\pdfoutput=1
\documentclass[phd, twoside, logo, notimes, electronic]{infthesis}

\usepackage{lipsum} 

\usepackage[square,sort,comma,numbers]{natbib}
\usepackage{enumitem}
\usepackage{multirow} 
\usepackage{adjustbox} 
\usepackage{subcaption} 
\usepackage{xcolor}
\usepackage{colortbl}
\usepackage{tabularx} 
\usepackage{tikz} 
\usepackage{placeins} 
\usepackage{pifont}
\newcommand{\cmark}{\ding{51}}%
\newcommand{\xmark}{\ding{55}}%

\usepackage{epigraph}

\usepackage{amsthm} 
\theoremstyle{definition}

\usepackage{helvet} 
\renewcommand{\familydefault}{\sfdefault} 
\usepackage[T1]{fontenc}

\usepackage{mathastext}
\MTfamily{\ttdefault}\Mathastext 

\definecolor{myBlue2}{RGB}{51,51,175}
\definecolor{myBlue}{RGB}{15,82,186}
\definecolor{myBlack}{RGB}{0,0,0}
\definecolor{myThemeColor}{RGB}{0,0,0}

\definecolor{myThemeColorTOCLink}{RGB}{10, 60, 183}

\usepackage{hyperref}
\hypersetup{colorlinks,
      linkcolor=myThemeColorTOCLink,
      citecolor=myThemeColorTOCLink,
      urlcolor=myBlue,
      linktoc=all,
      filecolor=myBlue}

\usepackage[capitalize, noabbrev, nameinlink]{cleveref}

\usepackage[tikz]{bclogo}

\usepackage{tcolorbox}
\tcbuselibrary{skins}

\definecolor{boxTitle}{gray}{0.85}
\definecolor{boxBackground}{gray}{0.95}
\definecolor{boxFrame}{gray}{0.75}

\tcbset{my box/.style={
    enhanced, fonttitle=\bfseries,
    colback=boxBackground, colframe=boxFrame,
    coltitle=black, colbacktitle=boxTitle,
    attach boxed title to top left={xshift=0.7cm,
                                    yshift*=-\tcboxedtitleheight/2},
    boxed title style={
      before upper=\hspace*{0.6cm}, 
      overlay={
       \node at ([xshift=0.5cm]frame.west)
         {\includegraphics[scale=0.9]{bc-lampe}};
      }
    }
  }
}

\newtcolorbox{mybox}[1][]{my box, #1}

\usepackage{marvosym} 
\usepackage[framemethod=tikz]{mdframed}

\newlist{myitem}{itemize}{1}
\setlist[myitem,1]{label=\Forward} 

\usetikzlibrary{shadows}
\newmdenv[
shadow=true,
frametitle=This chapter at glance...
]{mynote}

\usepackage{eso-pic}

\usepackage{pagecolor}

\usepackage{quotchap}

\makeatletter
\renewcommand\chapter{%
  \if@openright\cleardoublepage\else\clearpage\fi
  \thispagestyle{plain}%
  \global\@topnum\z@
  \null\hfill\@printcites\par
  \@afterindentfalse
  \secdef\@chapter\@schapter
}

\renewcommand{\@makechapterhead}[1]{%
  \chapterheadstartvskip%
  {\size@chapter{\sectfont\raggedright
  \vspace{-75pt}
    {\chapnumfont
      \ifnum \c@secnumdepth >\m@ne%
      \if@mainmatter\thechapter%
      \fi\fi
      \par\nobreak}%
    {\raggedright\advance\leftmargin10em\interlinepenalty\@M #1\par}}
  \nobreak\chapterheadendvskip}}
\makeatother

\colorlet{chaptergrey}{myThemeColor}
\renewcommand*{\sectfont}{\color{myThemeColor}\normalfont\fontfamily{\familydefault}\fontsize{21.5}{23}\bfseries}

\makeatletter
\renewcommand*{\chapnumfont}{%
  \usefont{T1}{\@defaultcnfont}{b}{n}\fontfamily{\familydefault}\fontsize{40}{50}\selectfont
  \color{chaptergrey}%
}
\makeatother

\usepackage{fancyhdr}
\renewcommand{\chaptermark}[1]{\markboth{\chaptername\ \thechapter.\ \ ~#1}{}}
\renewcommand{\sectionmark}[1]{\markright{\thesection.\ #1}}
\usepackage{etoolbox}
\makeatletter
\patchcmd{\f@nch@head}{\rlap}{\color{myThemeColor}\rlap}{}{}
\patchcmd{\headrule}{\hrule}{\color{myThemeColor}\hrule}{}{}
\makeatother

\usepackage{titlesec}
 
\titleformat*{\section}{\color{myThemeColor}\normalfont\fontfamily{\familydefault}\fontsize{17.5}{19}\bfseries}
\titleformat*{\subsection}{\color{myThemeColor}\normalfont\fontfamily{\familydefault}\fontsize{14}{17}\bfseries}
\titleformat*{\subsubsection}{\color{myThemeColor}\normalfont\fontfamily{\familydefault}\fontsize{14}{17}\selectfont}

\chapterfont{\color{myThemeColor}}  

\usetikzlibrary{shapes,arrows}
\usetikzlibrary{decorations.pathreplacing}

\usepackage[utf8]{inputenc}
\usepackage{alphabeta}

\usepackage[activate={true,nocompatibility},final,stretch=13,shrink=13]{microtype}

\usepackage{fontawesome5}

\usepackage[normalem]{ulem}
\usepackage{xpatch} %
\xpatchcmd{\sout}
  {\bgroup}
  {\bgroup}
  {}{}

\usepackage{mfirstuc}

\usepackage{listings}

\newif\ifshowcomment
\showcommenttrue

\ifshowcomment

\newcommand{\antonis}[1]{\noindent\textsf{\color{purple}{[Antonis: {\it#1}]}}}
\newcommand{\todo}[1]{\noindent\textsf{\color{orange}{[{Todo: \it #1}]}}}
\newcommand{\newtext}[1]{#1} 
\newcommand{\boris}[1]{\noindent\textsf{\color{Violet}{[Boris: {\it#1}]}}}

\newcommand{\yijun}[1]{\noindent\textsf{\color{OliveGreen}{[Yijun: {\it#1}]}}}
\newcommand{\zhaowei}[1]{\noindent\textsf{\color{brown}{[Zhaowei: {\it#1}]}}}

\else
\newcommand{\newtext}[1]{#1} 

\newcommand{\todo}[1]{}
\newcommand{\antonis}[1]{}
\newcommand{\boris}[1]{}
\newcommand{\vijay}[1]{}
\newcommand{\arpit}[1]{}
\newcommand{\yijun}[1]{}
\newcommand{\zhaowei}[1]{}

\fi

\newcommand{\beginbseq}[1]{\vspace{3pt}\noindent\textbf{#1? \vspace{0pt}}}
\newcommand{\beginfsec}[1]{\beginbsec{#1}}
\newcommand{\beginbsec}[1]{\vspace{3pt}\noindent\textbf{#1. \vspace{0pt}}}
\newcommand{\beginbsecBig}[1]{\noindent\textbf{\fontsize{13}{15}\selectfont #1}}
\newcommand{\beginbseceval}[1]{\beginbsec{#1}}

\newcommand{\CAP}[1]{\scalebox{0.95}{#1}}

\newcommand{\VAR}[1]{\texttt{#1}}

\newcommand{\papert}[1]{\textit{#1}}

\def\cid{\texorpdfstring{\VAR{c\textsubscript{id}}}{}} 

\newcommand*\circled[1]{\tikz[baseline=(char.base)]{
            \node[shape=circle,draw,inner sep=0.5pt] (char) {#1};}}

\newcommand*\circledZ[1]{\tikz[baseline=(char.base)]{
            \node[shape=circle, text=white, fill=black, draw,inner sep=0.5pt] (char) {#1};}}

\newcommand{\squishlist}{
 \begin{list}{$\bullet$}
  { \setlength{\itemsep}{2pt}
     \setlength{\parsep}{0pt}
     \setlength{\topsep}{2pt}
     \setlength{\partopsep}{0pt}
     \setlength{\leftmargin}{1em}
     \setlength{\labelwidth}{1em}
     \setlength{\labelsep}{0.5em} } 
}

\newcommand{\squishlistContrib}{ %
 \begin{list}{$\bullet$}
  { \setlength{\itemsep}{2pt}
     \setlength{\parsep}{0pt}
     \setlength{\topsep}{2pt}
     \setlength{\partopsep}{0pt}
     \setlength{\leftmargin}{1em}
     \setlength{\labelwidth}{1em}
     \setlength{\labelsep}{0.5em} }
}
\newcommand{\squishend}{ \end{list}  }

\newcommand{\squishenum}{\begin{enumerate}[itemsep=0.5pt,parsep=0pt,topsep=0pt,partopsep=0pt,leftmargin=1.5em,labelwidth=1em,labelsep=0.5em]{}}
\newcommand{\squishenumend}{\end{enumerate}}

\def\etal{et~al.~} 
\def\eg{e.g.,~} 
\def\ie{i.e.,~} 

\AtBeginEnvironment{savequote}{\quotefont\huge}

\newcommand{\nextlinepdf}[0]{\texorpdfstring{\\}{}}

\newcommand{\epoch}[0]{epoch \CAP{ID} }
\newcommand{\epochDot}[0]{epoch \CAP{ID}.}

\newcommand{\tocfont}[0]{
\fontsize{11pt}{12.5pt}\selectfont
}

\newcommand{\equote}[4]{

\vspace{#1 pt}
{\fontfamily{ptm}\selectfont 
\begin{flushright}
\fontsize{12pt}{12pt}\selectfont
#3
\end{flushright}

\begin{flushright}
\vspace{-22pt}
\textbf{#4}
\end{flushright}
\vspace{#2 pt}

\newcommand{\esquote}[3]{
\begin{savequote}[0.8\textwidth]
\normalsize\normalfont
\equote{#1}{#2}{#3}
\end{savequote}
}
}
}

\newcommand\capmystring[1]{\capmystringaux#1\relax}
\def\capmystringaux#1#2\relax{\texorpdfstring{\uppercase{#1}}{}\texorpdfstring{\lowercase{#2}}{}}

\newcommand{\captitle}[1]{\capmystring{#1}}

\newcommand{\tsection}[1]{
\section[\textcolor{black}{\captitle{#1}}]{\captitle{#1}}
}

\newcommand{\tsubsubsection}[1]{
\subsubsection[\textcolor{black}{\captitle{#1}}]{\captitle{#1}}
}

\newcommand{\tsubsection}[1]{
\subsection[\textcolor{black}{\captitle{#1}}]{\captitle{#1}}
}

\newcommand{\tnsection}[1]{
\section[\textcolor{black}{#1}]{#1}
}

\newcommand{\tnsubsection}[1]{
\subsection[\textcolor{black}{#1}]{#1}
}

\newcommand{\tchapter}[1]{
\chapter[\textcolor{black}{#1}]{#1}
}

\newcommand{\mscaption}[1]{\caption[\textcolor{black}{#1}]{#1}}
\newcommand{\mcaption}[2]{\caption[\textcolor{black}{#1}]{#2}}

\newcommand{\markedchapterTOC}[3]{\chapter[\textcolor{black}{#3}]{#2%
\chaptermark{#1}}
\chaptermark{#1}}

\newcommand{\markedsectionTOC}[3]{\section[\textcolor{black}{\captitle{#3}}]{\captitle{#2}%
\sectionmark{\captitle{#1}}}
\sectionmark{\captitle{#1}}}

\newcommand{\markedsection}[2]{\section[\textcolor{black}{\captitle{#2}}]{\captitle{#2}%
\sectionmark{\captitle{#1}}}
\sectionmark{\captitle{#1}}}

\def\INVNOSPACE{\CAP{R-INV}} 
\def\ACKNOSPACE{\CAP{R-ACK}}
\def\VALNOSPACE{\CAP{R-VAL}}
\def\ZINV{{\INVNOSPACE} } 
\def\ZACK{{\ACKNOSPACE} }
\def\ZVAL{{\VALNOSPACE} } 

\newcommand{\Zmylisting}[1]{\scalebox{0.9}{\texttt{#1}}}

\newcommand{\ecref}[2]{\cref{#1}\hyperref[#1]{#2}}

\newcommand{\tla}[0]{\CAP{TLA$^{+}$} }

\newcommand{\tstatement}[1]{
\vspace{10pt}
\centerline{\textbf{Thesis statement}}
\vspace{-10pt}
\begin{center} 
\parbox{0.7325\linewidth}{
#1}
\end{center} 
}

\definecolor{myBlue}{RGB}{0, 114, 178}
\definecolor{myOrange}{RGB}{204, 102, 0}

\title{Invalidation-Based Protocols\\ for Replicated Datastores}
\author{\normalfont Antonios Katsarakis}
\submityear{2021}

\AtBeginDocument{
  \addtocontents{toc}{\tocfont}
  \addtocontents{lof}{\tocfont}
  \addtocontents{lot}{\tocfont}
}

\abstract{
Distributed in-memory datastores underpin cloud applications that run within a datacenter and demand high performance, strong consistency, and availability. A key feature of datastores is data replication. The data are replicated across servers because a single server often cannot handle the request load. Replication is also necessary to guarantee that a server or link failure does not render a portion of the dataset inaccessible. A replication protocol is responsible for ensuring strong consistency between the replicas of a datastore, even when faults occur, by determining the actions necessary to access and manipulate the data. Consequently, a replication protocol also drives the datastore's performance.

Existing strongly consistent replication protocols deliver fault tolerance but fall short in terms of performance.  Meanwhile, the opposite occurs in the world of multiprocessors, where data are replicated across the private caches of different cores. The multiprocessor regime uses invalidations to afford strongly consistent replication with high performance but neglects fault tolerance.

Although handling failures in the datacenter is critical for data availability, we observe that the common operation is fault-free and far exceeds the operation during faults. In other words, the common operating environment inside a datacenter closely resembles that of a multiprocessor. Based on this insight, we draw inspiration from the multiprocessor for high-performance, strongly consistent replication in the datacenter. The primary contribution of this thesis is in adapting invalidating protocols to the nuances of replicated datastores, which include skewed data accesses, fault tolerance, and distributed transactions.

}

\begin{document}



\setlength{\parindent}{0em}
\setlength{\parskip}{0.8em} 

\begin{preliminary}

\maketitle 

{
\hypersetup{linkcolor=myThemeColorTOCLink}
\setlength{\parskip}{0em} 
\pagebreak

\papersdeclaration{
\begin{enumerate}
    \item[\cite{A&V:2018}] 
    {\small V. Gavrielatos,{\small${^{*}}$} A. Katsarakis,{\small${^{*}}$} A. Joshi, N. Oswald, B. Grot, and V. Nagarajan. \linebreak
    \papert{Scale-out ccNUMA: Exploiting skew with strongly consistent caching}.
    Proceedings of the Thirteenth European Conference on Computer Systems (EuroSys). ACM, 2018.
    {\small ${^{*}}$Equal contribution to this work}.\\}
  
    \item[\cite{Katsarakis:20}]
    \vspace{-22pt} {\small A. Katsarakis, V. Gavrielatos, S. Katebzadeh, A. Joshi, A. Dragojevic, B. Grot, and V. Nagarajan.
    \papert{Hermes: A fast, fault-tolerant and linearizable replication protocol}.
    Proceedings of the Twenty-Fifth International Conference on Architectural Support for Programming Languages and Operating Systems (ASPLOS). ACM, 2020.
    \textbf{Honorable mention in IEEE Micro Top Picks 2020}.}

    \item[\cite{Katsarakis:21}] \vspace{-2pt}
    {\small A. Katsarakis, Y. Ma, Z. Tan, A. Bainbridge, M. Balkwill, A. Dragojevic, B. Grot, B. Radunovic, and Y. Zhang.
    \papert{Zeus: Locality-aware distributed transactions}.
    Proceedings of the Sixteenth European Conference on Computer Systems (EuroSys). ACM, 2021.}
\end{enumerate}

\vspace{5pt}
In addition to the works mentioned above, which form the backbone of this thesis, 
\scalebox{1}{I also contributed to other relevant publications during my studies including:}

\begin{enumerate}

    \item[\cite{Katsarakis:21R}]
    {\small A. Katsarakis, Z. Tan, M. Balkwill, B. Radunovic, A. Bainbridge,  A. Dragojevic,  B. Grot, and  Y. Zhang. 
    \papert{rVNF: Reliable, scalable and performant cellular VNFs in the cloud}.
    Technical Report (MSR-TR-2021-7). Microsoft, 2021.}
    
    \item[\cite{Gavrielatos:21}] \vspace{-2pt} 
    {\small V. Gavrielatos, A. Katsarakis, and V. Nagarajan.
    \papert{Odyssey: The impact of modern hardware on strongly-consistent replication protocols}.
    Proceedings of the Sixteenth European Conference on Computer Systems (EuroSys). ACM, 2021.}
    
    \item[\cite{Gavrielatos:20}]
    \vspace{-2pt} {\small V. Gavrielatos, A. Katsarakis, V. Nagarajan, B. Grot, and A. Joshi.
    \papert{Kite: Efficient and available release consistency for the datacenter}.
    Proceedings of the Twenty-Fifth Symposium on Principles and Practice of Parallel Programming (PPoPP). ACM, 2020.}
    \textbf{Best paper nominee}.
    
\end{enumerate}
}

\microtypesetup{protrusion=false} 
\tableofcontents 

\bgroup

\begingroup
\let\cleardoublepage\relax
\listoffigures
\vspace{80pt}
\listoftables
\endgroup
\egroup

\microtypesetup{protrusion=true} 
}

\end{preliminary}


\tchapter{Introduction}

\equote{-30}{10}{Good ideas \dots want to connect, fuse, recombine.\\ They want to reinvent themselves by crossing conceptual borders.}{Steven Johnson}
Today's cloud applications deliver critical services to large audiences and are
underpinned by cavernous datastores that manage their ever-increasing data. 
A wide variety of applications rely on datastores, including social networks, e-commerce, telecommunications, and financial services~\cite{Bronson:2013, Corbett:2013, Venmo20, TATP:2009}.
Thus, datastores provide benefits to most people in numerous ways every day.

Datastores split and store application data across servers to leverage the 
in-memory speed and capacity of multiple nodes (i.e., servers) inside a datacenter.
\linebreak
To let applications access and manipulate their data, datastores provide a single-object read/write interface and occasionally, 
multi-object transactions. 
\linebreak A transaction 
is a series of reads and writes to one or more data objects 
treated as an indivisible unit, 
such that either all or none of the 
reads and writes 
occur.
\linebreak
It is common for modern services to generate several million such data queries per second~\cite{Xiao:20, Baker:2011}. Therefore, datastores must offer high performance. 
\linebreak
Furthermore, as datastores run on commodity failure-prone infrastructure~\cite{Forbes:2013}, it is essential that they also facilitate data availability in the case of faults.

Data replication is a fundamental feature of performant and resilient datastores. 
Datastores must replicate data 
across nodes to increase throughput because a single node often cannot keep up with the request load~\cite{Bronson:2013}. Replication is also necessary to guarantee that the failure of a node or network link does not render a portion of the dataset inaccessible. 
%
In such a \textit{replicated datastore}, consistency must be enforced across the data replicas.
Succinctly put, replicas should not arbitrarily diverge in the face of data updates, or it would be impossible to predict the behavior of the datastore and facilitate the correctness of the applications.

To ensure that the services running on the datastore operate correctly and intuitively, the data replicas must be \textit{strongly consistent}, 
providing the illusion of a single copy.
Maintaining the strong consistency of the replicas is a challenge, especially in the presence of failures. 
A \emph{replication protocol}\footnote{We use the term \textit{replication protocol} to refer to a wide range of protocols for accessing and manipulating replicated data, including protocols for data re-sharding and transactions.} 
is responsible for keeping
the replicas of a datastore strongly consistent, even when faults occur.
To achieve that, a replication protocol determines the exact actions necessary to perform reads, writes, or transactions on the data.
This includes the number of network exchanges as well as which and how many servers must be involved in completing each request.
Thus, besides ensuring strong consistency and fault tolerance,
replication protocols also define the datastore's performance.

Many applications that run on top of replicated datastores are highly sensitive to performance. High throughput is a common requirement. In addition, low latency is emerging as a critical design goal in the age of interactive services and machine actors. For instance, Anwar \etal\cite{Anwar:2018} note that a deep learning system running on top of a replicated datastore is profoundly affected by the latency of the datastore.


\markedsectionTOC
{Replication Protocols vs. Multiprocessors: Availability or Performance}
{Replication Protocols \textit{vs.} Multiprocessors:\\ Availability \textit{or} Performance}
{Replication Protocols \textit{vs.} Multiprocessors: Availability \textit{or} Performance}

In this thesis, we observe that the existing replication protocols which
support strongly consistent reads and writes and ensure data availability are unable to achieve high performance because they sacrifice either concurrency or speed.
Concurrency is typically diminished due to the serialization of reads or writes on a dedicated leader node (also called a primary or head node)~\cite{Alsberg:1976, Hunt:2010, Chandra:2016, Guerraoui:21, Aguilera:20, Terrace:2009, Poke:2015}.
Speed is jeopardized when writes need numerous network hops to complete~\cite{VanRenesse:2004, Terrace:2009} or when reads forfeit locality and require communication across multiple replicas to be served~\cite{Lamport:1998, Lynch:1997, Bolosky:2011, Moraru:2013, Ekstrom:2016, Liskov:2012, Ongaro:2014, Attiya:1995, Hadjistasi:17}. 

Strongly consistent replication is not a unique feature of datastores; it is also a well-established practice between the caches of a multiprocessor. While performance has been sacrificed in the name of fault tolerance and strong consistency in the distributed world of replication protocols, the story is entirely different for shared-memory multiprocessors. In the multiprocessor context, where fault tolerance is generally not a consideration, cache coherence protocols almost always enforce strong consistency while maintaining high performance.


Cache coherence protocols use \textit{invalidations} to efficiently guarantee strong 
consistency across replicated data in multiprocessor caches~\cite{Vijay:20}.
Their invalidation scheme ensures high performance for both reads and writes, compromising neither concurrency nor speed.
\textit{All}
data copies in a valid state across caches can be individually leveraged to perform \textit{local} reads.
Writes are also completed \textit{quickly} from \textit{any} cache after only a single round of 
invalidations to other caches.
Latency is further minimized through the use of a high-performance fabric with a fully hardened communication protocol.
Unfortunately, cache coherence invalidating protocols do not provide any fault-tolerance guarantees.

An analogous story unfolds when considering transactions.
Replication protocols in state-of-the-art datastores support strongly consistent transactions with data availability but sacrifice performance, as they cannot fully exploit locality. 
These protocols rely on static sharding, in which relevant data are randomly placed on fixed nodes. 
Thus, they cause excessive network traffic and require multiple network hops to complete each transaction, regardless of the access pattern~\cite{Dragojevic:2015, F-Kalia:2016}.
Meanwhile, the opposite is true for strongly consistent transactions in the multiprocessor.
The multiprocessor's transactional memory~\cite{htm} 
extends the invalidating coherence to afford
transactions that exploit 
access locality to boost performance.
For example, a core that has previously accessed and currently caches relevant data can perform a series of transactions on that data locally, eschewing remote access and coordination. 
Problematically, 
\linebreak
transactional memory is also not resilient to faults, 
\scalebox{0.99}{hence risking data availability.}


To summarize, in the world of datastores, replication protocols are fault tolerant but fall short in performance. While, in the multiprocessor world, invalidation protocols allow local reads with fast writes from all replicas and transactions that exploit locality to ensure high performance, but they are not fault tolerant.

\markedsectionTOC
{The common case of a replication protocol resembling a multiprocessor}
{The common case of a replication protocol resembling a multiprocessor}
{The common case of a replication protocol resembling a multiprocessor}

Although guaranteeing data availability in the presence of faults is critical for datastores, failures at the level of an individual server in a datacenter are relatively infrequent~\cite{Jha:2019}. Data from Google show that an average server fails at most twice per year~\cite{Barroso:2018}. Consequently, for  
a typical replica group comprised of a handful of datastore nodes, the amount of fault-free operation significantly dominates over the operation during failures.
Another observation is that modern datacenters aggressively enable high-performance networking. This trend includes user-space network stacks (e.g., 
\CAP{DPDK}~\cite{DPDK:online}) and fabrics featuring hardware offloading and remote direct memory access (\CAP{RDMA})~\cite{Guo:2016, Marty:2019, Gao:21}, which offer consistently low communication latencies. 
As such, our insight is that \textit{the common operating environment of a replication protocol inside a datacenter closely resembles that of a multiprocessor.} 
%
%

\begin{figure}[t]
  \centering
  \includegraphics[width=0.8\textwidth]{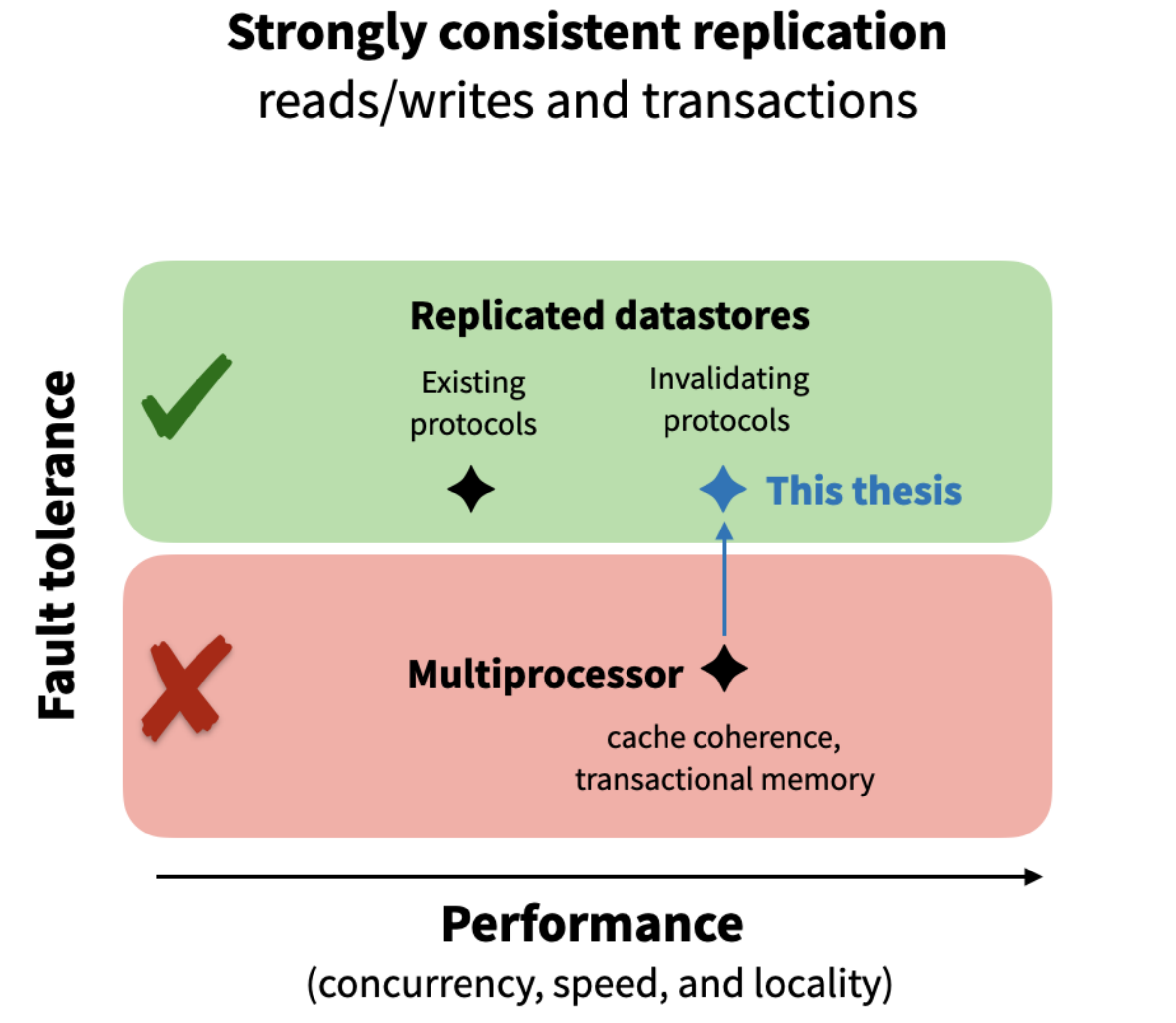}
  
  \mscaption{Performance and fault tolerance of strongly consistent replication.}
  \label{figB:motivation}
\end{figure}

\pagebreak
Based on this insight and the need for reliable yet 
performant 
replicated datastores, the thesis of this dissertation is as follows (also illustrated in \cref{figB:motivation}):


\tstatement{Adapting the multiprocessor-inspired invalidating protocols to intra-datacenter replicated datastores enables strong consistency with data availability and high performance.\linebreak}
\vspace{-30pt}
\tsection{Content and Primary Contributions}

To support our thesis statement, in this dissertation, 
we propose invalidating protocols that improve the three most common uses of data replication within intra-datacenter datastores.
These three use cases are listed below, followed by the name 
(in bold) of our associated proposal for each.
\squishenum
\item Replication for performance \hfill \textbf{Scale-out ccNUMA}
\item Replication for fault tolerance \hfill \textbf{Hermes}
\item Replicated distributed transactions \hfill \textbf{Zeus}
\squishenumend

For the first two use cases, we
consider the most typical setting for datastores, in which the data are 
statically sharded and accessed via a single-object read/write interface.
In the first use case,
we leverage replication 
to exploit
highly skewed data accesses, which are common in online services~\cite{Atikoglu:2012}. 
This use of 
replication aims to improve performance under strong consistency, but it does not cover fault tolerance. 
In the second use case, 
we demonstrate how an invalidation-based protocol enables fault-tolerant replication for data availability, with strong consistency and high performance.
Finally, in the third use case, 
we apply invalidation-based protocols to a more challenging datastore setting, with dynamic data sharding and fault-tolerant transactions for availability.
The main contextual differences considered in each use case are outlined in \cref{tabM:contexes}. 
Below, we briefly describe the primary contributions of this thesis.
%






\begin{table}[t]
\centering

\begin{adjustbox}{max width=0.8\textwidth}
\begin{tikzpicture}

\hspace{15pt}
\node (table) [inner sep=-0pt] {
\begin{tabular}{l|lll}
\rowcolor[HTML]{9B9B9B} 
                  & \textbf{{Scale-out ccNUMA}} 
                  & \textbf{{Hermes}} 
                  & \textbf{{Zeus}} \\ 
\textbf{\begin{tabular}[c]{@{}l@{}}Key drive \\ for replication\end{tabular}} 
  & Performance & Availability & Availability \\
\textbf{\begin{tabular}[c]{@{}l@{}}Access\\primitives\end{tabular}} 
  & Reads/writes & Reads/writes & Transactions\\
\textbf{Data sharding} 
  & Static & Static & Dynamic 
\end{tabular}
};
\draw [rounded corners=.3em] (table.north west) rectangle (table.south east);
\end{tikzpicture}
\end{adjustbox}

\mcaption
{Context differences in replicated datastores improved by this thesis.}
{\scalebox{0.98}{Contextual differences in replicated datastores improved by this thesis.}}
\label{tabM:contexes}
\end{table}

\begin{description}
    

    \item [1. Scale-out ccNUMA: Replication for performance under access skew]\hspace{-4pt}\footnote{\scalebox{0.9325}{This was a joint work with equal contributions from myself and my colleague, \textit{Vasilis Gavrielatos}}.}
    \hfill \\
    Data access skew is a prevalent workload characteristic of online services. 
    In short, a small number of data objects are widely more popular and likely to be accessed than the rest. 
    Thus, the datastore nodes holding these hot objects are overloaded while the majority of nodes remain underutilized.
    The resulting load imbalances inhibit the 
    performance of the datastore.
    
    To mitigate these load imbalances, 
    we propose a replication scheme that balances the request load over a pool of \CAP{RDMA}-connected servers (e.g., a rack). Each server is equipped with a small replicated software cache
    storing the (same) most popular objects in the pool, and client requests are spread across the pool.
    Requests for popular objects are served by the caches, filtering the skew.
    With the skew filtered, the remaining requests can leverage an uncongested \CAP{RDMA} network to complete quickly. 
    
    The key challenge, however, is ensuring strong consistency between the replicated hot objects.
    Existing protocols ensure strong consistency by serializing writes over a physical ordering point, which could itself easily become a hotspot under skewed accesses.
    %
    To resolve this issue, we introduce \textit{Galene}, a replication protocol that couples invalidations with logical timestamps to enable fully distributed write coordination from any replica and avoid hotspots.
    Our evaluation shows that for typical modest write ratios, the proposed scheme powered by Galene, improves throughput by 2.2$\times$ when compared with the state-of-the-art skew mitigation technique. 

    \item [2. Hermes: Strongly consistent and fault-tolerant replication made fast] \hfill \\
    Resilient datastores that guarantee data availability must replicate their data using fault-tolerant replication protocols. 
    Existing fault-tolerant \linebreak replication protocols that support strong consistency hinder datastore performance, as they compromise on speed or concurrency. 
    Briefly, these protocols fail to achieve both local reads and fast writes from all replicas.
    
    To address this shortcoming, we introduce \textit{Hermes}, an invalidation-based protocol that is strongly consistent and fault tolerant while exploiting the typical fault-free operation to enable local reads and fast writes from all replicas.
    We show that an invalidating protocol can be resilient and deliver high throughput with low latency.
    Five node replicas managed by Hermes afford hundreds of millions of reads and writes per second, resulting in significantly higher throughput than the state-of-the-art fault-tolerant protocols while offering at least 3.6$\times$ lower tail latency.
    
    \item [3. Zeus: Replicated and distributed locality-aware transactions] \hfill \\
    State-of-the-art datastores that provide multi-object transactions with data availability deploy protocols which cannot fully exploit the locality in access patterns that exist in several transactional workloads. Therefore, they incur excessive remote accesses and numerous network round-trips to commit each transaction, hence curtailing the datastore's performance.
    
    Inspired by the multiprocessor's transactional memory, we propose and implement \textit{Zeus}, a strongly consistent distributed transactional datastore that exploits and dynamically adapts to the locality of transactional workloads. To achieve this, we introduce a reliable ownership protocol for dynamic data sharding that quickly alters replica placement and access levels across the replicas and a fault-tolerant transactional protocol for fast, pipelined, reliable commit and local read-only transactions from all replicas. For workloads with data access locality, six Zeus nodes can achieve tens of millions of transactions per second with up to 2$\times$ the performance of state-of-the-art datastores while using less network bandwidth.
    

    
\end{description}
    
    \subsection*{Formally verified invalidation-based replication protocols} 
    
    \vspace{-12pt} 
    Overall, in this thesis we introduce and formally verify, in \tla\cite{Lamport:1994}, the \linebreak correctness of four invalidating protocols that provide strong consistency with high performance, advancing the state of affairs in replicated datastores.
    %
    \vspace{-10pt} 
    \begin{enumerate}
        \item \textbf{Galene}: A fully distributed replication protocol for high performance.
        
        \item \vspace{-5pt} 
        \textbf{Hermes}: A fast fault-tolerant replication protocol for reads and writes.
        
        
         
        \item \vspace{-5pt} 
        \textbf{Zeus ownership}: A fault-tolerant protocol for dynamic data sharding.
        
        \item \vspace{-5pt} 
        \textbf{Zeus reliable commit}: A locality-aware pipelined transaction commit.
        
        
    \end{enumerate}

\vspace{-10pt}
\tsection{Thesis Structure}
The remainder of this thesis is organized as follows.

\begin{description}

    \item [\cref{chap:background}] 
    provides background on replicated datastores, replication protocols, and consistency enforcement in the multiprocessor.
    
    \item [\textbf{\cref{chap:cckvs}} Scale-out ccNUMA] 
    reveals the benefits of aggressive replication backed by a fully distributed invalidating protocol for load balance and strong consistency in the presence of skewed data accesses.
    
    \item [\textbf{\cref{chap:hermes}} Hermes] 
    proposes a fault-tolerant invalidating protocol that affords strong consistency with local reads and fast writes from all replicas and demonstrates its throughput and latency advantages.
    
    \item [\textbf{\cref{chap:Zeus}} Zeus] 
    introduces and evaluates two invalidation-based protocols that enable fast dynamic sharding and distributed replicated transactions, with data availability and locality awareness.
    
    \item [\cref{chap:conclusion}] concludes the thesis by summarizing the key results and exploring possible directions for future work.
\end{description}

Supplementary material, including open-source code for the evaluated systems and detailed TLA$^{+}$ specifications of the proposed protocols, is available online:

\begin{table}[!ht]
\centering
\begin{tabular}{ll}
Scale-out ccNUMA &: \href{http://s.a-phd.com}{http://s.a-phd.com} \\
Hermes           &: \href{http://h.a-phd.com}{http://h.a-phd.com} \\
Zeus             &: \href{http://z.a-phd.com}{http://z.a-phd.com}
\end{tabular}
\end{table}



\tchapter{Background}
\label{chap:background}

\equote{-40}{5}{Make progress, and, before all else,\\ endeavor to be consistent.}{Seneca}

To put this work into context, we begin with background on replicated datastores, 
including details related to 
consistency, fault tolerance, and performance. 
We then outline existing replication protocols for datastores. 
Finally, we describe the multiprocessor's method of ensuring data consistency, which inspired the invalidation protocols developed in this thesis.

\vspace{-5pt}
\tsection{Replicated Datastores}
\vspace{-5pt}
\label{secB:datastores}
Replicated distributed datastores are the backbone of today's online services and cloud applications. They are responsible for storing application data while providing replica consistency, data availability, and high levels of performance.
One example of these datastores is
key-value stores (\CAP{KVS})~\cite{DeCandia:2007, Bronson:2013, Lim:2014}, which serve as the foundation of many of today's data-intensive online services, 
including e-commerce and social networks.
Another example is 
coordination services (e.g., Apache Zookeeper~\cite{Hunt:2010} and Google's Chubby~\cite{Burrows:2006}), which allow applications to maintain critical shared state, including configurations, metadata, and locks.
Yet another datastore example is 
shared-nothing transactional databases, such as those focusing on online transaction processing (\CAP{OLTP})~\cite{Dragojevic:2014, Kalia:2016}.


\beginbsec{Sharding and replication}
%
%
Replicated datastores partition
their data across multiple nodes inside a datacenter. Modern datastores statically partition the stored data into smaller pieces called \emph{shards} (i.e., static sharding)~\cite{Dragojevic:2014, Kalia:2016}.
Static sharding is typically achieved through consistent hashing~\cite{Karger:1997}, where a hash function is used to deterministically decide the home node of an object based on its unique identifier (e.g., a key or a memory address). 
Therefore, consistent hashing results in a uniformly random and fixed placement of objects among a datastore's nodes. 

Datastores also replicate each shard across multiple nodes to maintain data availability, even in the presence of faults.
A fault-tolerant replication protocol is deployed to enforce consistency and fault tolerance across all replicas of a given shard. 
The number of replicas of a shard is the \emph{replication degree}, and it presents a trade-off between cost and fault tolerance. More replicas increase fault tolerance but also increase the cost of the deployment. To facilitate data availability, a replication degree between three and seven replicas is commonly considered to offer a good balance between resilience and cost~\cite{Hunt:2010}. Overall, although a partitioned datastore may span numerous nodes, the replication protocol need only scale with the replication degree.





In this thesis, we focus on replication protocols deployed over datastores, \linebreak sharded, and replicated within 
a datacenter. 
Clients (i.e., application threads) interact with such a datastore by first establishing a session through which they \textit{invoke} requests and wait for \textit{responses}. 
The type of request is determined by the access primitives offered by the datastore.




\beginbsec{Access primitives}
Datastores provide fundamental primitives to access and modify data objects.\footnote{Throughout this thesis, we use the terms \textit{object} and \textit{key} interchangeably.}
Based on the number of objects involved, they can be classified as \textit{single-object} or \textit{multi-object} primitives.
Some of these primitives adhere to transactional semantics. 
Informally, a \textit{transaction} is an indivisible sequence of operations that access or modify at least one object. This sequence either completes in its entirety \textit{as if} executing without any other concurrent requests (i.e., \textit{commits}) or has no effect on the data (i.e., \textit{aborts}).


Most datastores provide a single-object interface that allows for \textit{read} and \textit{write} operations.
Occasionally, datastores afford another single-object operation called a \textit{read-modify-write} (\CAP{RMW})~\cite{Kruskal:88}, which is 
a single-object transaction that is equivalent to consensus~\cite{Nicolaou:16}.
An \CAP{RMW} facilitates arbitrarily powerful single-object procedures, such as compare-and-swap and other critical methods for locks and synchronization.
%
%

\begin{figure}[t]
  \centering
  \includegraphics[width=0.7\textwidth]{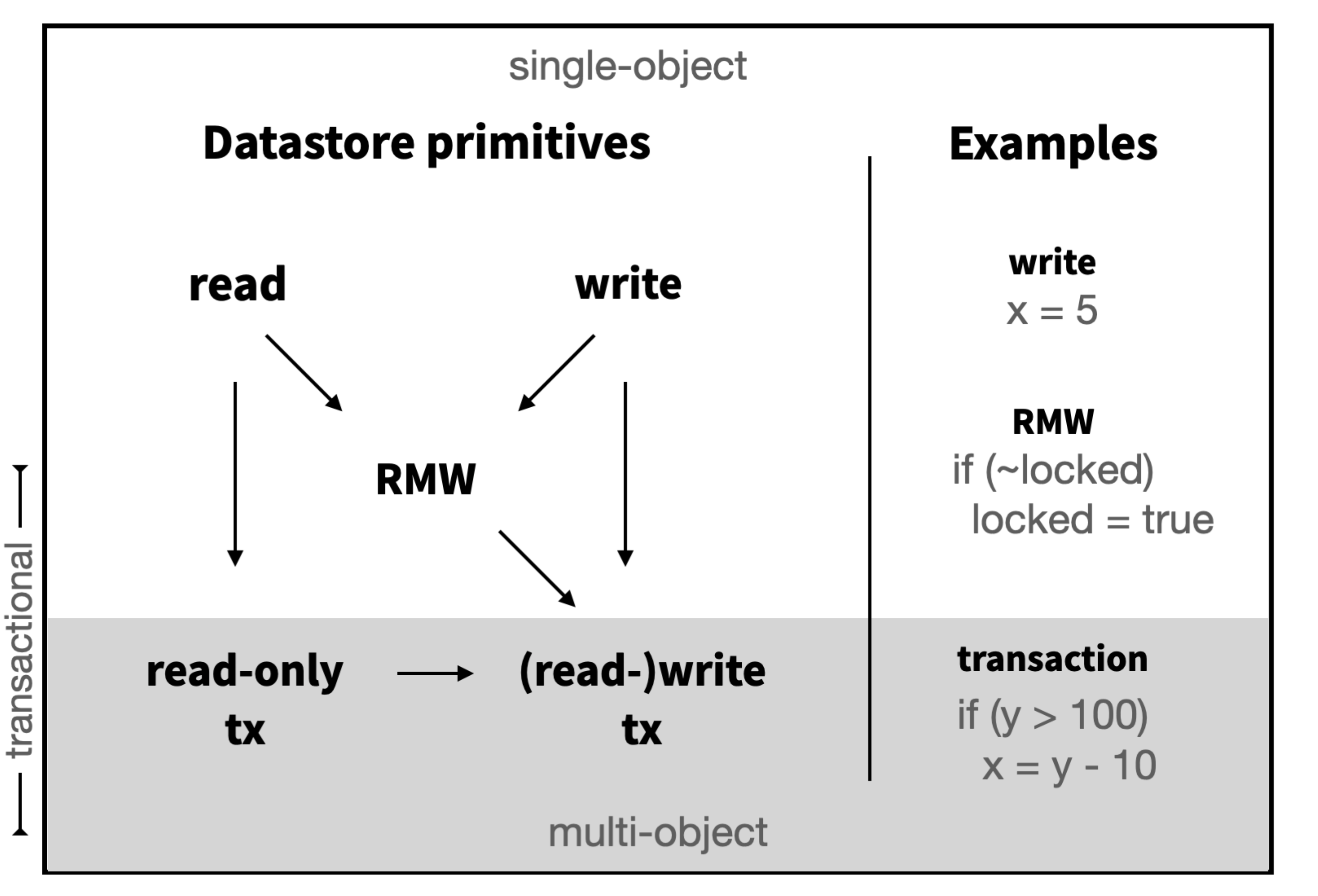}
  \mcaption
  {Access primitives offered by datastores.}
  {Access primitives offered by datastores and examples of primitives that update the state. Arrows point towards more general primitives. The variables \emph{x}, \emph{y}, and \emph{locked} represent objects stored in a datastore.}
  \label{figB:ops_n_example}
\end{figure}

Datastores with even richer interfaces support multi-object transactions. 
As illustrated by the examples on the right-hand side of \cref{figB:ops_n_example}, unlike in an \CAP{RMW}, 
each operation within a multi-object transaction may address a different object.
If these objects do not strictly need to be stored on the same node, the provided datastore primitive is called a \textit{distributed transaction}. Finally, \linebreak multi-object transactions can be further classified as \textit{read-only} if they only access and do not modify data (otherwise, they are classified as \textit{read-write}).
%

For brevity, in the remainder of this thesis, we refer to 
distributed transactions simply as transactions (also abbreviated as \textit{txs})
and read-write transactions as write transactions.

The left-hand side of \cref{figB:ops_n_example} summarizes the primitives offered by datastores, where \textit{A} $\rightarrow$ \textit{B} indicates that \textit{B} can implement (and is more general than) \textit{A}.
Although using more general primitives to implement less general ones results in the correct behavior, doing so 
comes at the expense of performance.
This is because the realization of a more general primitive fundamentally requires costlier protocol actions. For instance, a read can be served as an \CAP{RMW}; however, implementing an \CAP{RMW} is significantly more expensive than a read in a distributed setting with faults~\cite{Gavrielatos:20}.
As a result, a protocol should not simply focus on offering the most general access primitive, since this generalization would unnecessarily hinder the performance of the datastore.
%

In this context, the purpose of this dissertation is to provide invalidation-based protocols that support these fundamental datastore primitives over replicated datastores with strong consistency, fault tolerance, and high performance.






\tsubsection{Consistency}
\label{secB:consistency}

The problem of managing concurrent accesses over replicated data arises in many contexts, ranging from distributed replicated datastores to shared-memory multiprocessors.
To prevent the arbitrary divergence of replicas, which would render any system unusable, a \textit{consistency model} \footnote{A consistency model is also known as \textit{isolation level} in the database community.} must be enforced.
Informally, a consistency model is a set of rules that restricts the values a read may return when it is interleaved or overlapped with other operations executed over different replicas (and shards).

A plethora of weak consistency models exist that favor performance but incur hefty costs on programmability. 
Most weak models fall within the category of eventual consistency~\cite{Terry:95}. 
In such models, the only requirement is that all replicas must eventually converge on a value in the absence of new updates, allowing updates to be propagated asynchronously in any order. 
In terms of performance, weak models are beneficial, but they 
fall short in terms of providing adequate semantics to support all types of applications~\cite{Corbett:2013, Kakivaya:2018}.
In addition, weak models can be hard to reason about and can lead to nasty surprises for both developers and clients~\cite{Vogels:2009, Lu:15}.

\begin{figure}[t]
  \centering
  \includegraphics[width=\textwidth]{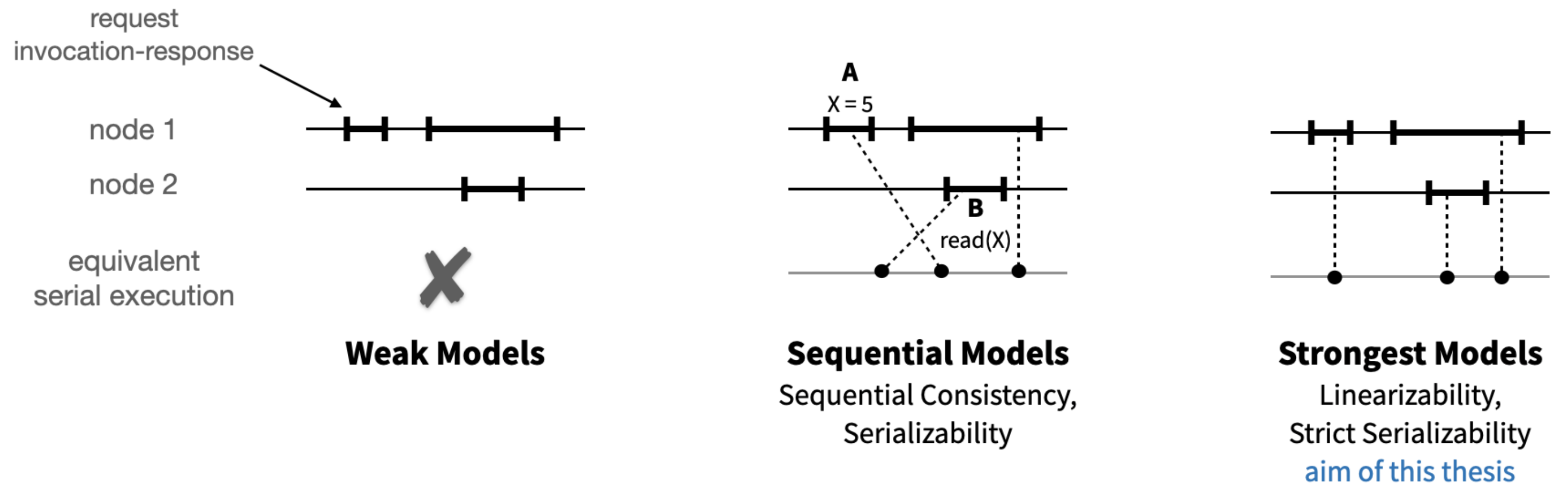}
  \mscaption{Operation ordering in different consistency models.} 
  \label{figB:consistency}
\end{figure}

More intuitive models are \textit{sequential}, such as sequential consistency for single-object operations~\cite{Lamport:1979} and serializability for transactions~\cite{Papadimitriou:79}.
As illustrated in \cref{figB:consistency}, unlike weak models, sequential models guarantee that the results of all operations are the same as if the operations on all the replicas (and shards) were executed in some sequential interleaving.
While sequential models are more intuitive than weak models, their operations are still not required to respect real-time boundaries. 
In other words, they permit operations to be sequenced outside of their invocation-response boundaries. 
For instance, as the execution example over the sequential models in \cref{figB:consistency} illustrates, a read operation (B) for an object that is invoked after the response to a write (A) to the same \linebreak object can be sequenced before A, thereby missing the value of the write.
Thus, problematically, operations may return stale values under a sequential model, which burdens programmers and confuses clients.
%



The strongest models provide the illusion of a single data copy and never return stale values.
More precisely, these models are sequential but also respect real time.
%
Consequently, they offer intuitive behavior to clients, permit the broadest spectrum of applications, and accommodate a simple programming interface. 
%
\linebreak 
Not surprisingly, many modern replicated datastores target the strongest 
\linebreak consistency~\cite{Baker:2011, Dragojevic:2015, F-Kalia:2016}.
For the above reasons, this thesis also focuses on guaranteeing the strongest consistency models, which are described next. 

\beginbsec{Linearizability}
For single-object operations, the strongest semantics are captured by \textit{linearizability}~\cite{Herlihy:1990}. In this model, as shown in \cref{figB:consistency}, each request appears to take effect globally and instantaneously at some point between its invocation and response.
Thus, a read invoked after the response of a write to the same object is guaranteed to 
be sequenced after the write and
return the value of the write (or a more recent value).
Besides 
its intuitive behavior, linearizability is also \textit{composable}~\cite{Herlihy:2008}.
Succinctly put, the union of individually linearizable entities results in a linearizable system.
Composability is important from a performance perspective, as it enables
independent, per-object linearizable, protocol instances to form a multi-object linearizable datastore in a highly concurrent fashion.
This modularity is also the reason why the linearizable protocols presented in this thesis can simply focus on a single object.


\beginbsec{Strict serializability}
The strongest consistency model for transactions 
is \textit{strict serializability}~\cite{sethi1982}. This model is equivalent to linearizability for transactions~\cite{Herlihy:1990}. 
Under strict serializability, all committed transactions appear as if they are atomically performed on all relevant shards and replicas at a single point between their invocation and response. 


For brevity, throughout this dissertation, we use the term \textit{strong consistency} (or the term \textit{safety}) to refer to linearizability for single-object operations and strict serializability for transactions.

\vspace{-10pt}
\tsubsection{Fault Tolerance}
\vspace{-5pt}
\label{secB:fault-model}
An essential feature of resilient datastores is ensuring data availability in the face of node and network failures. Availability requires data replication and \linebreak fault-tolerant protocols that 
enforce replica consistency even in the case of failures.
In this thesis, \cref{chap:hermes} and \cref{chap:Zeus} provide three such protocols based on the following failure model. 
\beginbsec{Failure model} This thesis primarily considers a partially synchronous system \cite{Dwork:1988} in which processes are equipped with loosely synchronized clocks (\CAP{LSC}s), as in~\cite{Chandra:2016}, and crash-stop or network failures may occur.
In this model, network failures can manifest as message reordering, duplication, or loss.
Although nodes
follow the replication protocol and do not act maliciously, they may fail due to a crash, and these crashes cannot be accurately detected.
We assume that only up to a minority of node replicas may crash, as we explain in \cref{secB:replication-protocols}.
Throughout this thesis, we describe datastores, protocols, or primitives as \textit{reliable} when they afford the strongest consistency and fault tolerance under this model.

%


%
While this thesis focuses on the partially synchronous model,
\cref{secH:async} demonstrates how the proposed invalidation-based protocols can be made \textit{indulgent} (i.e., safe under non-reliably detected crash faults and asynchrony --- that is, without \CAP{LSC}s)~\cite{Guerraoui:00}.
Note that in accordance with the seminal \CAP{FLP} impossibility result~\cite{Fischer:1985}, indulgent protocols \newtext{with primitives equivalent to consensus (e.g., \CAP{RMW}s)} may not always provide progress.

Datacenter network topologies are highly redundant~\cite{Gill:2011, Singh:2015, Filer:19}. Therefore, link failures leading to network partitions are not common inside a datacenter. This renders partitions outside the main scope of this dissertation. Nevertheless, we discuss how our approach maintains safety under network partitions in~\cref{secH:hermes-discussion}.

\beginbsec{Common operation $=$ fault-free}
Although ensuring data availability and correctness under faults is of the utmost importance, the failure-free operation 
considerably prevails
for a replica group.
Failures at the level of an individual node inside a datacenter are relatively infrequent.
An average server fails at most twice per year, according to data from Google~\cite{Barroso:2018}, and other independent studies over large clusters have reported similar numbers~\cite{Sato:12, Snir:14}. With a typical replication degree spanning 3--7 nodes for fault tolerance, one can generally expect well over 20 days of crash-free operation within each replica group.
Unsurprisingly, researchers have advocated for leveraging the common fault-free operation to increase the performance of reliable datastores~\cite{Birman:10, Jha:2019}. 

\vspace{-7pt}
\tsubsection{Performance}
\vspace{-3pt}

\beginbsec{Modern hardware and workload characteristics matter}
Cloud applications and online services supported by datastores are characterized by numerous concurrent and latency-sensitive requests~\cite{Bronson:2013, Anwar:2018}.
To satisfy the application demands, datastores must deliver high performance. This translates to serving requests in a high-throughput and low-latency manner. 
Leveraging modern hardware~---~for example, keeping the dataset in-memory, employing highly multithreaded designs, and exploiting fast networking (e.g., \CAP{DPDK} or \CAP{RDMA})~---~ is necessary but insufficient for performance.

Workload characteristics play a significant role in performance.
Online services present data accesses that are highly skewed in popularity, following a power-law distribution~\cite{Atikoglu:2012}. 
For instance, posts by state leaders in a social network are exponentially more likely to be accessed than the vast majority of other posts.
Handling workloads with very popular objects requires extra care to avoid hotspots.
Another workload characteristic is that several transactional applications
exhibit a high degree of locality in their access patterns~\cite{TPC-C, Venmo20}. In other words, transactions tend to repeat, accessing the same or adjacent sets of objects. For example, in a cellular control plane application, each phone user repeats transactions, accessing the same phone context and its nearest base station. 
Despite the locality in the workload, if the datastore does not strive to keep relevant objects on the same node, costly network communication is inevitable to execute transactions.
In short, ignoring these important workload characteristics leads to load imbalances and excessive network traffic, which adversely affect performance.


\beginbsec{The protocol is essential} 
The replication protocol determines the actions necessary to execute each request, thus defining the datastore's performance.
To achieve high performance, a replication protocol should strive to follow two high-level principles: 
(1) maximize concurrency and load balance 
and 
(2) complete operations as fast as possible.
For concurrency, the protocol must allow the execution of operations from all replicas in the deployment. The protocol should not neglect the skewed data access nature of online services, which are susceptible to load imbalances. 
For instance, a dedicated replica that acts as a physical ordering point for writes limits concurrency and is prone to hotspots, thereby significantly inhibiting the performance of the datastore.

The second principle~--~namely, completing operations fast~--~calls for minimizing coordination (i.e., network exchanges between replicas) in the critical path of a request, which is challenging under strong consistency and fault tolerance.
Intuitively, under strong consistency, a read on an object replica that executes after the completion of a write to another object replica must return the value of the write. 
However, if both the write and read execute locally to their respective replicas, it would be impossible for the read to know and return the latest value.
Therefore, a coordination round-trip among replicas is inevitable under strong consistency for either reads or writes when served by different replicas.

When it comes to fault tolerance, coordination is necessary for writes. Each write must ensure that it has replicated its value before its completion. Otherwise, if the coordinating replica of the write fails, the datastore will permanently lose that committed value.
Undesirably, losing the latest committed value means that future reads on the affected object, even if they repeatedly contact all alive replicas, would 
remain indefinitely blocked or return stale values. 

\begin{table}[t]
\centering
\begin{adjustbox}{max width=0.8 \textwidth}

\begin{tikzpicture}

\node (table) [inner sep=-0pt] {
\begin{tabular}{l|cc}
\rowcolor[HTML]{9B9B9B}

  \begin{tabular}[c]{@{}c@{}}\textbf{Coordination} \end{tabular}
 &
  \begin{tabular}[c]{@{}c@{}}\textbf{Strong consistency} \end{tabular} &
  \begin{tabular}[c]{@{}c@{}}\textbf{Fault tolerance} \end{tabular} \\ \hline
\begin{tabular}[c]{@{}l@{}}
Reads  ~~$=$~~ 0\\ 
Writes ~~$=$\hspace{7pt} 0\end{tabular} &
\xmark & \xmark \\ \hline
\begin{tabular}[c]{@{}l@{}}
Reads  ~~$=$~~ 1\\ 
Writes ~~$=$\hspace{7pt} 0\end{tabular} & 
\cmark & \xmark \\ \hline
\begin{tabular}[c]{@{}l@{}}
Reads  ~~\raisebox{2pt}{\scalebox{0.7}{$\geq$}}\hspace{6pt} 0\\
Writes ~~\raisebox{2pt}{\scalebox{0.7}{$\geq$}}~~ 1\end{tabular} & 
\cmark & \cmark
\end{tabular}

};
\draw [rounded corners=.3em] (table.north west) rectangle (table.south east);
\end{tikzpicture}

\end{adjustbox}

\mcaption
{Coordination of reads and writes for fault tolerance and consistency.}
{Fault tolerance and consistency of reads and writes from all replicas based on coordination (i.e., round-trips to other replicas in the critical path).}
\label{tabM:read-write}
\end{table}

Overall, as outlined in \cref{tabM:read-write}, 
when all replicas can execute reads and writes,
the following two conditions apply on coordination to attain strong consistency and fault tolerance. For strong consistency, either reads or writes must contact other replicas (one or more times). For fault tolerance, writes necessarily need to contact other replicas at least once before completion to replicate their value. Thus, the best result a reliable protocol can aim for is to allow all replicas to serve local reads and fast writes that complete after a round-trip to other replicas, as this combination yields the maximum concurrency with the least amount of coordination for strong consistency and fault tolerance.

However, achieving this "holy grail" of local reads and fast writes is not always feasible. For example, multi-object transactions or failures in the middle of a single-object primitive may necessitate further coordination~\cite{Schiper:04}.
Nevertheless, protocols can optimize performance during the standard fault-free operation, which is far more common than the operation during faults. When it comes to transactions, protocols can exploit the locality exhibited by several workloads to avoid unnecessary network hops and traffic, which significantly reduce the overall performance.

In summary, high performance under fault tolerance and strong consistency, demands 
--- in addition to exploiting modern hardware ---
replication protocols that strive for local reads and fast writes from all replicas. To approach this performance ideal, protocols should leverage the common fault-free operation without neglecting critical workload characteristics, such as data access skew and locality in transactional workloads.

 \vspace{-15pt}
\tsection{Existing Replication Protocols}
\label{secB:replication-protocols}

Replication protocols that guarantee strongly consistent reads and writes and are capable of dealing with failures under our fault model can be classified into two categories: \textit{majority-based} protocols, which are typically variants of Paxos~\cite{Lamport:1998}, and \textit{membership-based} protocols, which require
a stable
\linebreak
membership of live nodes, such as the seminal primary-backup protocol~\cite{Alsberg:1976}.

\beginbsec{Majority-based protocols}
This class of protocols requires the majority of replicas to respond in order to commit a write. 
As a result,
majority-based protocols are naturally available, provided that a majority is responsive.
However, majority-based protocols pay the price in performance. To commit writes from all replicas, they 
must make multiple round-trips to a majority~\cite{Lamport:1998, Georgiou:11, Hadjistasi:17, Georgiou:19, Burke:20, Lynch:1997}.
More importantly, in the absence of responses from all replicas, there is no guarantee that a given write has reached all replicas, which makes linearizable local reads fundamentally challenging. 
Myriad of protocols are majority-based and cannot afford linearizable local reads from all replicas~\cite{Lamport:1998, Lamport:2001, Lamport:2005, Lynch:1997, Ongaro:2014, Moraru:2013, Burke:20, Skrzypczak:20, Enes:20, Georgiou:11, Hadjistasi:17, Georgiou:19, Gavrielatos:21Arxiv}.
In short, majority-based protocols seamlessly handle failures but hurt the performance of writes and reads, even in the absence of faults.

\beginbsec{Membership-based protocols}
Protocols in this class require {\em all operational} nodes in the replica group to acknowledge a write (also called read-one/write-all protocols~\cite{Jimenez:2003}). 
In doing so, they ensure that a committed write has reached all replicas in the ensemble, which naturally facilitates local reads. 

Membership-based protocols are supported by a \textit{reliable membership} (\CAP{RM})~\cite{Lamport:2009, Birman:1987, Chockler:01}. 
Modern \CAP{RM} implementations use a majority-based protocol to reliably maintain a stable membership of \textit{live} nodes guarded by leases~\cite{Gray:89, Dragojevic:2015, Kakivaya:2018}.
Specifically, each node locally stores a membership variable with an \epoch and a lease. 
The membership variable indicates the set of live nodes. 
Live nodes remain operational (i.e., execute reads and writes) as long as their lease has not expired.
Protocol messages are tagged with the \epoch of the sender at the time of the message creation, and a receiver drops any message that is tagged with an \epoch that differs from its local \epochDot

Membership-based protocols favor performance in the standard fault-free \linebreak operation in exchange for a reconfiguration, causing a performance hiccup when nodes crash.
During failure-free operation, membership leases are 
\linebreak
regularly renewed, facilitating local reads. 
When a failure is suspected, the membership variable is reliably updated (and the \epoch is incremented) through a majority-based protocol, but only after the expiration of the membership leases.
This guarantees safety by circumventing the potential false positives of unreliable failure detection.
Simply put, updating the membership variable only after the lease expiration ensures that unresponsive nodes cannot compromise consistency, since they have stopped serving requests before they are removed from the membership. 
As a result, when a fault occurs, the lease duration translates to a short 
unavailability for the affected shard.
Nevertheless, given the common failure-free operation, this is a fair trade-off 
in comparison with
the majority-based protocols, which avoid reconfiguration but forfeit local reads, even in the absence of faults.

Another benefit of membership-based protocols is that they require a fewer number of replicas to sustain the same number of node crashes. Unlike majority-based protocols that need $2f+1$ node replicas to sustain $f$ node crashes, membership-based protocols need just $f+1$ node replicas to sustain $f$ node crashes if the reliable membership is maintained by an external set of nodes~\cite{Lamport:2009}. However, to facilitate a fair performance comparison, in this thesis the membership is maintained by the same set of nodes (i.e., the replicas themselves). 
Consequently, we assume that both majority- and membership-based protocols can sustain only a minority of replica failures, even though this is not in favor of the protocols we propose.

Although existing membership-based protocols maximize performance on reads, they still hinder performance on writes in the steady state. 
The state-of-the-art membership-based protocol~\cite{Terrace:2009} improves upon the primary-backup protocol and allows for linearizable local reads from all replicas. Unfortunately, as we detail in \cref{secH:craq}, writes in this protocol always serialize on a dedicated replica and are propagated to one replica at a time.
%
This deficiency adversely affects the throughput and latency of writes.

\pagebreak

\beginbsec{Reliable transactions via distributed commit}
Modern resilient datastores afford reliable transactions~\cite{Dragojevic:2014, F-Kalia:2016,wei2015fast}. 
To implement transactions, they rely on statically sharded data.
In
static sharding, objects are placed randomly on fixed nodes, making it
easy to locate and access objects.
Because related objects reside on different shards across nodes, a \textit{distributed commit} must take place to accommodate transactions. 
In other words, multiple nodes, i.e., those storing data relevant to the transaction, must reach a unanimous agreement on whether a transaction can commit; otherwise, it should abort (e.g., due to conflicts)~\cite{Gray:78}.
The distributed commit is typically 
implemented over primary-backup replicated shards for fault tolerance~\cite{Dragojevic:2014}.
This combination of static sharding and distributed commit supports reliable transactions, regardless of the workload access patterns.


%
%
%
However, a distributed commit over static sharding cannot fully exploit the data access locality in transactional workloads and requires several round-trips to execute and commit each transaction reliably.
It is most likely that during
execution under static sharding, 
the node executing a transaction must fetch one or more objects that are stored remotely, potentially in a serial manner (\eg due to control flow or pointer chasing). 
Moreover, since a transaction can be aborted by remote participants, which may fail, committing a transaction also fundamentally mandates several round-trips for a reliable agreement~\cite{Dragojevic:2014, F-Kalia:2016}. 
This overhead is embedded in the commit of each transaction, even when failures do not occur.
To make matters worse,
even if an identical transaction immediately repeats on the same node on which it was previously completed, it will cause as much network traffic to be executed and committed again.
In summary, the inability of the distributed commit over static sharding to exploit access locality in transactions results in numerous network round-trips, drastically affecting the performance of the datastore. 

\tsection{Multiprocessor Consistency Enforcement}


Replication is not a unique feature of datastores; it has long been practiced in the world of shared-memory multiprocessors.
In the world of multiprocessors, in which the memory system is the "datastore", every processing core with its local private cache is a "node", and the cache blocks are the "objects" --- a cache block may be replicated in one or more caches, each private to a core.

While in the distributed world of datastores, strongly consistent replication protocols sacrifice performance for fault tolerance, in the world of shared-memory multiprocessors, the story is exactly the opposite.
In the multiprocessor, fault tolerance is generally not a consideration, and cache consistency protocols (also known as coherence protocols) almost always enforce strong consistency with high performance~\cite{Vijay:20}.

\beginbsec{Cache coherence protocol}
A multiprocessor cache coherence protocol ensures that, at any given time, a cache block can be either written or read.
To maintain this invariant, a typical coherence protocol allows a reader to return the local cache block value if and only if the block's local state is valid and forces a writer to invalidate all copies of the block before completing an update.
Therefore, a read always returns the most up-to-date value, thus providing linearizability~\cite{Vijay:20}.\footnote{Scheurich and Dubois~\cite{Scheurich:1987} presented this approach as a sufficient condition for enforcing sequential consistency. In reality, however, it satisfies the stronger linearizability property, a model that was later formalized by Herlihy and Wing~\cite{Herlihy:1990}.} 
We call such a protocol that invalidates all operational replicas before completing an update as an \textit{invalidating} (or \textit{invalidation-based}) protocol.  
The performance benefits of an invalidating cache coherence protocol are twofold.
First, this protocol allows \textit{each} core with an object replica to perform linearizable \textit{local} reads against its copy.
Second, the invalidating coherence protocol allows for high-performance writes.
\textit{Any} core can \textit{quickly} perform a write after a single broadcast round of invalidations to all other object replicas (i.e., cores caching the block).
Succinctly put, invalidating coherence protocols allow for concurrency and speed in both reads and writes.

\beginbsec{Hardware transactional memory}
Modern multiprocessors are enhanced with hardware transactional memory (\CAP{HTM})~\cite{htm}. Architectures that support \CAP{HTM} afford arbitrary transactions over independent cache blocks in an efficient manner.
Akin to distributed transactions in datastores, which typically build on top of replication protocols, such as primary-backup protocols, an \CAP{HTM} implementation extends the invalidation-based cache coherence protocols.

Unlike datastore transactions over static sharding, transactions in \CAP{HTM} are not executed or committed in a costly distributed way. Instead, a core coordinating a transaction leverages the dynamic sharding ability of the underlying coherence protocol to gather all involved objects in its local cache with exclusive write 
permissions (i.e., it dynamically acquires \textit{ownership}). This dynamic scheme allows for local transaction execution and commit. 
Crucially, subsequent transactions to those blocks eschew any remote coherence actions until another core takes over the ownership. As a result, this approach tremendously benefits workloads with locality in their transactional access patterns.

\beginbsec{Performant but not fault tolerant}
The multiprocessor's approach to strong consistency is highly efficient,
offering both local reads with fast writes from all cores and performant transactions once the workload's locality is captured.
However, the multiprocessor's consistency protocols are tailored for a single machine, and are therefore, not readily applicable to the distributed setting of replicated datastores. 
For example, scalable coherence protocols arbitrate concurrent writes through a directory-based design, which has two performance drawbacks in the distributed setting. First, writes must consult a directory before invalidating other copies. This can be fast within a multiprocessor but costly in a distributed setting, where it translates to across-server network hops.
Second, writes to the same object serialize on the directory, which is not a good fit for online services with highly skewed data accesses, as it is prone to hotspots.


However, the most critical issue is that neither coherence protocols nor \CAP{HTM} provide any fault tolerance guarantees.
To begin with, the centralized directory on which they rely is a single point of failure. 
In addition, invalidating all copies of a block on a write is impossible if one of the cores with a copy fails, as in this case the writer will endlessly wait for the failed core to acknowledge its invalidation.
More subtly, the entire system is vulnerable when a core with ownership to a block crashes; as the sole up-to-date replica of the block is permanently lost, and any other cores attempting future access to that block will be indefinitely stalled.
\tsection{Summary}
In summary, distributed datastores need high performance, strong consistency, and fault tolerance. To guarantee fault tolerance and strong consistency, they replicate data and rely on replication protocols. Replication protocols define the exact actions necessary to execute all operations and thus the datastore's performance. Traditional datastore protocols for reads/writes and distributed transactions are designed for the distributed setting but 
do not adequately address performance.
In contrast, the multiprocessor's consistency protocols provide high performance 
but cannot handle the challenges of the distributed setting, such as highly skewed workloads or fault tolerance.
%

To resolve this tension, the remainder of this dissertation primarily proposes multi\-processor-inspired invalidating protocols tailored for performance, load balance, and fault tolerance to accelerate both single-object operations and distributed transactions in replicated datastores. 

\markedchapterTOC
{Scale-out ccNUMA}
{Scale-out ccNUMA:\nextlinepdf Replication for Performance}
{Scale-out ccNUMA: Replication for Performance}

\equote{-40}{10}{The whole is greater than the sum of its parts.}{Aristotle}

\label{chap:cckvs}


In this chapter, we explore data replication solely to improve performance 
\linebreak
without considering fault tolerance.
We examine popularity skew in data access as this is a prevalent characteristic of online services and is responsible for load imbalances which can greatly degrade the 
datastore performance.
We propose a novel caching strategy that exploits skew to increase performance by aggressively replicating the hottest objects. 
At the core of this strategy is a new replication protocol that combines multiprocessor-inspired invalidations with logical timestamps to enforce strong consistency while balancing writes across all replicas and avoiding hotspot-prone physical serialization points.

\tsection{Overview}
\label{secC:intro}

Today's online services, such as search, e-commerce, and social networking, are backed by distributed key-value stores (\CAP{KVS}).
Such datastores must provide high throughput in order to serve millions of user requests simultaneously while meeting online response time requirements. 
To sustain these performance objectives, the application datasets are typically kept in-memory and sharded across multiple servers using techniques such as consistent hashing.


Although sharding data across individual servers enables massive parallelism, such a datastore design can suffer from hotspots.  This is because the popularity distribution of objects is highly skewed in these workloads, typically following power-law distributions~\cite{Atikoglu:2012,Novakovic:2016,Huang:2014,Volos:2017}. In other words, in the presence of skew, the server(s) serving the most popular objects will become saturated, turning into a bottleneck and limiting the throughput of the entire 
\CAP{KVS}. 

The skew problem is well established, and a number of techniques have been proposed for mitigating it. These techniques can be classified into two 
\linebreak
categories. 
The first class of techniques~\cite{Fan:2011,Li:2016,Jin:2017} uses a dedicated cache for storing popular keys to filter the skew.
The second class of techniques (FaRM~\cite{Dragojevic:2014} and RackOut~\cite{RNovakovic:2016}) mitigate skew by evenly distributing read 
\linebreak
requests across all servers of the \CAP{KVS}, regardless of the object's location.
To ensure low latency, the servers use an \CAP{RDMA}-enabled interconnect to 
\linebreak
access objects that reside at other servers. 
In essence, this class of techniques exposes a non-uniform memory access (\CAP{NUMA}) shared memory abstraction across the \CAP{KVS} servers.

The first approach is not scalable because the limited computational resources of a single cache node may not be able to keep up with the load. In contrast, the second approach is scalable in its processing capability but is network bound because the vast majority of accesses are serviced by remote nodes.

In this chapter, we view skew as an opportunity and leverage it to improve \CAP{KVS} performance. Taking inspiration from the effectiveness of caches in shared memory multiprocessors, we propose a \emph{Scale-out ccNUMA} architecture that augments \emph{each} server node in a distributed KVS deployment with a small cache of hot objects. Because object popularity is a function of the entire dataset, and not of individual shards, all cache instances maintain an identical set of objects, which are the most popular objects in the dataset. This {\em symmetric cache} not only ensures a high hit rate, but also relieves the clients from needing to know which caches maintain what objects, and avoids the need for costly metadata to track sharers on the KVS side.

Replicating the hottest data in multiple caches raises the problem of ensuring consistency in the presence of writes.
Traditional strongly consistent replication protocols, which allow for local reads and are suitable for caching, mandate that writes must serialize on a physical ordering point which is prone to hotspots in the presence of skew.
To address this issue, we propose \textit{Galene}, a novel replication protocol 
that couples cache-coherence-inspired invalidations with logical timestamps. This combination affords linearizability and fully distributed write serialization. Thus, writes can be coordinated by any cache, equally spreading the cost of consistency actions across the datastore nodes.

We then develop {\em ccKVS}, a distributed \CAP{RDMA}-based \CAP{KVS} that 
employs a Scale-out ccNUMA architecture, featuring symmetric caching with the strongest 
\linebreak
consistency enforced via the Galene protocol.
Our evaluation on a nine-node rack-based cluster shows that in comparison with a state-of-the-art NUMA-approach \CAP{KVS},
ccKVS achieves a 2.2$\times$ improvement in throughput for a
\linebreak
typical skewed workload with a modest write ratio while satisfying the strongest
\linebreak
consistency.

\noindent In short, the main contributions of this chapter are as follows:

\squishlistContrib

    \item \vspace{15pt}
    \textbf{We introduce symmetric caching} (\cref{secC:cache}), a transparent caching strategy that replicates the most popular objects in all caches, thus enabling high throughput and load balance
    while eliminating the costly requirement of tracking sharers. 

    \item \vspace{20pt}
    \textbf{To keep the caches consistent, we propose Galene} (\cref{secC:protocols}), a fully distributed protocol that couples invalidations with logical timestamps. Galene avoids hotspot-prone serialization points and enables write coordination from any replica to equally spread the cost of consistency actions across the deployment. We also verify Galene for safety and deadlock freedom in \CAP{TLA$^{+}$}.
   
    \item \vspace{20pt}
    \textbf{We build and evaluate ccKVS} (\cref{secC:design} and \cref{secC:eval}), an \CAP{RDMA}-based \CAP{KVS} that implements symmetric caching with the Galene protocol. Our evaluation shows that ccKVS achieves a throughput improvement of 2.2$\times$ over a state-of-the-art RDMA-based skew mitigation scheme for a workload with modest write ratios while satisfying linearizability.
    
    
\squishend

\break
\vspace{-10pt}
\tsection{Motivation}
\label{secC:motiv}
\vspace{-10pt}

\tsubsection{Skew and Load Imbalance}
\label{secC:skew}
\vspace{-7pt}

Prior research characterizing data access patterns in real-world datastore settings has shown that the popularity of individual objects in a dataset often follows a power-law distribution~\cite{Atikoglu:2012,Novakovic:2016,Huang:2014,Bodik:2010,Yang:2016,Sharma:2011}. In such a distribution, a small number of hot objects receive a disproportionately high share of accesses, while the majority of the dataset observes relatively low access frequency. The resulting {\em skew} can be accurately represented using a Zipfian distribution, in which an object's popularity $y$ is inversely proportional to its rank
$r$: $y ~ r^{-\alpha}$. The exponent $\alpha$ is a function of the dataset and access pattern, and has been shown to lie close to unity. The most common value for $\alpha$ in the recent literature is 0.99~\cite{Dragojevic:2014,RNovakovic:2016,Jin:2017,Hong:2013,Li:2016}, with 0.90 and 1.01 also frequently used and cited in KVS research~\cite{Armstrong:2013, Fan:2011}.

\begin{figure}[t!]
  \centering
  \includegraphics[width=0.7\textwidth]{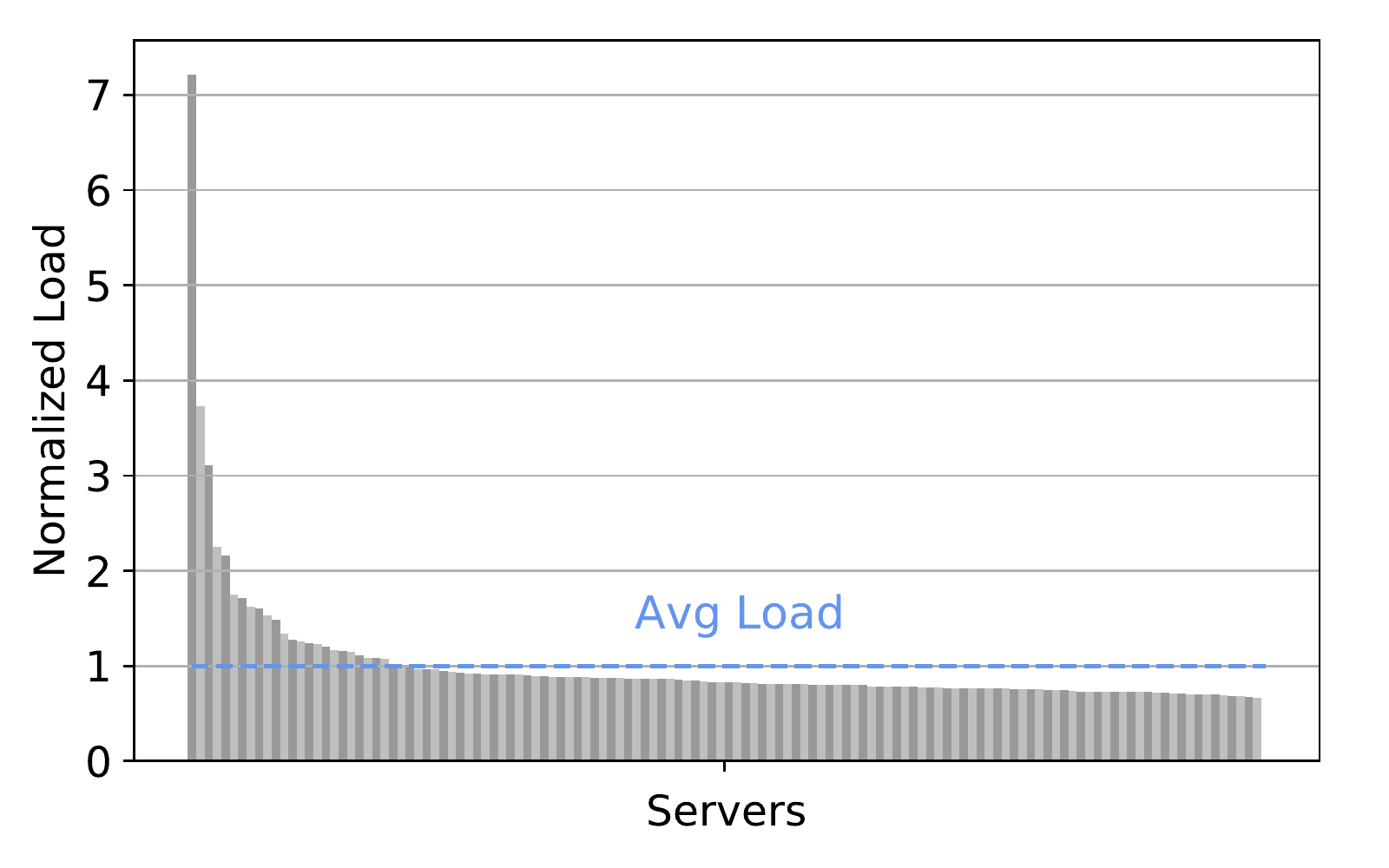} 
  \mcaption
  {Load imbalance in a server cluster caused by skewed accesses.}
  {Load imbalance in a cluster of 128 servers caused by skewed workload with $\alpha=0.99$.}
  \label{figC:server_load}
\end{figure}

\begin{figure}[t]
  \centering
  \includegraphics[width=0.7\textwidth]{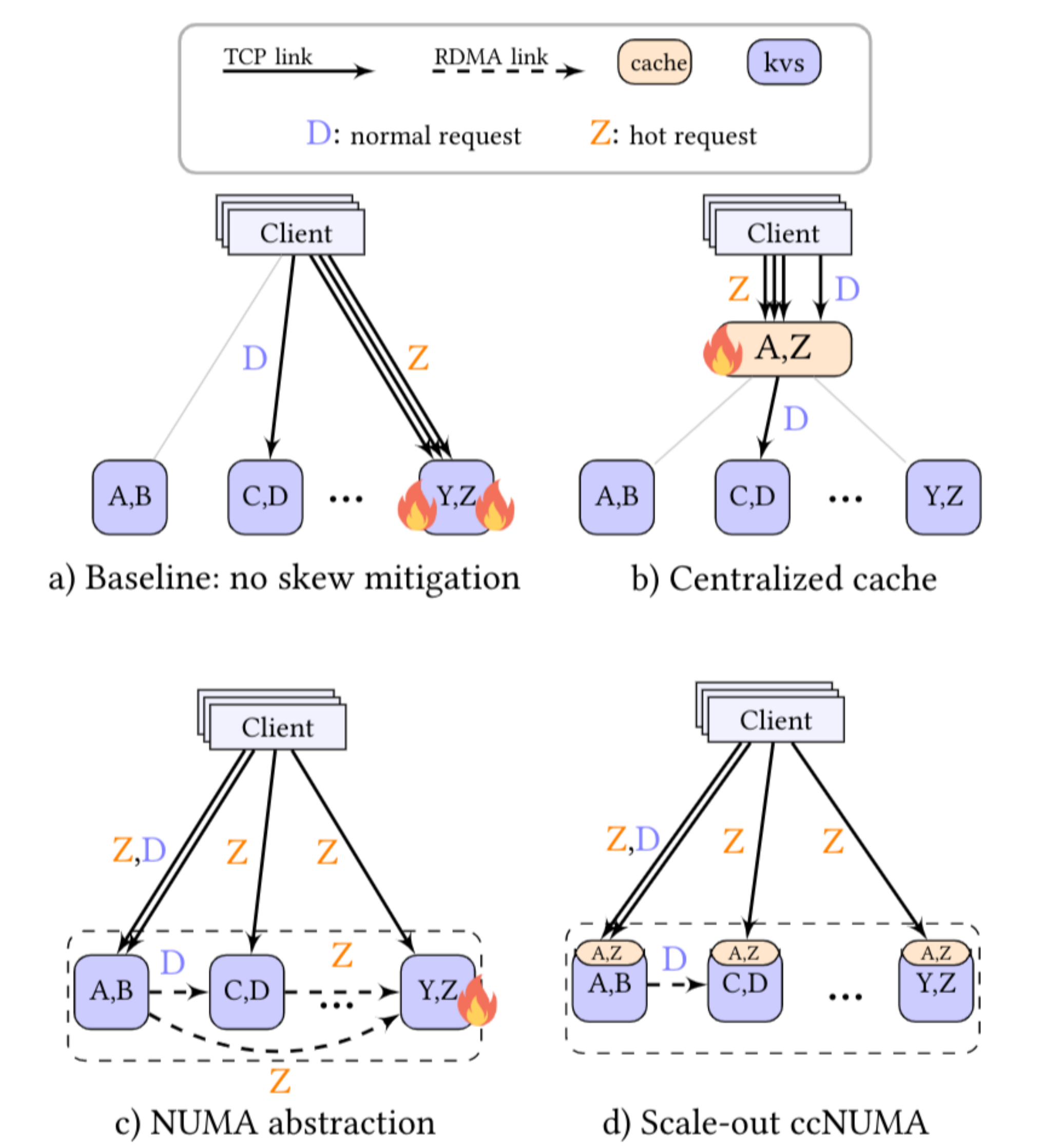}
  \mscaption{Design space for skew mitigation techniques.}
  \label{figC:blackbox}
\end{figure}

An important implication of popularity skew is the resulting load imbalance across the set of servers maintaining the dataset. As shown in \ecref{figC:blackbox}{a}, the server(s) responsible for the hottest keys may experience several times more load than an average server storing a slice of the dataset~\cite{Novakovic:2016}. For instance, \cref{figC:server_load} shows an example deployment of 128 servers and a data-serving workload with an access skew of $\alpha=0.99$. 
In this scenario,
the server storing the hottest key receives over 7$\times$ the average load in the system. 

\tsubsection{Existing Skew Mitigation Techniques}
\label{secC:existing}


\ecref{figC:blackbox}{b} and \ecref{figC:blackbox}{c} depict two approaches for skew mitigation that have emerged in the recent literature: caching and the \CAP{NUMA} abstraction. 

\beginbsec{Caching} 
Noting that a small fraction of keys is responsible for the load imbalance, recent work has suggested using a dedicated cache to filter the skew from the access stream before it hits the data-serving nodes (\ecref{figC:blackbox}{b}). 
Several variants of this
idea have been proposed: (i) placing a cache at the front-end load balancer~\cite{Fan:2011}; (ii) using a programmable switch to steer requests for hot objects to the cache node~\cite{Li:2016}; and (iii) using a programmable switch as a cache node~\cite{Jin:2017}. 

These caching approaches suffer from two important limitations. First, they usually target storage clusters where the back-end nodes are limited by the performance of the storage I/O~\cite{Li:2016}. Thus, a powerful server with an 
in-memory object cache is sufficient to keep pace with the load. 
The same is not true if the datastore is in-memory, in which case the high request rate it can sustain would overwhelm a single cache node. 
Second, these approaches do not offer a viable strategy for scaling the cache beyond a single node to accommodate larger deployments. 
While it is true that
simple partitioning of hot keys across servers is one way to scale to multiple cache nodes; in the limit, however, this strategy is fundamentally limited by the ability of the cache node with the hottest key to keep up with the load.

\beginbsec{NUMA abstraction}
This approach, which was pioneered in FaRM~\cite{Dragojevic:2014} and leveraged in RackOut~\cite{RNovakovic:2016}, offers a \CAP{NUMA}-like shared memory abstraction across the nodes storing the dataset via remote access primitives over a low-latency \CAP{RDMA}-enabled network, as shown in \ecref{figC:blackbox}{c}. More specifically, the one-sided \CAP{RDMA} reads allow any node to directly access the memory of any other node in the deployment. The design exploits this remote access capability to offer a {\em black-box abstraction} 
to the outside world wherein a client can send a request to any node in the deployment regardless of the data's location. By allowing requests for any object to be evenly distributed across the entire deployment, load imbalance is mitigated in the face of a skewed access distribution.


The key limitation of this approach is that the vast majority of requests require remote access. Indeed, the fraction of requests satisfied locally is inversely proportional to the number of servers in a deployment. Subsequent work (FaSST~\cite{F-Kalia:2016}) improved on network performance by replacing the one-sided primitives with two-sided \CAP{RDMA} communication, reducing the overall network overhead of the approach.
Novakovic \etal\cite{Novakovic:2014} demonstrated that integrated on-chip \CAP{NIC}s can further enhance performance by lowering the remote access latency.
Nevertheless, network bandwidth has persisted as the main performance limiter of the \CAP{NUMA} shared memory abstraction~\cite{F-Kalia:2016}. 

\vspace{0.08in}
\noindent
To summarize, existing skew mitigation techniques either (1) use a powerful cache node to filter the skew from the access stream before it hits the storage nodes or (2) exploit a \CAP{NUMA}-like shared memory abstraction that relies on remote access primitives to distribute the load across all servers. The first approach is processing bound because a single cache node may not be able to keep pace with the load, which makes it applicable mainly in a disk-based cluster environment. Meanwhile, the latter approach is scalable in its processing capability, but is network bound because the vast majority of requests require a remote access.

\tnsection{Scale-out ccNUMA}
\label{secC:case}

The central thesis of this chapter is that \emph{
a small cache of hot objects at each data serving node can effectively filter the skew while scaling cache throughput with the number of servers}. \ecref{figC:blackbox}{d} demonstrates the proposed approach, which combines the best features of caching and the \CAP{NUMA} abstraction in an architecture we call {\em Scale-out ccNUMA}. As shown in the figure, Scale-out ccNUMA augments each node in a pure \CAP{NUMA} deployment with a cache of hot objects. Whenever a client request hits in a server's cache, that node can immediately return the data, thus avoiding a remote access to the node containing the corresponding shard. 

The proposed approach has the following benefits:

\squishlist

\item Compared with existing cache proposals, which have a centralized cache at a load balancer or a network switch~\cite{Fan:2011, Li:2016, Jin:2017} and are thus limited by the throughput of that cache, the per-node cache naturally scales its throughput with the size of the deployment. Moreover, the per-node cache avoids the need for heterogeneous or exotic hardware required by prior work, such as a more powerful server in the cache node~\cite{Fan:2011, Li:2016} and/or programmable 
network switches~\cite{Li:2016, Jin:2017}. Avoiding hardware heterogeneity in a datacenter setting is beneficial from a cost, maintenance, and engineering (programmability) perspectives. 

\begin{figure}[t]
  \centering
  \includegraphics[width=0.7\textwidth]{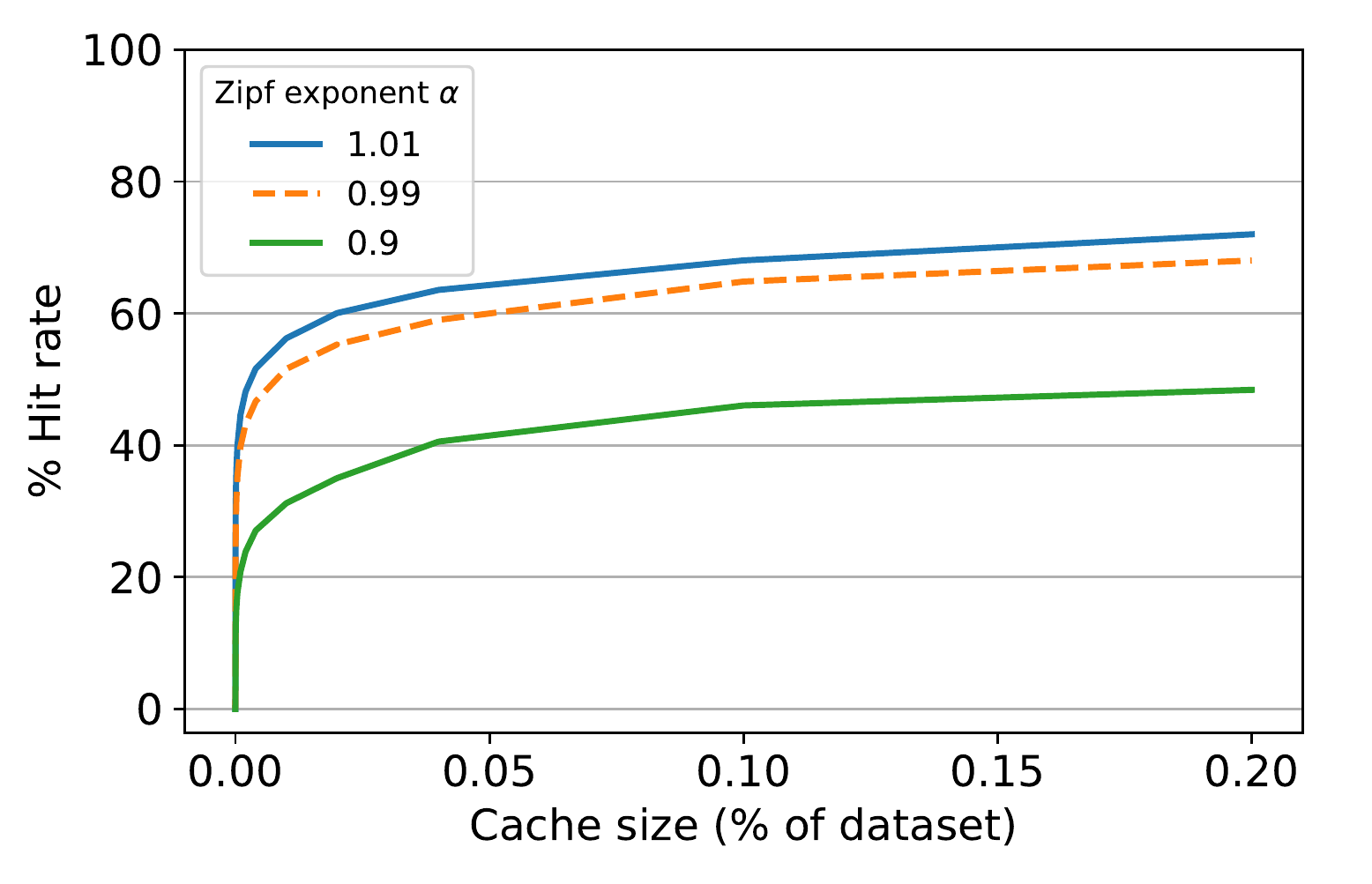}
  \mscaption{Effectiveness of caching under popularity skew.}
  \label{figC:cache-opportunity}
\end{figure}

\item Compared to a pure \CAP{NUMA} abstraction (e.g., FaRM~\cite{Dragojevic:2014}, RackOut~\cite{RNovakovic:2016}, FaSST~\cite{F-Kalia:2016}),  
adding a cache to each node can significantly lower the 
\linebreak
incidence of remote accesses. As \cref{figC:cache-opportunity} shows, for a Zipfian skew with an exponent $\alpha=0.99$ and a cache storing as little as 0.1\% of the hottest data, 65\% of requests will hit in the cache. Thus, only the remaining 35\% of the accesses (i.e., cache misses) may require remote access. 

Critically, the use of caching does not compromise the black-box abstraction presented by the \CAP{NUMA} shared memory architecture. Thus, any client can send a request to any server in the deployment without knowing the data's location. By load balancing the requests across the nodes and avoiding the majority of remote accesses, co-locating a cache with each node naturally improves the scalability of the shared memory architecture. 



\squishend

Despite these benefits, the proposed approach introduces a significant 
\linebreak
challenge in that it requires the caches to be consistent with respect to each other whenever a write occurs. This consistency-related challenge can further be broken down into two components. 

The first is how to determine which caches store what objects. This is
\linebreak
necessary to find the set of replicas, which is needed for consistency-preserving actions (e.g., invalidations or updates). The consistency protocols used in scalable multiprocessors use a directory to track replicas; however, the node holding the directory can potentially become a performance bottleneck.
While a directory can be distributed, a skewed access distribution naturally makes certain directory nodes more loaded than others, likely negating the benefits of caching. 

The second aspect of the challenge is related to write serialization, which is an important consistency requirement: all sharers must agree on the order of writes. Scalable multiprocessors accomplish this by physically serializing at the directory, which can, again, cause a bottleneck in our setting. 

Finally, we note that in addition to the consistency challenge, Scale-out 
\linebreak
ccNUMA also introduces the need for \emph{push-based} protocols. Protocols used in multiprocessors tend to employ invalidating {\em pull-based} protocols, meaning that a writer invalidates all sharers, which then must re-read the object to bring it back into the cache. This strategy is intended for parallel workloads where, for example, a variable can be written multiple times before being read by another thread.
In contrast, with read-intensive workloads that are the target of this work, an object that was updated will very likely be read in the nearest future at other nodes. 
This motivates the need for a push-based protocol that proactively pushes the updated object to all caches.


In the next two sections, we describe a cache organization, followed by the Galene consistency protocol, which address the challenges outlined above. 

\tsection{Symmetric Caching}
\label{secC:cache}

We exploit a simple insight in designing a scalable cache architecture that helps address the concerns outlined in the previous section. Specifically, we observe that the most popular objects are by nature the most likely ones to be accessed; hence, even though there are multiple cache nodes, they should all cache the same set of objects --- namely, the most popular ones. This idea, which we call {\em symmetric caching}, is illustrated in \ecref{figC:blackbox}{d}.

Despite its apparent simplicity, the symmetric cache architecture is extremely powerful, as it naturally resolves a number of challenges. For one, because all caches keep the same set of objects, there is no need to inform clients of which node caches what objects. Thus, clients can leverage the black-box abstraction and send requests to any node in the data serving deployment, with the probability of a cache hit being dependent solely on the requested key and not the choice of the node. This ensures both a load-balanced request distribution and a high cache-hit rate. 

Another advantage of the symmetric cache is that a node can determine which, if any, nodes cache an object solely by querying its local cache; if an object is found there, then {\em all} nodes have it; otherwise, none do. 
The ability to query a local cache to learn the status of an object naturally avoids the need for a directory, whose role in  cache-coherent multiprocessors is to track the set of caches that have a copy of a cache block. 
By not having a directory through which consistency actions would need to serialize,
the symmetric cache eliminates a potential serialization bottleneck and enables fully distributed consistency, as we describe in the next section.

An important feature of symmetric caching is that the caches are write-back. This means that writes to an object residing in the symmetric cache do not update the underlying \CAP{KVS} until the object is evicted from the cache. This feature is critical in avoiding throughput degradation at the home node of a popular object, whenever writes follow a skewed distribution. Because all caches maintain the same set of objects in the cache, on eviction, only the node containing the shard with the evicted key needs to check whether the object has been modified and, if so, update the underlying \CAP{KVS}. 

In order for the symmetric cache to be effective, it is essential to be able to identify the most popular objects with minimal overhead.
This problem has been well researched, with highly efficient solutions proposed in recent work. 
A particularly attractive approach for symmetric caching is one
proposed by Li \etal~\cite{Li:2016}, which relies on memory-efficient top-k algorithms~\cite{Cormode:2008, Metwally:2005} to dynamically learn the popularity distribution. In the algorithm proposed by Li et al., each server maintains a key-popularity list with $k$ entries, approximating the popularity of the $k$ hottest keys, and a frequency counter that keeps track of recently visited keys, such that newly popular keys can be detected.
The scheme uses an epoch-based approach, whereby the key popularity list gets updated and propagated to the cache at the end of each epoch. Finally, request sampling is used to alleviate the performance impact of updating the frequency counter upon each request.

Conveniently, because symmetric caching exposes a \CAP{NUMA} abstraction, whereby clients spread their requests across all servers, each server sees the same access distribution as do the other servers in the deployment. Therefore, in our setting (and in contrast to~\cite{Li:2016}), it is sufficient 
for just a single server to act as the cache orchestrator, responsible for identifying the most popular objects and informing the other nodes.
Centralizing the process of classifying an object as popular not only reduces the overhead of tracking hot objects but also naturally alleviates the burden of reaching a consensus on which objects are popular, thus simplifying the entire process. While our evaluation does not consider shifts in popularity skew, we expect the set of most popular keys to evolve slowly, with only a handful of keys removed or added to the cache every few seconds~\cite{Li:2016}.

\markedsection
{Galene: \uppercase{F}ully distributed Strong Consistency}
{Galene: \uppercase{F}ully distributed Strong Consistency}
\label{secC:protocols}


With symmetric  caching, while a significant portion of read requests (the cache hits) can be served locally,
ensuring consistency in the presence of writes is  challenging. 
%
%
To facilitate strongly consistent local reads, a replication protocol must propagate writes that hit in one cache to all other symmetric caches.
Our targeted consistency model of linearizability mandates that writes must be atomically reflected across the replicas at a point within their invocation and response. 
This implies
\textit{write serialization}: all replicas must agree on the order of writes to a key.\footnote{Because linearizability is composable, ensuring linearizability for each individual key guarantees that the entire datastore is linearizable.}

\tsubsection{Hotspots in Write Serialization}

\begin{figure}[t]
  \centering
  \includegraphics[width=0.7\textwidth]{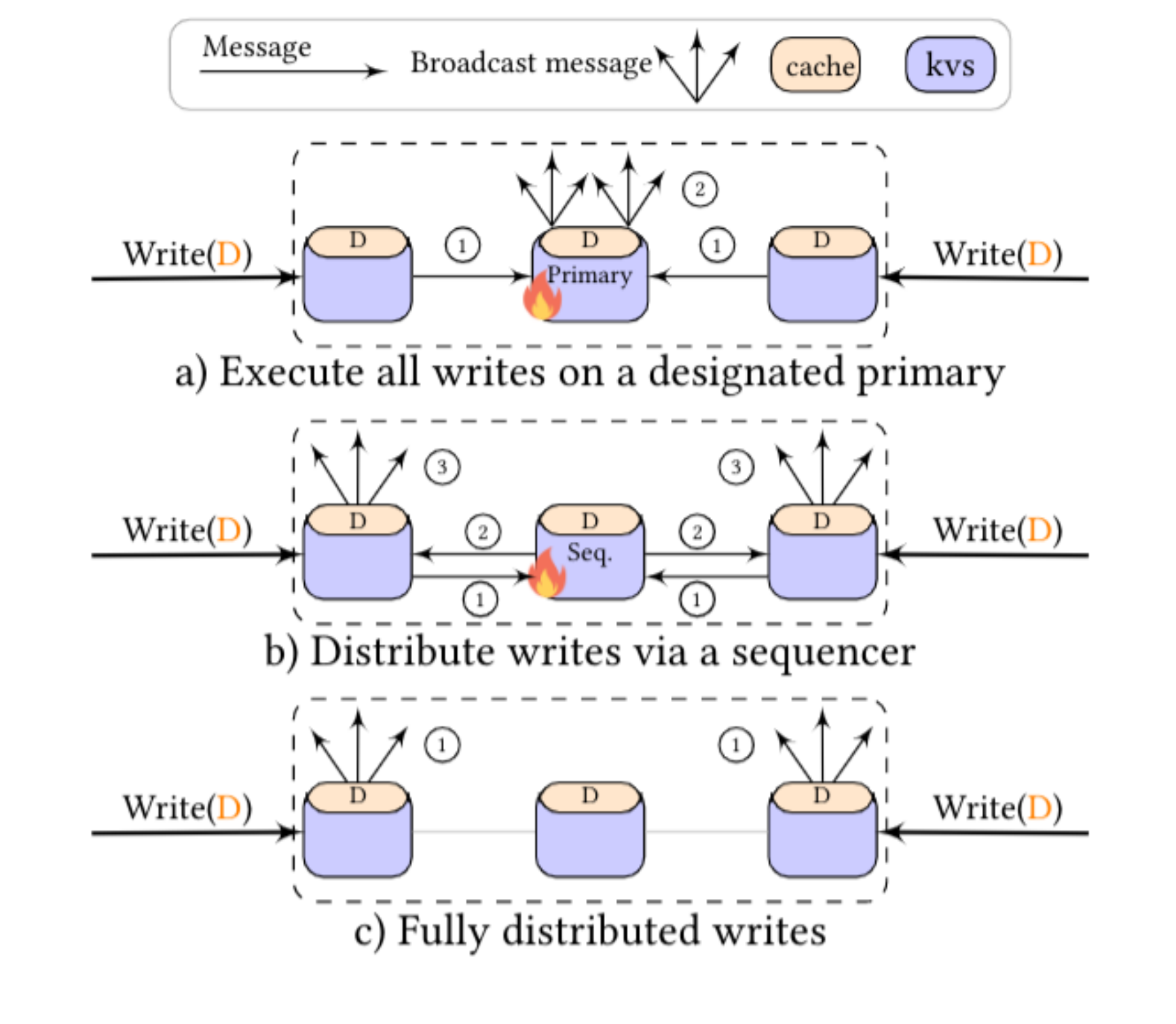}
  \mcaption{Design space for ensuring a global order of writes.}{Design space for ensuring a single global order for writes to a key. }
  \label{figC:protocols-design}
\end{figure}

Enforcing write serialization in a high-throughput fashion is challenging under skew. One natural way to enforce write serialization is to constrain writes to a single replica (e.g., via a primary-backup protocol), where all writes to a specific key must occur at a designated primary, as shown in \ecref{figC:protocols-design}{a}. Such protocols are commonly used in 
distributed datastores that demand strong consistency~\cite{Dragojevic:2015, VanRenesse:2004, Terrace:2009}. 
In the same spirit, distributed datastores can opt to achieve write serialization through a sequencer (shown in \ecref{figC:protocols-design}{b}), which assigns monotonically increasing timestamps to writes and their consistency actions (e.g., update messages). 
%
%
However, in the presence of skew, the primary (or sequencer) in the two approaches could easily become a hotspot on writes to a popular object, as consistency actions related to that object must serialize through it. 
%
The same flaw arises in coherent multiprocessors where write serialization to a block is performed at a physical point (i.e., a directory).

We address this issue by employing Galene, a multiprocessor-inspired 
invalidation protocol with a twist. 
As in the multiprocessor's coherence, Galene invalidates all replicas before performing a write.
Unlike the multiprocessor 
protocols, however, Galene eschews the directory and achieves write serialization in a fully distributed manner through \textit{logical timestamps}.
\ecref{figC:protocols-design}{c} shows the fully distributed nature of the protocol,
which completely avoids physical serialization points that are prone to load imbalances.


\begin{figure}[t]
  \centering
  \includegraphics[width=0.9\textwidth]{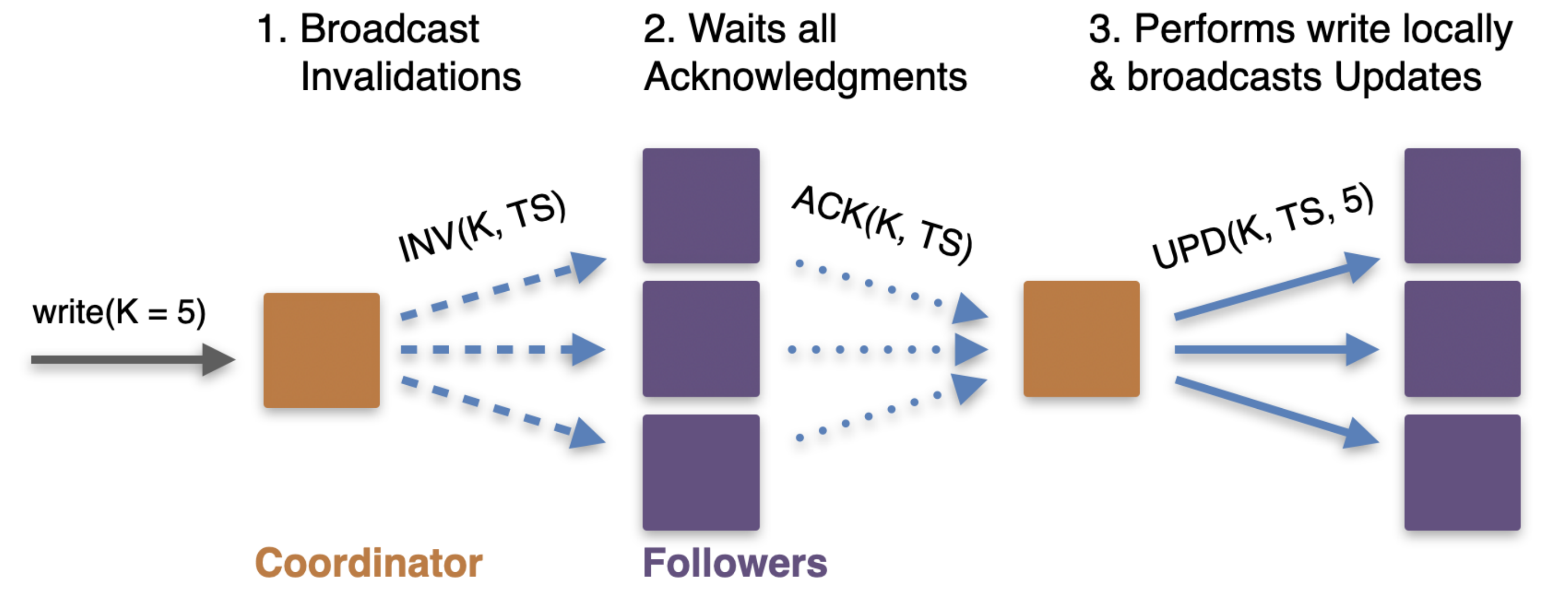}
  \mcaption
  {Write actions in Galene.}
  {Write actions in Galene when writing the value \emph{5} to a hot (cached) key \emph{K}. \emph{TS} is the write's logical timestamp.}
  \label{figC:consistency-actions}
\end{figure}

\tsubsection{Galene Overview}
\label{secC:dist-protocols}
Galene is a fully distributed push-based invalidating protocol that guarantees linearizability. 
Read cache hits are executed locally on any of the symmetric cache nodes. 
A write cache hit also proceeds on any node. As illustrated in \cref{figC:consistency-actions}: the cache coordinating the write (called the \textit{coordinator}) broadcasts an \textit{Invalidation} (\CAP{INV}) message to all other caches (called the \textit{followers}) and waits on \textit{Acknowledgments} (\CAP{ACK}s). Once all \CAP{ACK}s have been received, the write completes via an \textit{Update} (\CAP{UPD}) message broadcast where the coordinator proactively pushes the new value to the followers.

Before describing the protocol, we briefly outline the two essential mechanisms behind Galene's fully distributed method of ensuring strong consistency:

\beginbsec{Invalidations} 
When an \CAP{INV} message is received, the target key is placed in an Invalid state, meaning that reads to the key cannot be served. While conceptually similar to a lock, the main difference is that with invalidations, concurrent writes
to the same key do not fail and are resolved in place through the use of logical timestamps, as discussed below. The use of invalidations is inspired by cache coherence protocols, where a cache line in an Invalid state informs the readers that they must wait for an updated value. 

\beginbsec{Logical timestamps} 
Each write in Galene is tagged with a monotonically-increasing per-key logical timestamp, \newtext{implemented using logical clocks (as in the seminal \CAP{ABD} protocol~\cite{Attiya:1995})} and computed locally at the coordinator cache. The timestamp is a lexicographically ordered tuple <\VAR{v}, \cid>
that combines a key's version number (\VAR{v}), which is 
incremented on every write, with the node \CAP{ID} of the coordinator 
(\cid).
Two or more writes to a key are \textit{concurrent} if 
their execution is initiated by different caches holding the same timestamp.
Non-concurrent writes to a key are 
ordered based on their timestamp version, while concurrent writes from different coordinators (same version) are ordered via their 
\texorpdfstring{\VAR{c\textsubscript{id}}}{}.\footnote{
  More precisely, a timestamp A: <\texorpdfstring{\VAR{v\textsubscript{A}}}{}, \texorpdfstring{\VAR{c\textsubscript{idA}}}{}> is higher than a timestamp B: <\texorpdfstring{\VAR{v\textsubscript{B}}}{}, \texorpdfstring{\VAR{c\textsubscript{idB}}}{}>, if either $\texorpdfstring{\VAR{v\textsubscript{A}}}{} > \texorpdfstring{\VAR{v\textsubscript{B}}}{}$ or $\texorpdfstring{\VAR{v\textsubscript{A}}}{} = \texorpdfstring{\VAR{v\textsubscript{B}}}{}$ and $\texorpdfstring{\VAR{c\textsubscript{idA}}}{} > \texorpdfstring{\VAR{c\textsubscript{idB}}}{}$.
} 
Uniquely tagged writes allow each node to locally establish the same global order of writes  to a key. 


\begin{figure}[t]
  \centering
  \includegraphics[width=0.7\textwidth]{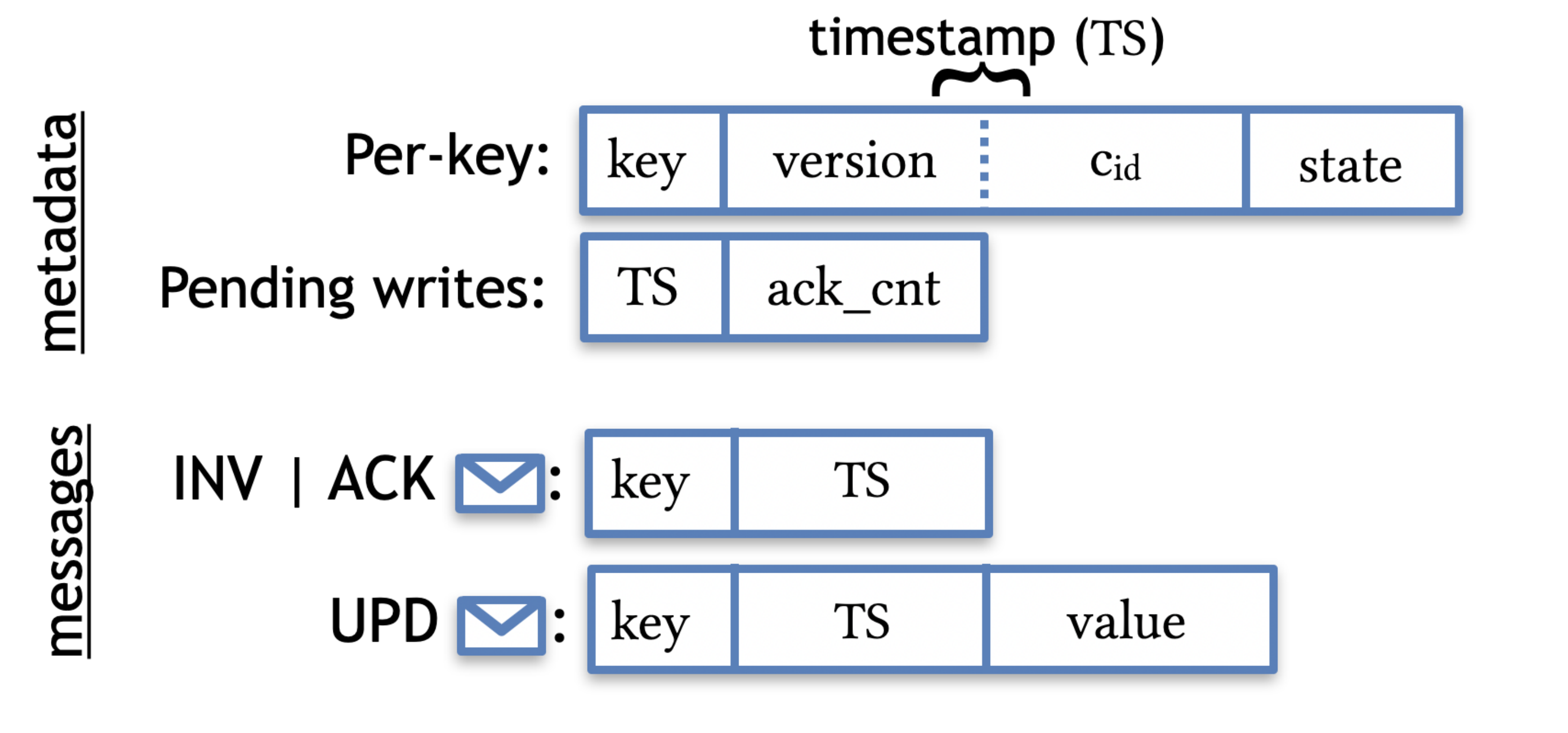}
  \mscaption{Metadata stored and messages sent by Galene.}
  \label{figC:galene_meta}
\end{figure}


\tsubsection{Galene Protocol}
Each cached object replica maintained by Galene can be in one of three states: \textit{Valid}, \textit{Invalid}, or \textit{Write}. 
\cref{figC:galene_meta} illustrates the format of protocol messages and the metadata stored at each cached replica of an object. Notice that a write's associated messages (i.e., \CAP{INV}s, \CAP{ACK}s, and \CAP{UPD}s) are tagged with the logical timestamp of that write.
The exact protocol steps to execute a read and a write are described below.

%
%


\vspace{0.02in}
\noindent\textbf{\textit{Read}}: 
Any symmetric cache node can service a read request of a cached object by returning the local value of the requested key if it is in the Valid state. If the key is in any other state, the request is temporarily stalled.

\noindent\textbf{\textit{Write}}:\\
\centerline{\underline{Coordinator}}
Any node can be a coordinator and issue a write to a cached key, but only when its local replica of the targeted key is in the Valid state. Otherwise, the write is temporarily stalled. To issue and complete a write, the coordinator node: 
\begin{itemize}[leftmargin=*]
\item\textbf{C\textsubscript{\CAP{TS}}}: Updates the key's local timestamp by incrementing its \VAR{version} and appending its node \CAP{ID} as the 
\cid, then assigns this timestamp to the write.

\item\textbf{C\textsubscript{\CAP{INV}}}: Promptly broadcasts an \CAP{INV} message consisting of the \VAR{key} and the new timestamp (\VAR{\CAP{TS}}) to all followers and transitions the key to the Write state. 

\item\textbf{C\textsubscript{\CAP{ACK}}}: Once the coordinator receives \CAP{ACK}s from all the followers, it checks whether the key's local timestamp is unchanged (i.e., is the same as its write's timestamp).
If the timestamp remained the same, it \textit{applies} the write in its local cache by updating the key to the new value and transitioning the key's state back to Valid.
Otherwise, it leaves the key unchanged.
\item\textbf{C\textsubscript{\CAP{UPD}}}: Finally, if the coordinator applied the write locally, it completes the write by broadcasting a \CAP{UPD} message which consists of the key, the write's value, and timestamp to all the followers. If it did not apply the write locally, then another concurrent write with a higher timestamp took precedence, covering its write. In this case, the coordinator completes its write without broadcasting a \CAP{UPD} message. 
\end{itemize}

\vspace{5pt}
\centerline{\underline{Follower}}
\vspace{-20pt}
\begin{itemize}[leftmargin=*]
\item\textbf{F\textsubscript{\CAP{INV}}}: Upon receiving an \CAP{INV} message, a follower compares the timestamp of the incoming message with its local timestamp of the key.
If the received timestamp is lower than the local timestamp, the follower simply ignores the message. Otherwise,
the follower performs the \CAP{INV} to its local cache by transitioning the key to the Invalid state and updating the key's local timestamp (both its \VAR{version} and the \cid).


\item \textbf{F\textsubscript{\CAP{ACK}}}: Regardless of the result of the timestamp comparison, a follower always responds with an \CAP{ACK} containing the same timestamp as that in the \CAP{INV} message of the write. 

\item \textbf{F\textsubscript{\CAP{UPD}}}: When a follower receives a \CAP{UPD} message, it updates the key's local value with the value from the message and transitions the key to the Valid state if and only if the received timestamp is equal to the key's local timestamp. Otherwise, the \CAP{UPD} message is simply ignored. 
\end{itemize}

\vspace{-5pt}
\beginfsec{Formal verification} 
We expressed Galene in \tla\cite{Lamport:1994} and verified its reads and writes for safety and the absence of deadlocks. 
For safety, we verified against
the \emph{data value invariant}: if an object copy is in the Valid state (i.e., can be read), then it must hold the most recent value written to that object. 
Our \tla model allows for the number of caches and number of total writes to be configured. We have verified with up to 5 caches and 4 writes. A detailed state transition table as well as the \tla specification are available online.\footnote{ 
\label{footC:link}\href{http://s.a-phd.com}{http://s.a-phd.com}} 
\newtext{We sketch why this protocol specification provides linearizability in \cref{Apendix}.}

\beginbsec{Enhancing fairness}
Galene linearizes writes based on their unique 
\linebreak
timestamps, consisting of a version and a node \CAP{ID}.
In the event that the versions are the same
(i.e., concurrent writes), the linearization is resolved based on the node \CAP{ID}s, which might raise concerns about ordering fairness. 
This is easily mitigated by assigning several \textit{virtual node \CAP{ID}s} to each physical node. With this scheme, before issuing a write, a node randomly picks one of its virtual node \CAP{ID}s to be used for the write's logical timestamp. Of course, to maintain correctness, the same virtual node \CAP{ID} cannot be assigned to more than one physical node. 
For example, given three nodes \textit{A, B, and C}, the sets of virtual \CAP{ID}s \textit{A}: $\{$1, 4, 7, 10$\}$, \textit{B}: $\{$2, 5, 8, 11$\}$, and \textit{C}: $\{$3, 6, 9, 12$\}$ are safe and would increase fairness.

\begin{figure}[t]
  \centering
  \includegraphics[width=0.9\textwidth]{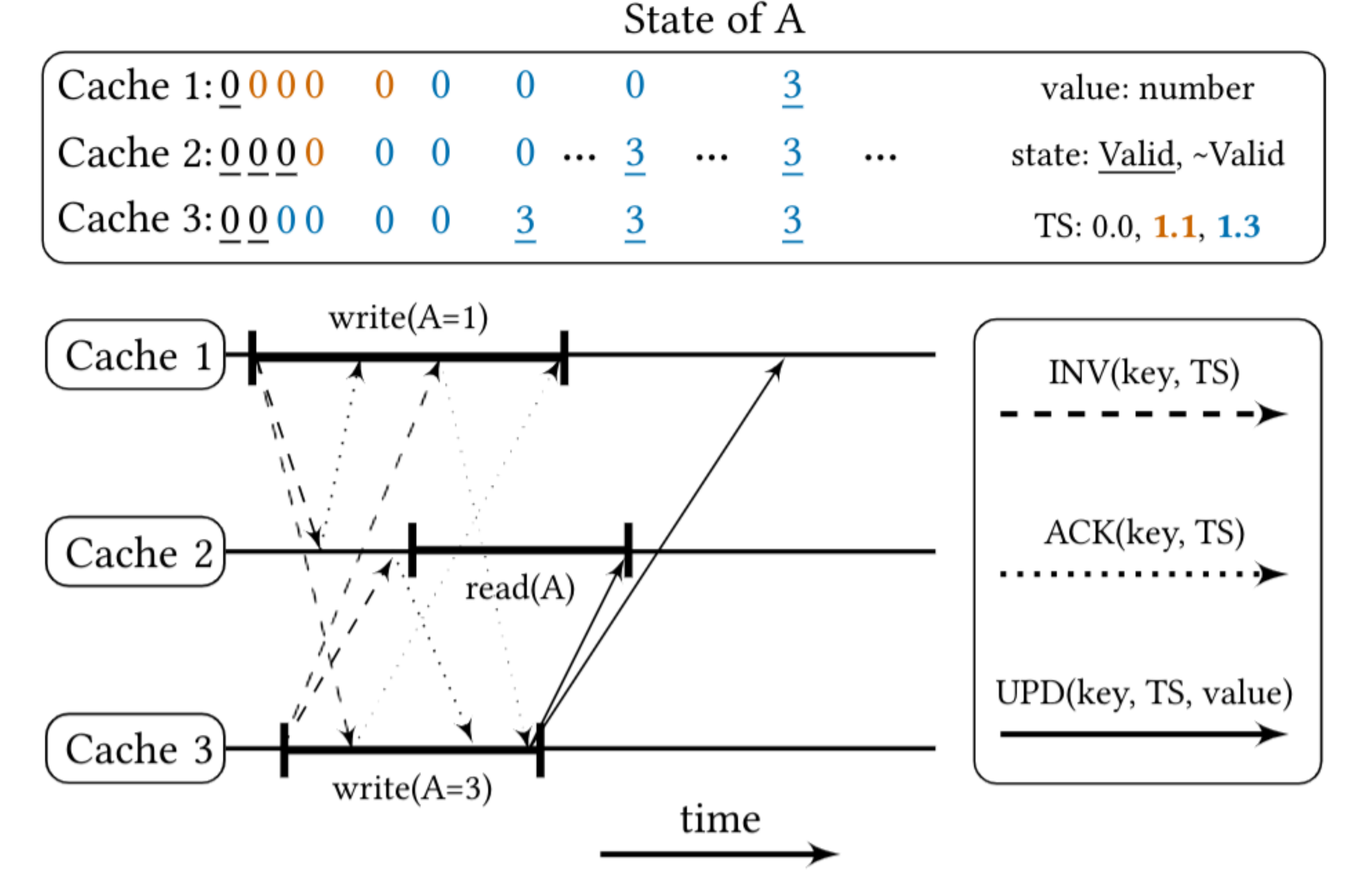}
  \mcaption{Execution example with concurrent writes in Galene.}
  {Concurrent writes to a hot key A, followed by a read, each hitting a different instance of a three-node symmetric cache. State of A shows the values of the replicas in each cache; underlined values represent Valid state, non-underlined represent other states. The color indicates the timestamp value.}
  \label{figC:galene_example}
\end{figure}
\tsubsection{Operational Example}

In this subsection, we discuss \cref{figC:galene_example}, which illustrates an example of 
Galene's execution with reads and writes to a cached key $A$. 
The purpose is to demonstrate the operation of Galene in the presence of concurrent reads and writes to a key.
For clarity, we consider a symmetric cache over three nodes, no use of virtual node \CAP{ID}s, and use the notation \VAR{v}.\cid~instead of <\VAR{v}, \cid> for the timestamp.
We assume that key $A$ is initially stored in the Valid state, with the same value (zero) and timestamp (0.0) in all three nodes. 

First, node 1 initiates a write ($A = 1$) by incrementing its key's local timestamp to {\color{myOrange}1.1}, broadcasting \CAP{INV} messages (dashed lines), and transitioning key $A$ to the Write state but without yet updating its local value. Similarly, node 3 initiates another concurrent write ($A = 3$, with timestamp {\color{myBlue}1.3}). 
Recall that \CAP{INV}s in Galene contain the key and timestamp (including the \cid).

Node 2 \CAP{ACK}s the \CAP{INV} message from node 1 (dotted line), transitions the key $A$ to the Invalid state and updates its timestamp to {\color{myOrange}1.1}. Node 3 \CAP{ACK}s the \CAP{INV} of node 1 but does not modify the local copy of $A$ (or its metadata) because its local timestamp is higher (same version, but higher \cid).
Subsequently, node 2 receives the \CAP{INV} from node 3, which has a higher timestamp than the locally stored timestamp, resulting in an update to its local timestamp (from {\color{myOrange}1.1} to {\color{myBlue}1.3}), all while remaining in the Invalid state. 
Likewise, node 1 \CAP{ACK}s the \CAP{INV} of node 3 by updating the timestamp and remaining in the Invalid state. 

Meanwhile, node 2 starts a read, but it is stalled because its local copy of $A$ is invalidated. 
Once node 3 receives all of the \CAP{ACK}s, it completes its own write by updating the local value, transitioning $A$ to the Valid state, and broadcasting a \CAP{UPD} message (solid lines) to the other replicas that includes the write's timestamp and value. 
When node 2 receives node's 3 \CAP{UPD} message, it updates $A$'s local value, transitions its state to Valid and completes its stalled read. 

Once node 1 receives all of the \CAP{ACK}s, it completes its write. However, its key remains in the Invalid state. This occurs because the write from node 3 took precedence over node's 1 own write due to its higher timestamp, but the \CAP{UPD} from node 3 has not yet been received. Note that although the write from node 1 completes later than the concurrent write from node 3, it is linearized directly before the write of node 3 due to its lower timestamp (\cid). Finally, node 1 receives node's 3 \CAP{UPD} message, which results in updating the value of $A$ and transitioning its state back to Valid.

\tnsection{ccKVS}
\label{secC:design}

To understand the benefits and limitations of the proposed Scale-out ccNUMA architecture, we build ccKVS, an in-memory \CAP{RDMA}-based distributed KVS that combines a \CAP{NUMA} abstraction~\cite{Dragojevic:2014} 
with symmetric caching and the Galene protocol.
The code of ccKVS is available online.\footnote{\href{http://s.a-phd.com}{http://s.a-phd.com}} 

Each node in ccKVS is composed of two entities: a shard of the \CAP{KVS} and an instance of the cache. Each entity has an object store and a dedicated pool of threads for request processing. As described in \cref{secC:cache}, the content of all caches is identical, composed of the most popular objects in the dataset. The caches are kept consistent using the Galene protocol (described in \cref{secC:protocols}). The nodes of a ccKVS deployment are connected via \CAP{RDMA} with two-sided primitives used for communication. Clients load balance their requests (both reads and writes) across all nodes in a ccKVS deployment --- for example, by picking a server at random or in a round-robin fashion.

\tsubsection{Functional Overview}
\label{secC:overview}
\beginbsec{Reads}
When a client request arrives at a ccKVS server, the server probes its instance of the symmetric cache. If the requested key is found, the associated object is retrieved from the cache and the server directly responds to the client. In case of a miss, the server determines whether the key belongs to a local or remote \CAP{KVS} partition. If remote, the server issues a remote access to the server containing the requested key using a two-sided \CAP{RDMA} primitive. On the destination side, the server picks up the remote access and responds with the data to the requesting server. Once the object is available, either by virtue of being in the local partition or through a remote access, the server handling the request responds to the client.

\beginbsec{Writes}
Similar to reads, write requests are load-balanced across all nodes in a ccKVS deployment, thus avoiding write-induced load imbalance. If the write request hits in the cache, the server handling the request (i.e., the coordinator of the write) executes the steps necessary to maintain consistency across all symmetric caches in accordance with the Galene protocol. Briefly, this means first invalidating all the caches and only then performing the write locally and propagating the new data to the other caches.
The communication required for maintaining consistency also occurs via two-sided \CAP{RDMA} primitives. 
If the write request misses in the cache, the server forwards it to the home node (if remote), which directly performs the write. 




\tsubsection{Cache and \uppercase{KVS} Implementation}
\label{secC:design-cache-kvs}


\beginbsec{Thread partitioning}
The threads inside a machine in ccKVS are partitioned into two pools: cache threads and \CAP{KVS} threads. 
The cache threads receive the requests from outside clients and are responsible for the cache accesses. 
The back-end \CAP{KVS} is handled by the \CAP{KVS} threads; thus, in case of a cache miss, the request must be propagated from a cache thread to a \CAP{KVS} thread (local or remote). Finally, the cache threads also communicate with each other to exchange consistency messages: invalidations, acknowledgments and updates. Notably, the \CAP{KVS} threads do not communicate with each other.

\beginbsec{Concurrency control}
Among the cache threads, which are responsible for servicing requests to the most popular objects in the request stream, ccKVS leverages the concurrent-read-concurrent-write (\CAP{CRCW}) model, whereby any cache thread can read or write any object in the cache. Despite the mandatory synchronization overheads, we find that this design maximizes throughput given the demand for the most popular keys in the dataset. 

The \CAP{KVS} design is more involved. The conventional wisdom~\cite{Lim:2014} is that when the requests are load-balanced across all machines, it is beneficial to 
partition the \CAP{KVS} at a core granularity (i.e., exclusive-read-exclusive-write (\CAP{EREW}) model) to avoid inter-thread synchronization on data accesses. Our design, however, employs the \CAP{CRCW} model for the \CAP{KVS}, even though with the skew 
\linebreak
filtered by the caches, \CAP{KVS} accesses observe an access distribution that closely approaches uniform. 

We choose \CAP{CRCW} because 
it allows us to minimize the connections among the cache threads and the \CAP{KVS} threads in the deployment. Our experiments show that this is a favorable design choice, as the benefits of limiting the connectivity among threads on different machines outweigh the overhead of the concurrency control in \CAP{CRCW}. We elaborate on these benefits in \cref{secC:design-perf}. Finally, the \CAP{CRCW} concurrency model in the \CAP{KVS} increases the ability of cache threads to batch multiple requests in a single packet, alleviating  network-related bottlenecks. We explore the benefits of this optimization in \cref{secC:coalesc-perf}.

To ensure high read and write performance under the \CAP{CRCW} model, ccKVS synchronizes accesses using \textit{sequential locks} (seqlocks)~\cite{Hemminger:2002, Lameter:2005}, which allow lock-free reads without starving the writes. 
The seqlock is composed of a spinlock and a version. The writer acquires the spinlock and increments the version, goes through its critical section,  increments the version again, and releases the lock. Meanwhile, the reader never needs to acquire the spinlock; it simply checks the version immediately before entering the critical section and immediately after exiting.
If the version has changed or is an odd number, 
then a write has happened concurrently with the read, and the reader retries. 

The seqlocks are implemented in the header of each object. The header contains a version number that has a dual role: it is used to implement both seqlocks and the version of the logical timestamps for the Galene protocol. Therefore, we only need to add one byte to the header to implement the spinlock. 
Our seqlock implementation is inspired by the \CAP{OPTIK} design pattern~\cite{Trigonakis:2016}.

All consistency messages are treated as writes, as they must modify metadata in the header of the key-value pair. Meanwhile, reads to the cache do not modify state and thus happen ``lock-free'' and in parallel.

\beginbsec{KVS}
We use \CAP{MICA}~\cite{Lim:2014} as a state-of-the-art \CAP{KVS} and leverage the source code for \CAP{EREW} found in \cite{Kalia:2016} to build our \CAP{KVS}. Since ccKVS adopts the \CAP{CRCW} model, the \CAP{KVS} is concurrently accessed by all \CAP{KVS} threads; therefore, we implement seqlocks over \CAP{MICA}. Our evaluation considers both \CAP{EREW} and \CAP{CRCW} design choices.
Finally, we note that symmetric caching and Galene are not tied to any particular \CAP{KVS}.

\beginbsec{Symmetric cache}
The symmetric cache is a data structure that is concurrently accessed by all the cache threads within a node. It inherits its structure from our \CAP{KVS}. 
We extend the \CAP{KVS}'s \CAP{API} and functionality to provide support for Galene's consistency-related operations (i.e., Invalidations, Acknowledgments, and Updates) and object states (i.e., Valid, Invalid, and Write). For example, a read request may hit in the cache but not immediately return, since the key-value pair could be in the Invalid state.

Each key-value pair stored in the cache has an 8B header, where the necessary metadata for synchronization and consistency are efficiently maintained. These metadata include: the consistency state (1B), the logical timestamp (i.e., the version (4B) and \cid~(1B)), a counter for the received acknowledgments (1B), and the spinlock required to support the seqlock mechanism (1B). 


\tsubsection{Communication Layer} \label{secC:communication}

\beginbsec{RDMA}
There are two prevalent techniques for building an \CAP{RDMA}-based \CAP{KVS}: (i) using 
one-sided primitives such as \CAP{RDMA} reads in FaRM~\cite{Dragojevic:2014} and (ii) using remote procedure calls over unreliable datagram (\CAP{UD}) sends, similar to FaSST~\cite{F-Kalia:2016}. We choose the more general remote procedure calls over the \CAP{UD} sends approach but note that the Scale-out ccNUMA paradigm is not 
\linebreak
constrained by the choice of the communication primitive and could equally work with one-sided accesses.

\beginbsec{Flow control}
The communication between cache and \CAP{KVS} threads is
\linebreak
facilitated by a credit-based flow control mechanism~\cite{Kung:1994}. The cache threads have a number of credits for each remote \CAP{KVS} thread, and the \CAP{KVS} threads have a matching amount of buffer space for each remote cache thread.  Each time a cache thread sends a request, the credits for the receiving \CAP{KVS} thread are decremented. Similarly, the credits are incremented whenever the \CAP{KVS} 
\linebreak
responds.
Because a request always receives a response, the flow control does not require additional credit update messages; the responses to the 
\linebreak
requests are implicitly used as credit update messages.

In contrast, the communication between cache threads on consistency actions requires explicit credit updates because not all messages receive a response. For example, a cache thread that broadcasts updates to all other machines does not receive acknowledgments for those updates. Thus, ccKVS uses explicit credit update messages to inform cache threads of buffer availability across the symmetric cache nodes. 
\cref{secC:design-perf} describes optimizations to alleviate the network bandwidth overhead of credit updates.

\beginbsec{Broadcast primitive} 
To facilitate Galene's write actions, we implement a software broadcast where the sender prepares and sends a separate message to each receiver. The application sends a linked list of work requests (i.e., packets) to the \CAP{NIC} as a batch; all work requests point to the same payload but each work request points to a different destination. When a cache thread intends to send more than one broadcast, we batch these broadcasts together to the \CAP{NIC} to amortize the PCIe overheads.

\tsubsection{Performance Optimizations}
\label{secC:design-perf}
\beginbsec{Reducing connections}
One of our goals in implementing ccKVS is to 
\linebreak
maintain \CAP{RDMA} scalability by limiting the number of threads that communicate with each other. Despite using the more scalable \CAP{UD} transport, all-to-all 
\linebreak
communication at the thread level can still prove challenging to scale because of the required buffer space,
which scales linearly with connection count~\cite{F-Kalia:2016}.

Partitioning threads helps to limit the extent of all-to-all communication, as the KVS threads of different nodes do not need to communicate with each other. Additionally, we bind each cache thread to exchange messages with just two threads in each remote machine: one cache thread and one KVS thread. This optimization is enabled by the use of the \CAP{CRCW} model in both the symmetric cache and KVS, since each thread has full access to the dataset (cache or KVS, respectively).

Reducing the connections minimizes the
buffer space that needs to be registered with the NIC. As discussed in \cref{secC:design-cache-kvs},  transitioning the KVS from the \CAP{EREW} to the \CAP{CRCW} model incurs a concurrency control overhead. However, in our experiments, we measure a performance increase of up to 10\% when employing \CAP{CRCW} rather than \CAP{EREW}, which we attribute to the reduction of the connections between cache and KVS threads. 

\beginbsec{RDMA optimizations}
Using the \CAP{UD} transport allows us to perform opportunistic batching in all communications with the \CAP{NIC} to amortize the \CAP{PCI}e overheads. We post 
work requests as linked lists and notify the \CAP{NIC} about their existence.
The \CAP{NIC} can then read these requests in bulk, amortizing the \CAP{PCI}e overheads. To further alleviate \CAP{PCI}e overheads, we inline payloads inside their respective work requests, whenever the payloads are small enough (less than 189 Bytes), such that the \CAP{NIC} does not need a second round of \CAP{DMA} reads to fetch the payloads after reading the work requests.

We follow the guideline to use multiple queue pairs per thread~\cite{Kalia:2016}; for example, a cache thread uses different queue pairs for remote requests, consistency messages, and credit updates.  
Moreover, we leverage selective signaling when sending messages: the sender polls for only one completion every time it sends a fixed-size batch of messages. 

\beginbsec{Flow control optimizations}
To prevent flow control from becoming an important factor in network bandwidth consumption, we apply a batching optimization on the credit updates. We do not send a credit update for each received 
message; instead, we send a credit update after receiving a number of consistency messages to amortize the network cost 
of the credits. Additionally, the credit update messages have no payload (i.e., are header-only 
messages), reducing the required PCIe transactions and network traffic for sending and receiving them. In \cref{secC:eval}, we show that through these optimizations, the overhead of the credit update messages becomes trivial.
\tsection{Experimental Methodology}
\label{secC:method}

In this section, we first present the designs that we evaluate, then describe our evaluation infrastructure. 

\tsubsection{Evaluated Systems}
%

We evaluate Scale-out ccNUMA by comparing it with a state-of-the-art skew mitigation approach based on FaSST~\cite{Kalia:2016}. 
Although FaSST is designed for transaction processing, it has two key attributes that make it a good baseline for a system to tackle skew. Namely, it offers a \CAP{NUMA} abstraction like FaRM~\cite{Dragojevic:2014} and RackOut~\cite{RNovakovic:2016}, and it leverages several design techniques to achieve high performance using \CAP{RDMA}~\cite{F-Kalia:2016}.
We implement FaSST over our datastore with efficient single-object read and write operations by stripping all of the transaction processing overheads.
We apply all of the optimizations discussed in \cref{secC:design-perf} to maximize the performance of this baseline and negate any implementation-specific advantages of ccKVS. The performance of our baseline system is on par with the reported FaSST results (subject to different evaluation setups).
%
%
%

We evaluate three flavors of the FaSST-based baseline design:

\begin{itemize}
\item \textbf{Base-EREW} has its \CAP{KVS} partitioned at a core granularity similarly to \CAP{MICA}. We expect this system to suffer under a skewed distribution, as the performance will be limited by the core responsible for the hottest shard.

\item \textbf{Base} has its \CAP{KVS} partitioned at a server granularity (\CAP{CRCW}). Compared with Base-\CAP{EREW}, we expect this system to perform better under skew while still being bottlenecked by the server with the hottest shard.  
\item \textbf{Uniform} represents the performance of Base under a uniform distribution. 
This establishes an upper bound on the performance of baseline designs.
\end{itemize}

We build ccKVS by adding symmetric caches on top of Base. More specifically, we add a cache to each node and implement a system as described in \cref{secC:design} that supports the Galene protocol specified in \cref{secC:protocols}. 
We configure the symmetric cache size to 0.1\% of the total dataset (250K objects of up to 1KB each, with an overall memory footprint of up to 1GB). In accordance with \cref{figC:cache-opportunity}, the expected cache-hit ratio is 46\%, 65\%, and 69\% for skew exponents of $\alpha$ equal to $0.9$, $0.99$, and $1.01$, respectively.  

\tsubsection{Testbed} \label{{secC:method-setup}}
\beginbsec{Infrastructure} 
We conduct our experiments on an isolated cluster of 9 servers interconnected via a 12-port Infiniband switch (Mellanox MSX6012F-BS). Each machine runs Ubuntu server 14.04 and is equipped with two 10-core \CAP{CPU}s (Intel Xeon E5-2630) with 64 GB of system memory and a single-port 56Gb NIC (Mellanox \CAP{CX4} \CAP{FDR} \CAP{IB} \CAP{PCI}e3 x16) connected on socket 0. Each \CAP{CPU} has 25 MB of L3 cache and two hyper-threads per core. We disable turbo-boost, pin threads to cores, and use huge pages (2MB) for both the \CAP{KVS} and the cache.


\beginbsec{Workloads} 
Our evaluation is performed on workloads following a Zipfian access distribution. We use the skew exponent $\alpha=0.99$ as the default value (as in \CAP{YCSB}~\cite{Cooper:2010}) and also study $\alpha=\{0.90,1.01\}$. For comparison purposes, we also assess a uniform access distribution. We evaluate both a read-only workload and workloads with modest write ratios,
which are representative of large-scale data serving deployments with high skew (e.g., Facebook reports a write ratio of 0.2\%~\cite{Bronson:2013}).
The \CAP{KVS} consists of 250 million distinct key-value pairs, making each node responsible for nearly 28 million keys. Unless stated otherwise, we use keys and values of 8 and 40 bytes, respectively, thus allowing a direct comparison with FaSST~\cite{F-Kalia:2016}. Finally, we apply request coalescing optimization in \cref{secC:bottlenecks}, \cref{secC:coalesc-perf}, and \cref{secC:perf-vs-latency}.
\tsection{Evaluation}

\label{secC:eval}

\tsubsection{Read-Only Performance}
\label{secC:read-only}

\begin{figure}[t]
  \centering
  \includegraphics[width=0.7\textwidth]{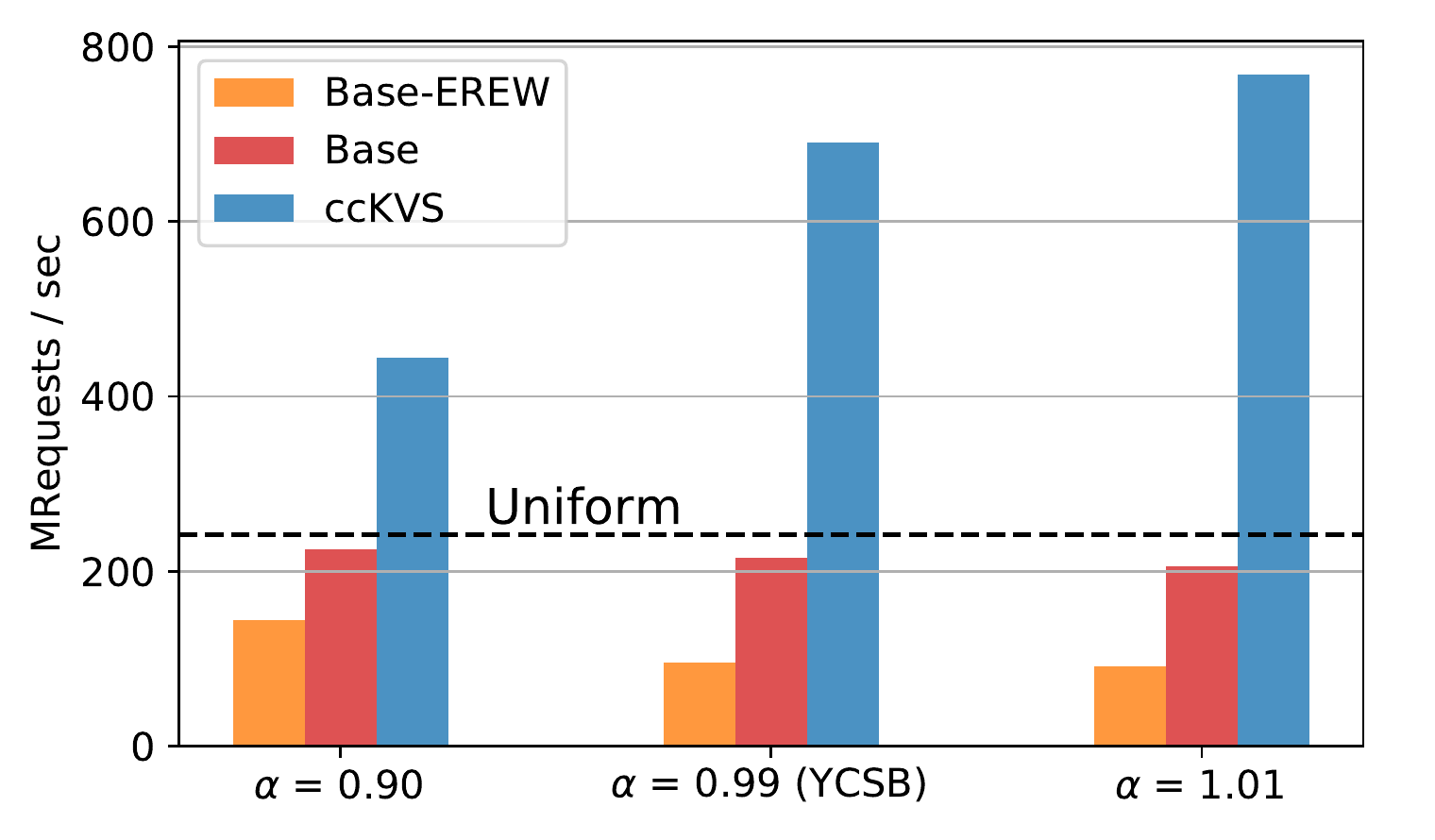}
  \mcaption{Throughput of a read-only workload while varying skew.}
  {Throughput of a read-only workload while varying skew. [9 nodes]}
  \label{figC:varying-skew}
\end{figure}

We first evaluate the performance of all the designs for a read-only workload. \cref{figC:varying-skew} shows the performance of Base-\CAP{EREW}, Base, and ccKVS under three different skewed distributions ($\alpha = \{0.9, 0.99, 1.01\}$). As the results are similar for all three distributions, we focus our discussion on $\alpha=0.99$.

As expected, Base-EREW has poor performance and achieves only 95 million requests per second (MReq/s), as the whole system is bottlenecked by the throughput of the core responsible for the hottest shard. On the other hand, Base achieves 215 MReq/s, significantly mitigating the skew, as the bottleneck shifts from the hottest core to the hottest server. 
In fact, the performance of Base is within 10\% of Uniform, which achieves 240 MReq/s.
It is worth noting that this performance gap is strongly correlated with the skew exponent ($\alpha$) and the number of servers in the deployment. 

ccKVS achieves 690 MReq/s, which is 3.2$\times$ higher than the throughput of Base and 2.85$\times$ higher than Uniform. 
The significantly higher throughput of ccKVS compared to Uniform highlights the fact that the baseline systems are network limited.
ccKVS is able to achieve considerably higher throughput by avoiding the need to access remote nodes for cached objects, thus reducing network bandwidth pressure. 
ccKVS also benefits from the fact that symmetric caches allow all the nodes in the \CAP{KVS} to serve requests for hot objects, thus distributing the load evenly among them. 

\begin{figure}[t]
  \centering
  \includegraphics[width=0.7\textwidth]{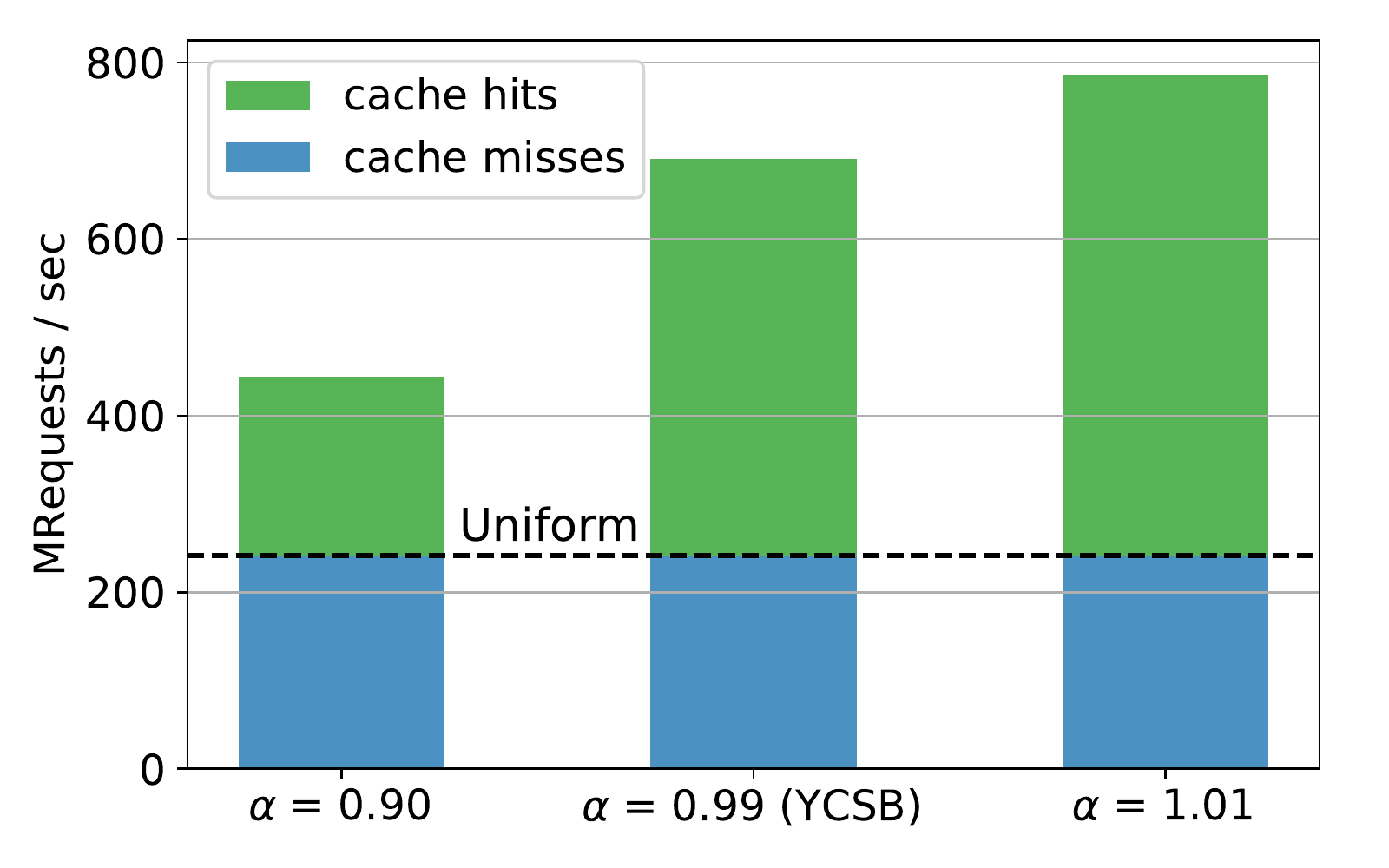} 
  \mcaption{Request breakdown for a read-only workload.}
  {Breakdown of completed requests in ccKVS for a read-only workload with varying skew. [9 nodes]}
  \label{figC:break-down}
\end{figure}

To better understand the reasons behind the significant performance improvement provided by ccKVS, we analyze its throughput. \cref{figC:break-down} shows the breakdown of ccKVS throughput in terms of the number of cache hits and misses for a read-only workload with varying skew.
In general, as the skew increases, the cache-hit rate also increases. 
Cache hits require compute 
\linebreak
resources, whereas cache misses mostly require network resources due to remote \CAP{KVS} access. 
We observe that the cache-miss throughput of ccKVS is equal to the entire throughput of Uniform and 
stays constant, even though the cache-miss rate is higher with lower skew exponents. This leads to the conclusion that both ccKVS and Uniform are network bound. Meanwhile, the cache-hit throughput increases as the cache-hit rate increases, indicating that the \CAP{CPU} is not the bottleneck. We confirm these hypotheses in \cref{secC:bottlenecks}. 
\tsubsection{Performance under Writes}
\label{secC:write-perf}

\begin{figure}[t]
  \centering
  \includegraphics[width=0.7\textwidth]{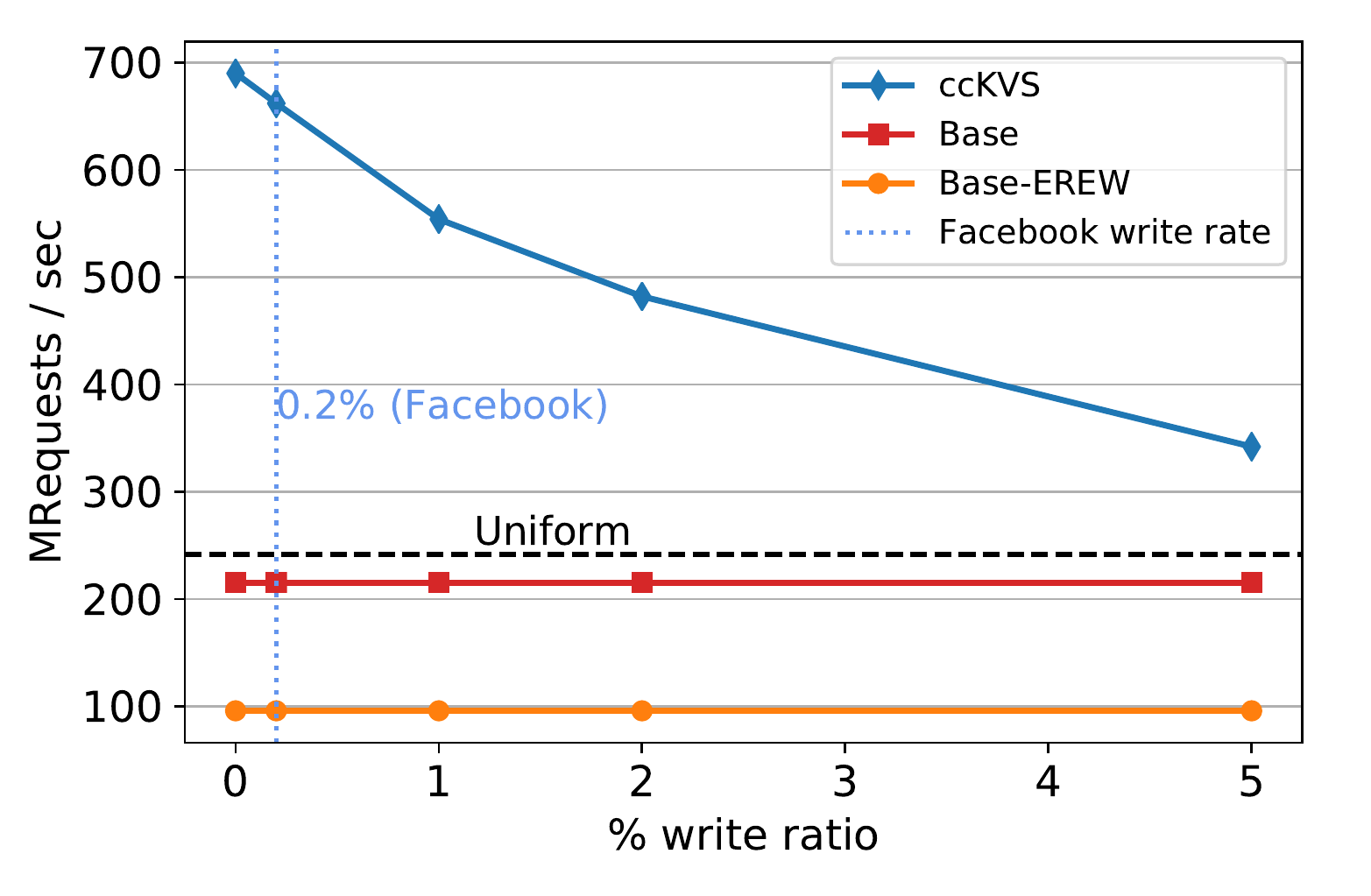} 
  \mcaption{Sensitivity to write ratio.}{Sensitivity to write ratio. [9 nodes, $\alpha$ = 0.99]}
  \label{figC:writes}
\end{figure}

We now analyze the performance of ccKVS in the presence of writes.
\cref{figC:writes} shows the throughput of the evaluated systems for varying write ratios with $\alpha = 0.99$. 
None of the baselines are sensitive to the write ratio, as they are all bottlenecked by the network. Note that in the baseline design, the network traffic does not change with varying write ratios, as remote read and 
\linebreak
remote write requests both consume the same amount of network bandwidth.
In 
\linebreak
contrast, the throughput for ccKVS decreases with increasing write ratios. This decrease is caused by the additional consistency actions required for every cache write, such as broadcasting updates over the network. These actions consume network resources and thus diminish the throughput of the system, which is network bound even in the read-only scenario.

\begin{figure}[t]
  \centering
  \includegraphics[width=0.7\textwidth]{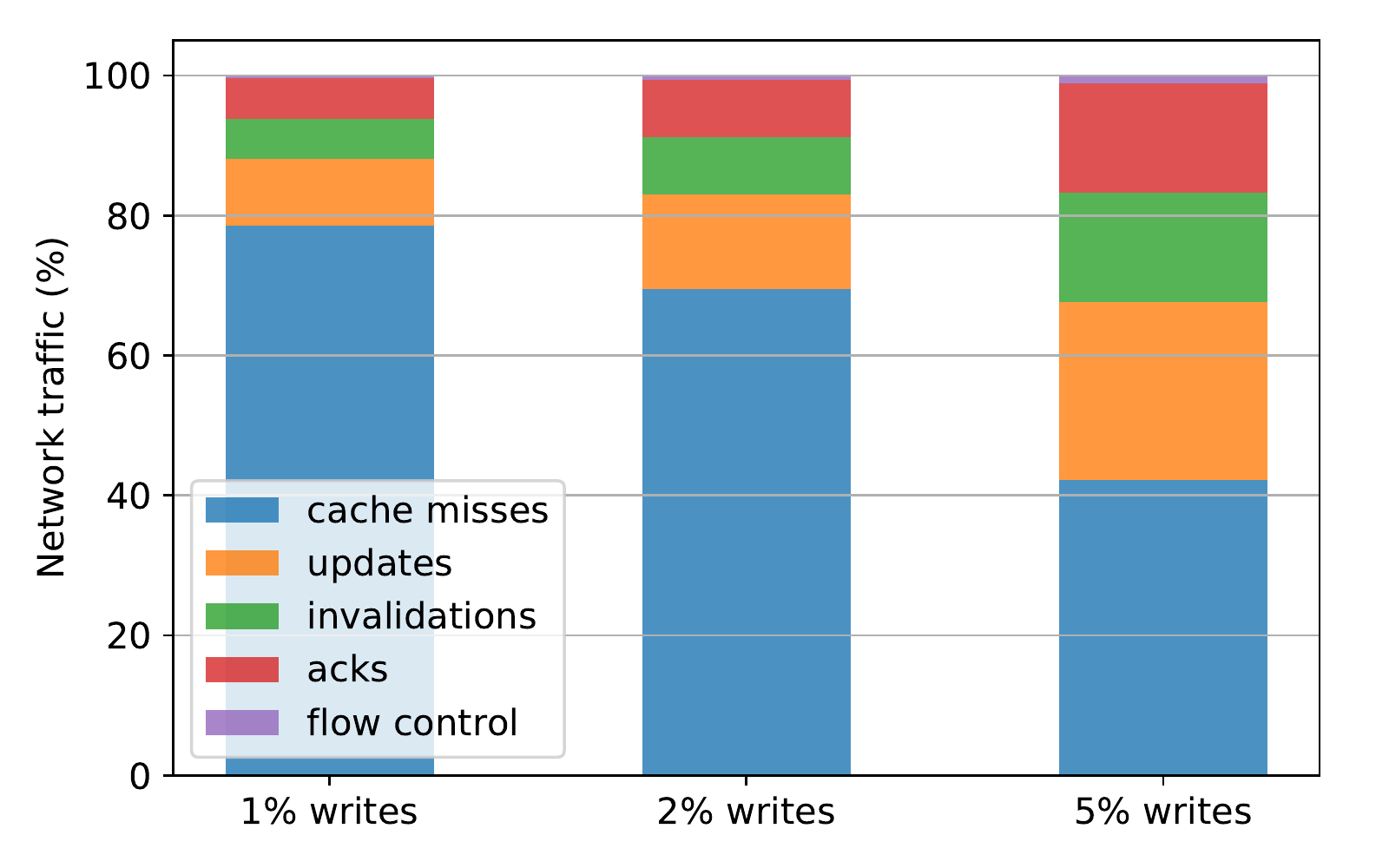} 
  \mcaption{Network traffic breakdown.}
  {Network traffic breakdown for ccKVS. [9 nodes, $\alpha$ = 0.99]}
  \label{figC:network-break-down}
\end{figure}

However, for realistic write ratios in skewed workloads, such as 0.2\% for Facebook's workload~\cite{Bronson:2013}, ccKVS provides throughput within 3\% of a read-only workload. In fact, ccKVS outperforms Base even for write ratios as high as 5\% while providing the strongest consistency guarantee (linearizability). This is a particularly important result, as it shows that, \emph{contrary to conventional
\linebreak
wisdom, it is possible to achieve high throughput in the presence of aggressive replication under strong consistency guarantees.}

To further analyze the throughput of ccKVS with increasing write ratios, we show the breakdown of the network traffic 
for 1\%, 2\%, and 5\% write ratios in \cref{figC:network-break-down}.  As the write ratio increases, consistency actions (i.e., updates, invalidations and acks) claim an increasingly large percentage of the available network bandwidth. As a result, less bandwidth is available for remote \CAP{KVS} accesses triggered by cache misses. Since the system is network bound, a reduction in available bandwidth for remote \CAP{KVS} accesses proportionately lowers total system throughput. 
Finally, we note that flow control consumes a negligible amount of bandwidth thanks to batching of credits (\cref{secC:design-perf}).

\begin{figure}[t]
  \centering
  \includegraphics[width=0.7\textwidth]{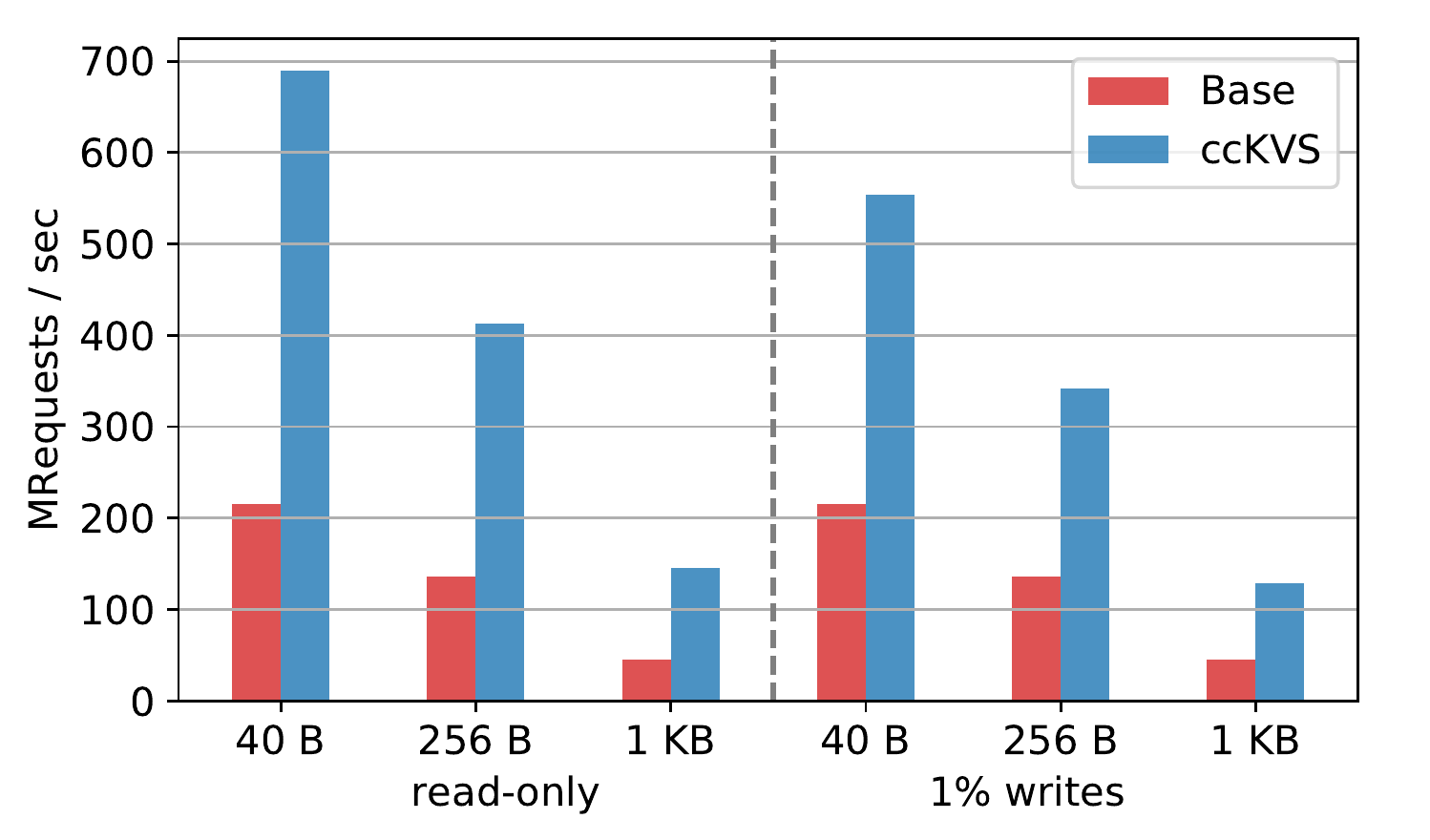} 
  \mcaption{Varying object size study.}
  {Read-only and 1\% writes, varying object size. [9 nodes, $\alpha$ = 0.99]}
  \label{figC:varying-value-size}
\end{figure}

\tsubsection{Sensitivity to Object Size}
\label{secC:value-perf}
We next study the performance of ccKVS with varying object sizes.
\cref{figC:varying-value-size} shows the throughput of ccKVS in comparison to Base for various object sizes over read-only and 1\% writes with $\alpha = 0.99$. 

In the read-only scenario, the relative performance of Base and ccKVS follow the same trend, irrespective of object size. ccKVS still outperforms Base by over 3$\times$ for larger objects. 
The trend is similar with writes.
As expected, the performance of ccKVS is lower than read-only due to the bandwidth spent on consistency messages (i.e., invalidations, acknowledgments, and validations). 
Nevertheless, ccKVS still outperforms Base by more than 2.2$\times$ across all object sizes.

\tsubsection{System Bottlenecks} \label{secC:bottlenecks}
In order to identify the system bottlenecks,
we analyze the hardware counters for the \CAP{NIC}, \CAP{PCI}e, and memory.\footnote{We used Mellanox's NEO-Host suite \cite{NeoHost} for \CAP{NIC} profiling and Intel's pcm \cite{pcm} for the \CAP{PCI}e and memory measurements.}
We also profile ccKVS (using the Zoom profiler~\cite{Zoom}) and use busy-wait counters within the ccKVS. After inspecting all measurements, we observe that  bottleneck shifts depending on the network packet size. We identify two distinct cases: large objects that result in large packet sizes and small objects that result in small packet sizes. 

For large objects, network utilization in ccKVS closely approaches the available network bandwidth, while the rest of the resources remain underutilized. Thus, we can safely infer that the bottleneck, in this case, is the available network bandwidth. 
In contrast, with small objects, \CAP{CPU}, \CAP{PCI}e, memory bandwidth, and network bandwidth are all underutilized. To our surprise, the bottleneck for small object sizes appears to be the packet processing rate of the switch.

To validate our claim, we conduct the following experiment: we measure the maximum packet rate using Mellanox's micro-benchmark (ib\_send\_bw) when connecting two machines directly (i.e., without the switch) and when connecting them through the switch. We observe that the maximum rate of sent/received packets per second is significantly higher (up to 25\%) when the servers are connected directly.\footnote{Our findings were confirmed by the manufacturer of the switch.} These results hold for ccKVS, as well.

For simplicity, throughout the evaluation section, we assume that the
bottleneck is in the network in both cases, as the limited switch processing rate for small packets can be viewed as an artificial network bandwidth limitation. 
We measure the maximum achievable bandwidth to be around 21.5 Gbps for small packets, while the \CAP{NIC} nominally supports 54 Gbps. 

\vspace{-10pt}

\tsubsection{Request Coalescing}
\label{secC:coalesc-perf}
In order to demonstrate and alleviate the bottlenecks imposed by transmitting small packets, we perform \textit{request coalescing}, whereby multiple requests 
destined for the same node are opportunistically coalesced into a single network packet.
We only apply request coalescing to cache misses (requests and the associated responses), since these dominate the network traffic in ccKVS at modest write ratios.

\begin{figure*}[!ht]
  \centering
  \begin{subfigure}[b]{0.655\textwidth}
    \includegraphics[width=0.99\textwidth]{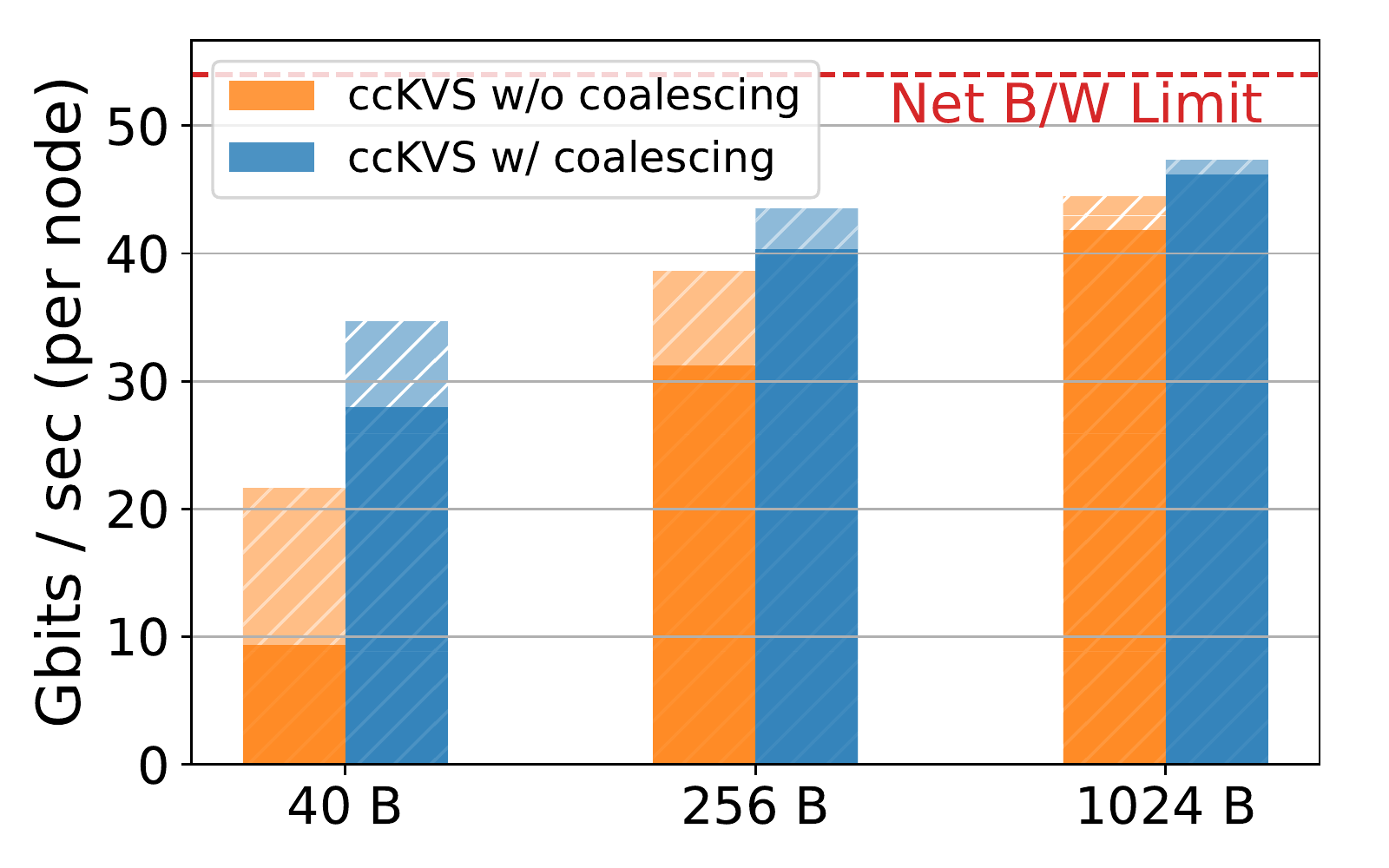}
    \caption{Network utilization of read-only workload with different object sizes. Solid bars: data payload; stripes: packet headers.}
    \label{figC:coalescing-net}
  \end{subfigure}
  \begin{subfigure}[b]{0.655\textwidth}
    \includegraphics[width=0.99\textwidth]{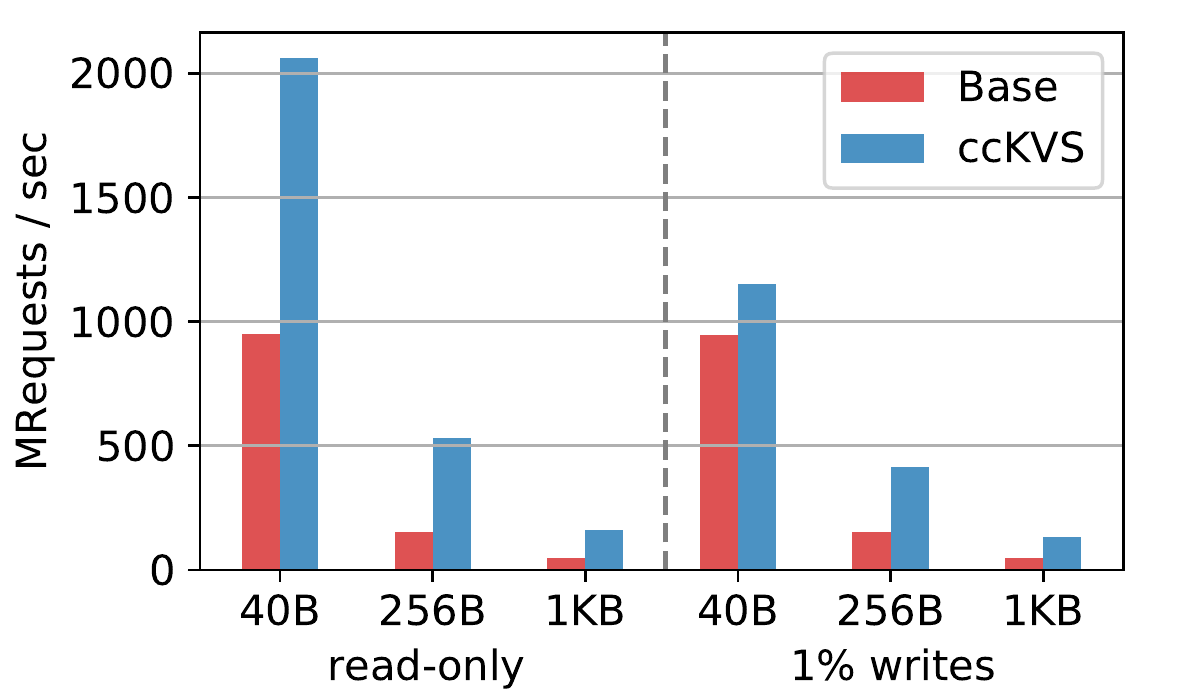}
    \caption{Performance impact of coalescing for read-only and 1\% writes while varying object size. \\}
    \label{figC:coalescing-perf}
  \end{subfigure}
  \begin{subfigure}[b]{0.655\textwidth}
    \includegraphics[width=0.99\textwidth]{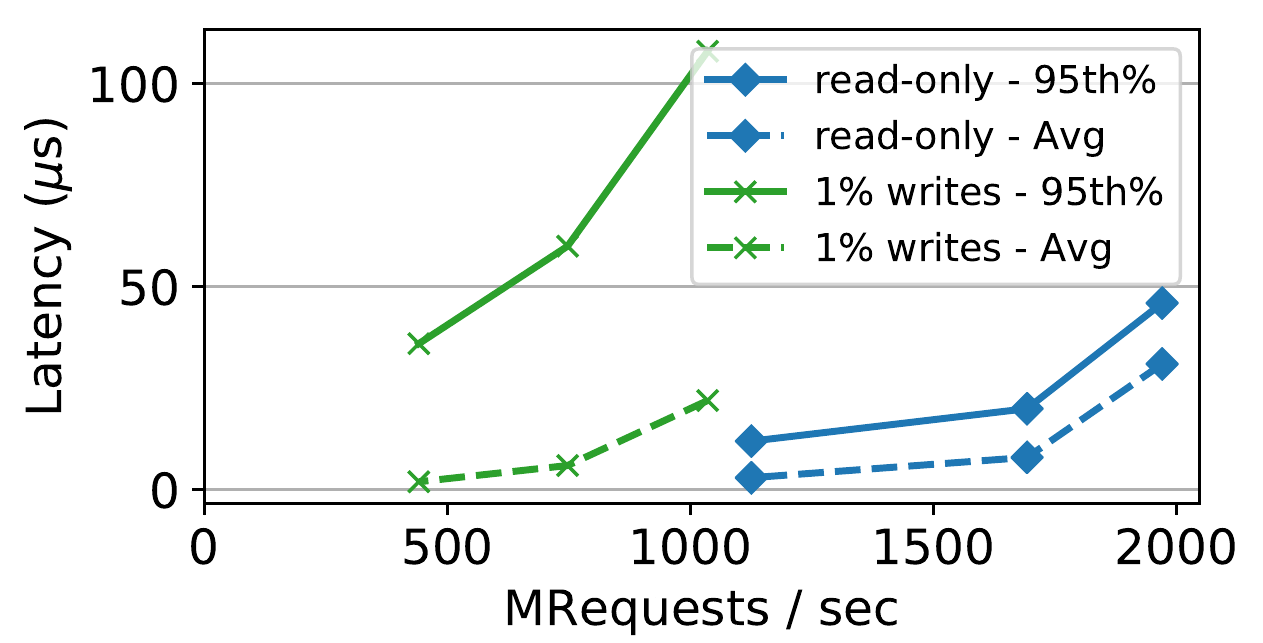}
    \caption{ccKVS average and 95th percentile latencies for 40B objects at various load levels for read-only and 1\% writes with coalescing.}
    \label{figC:latency-perf}
  \end{subfigure}
  \mcaption{Analysis of coalescing and latency.}
  {Analysis of coalescing and latency. [9 nodes, $\alpha$ = 0.99]}
\end{figure*}

\cref{figC:coalescing-net} shows the network utilization of ccKVS for a read-only workload with and without request coalescing. This figure breaks down the network utilization into packet header and payload (i.e., data traffic), illustrated by striped and solid bars, respectively.
Coalescing multiple requests results in larger network packets, shifting the bottleneck from the switch's packet processing rate to network bandwidth. As a result, the optimized ccKVS that supports coalescing  increases throughput by almost 3$\times$ for the 40B values.

To ensure a fair comparison, we also add support for coalescing to Base and present the optimized performance of ccKVS and Base in \cref{figC:coalescing-perf} for read-only and 1\% writes while varying the object size. As expected, when comparing the results presented in \cref{figC:varying-value-size} and \cref{figC:coalescing-perf}, both ccKVS and Base enjoy increased throughput for small object sizes when coalescing is applied. However, coalescing is less beneficial for larger objects, as the system is already bottlenecked by the network bandwidth. 

Specifically, when examining the effects of coalescing on small (40B) objects, we observe that the performance of Base is almost 950 MReq/s for both the read-only and 1\% writes workloads, which yields an improvement of over 4$\times$ relative to the no-coalescing Base. In turn, ccKVS achieves a 3$\times$ improvement in performance with coalescing enabled, delivering over 2 billion requests per second, which is more than twice the performance of Base with coalescing. 

The benefits of coalescing diminish in ccKVS on the 1\% writes workload
\linebreak
because a fraction of network traffic carries consistency messages, which we do not coalesce. Nonetheless, even with writes, request coalescing improves the performance of ccKVS by 2$\times$ over no-coalescing.

\vspace{-10pt}
\tsubsection{Latency analysis}
\label{secC:perf-vs-latency}
\vspace{-10pt}

\cref{figC:latency-perf} illustrates the average and the 95th-percentile latency of ccKVS for a read-only workload and a workload with 1\% writes while varying the load and with request coalescing.
We observe that, even at high loads, the tail latency is about an order of magnitude lower than the target of 1ms for a typical \CAP{KVS} service~\cite{Leverich:2014}. 
In fact, at maximum load, the 95th-percentile read-only latency is quite close to the average latency. 
However, for 1\% writes when ccKVS is at high load, its 95th-percentile latency is noticeably higher than its average latency. This is expected, since writes in ccKVS are blocking (i.e., send invalidations and wait for acknowledgments in the critical path).
\newtext{
Note we do not optimize ccKVS for latency (or compare it with the baseline), as this chapter focuses on the throughput benefits of replication and invalidating protocols. We thoroughly study latency and compare with existing works in the next chapter.}

\FloatBarrier

\tsubsection{Analytical Model} \label{secC:analytical}

Since our 9-machine deployment prevents us from directly evaluating the 
\linebreak
scalability of ccKVS, 
we build an analytical model that models the throughput 
\linebreak
of ccKVS.
This model leverages the fact that ccKVS is bottlenecked by the 
\linebreak
network bandwidth (\cref{secC:bottlenecks}). Therefore, the throughput of ccKVS is inversely proportional to the overall network traffic.

There are two sources of network traffic. The first is due to requests that miss in the cache targeting keys mapped to a remote node.  
A request is a cache miss with probability $(1-h)$, where $h$ denotes the cache hit ratio. The cache miss targets a remote node with probability $1 - \frac{1}{N}$, where $N$ denotes the number of servers. 
A remote request generates two messages: one request and one reply. The total size of these two messages (in bytes) is denoted as $B_{RR}$.
On average, the cache miss-related traffic ($TR_{M}$) generated per request is given by:
   \begin{equation}
    \label{Ccache-miss-traffic}
    TR_{M} = (1-h) * (1 - \frac{1}{N}) * B_{RR}
    \end{equation}
    
The second source of traffic is the messages for consistency actions, which are generated by hot writes (i.e., writes that hit in the cache). 
Consistency actions in ccKVS include invalidations, acknowledgments, and updates. These three messages amount to $B_{G}$ bytes, with each hot write generating $(N-1)$ of each of these types of messages.
The probability of a hot write is given by $h * w$, where $w$ denotes the write ratio. 
Therefore, the overall consistency-related traffic ($TR_{C}$) generated per request in ccKVS is given by:
\begin{equation}
    \label{CLin-traffic}
    TR_{C} = h * w * (N - 1) * B_{G}
    \end{equation}

From \cref{Ccache-miss-traffic} and \cref{CLin-traffic}, each request in ccKVS generates $TR_{M} + TR_{C}$ bytes worth of traffic.
Because ccKVS is network bound, the throughput (i.e., number of requests per second) of a server can be computed as the available network bandwidth ($BW$) divided by the bytes required per request.  To compute the total throughput of ccKVS ($T_{ccKVS}$), we multiply by the number of servers, as shown in \cref{eqC:Lin_throughput}:
\begin{equation}
T_{ccKVS} =  N * \frac{BW}{TR_{M} + TR_{C}} 
\label{eqC:Lin_throughput}
\end{equation}

In the Uniform design, network traffic is generated for requests that map to a remote node. Requests map to a remote node with the probability $1 - \frac{1}{N}$ and such requests generate a request and a reply message (similar to cache misses in ccKVS) that amount to $B_{RR}$ bytes transferred over the network. Therefore, the total traffic $(TR_{U})$ generated by a request in Uniform is given by:
\begin{equation}
    \label{CUni-traffic}
    TR_{U} = (1 - \frac{1}{N}) * B_{RR}
    \end{equation}
And the total throughput ($T_U$) of Uniform is as shown in \cref{eqC:Uni_throughput}.
\begin{equation}
T_{U} =  N *\frac{BW}{TR_{U}} \label{eqC:Uni_throughput}
\end{equation}

\begin{figure}[t]
  \centering
  \includegraphics[width=0.7\textwidth]{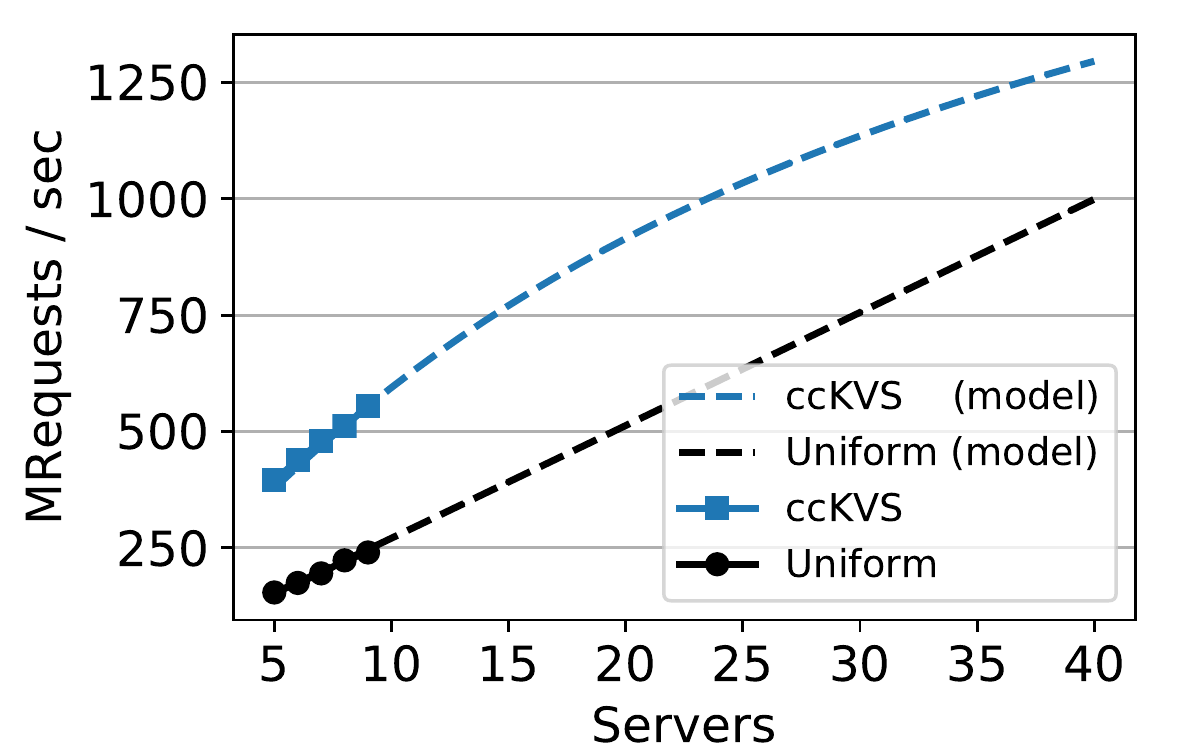}
  \mcaption
  {Scalability study.}
  {ccKVS scalability study using the model (dashed) and real-system validation (solid). [1\% writes, $\alpha= 0.99$]}
  \label{figC:analytical-server}
\end{figure}

\subsubsection*{\textbf{Scalability study}}
In the presence of writes, we anticipate the per-server throughput of ccKVS to degrade as the number of servers increases due to the proportional increase in consistency traffic. 
We employ the proposed analytical model to conduct a scalability study and understand the extent of this degradation.

To validate the model with our existing setup, we feed the model with the same parameters as in our implementation with request coalescing disabled. We set the cache hit ratio ($h$) to $65\%$ and the message sizes with the exact numbers used in our evaluation for small objects:
$B_{RR} = 113 $ bytes and $B_{G} = 183$ bytes (including network headers). Finally, we set the available network bandwidth ($BW$) at $21.5$ Gbps, which is the network bandwidth observed for the configuration with small objects.

\cref{figC:analytical-server} shows the estimated throughput of Uniform and ccKVS when 
\linebreak
scaling the number of servers of the deployment from 5 to 40 while fixing the write ratio at 1\%. 
As expected, the scaling of Uniform is almost perfectly linear. However, ccKVS scales sublinearly with the number of servers; this is  because, as the number of servers increases, the consistency traffic increases too. 

We also plot the measured throughput of our system for up to nine machines (i.e., the size of our deployment). As we can see, the analytically computed throughput is similar to the measured throughput for both ccKVS and Uniform. 
With nine servers, ccKVS is estimated to achieve 554 MReq/s, which is within 1\% of the measured throughput in our implementation (558 MReq/s).

In general, we find that the analytical model predicts the performance of ccKVS with sufficient accuracy. Using the validated model, we find that the performance of ccKVS is significantly better than the upper bound for the baseline (i.e., Uniform) for moderately sized deployments with 1\% write ratio.

\begin{figure}[t]
  \centering
  \includegraphics[width=0.7\textwidth]{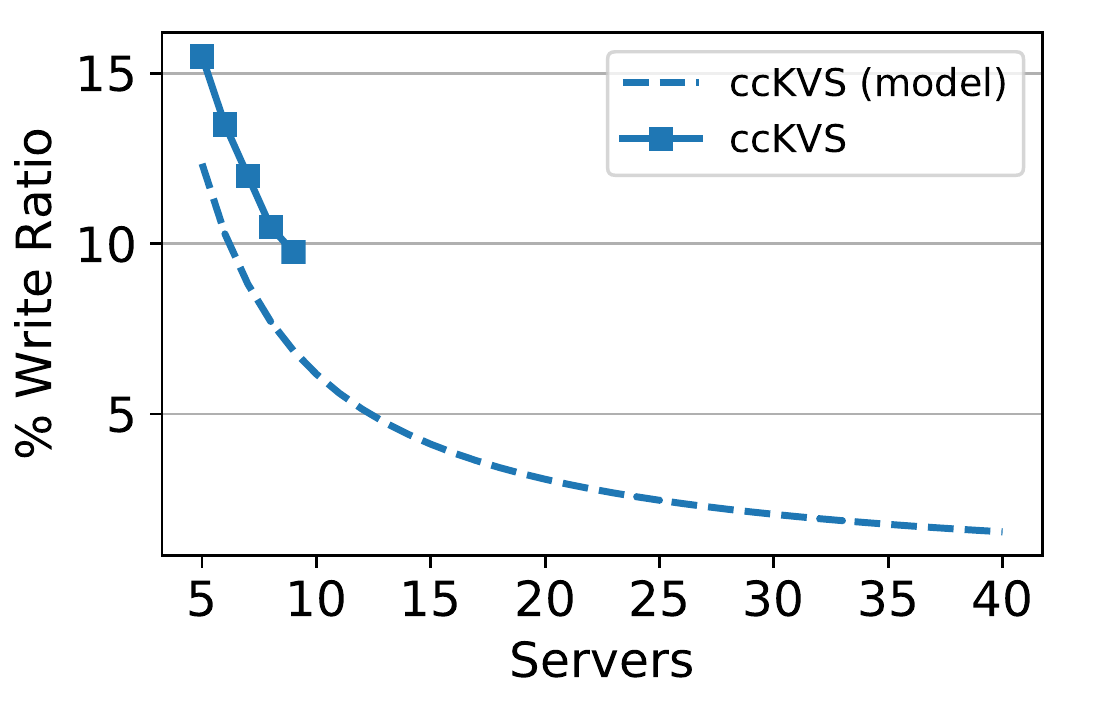}
  \mcaption
  {Break-even write ratio.}
  {Break-even write ratio model (dashed) and real-system validation (solid) for up to 9 nodes. [$\alpha$ = 0.99]}
  \label{figC:break-even}
\end{figure}

\subsubsection*{\textbf{When does symmetric caching break even?}}  

Next, we use our analytical model to answer the following question: for a 
\linebreak
deployment of X servers, what is the write ratio at which ccKVS yields the same throughput as Uniform?
We call this write ratio the \textit{break-even write ratio}. 
To calculate the break-even write ratio for ccKVS, we equate the throughput of Uniform, $T_U$ (\cref{eqC:Uni_throughput}), with the throughput of ccKVS, $T_{ccKVS}$ (\cref{eqC:Lin_throughput}), and solve for the write ratio.

\cref{figC:break-even} illustrates the break-even write ratio for ccKVS deployments of up to 40 servers. For example, a ccKVS deployment with 15 servers yields the same performance as Uniform at a write ratio of 4\%. Therefore, a 15-server deployment with a write ratio below 4\% can benefit from employing ccKVS. 

To validate the model, \cref{figC:break-even} also depicts the measured break-even write ratios for actual deployments of up to nine nodes. We observe that the trend is similar for both the model and actual measurements; however, the real system can sustain slightly higher break-even write ratios than what the model predicts. 
The reason for this slight discrepancy is that, as noted in \cref{secC:bottlenecks}, the actual bottleneck for small packets is in the switch packet processing;
because the update messages in Galene are large (i.e., contain both key and value), ccKVS achieves higher network bandwidth than predicted for high write ratios. 

As expected, the break-even write ratio decreases when the number of servers increases. This occurs because the consistency traffic increases linearly with the number of servers, since a write to a hot object must be propagated to all servers. With 40 servers, the break-even write ratio is 1.7\% for ccKVS. This indicates that in a moderately sized deployment with low write ratios, ccKVS should outperform the baseline while maintaining strong consistency guarantees. However, at higher write ratios or in larger deployments, the performance benefit of ccKVS may vanish.


\beginbsec{Note on scalability}
We have established that the benefits of symmetric caching decrease with increasing size of the deployment. 
However, this constraint does not strictly prohibit the application of symmetric caching in large deployments.
To scale beyond a rack-scale or small-sized cluster deployment, we believe our ideas can be applied by simply partitioning bigger deployments into smaller Scale-out ccNUMA clusters, each of which can independently 
\linebreak
apply symmetric caching. For example, a \CAP{KVS} spanning 100 nodes can be split into five 20-machine groups (similarly to~\cite{RNovakovic:2016}) where each group employs symmetric caching for its portion of the \CAP{KVS}. 

\tsection{Related Work} 
\label{secC:Related}

\beginbsec{Data replication} Service providers 
often use data replication to improve system performance, particularly to provide load balancing. While conceptually straightforward, replicating hot data across some number of servers~\cite{Hong:2013,Huang:2014}, 
it entails a number of practical shortcomings, as detailed in~\cite{RNovakovic:2016}. These include determining the appropriate level of replication granularity (object, partial shard, or entire shard), tracking replicas, maintaining replica consistency, and informing clients of the replica's locations. The latter can be particularly onerous if the number of clients is much greater than the number of servers, which is often the case. In practice, these problems tend to have ad-hoc solutions requiring complex engineering and with significant system-level overheads, hence spurring the recent work on alternative approaches using fast remote access and caching (as discussed in \cref{secC:existing}). 

Our work takes the best features of replication, caching, and fast remote 
access. Compared to traditional replication, our solution allows for fine-granularity 
replication of individual keys and does not require client-side knowledge of replicas while affording strong consistency across all replicas.

\beginbsec{Distributed shared memory (DSM)} In principle, a distributed \CAP{KVS} is not all that different from a \CAP{DSM}~\cite{Stets:1997, Keleher:1994, munin, Li:1989}.
The underlying problem boils down to enforcing a consistency model in the presence of replication.  However, there is one important difference: the workloads. The goal of \CAP{DSM} is to support scalable parallel programs, whereas the goal of \CAP{KVS} is to support data-serving workloads. The former is characterized by CPU-intensive programs that ideally do not spend all their time waiting for memory, while the latter does little more than perform data accesses to main memory. 
Whereas locality in \CAP{DSM} arises from program working sets, locality in a \CAP{KVS} can be explained by a skewed access distribution. 

These workload and sharing pattern differences translate into significant 
\linebreak
divergence in the design of caches and consistency protocols. In particular, the popularity skew naturally dictates that only the popular objects should be cached. Similarly, it dictates that there is no need to have different objects in different caches, thus avoiding the need to track sharers (e.g., through a directory) or migrate pages. 

\beginbsec{Cache coherence} In shared-memory multiprocessors, the local caches of each processor are typically kept coherent using hardware-based coherence protocols~\cite{Vijay:20}. Our approach is inspired by the effectiveness of coherent caches in such architectures. 
However, as discussed in \cref{secC:case}, the 
protocol we employ (push-based and fully distributed) is quite different from those typically used in multiprocessors (pull-based and serializing at a directory). 

\tsection{Summary}
\label{secC:conclusion}
Popularity skew is a well-known bottleneck in existing \CAP{KVS} deployments. 
\linebreak
Existing skew-mitigation techniques are limited in their efficacy when applied to a distributed in-memory \CAP{KVS}. 
This chapter 
embraces skew as an opportunity through aggressive caching of popular objects across all nodes of the \CAP{KVS}. While aggressive replication is generally thought to be a challenge in distributed datastores due to the perceived cost of keeping replicas consistent, our work shows otherwise. Using a low-overhead {\em symmetric cache} architecture powered by a fully distributed strongly consistent replication protocol, we demonstrate that our prototype ccKVS outperforms a state-of-the-art \CAP{KVS} on workloads with a moderate write ratio.

\markedchapterTOC
{Hermes}
{Hermes:\nextlinepdf Fast Fault-Tolerant Replication}
{Hermes: Fast Fault-Tolerant Replication}
\label{chap:hermes}

\equote{-40}{0}{The greatest glory in living lies not in never falling,\\ but in rising every time we fall.}{Nelson Mandela}

In the previous chapter, replication was used to improve performance, but it did not ensure fault tolerance. 
In this chapter, we introduce a fault-tolerant replication protocol with very high throughput and significantly lower latency than the state of the art by extending the cache-coherence-inspired invalidating protocols in a setting with failures. 

\tsection{Overview}
\label{secH:introduction}
Modern reliable datastores are expected to keep strongly consistent replicas for data availability despite failures while also delivering high performance.
%
When it comes to performance, recent works on 
reliably replicated datastores focus on 
throughput and tend to ignore latency~\cite{Terrace:2009}. Meanwhile, latency is emerging as a critical design goal in the age of interactive services and machine actors~\cite{Barroso:2017}. For instance, a recent work~\cite{Anwar:2018} notes that a deep learning system running on top of a reliable datastore is greatly affected by the latency of the datastore.

Today's replication protocols are not designed to handle the latency challenge of in-memory reliable datastores. Chain replication (\CAP{CR})~\cite{VanRenesse:2004}, a state-of-the-art high-performance reliable replication protocol, is a striking example of trading latency for throughput. Our detailed study of \CAP{CRAQ}~\cite{Terrace:2009}, the state-of-the-art \CAP{CR} variant, reveals that while \CAP{CRAQ} can offer very high throughput, it is 
ill-suited for latency-sensitive workloads. \CAP{CRAQ} organizes the replicas in a chain. Although reads can be served locally by each of the replicas, writes expose the entire length of the chain. Moreover, when a read hits a key for which a write is in progress, the read incurs an additional latency as it waits for the write to be resolved. With high-latency writes and mixed-latency reads, \CAP{CRAQ} fails to provide predictably low latency.


\begin{table}[t]
\centering

\begin{adjustbox}{max width=0.8 \textwidth}

\begin{tikzpicture}
\renewcommand{\arraystretch}{1.5}
\node (table) [inner sep=-0pt] {
\begin{tabular}{
>{\columncolor[HTML]{C0C0C0}}c l}
\cellcolor[HTML]{C0C0C0}                                 & local         \\ \cline{2-2} 
\multirow{-2}{*}{\cellcolor[HTML]{C0C0C0}\textbf{reads}} & load-balanced
\end{tabular}
};
\draw [rounded corners=.3em] (table.north west) rectangle (table.south east);
\end{tikzpicture}

\begin{tikzpicture}
\node (table) [inner sep=0pt] {
\begin{tabular}{
>{\columncolor[HTML]{C0C0C0}}l l}
\cellcolor[HTML]{C0C0C0}                                  & decentralized                         \\ \cline{2-2} 
\cellcolor[HTML]{C0C0C0}                                  & inter-key concurrent                  \\ \cline{2-2} 
\multirow{-3}{*}{\cellcolor[HTML]{C0C0C0}\textbf{writes}} & fast (\eg few \CAP{RTT}s)
\end{tabular}
};
\draw [rounded corners=.3em] (table.north west) rectangle (table.south east);
\end{tikzpicture}
\end{adjustbox}

\mcaption
{Replication protocol features for high performance.}
{Reliable replication protocol features for high performance.}
\label{tabH:hp-features}
\end{table}

This chapter addresses the challenge of designing a reliable replication protocol that provides both high throughput and low latency within a datacenter. To that end, we identify essential features necessary for high performance, which are summarized in \cref{tabH:hp-features}.
For reads, this means the ability to
execute a read locally 
on any of the replicas. For writes, high performance mandates the ability to execute writes 
in a decentralized manner (i.e., any replica can initiate and drive a write to completion without serializing it through another node), 
concurrently execute writes to different keys, and  
complete writes fast
(e.g., by minimizing round-trips).

Based on these insights, we introduce \textit{Hermes}, a strongly consistent fault-tolerant replication protocol for in-memory datastores that provides high throughput and low latency. At a high level, Hermes is a broadcast-based protocol for single-object reads, writes, and \CAP{RMW}s that resembles two-phase commit (\CAP{2PC})~\cite{Gray:78}. However, \CAP{2PC} is not reliable (\cref{secH:related-work}) and is overkill for replicating single-object writes. In contrast, Hermes is highly optimized for single-object operations and is reliable.

Hermes builds upon the two main ideas of Galene to achieve high performance while also ensuring fault tolerance. Galene's first idea is the use of {\em invalidations}: a form of lightweight locking inspired by cache coherence protocols. The second is per-key \textit{logical timestamps} implemented as Lamport clocks~\cite{Lamport:1978}. Together, these enable linearizability; local reads; and fully concurrent, decentralized, and fast writes. Logical timestamps further allow each node to locally establish a single global order of writes to a key, which enables conflict-free write resolution (i.e., writes never abort\footnote{\CAP{RMW} in Hermes may abort (\cref{secH:hermes-RMWs}).} -- another difference from \CAP{2PC}) and 
\textit{write replays} to handle faults.

\noindent In short, the contributions we make in this chapter are as follows:

\begin{itemize}[leftmargin=*]
    \item \textbf{We introduce Hermes, a reliable replication protocol} (\cref{secH:hermes}) that utilizes invalidations and logical timestamps to achieve high performance and linearizability. 
    In the common failure-free operation, any replica in Hermes
\linebreak
    affords efficient local reads and fast fully concurrent writes.
Hermes handles message loss and node failures by guaranteeing that any write can always be safely replayed.  
We also detail efficient \CAP{RMW} support in Hermes and propose a variant of Hermes for safe operation under asynchrony that does not compromise on throughput.
    
    \item \textbf{We formally verify Hermes} (\cref{secH:hermes}) 
    reads, writes, and \CAP{RMW}s
    in \tla for safety and absence of deadlocks in the presence of crash-stop failures as well as message reordering and duplicates. 
    
    
    
    \item \textbf{We implement and evaluate Hermes} (\cref{secH:methodology} and \cref{secH:evaluation}) over a high-performance \CAP{RDMA}-based \CAP{KVS}.
    Our evaluation shows that Hermes outperforms the state-of-the-art \CAP{RDMA}-enabled virtual Paxos protocol~\cite{Jha:2019} 
    by an order of magnitude.
    Moreover, Hermes achieves higher throughput than the highly optimized \CAP{RDMA}-based state-of-the-art \CAP{ZAB}~\cite{Junqueira:2011} and \CAP{CRAQ}~\cite{Terrace:2009} replication protocols across all write ratios while significantly reducing the tail latency. At 5\% writes, the tail latency of Hermes is at least 3.6$\times$ lower than that of \CAP{CRAQ} and \CAP{ZAB}.
\end{itemize}

\break
\tsection{Motivation}
\label{secH:background}

\vspace{-5pt}
\tsubsection{High-Performance Reads and Writes}
\vspace{-5pt}
\label{secH:high-perf}
Maintaining high performance under strong consistency and fault tolerance is an established challenge~\cite{VanRenesse:2004, Baker:2011}. In the context of in-memory datastores, high performance is accepted to mean low latency and high throughput. 
Requirements for achieving high performance differ for reads and writes.

\beginbsec{Reads}
The key to achieving both low latency and high throughput on reads is (1) being able to service a read on any replica, which we call {\em load-balanced reads}, and (2) completing the read locally (i.e., without engaging other replicas). 
%
While seemingly trivial, load-balanced local reads (thereafter simply {\em local reads}) present a challenge for many reliable protocols, which may require 
\linebreak
communication among nodes to agree on a read value (e.g., \CAP{ABD}~\cite{Attiya:1995,Lynch:1997} and Paxos~\cite{Lamport:1998}) or mandate that only a single replica serve linearizable reads for a given key (e.g., primary-backup~\cite{Alsberg:1976} or Zookeeper's \CAP{ZAB} protocol~\cite{Junqueira:2011}).

\beginbsec{Writes}
Achieving high write performance under strong consistency and fault tolerance is notoriously difficult. We identify the following requirements necessary for low-latency high-throughput writes:

\noindent $\succ$ \underline{{\em Decentralized}} : In order to reduce network hops and preserve load balance across the replica ensemble, any replica must be able to initiate a write and drive it to completion (by communicating with the rest of the replicas) while avoiding centralized serialization points. For instance, both \CAP{ZAB} and \CAP{CR} require  writes to initiate at a particular node, thus failing to achieve decentralized writes.

\noindent $\succ$ \underline{{\em Inter-key concurrent}} : Independent writes on different keys should be able to proceed in parallel, enabling intra- and multi-thread parallel request execution.
For example, \CAP{ZAB} requires all writes to be serialized through a leader, thus failing to provide inter-key concurrency.
\newtext{
Linearizable protocols, like \CAP{CR}, can offer inter-key concurrency but some need costly per-key leases
~(see~\cref{secH:discussion}).}

\noindent $\succ$ \underline{{\em Fast}} : Fast writes require minimizing the number of message round-trips, avoiding long message chains (e.g., contrary to \CAP{CR}), and shunning techniques that otherwise increase latency (e.g., 
performing writes in lockstep~\cite{Mao:2008,Poke:2017}).

\tsubsection{Reliable Replication Protocols}
\label{secH:reliable-protocols}
%
As stated in the background section, reliable replication protocols capable of dealing with failures under our fault model are either majority-based or membership-based protocols.
%
Majority-based protocols require only a 
\linebreak
majority of nodes to respond in order to commit a write. They therefore tend to give up on local reads but may support decentralized or inter-key concurrent writes~\cite{Lamport:1998, Lynch:1997, Moraru:2013}. Majority-based protocols that afford local reads either relax consistency and serialize independent writes on a master (e.g., \CAP{ZAB}) or require communication-intensive per-key leases (detailed in \cref{secH:related-work}).
Problematically, both approaches hurt performance even in the absence of faults.

In contrast, membership-based protocols ensure that a committed write reaches all replicas in the ensemble.
Thus, in the absence of faults, membership-based protocols are naturally free of performance limitations associated with majority-based protocols and can easily facilitate local reads.

\begin{figure}[t]
  \centering
  \includegraphics[width=0.8\textwidth]{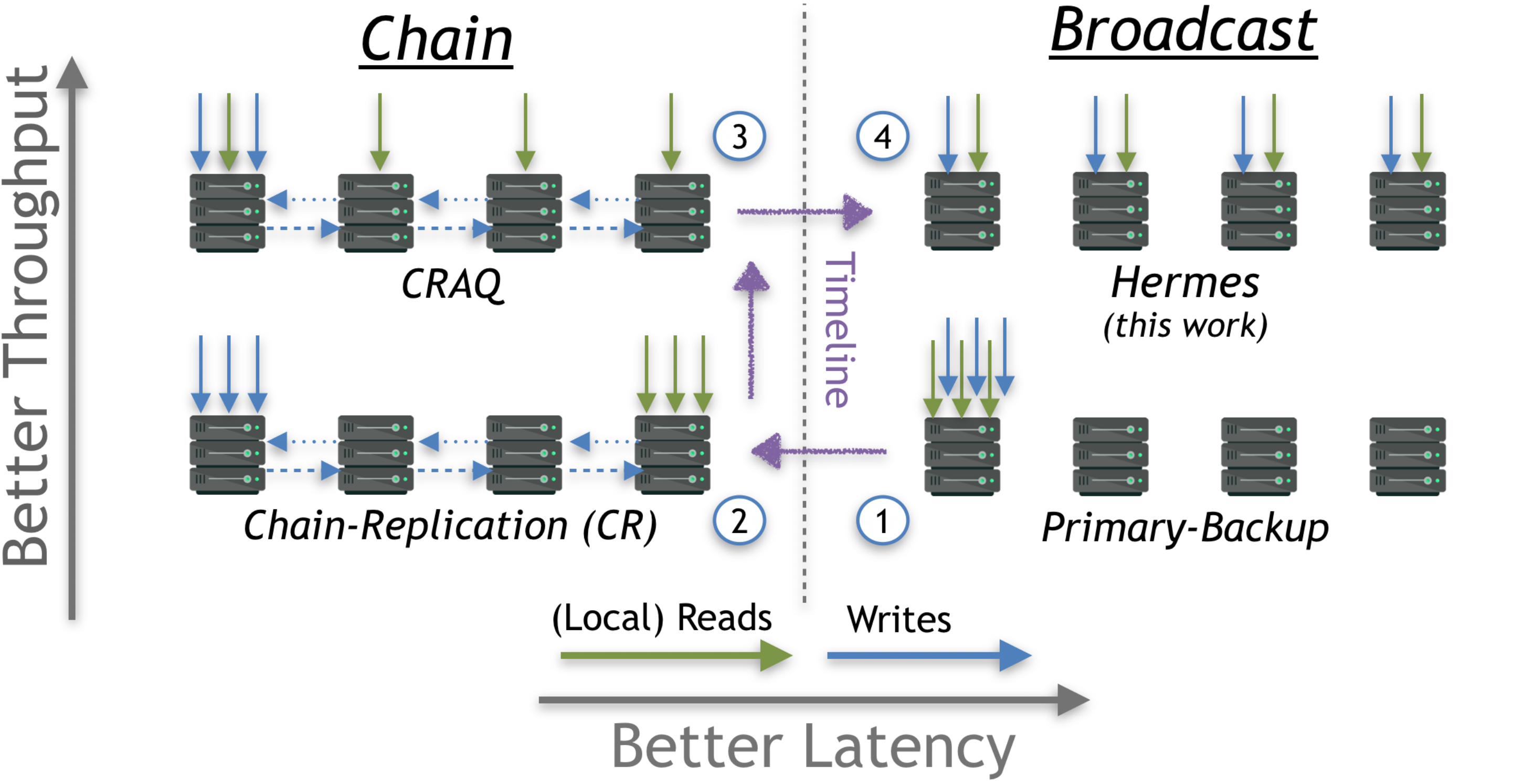}
  \mcaption{Throughput and latency of membership-based protocols.}
  {Comparison of reliable membership-based protocols in terms of throughput and latency.} 
  \label{figH:design}
\end{figure}

A common practice for high-performance replication 
is to optimize for the 
\linebreak
typical case of failure-free operation by harnessing the performance benefits of membership-based protocols  
and limiting the use of majority-based protocols to \CAP{RM} reconfiguration~\cite{Lamport:2009, Jimenez:2003, Dragojevic:2015}.
In fact, major datacenter operators, such as Microsoft, not only exploit mem\-ber\-ship-based protocols in their datastores~\cite{Dragojevic:2015, Shamis:2019}, but they also provide \CAP{LSC}s~\cite{Microsoft-time:2018, Corbett:2013} 
and \CAP{RM}~\cite{Kakivaya:2018} as datacenter services to ease the deployment of member\-ship-based protocols by third parties. 

The earliest membership-based protocol is primary-backup \cite{Alsberg:1976}, which serves all requests at a primary node and does not leverage the backup replicas for performance. 
Chain replication \CAP{(CR)}~\cite{VanRenesse:2004} improves upon primary-backup by organizing the nodes in a chain and dividing the responsibilities of the primary between the \textit{head} and the \textit{tail} of the chain, as shown in \cref{figH:design} (bottom left). 
\CAP{CR} is a common choice for implementing high-performance reliable replication~\cite{Jin:2018, Anwar:2018, Terrace:2009, Balakrishnan:2012, Wei:2017}.
We next discuss \CAP{CRAQ}~\cite{Terrace:2009}, a highly optimized variant of \CAP{CR}.

\subsection*{CRAQ} 
\label{secH:craq} 

\CAP{CRAQ} is a state-of-the-art membership-based protocol that offers high throughput and strong consistency (linearizability). In \CAP{CRAQ}, nodes are organized in a chain and writes are directed to its head, as in \CAP{CR}. The head propagates the write down the chain, 
which completes 
once it reaches the tail. Subsequently, the tail propagates acknowledgment messages upstream toward the head, informing all nodes of the write's completion.

\CAP{CRAQ} improves upon \CAP{CR} by enabling read requests to be served locally from all nodes, as shown in \cref{figH:design} (top left). However, if a non-tail node attempts to serve a read for which it has seen a write message propagating downstream from head to tail, but has not seen the acknowledgment propagating upstream, then the tail must be queried to determine whether the write has been applied or not. 

\CAP{CRAQ} is the state-of-the-art reliable replication protocol that achieves high throughput via a combination of local reads and inter-key concurrent writes. However, \CAP{CRAQ} fails to satisfy the low latency requirement: while reads are typically local and thus very fast,  writes must traverse multiple nodes sequentially, incurring a prohibitive latency overhead.

\tsection{Hermes} 
\label{secH:hermes}
Hermes
is a reliable mem\-ber\-ship-based broadcasting protocol that offers high throughput and low latency
while providing linearizable reads, writes, and \CAP{RMW}s. 
Hermes optimizes for the common case of no failures~\cite{Barroso:2018} and targets intra-datacenter in-memory datastores 
with a replication degree typical of 
\linebreak
today's deployments (3--7 replicas)~\cite{Hunt:2010}.
As noted in \cref{secB:datastores}, the replica count does not constrain the size of 
a sharded datastore, since each
shard is
replicated independently of other shards.
%
Example applications that can benefit from Hermes include reliable datastores~\cite{Baker:2011, Balakrishnan:2012, Edmund:2012, Wei:2017}, lock-services~\cite{Hunt:2010,Burrows:2006} and applications that require high performance, strong consistency, and availability (e.g.,~\cite{Adya:2016,Botelho:2013, Woo:18}).

\begin{figure}[t]
  \centering
  \includegraphics[width=0.8\textwidth]{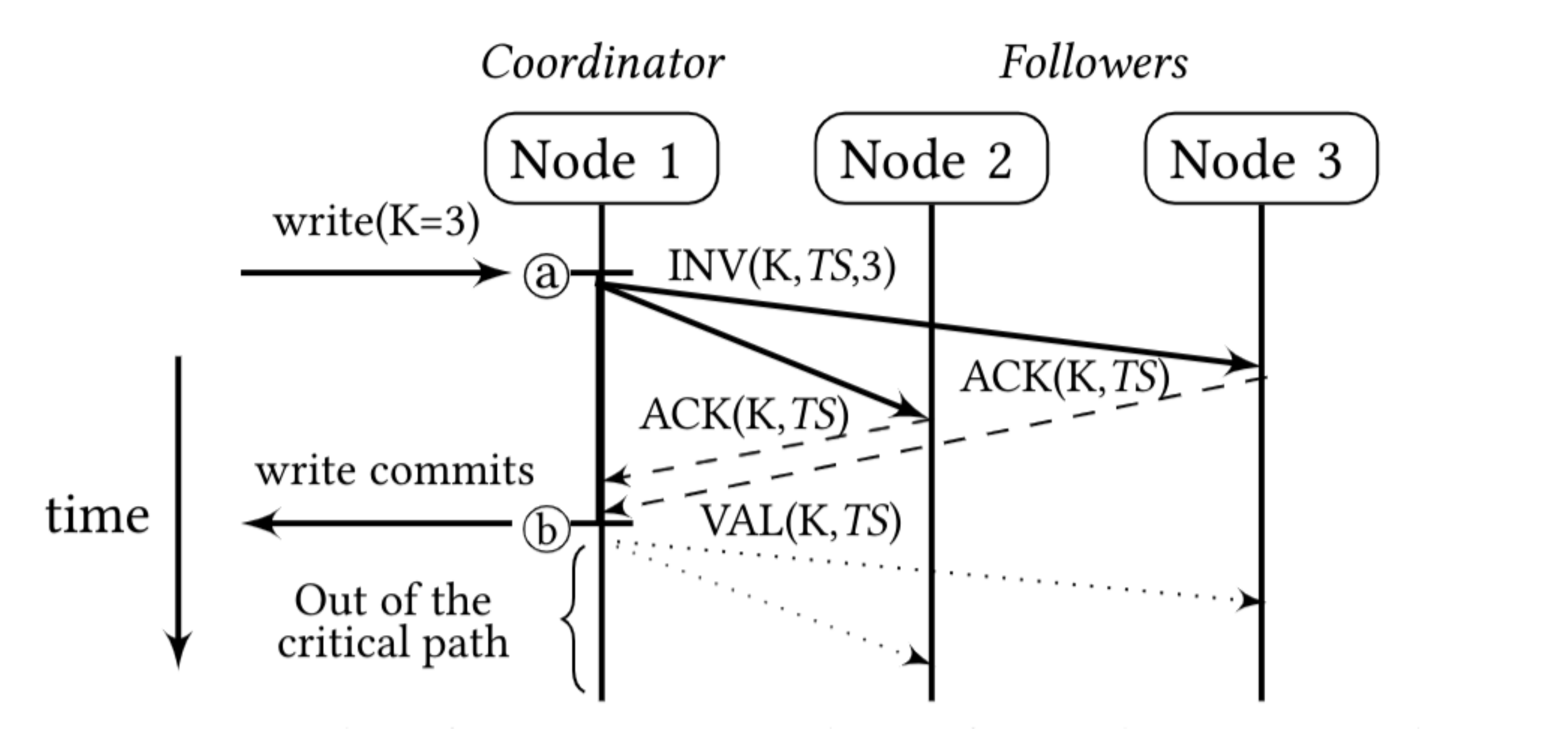}
  \mcaption{Write actions in Hermes.}
  {Example of writing a value of \emph{3} to key \emph{K}. Nodes 1, 2, and 3 hold a replica of \emph{K}. \emph{\CAP{TS}} is the timestamp.}
\label{figH:Hermes-write}
\end{figure}

\tsubsection{Protocol Overview}

In Hermes, reads complete locally. Writes can be initiated by any replica and complete fast regardless of conflicts. 
As illustrated in \cref{figH:Hermes-write}, a write to a key proceeds as follows. The replica initiating the write (called \textit{coordinator}) broadcasts an \textit{Invalidation} (\CAP{INV}) message to the rest of the replicas (called {\em followers}) and waits on \textit{Acknowledgments} (\CAP{ACK}s). Once all \CAP{ACK}s have been received, the write completes via a \textit{Validation} (\CAP{VAL}) message broadcast by the coordinator replica. 

We now briefly overview the salient features of Hermes and discuss the specifics in the following subsections. 

\beginbsec{Invalidations and logical timestamps} 
Similar to Galene, Hermes leverages the combination of invalidations and logical timestamps\footnote{Recall that the timestamp is a lexicographic tuple of <\VAR{v}, \cid> combining a key's version number (\VAR{v}), which is incremented on every write, with the node \CAP{ID} of the coordinator (\cid).} for high-performance reads and writes.
%
Given that a write invalidates all replicas prior completion, strongly consistent local reads in Hermes are simple. The very fact that an object replica is in a valid state implies that it contains the most up to date value and thus is safe to read. Writes are more involved as we discuss next.

\beginbsec{High-performance non-conflicting writes}
\label{secH:conc-writes}Hermes affords high-performance writes (\cref{secH:high-perf}) by maximizing concurrency while maintaining low latency.
First, writes in Hermes are executed from any replica in a decentralized manner, eschewing the use of a serialization point (e.g., a leader) and thus reducing the number of network hops and ensuring load balance. 
In contrast to 
approaches that globally order independent writes for strong consistency (e.g., \CAP{ZAB} -- \cref{secH:meth-protocols}), Hermes allows writes to different keys to proceed in parallel, hence achieving inter-key concurrency. 
This is accomplished through Hermes' approach of invalidating all operational replicas to achieve 
linearizability.
When combined with per-key logical timestamps, invalidations permit concurrent writes to the same key to be correctly linearized at the endpoints; thus, writes do not appear to conflict, making aborts unnecessary. 

Finally, in the absence of a failure, writes in Hermes cost one-and-a-half round-trips (\CAP{INV$\rightarrow$ACK$\rightarrow$VAL}); however, the exposed latency is only a single round-trip for each node. From the perspective of the coordinator, once all \CAP{ACK}s are received, it is safe to respond to a client because, at this point, the write is guaranteed to be visible to all live replicas and any future read cannot return the old value (\ie the write is \textit{committed} -- \cref{figH:Hermes-write}\scalebox{0.7}{\circled{b}}).
The followers also observe only a single round-trip (further optimized in \cref{secH:optimizations}), which starts once an \CAP{INV} arrives. At that point, each follower responds with an \CAP{ACK} and completes the write when a \CAP{VAL} is received.

\beginbsec{Safely replayable writes}
Hermes takes the ideas of Galene a step further by ensuring fault tolerance.
In Galene, node and network faults during a write to a key may leave the key in a permanently Invalid state in some or all nodes. To prevent this, Hermes allows any invalidated operational replica to replay the write to completion without violating linearizability. This is accomplished using two mechanisms. First, the new value for a key is propagated to the replicas in \CAP{INV} messages (\cref{figH:Hermes-write}\scalebox{0.85}{\circled{a}}).
Such \textit{early value propagation}
guarantees that every invalidated node is aware of the new value. Second, logical timestamps enable a precise global ordering of writes in each replica, facilitating idempotence. By combining these ideas, a node that finds a key in an Invalid state for an extended period 
can safely replay a write by taking on a coordinator role and retransmitting \CAP{INV} messages to the replica ensemble with the {\em original} timestamp (i.e., original version number and 
\cid),
hence preserving the global write order.

The above features afford the following properties:

\noindent $\succ$ \underline{{\em Strong consistency}}:
By invalidating all replicas of a key at the start of a write, Hermes ensures that a key in a Valid state is guaranteed to hold the most up-to-date value. Hermes enforces the invariant that a read may complete if and only if the key is in a Valid state, which provides linearizability. 

\noindent $\succ$ \underline{{\em High performance}}:
Local reads, in concert with high-performance broadcast-based non-conflicting writes from any replica, help ensure both low latency and high throughput. 

\noindent $\succ$ \underline{{\em Fault tolerance}}:
Hermes uses safely replayable
writes to tolerate a range of faults, including message loss, node failures, and network partitions. As a membership-based protocol, Hermes is aided by \CAP{RM} (\cref{secB:replication-protocols}) to provide a stable membership of live nodes in the face of failures and network partitions.

\begin{figure}[t]
  \centering
  \includegraphics[width=0.8\textwidth]{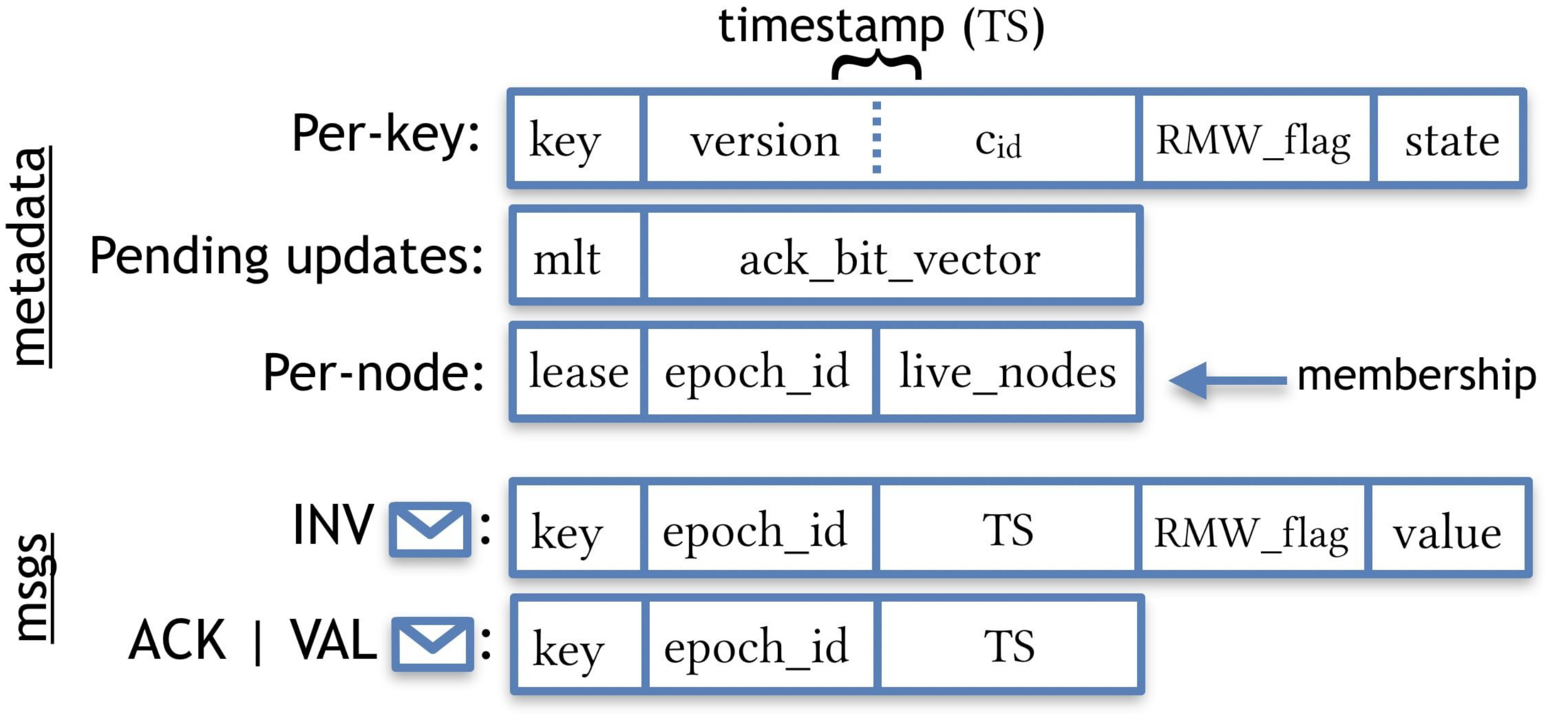}
  \mscaption{Metadata stored and messages sent by Hermes.}
  \label{figH:metadata}
\end{figure}

\tsubsection{Read/Write Protocol}
\label{secH:protocol}
The Hermes protocol consists of four stable states (\textit{Valid}, \textit{Invalid}, \textit{Write}, and \textit{Replay}) and a single transient state (\textit{Trans}).
\cref{figH:metadata} illustrates the format of protocol messages and the metadata stored at each replica.
A detailed protocol transition table, as well as the \tla specification, are available online.\footnote{\href{https://hermes-protocol.com}{https://hermes-protocol.com}}

The following protocol is slightly simplified in that it focuses only on reads and writes (i.e., omits \CAP{RMW}s) and deals only with node failures (not network faults). 
Resilience to network faults and \CAP{RMW}s are described in \cref{secH:hermes-discussion} and \cref{secH:hermes-RMWs}, respectively. 

\vspace{0.02in}
\noindent\textbf{\textit{Reads}}: A read request is serviced on an \textit{operational} replica (\ie one with an \CAP{RM} lease) by returning the local value of the requested key if it is in the Valid state. If the key is in any other state, the request is stalled.

\noindent\textbf{\textit{Writes}}:\\
\centerline{\underline{Coordinator}}
A coordinator node issues a write to a key only if it is in the Valid state; otherwise, the write is stalled. To issue and complete a write, the coordinator node: 
\begin{itemize}[leftmargin=*]
\item\textbf{C\textsubscript{\CAP{TS}}}: Updates the key's local timestamp by incrementing its \VAR{version} and appending its node \CAP{ID} as the 
\cid,
then assigns this timestamp to the write.
\item\textbf{C\textsubscript{\CAP{INV}}}: Prompt\-ly broadcasts an \CAP{INV} message consisting of the \VAR{key}, the new timestamp (\VAR{\CAP{TS}}), and the \VAR{value} to all replicas and transitions the key to the Write state while applying the new value locally. 
\item\textbf{C\textsubscript{\CAP{ACK}}}: Once the coordinator receives \CAP{ACK}s from all \textit{live} replicas, the write is completed by transitioning the key to the Valid state (Invalid state if the key was in the Trans state\footnote{The Trans state indicates a coordinator with a pending write that got invalidated. While not required, the Trans state helps track when the coordinator's original write completes, hence allowing the coordinator to notify the client of the write's completion.}).
\item\textbf{C\textsubscript{\CAP{VAL}}}: Finally, the coordinator broadcasts a \CAP{VAL}
consisting of the key and the same timestamp to all followers.
\end{itemize}

There are two simple yet subtle differences when comparing the coordinator actions of a write in Hermes and Galene. First, in Hermes, the value of a write is sent eagerly with the \CAP{INV} message. Second, the coordinator in Hermes waits for \CAP{ACK}s only from the live replicas, 
as indicated in the membership variable. If a follower fails after an \CAP{INV} has been sent, the coordinator waits for the \CAP{ACK} from the failed node until the membership is reliably updated (after the node is detected as failed and the membership leases expire -- \cref{secB:replication-protocols}). 
Once the coordinator is no longer missing any \CAP{ACK}s, it can safely continue the write.

\vspace{10pt}
\centerline{\underline{Follower}}
\vspace{-20pt}
\begin{itemize}[leftmargin=*]
\item\textbf{F\textsubscript{\CAP{INV}}}: Upon receiving an \CAP{INV} message, a follower compares the timestamp of the incoming message to its local timestamp of the key. If the received timestamp is higher than the local timestamp, the follower transitions the key to the Invalid state (Trans state if the key was in the Write or Replay state) and updates the key's value and local timestamp (both its \VAR{version} and the~\cid).
\item \textbf{F\textsubscript{\CAP{ACK}}}: Regardless of the result of the timestamp comparison, a follower always responds with an \CAP{ACK} 
containing
the same timestamp as that in the \CAP{INV} message of the write. 
\item \textbf{F\textsubscript{\CAP{VAL}}}: When a follower receives a \CAP{VAL} message, it transitions the key to the Valid state if and only if the received timestamp is equal to the key's local timestamp. Otherwise, the \CAP{VAL} message is simply ignored. 
\end{itemize}

\noindent\textbf{\textit{Write replays}}:
A request that finds a key in the Invalid state for an extended period of time (determined via the {\em mlt} timer, described in~\cref{secH:hermes-discussion}) triggers a write replay. The node servicing the request takes on the coordinator role, transitions the key to the Replay state, and begins a write replay by re-executing steps \textbf{C\textsubscript{\CAP{INV}}} through \textbf{C\textsubscript{\CAP{VAL}}} 
using the timestamp and value received with the \CAP{INV} message.
Note that the original timestamp is used in the replay (\ie the
version and \cid~
are the same as that of the original coordinator) to allow the write to be correctly linearized.  
Once the replay is completed, the key transitions to the Valid state, after which the initial request is serviced.

\beginfsec{Formal verification}
We expressed Hermes in \tla and model-checked the protocol's reads, writes, \CAP{RMW}s and replays for safety and the absence of deadlocks in the presence of 
message reordering and duplicates, as well as membership reconfigurations due to crash-stop failures.

\subsection*{Protocol Optimizations}
\label{secH:optimizations}

\beginbsec{[O\textsubscript{1}] Enhancing fairness} Like Galene, Hermes can utilize virtual node \CAP{ID}s to increase fairness in the order of concurrent writes to the same key from different replicas. 

\beginbsec{[O\textsubscript{2}] Eliminating unnecessary validations} When the coordinator of a write gathers all of its \CAP{ACK}s 
but discovers a concurrent write to the same key with a higher timestamp (i.e., the key was in the Trans state), it does not need to broadcast \CAP{VAL} messages (\textbf{C\textsubscript{\CAP{VAL}}}), thus saving valuable network bandwidth. 

\beginbsec{[O\textsubscript{3}] Reducing blocking latency} In failure-free operation, and during a write to a key, followers block reads to that key for up to a round-trip (\cref{secH:conc-writes}). This blocking latency can be reduced to half of a round-trip if followers broadcast \CAP{ACK}s to all replicas rather than responding only to the coordinator of the write (\textbf{F\textsubscript{\CAP{ACK}}}). 
Once all \CAP{ACK}s have been received by a follower, it can service the reads to that key without waiting for the \CAP{VAL} message. While this optimization increases the number of \CAP{ACK}s, the actual bandwidth cost is minimal as \CAP{ACK} messages have a small constant size. The bandwidth cost is further offset by avoiding the need to broadcast \CAP{VAL} messages. Thus, under the typical small replication degrees, this optimization comes at a negligible cost in bandwidth.

\tsubsection{Network Faults, Reconfiguration, and Recovery}
\label{secH:hermes-discussion}
This section presents Hermes' operation under imperfect links, network partitions,
and the transient period of membership reconfiguration on a fault. 
It then 
provides an overview of the mechanism to add new nodes to the replica group.

\beginbsec{Imperfect links}
In typical multi-path datacenter networks, messages can be reordered, duplicated, or lost~\cite{Farring:2009,Gill:2011,Lu:2018}. Hermes operates correctly in all of these scenarios, as described below. In Hermes, the information necessary to linearize operations is embedded with the keys and in messages in the form of logical timestamps. Thus, the protocol never violates linearizability even if messages get delayed, reordered, or duplicated in the network.

Hermes uses the same idea of replaying writes if any \CAP{INV}, \CAP{ACK}, or \CAP{VAL} message is suspected to be lost. A message is suspected to be lost for a key if the request's \textit{message-loss timeout (mlt)} -- within which every write request is expected to be completed -- is exceeded. To detect the loss of an \CAP{INV} or \CAP{ACK} for a particular write, the coordinator of the write resets the 
request's mlt 
once it broadcasts \CAP{INV} messages. If a key's 
mlt
is exceeded before its write completion, the coordinator suspects a potential message loss and 
resets the request's mlt before retransmitting the write's \CAP{INV} broadcast.

In contrast, the loss of a \CAP{VAL} message is handled by the follower using a write replay. Once a follower receives a request for a key in the Invalid state, it resets the request's message-loss timeout.
If the timestamp or state has not been updated within the 
mlt
duration, it suspects the loss of a \CAP{VAL} message and triggers a write replay. Although a write replay will never compromise the safety of the protocol, a carefully calibrated timeout will reduce unnecessary replays (e.g., when messages are not lost).

\beginbsec{Network partitions}
Datacenter network topologies are highly redundant~\cite{Gill:2011,Singh:2015}; however, in rare cases, link failures might result in a network partition. According to the \CAP{CAP} theorem~\cite{Brewer:2000,Gilbert:2002},
either consistency or availability must be sacrificed in the presence of network partitions. 
Hermes follows the guidelines of Brewer~\cite{Brewer:2012} to permit the datastore to continue serving requests only in its \textit{primary partition}: a partition with the majority of replicas. 
\newtext{Recall that we consider a membership service using a majority-based protocol.
Thus, although failure detectors cannot differentiate between node failures and network partitions,
the membership can only be reliably updated in the primary partition and does so only after the expiration of the membership leases}. As a result, replicas in a minority partition stop serving requests before the membership is updated and new requests are able to complete only in the primary partition.
Updating the membership in the primary partition is always feasible because the \CAP{RM} protocol is run by the datastore replicas, not external nodes (e.g., an external service).\footnote{If Hermes is deployed over an external \CAP{RM} service, only the nodes that remain connected with the service would continue to be operational under a network partition.}
Once network connectivity is restored, nodes previously on a minority side can rejoin the replica group via a recovery procedure explained below. 


\beginbsec{Membership reconfiguration after a failure} Following a network partition or a node failure and expiration of the leases for all of the nodes in a membership group, a majority-based protocol is used to reliably update the membership. We refer to this update as \textit{m-update}, which consists of a lease renewal, a new list of live nodes, and an incremented \epoch (\textit{epoch\_id}). Although the m-update is consistent even in the presence of faults, it does not reach all live replicas instantaneously. 
Rather, there is a transient period in which some replicas that are considered live, according to the latest value of the membership, have received the m-update while others have not and are still non-operational.

Hermes seamlessly deals with the transition of m-update without violating safety. Hermes' replicas, which have received the m-update, are able to act as coordinators and serve new requests. Thus, reads that find the target key in the Valid state can immediately be served as usual.
In contrast, writes or reads that require a replay (\ie the targeted key is Invalid) are effectively stalled 
until all live nodes (as indicated by the membership variable) receive the m-update. This is because writes and write replays do not commit until all live replicas become operational and acknowledge their \CAP{INV} messages.

During this transition period, any live follower that has not yet received the latest m-update will simply drop the \CAP{INV} messages, because those messages are tagged with an epoch\_id greater than the follower's local epoch\_id. This manifests as a simple message loss to a coordinator, which triggers retransmission of the \CAP{INV}s (\cref{secH:hermes-discussion}). The coordinator eventually completes its writes once all live followers have received the latest membership and become operational.

\beginbsec{Recovery}
Hermes' fault tolerance properties enable a datastore to continue operating even 
in the presence of failures. 
However, as nodes fail, new nodes need to be added to the datastore to continue operating at peak performance. To add a new node, the membership is reliably updated, following which all other live replicas are notified of the new node's intention to join the replica group. Once all replicas acknowledge this notification, the new node begins to operate as a \textit{shadow replica} that participates as a follower for all 
writes but does not serve any client requests. Additionally, it reads chunks (multiple keys) from other replicas to fetch the latest values and reconstruct the shard for which it is responsible, similar to existing approaches~\cite{Dragojevic:2015,Ongaro:2011}. After reading the entire shard, the shadow replica is up to date and transitions to an operational state, in which it is able to serve client requests.

\begin{figure}[t]
  \centering
  \includegraphics[width=0.9\textwidth]{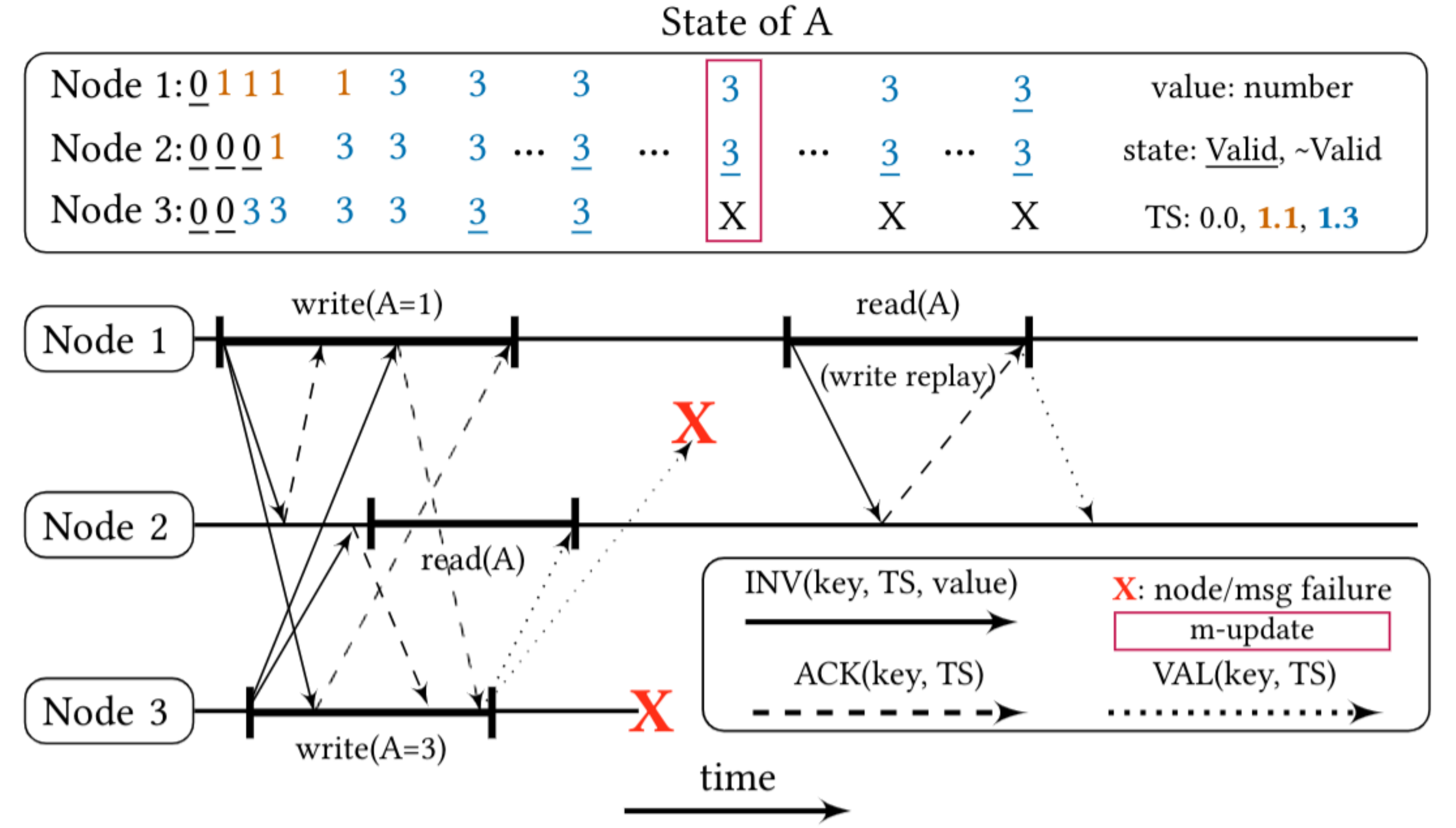}
\mcaption{Execution example with faults in Hermes.}
{Concurrent writes to key A, then a read, followed by a node and a message failure which trigger a write replay on the last read. 
State of A shows the value on each object replica; underlined represents Valid state, non-underlined represents other states. The color indicates the timestamp's value.}
\label{figH:Hermes-example}
\end{figure}
\tsubsection{Operational Example}
In this subsection, we discuss \cref{figH:Hermes-example}, which illustrates an example of Hermes' execution with reads and writes to key $A$ that is replicated in three nodes. 
The key $A$ is initially stored in the Valid state with the same value (zero) and timestamp in all three nodes. 
The purpose 
is to demonstrate the operation of Hermes while shedding light on some of its corner cases in the presence of concurrency and failures. 
For simplicity, we assume no use of virtual node \CAP{ID}s or any latency optimizations (\cref{secH:optimizations}).

First, node 1 initiates a write ($A = 1$) by incrementing its local timestamp, broadcasting \CAP{INV} messages (solid lines), transitioning key $A$ to the Write state, and updating its local value. Likewise, node 3 initiates another concurrent write ($A = 3$). 
Recall that \CAP{INV}s in Hermes contain the key, timestamp (including the \cid), and value to be written. 

Node 2 \CAP{ACK}s the \CAP{INV} message from node 1 (dashed line), updates its timestamp and value, and transitions key $A$ to the Invalid state. Node 3 \CAP{ACK}s the \CAP{INV} of node 1, but it does not modify $A$ or its state because its local timestamp is higher (same version, but higher 
\cid).
Subsequently, node 2 receives the \CAP{INV} from node 3, which has a higher timestamp than the locally stored timestamp, resulting in an update to its local value and timestamp, all while remaining in the Invalid state. 
Similarly, node 1 \CAP{ACK}s the \CAP{INV} of node 3 by updating the value and the timestamp before transitioning to the Trans state. 

Meanwhile, node 2 starts a read, but it is stalled because its local copy of $A$ is invalidated. 
Once node 3 receives all \CAP{ACK}s, it completes its own write by transitioning $A$ to the Valid state and broadcasting a \CAP{VAL} message (dotted lines) to the other replicas. 
When node 2 receives node's 3 \CAP{VAL} message, it transitions $A$ to Valid state and completes its stalled read. 

Once node 1 receives all of the \CAP{ACK}s, it completes its write but transitions to the Invalid state.  This occurs because the write from node 3 took precedence over node 1's write due to its higher timestamp, but the \CAP{VAL} from node 3 has not yet been received. Note that, although the write from node 1 completes later than the concurrent write from node 3, it is linearized before the write of node 3 due to its lower timestamp 
(\cid).

As a last step, we consider a failure scenario wherein the \CAP{VAL} message from node 3 to node 1 is dropped, node 3 crashes, and key $A$ in node 1 thus remains in the Invalid state.
Once the leases expire and node 3 is detected as failed, the membership variable is reliably updated.
Subsequently, node 1 receives a read for $A$ but finds $A$ invalidated by a failed node.
Thus, node 1 triggers a write replay by broadcasting \CAP{INV} messages with the key's locally stored timestamp and value (i.e., replaying node 3's original write). Crucially, the fact that \CAP{INV} messages contain both the timestamp and value to be written allows node 1 to safely replay node 3's write. Node 2 \CAP{ACK}s the \CAP{INV} from node 1 without applying it, since it already has the same timestamp. Once node 1 gets the \CAP{ACK} from node 2, it can unblock itself. Lastly, node 1 completes the replay of the write by broadcasting a \CAP{VAL} message to all live nodes (i.e., node 2, in this example).


\tsubsection{Read-Modify-Writes}
\label{secH:hermes-RMWs}

So far, we have focused on read and write operations. However, Hermes also supports read-modify-write (\CAP{RMW}) atomics that are useful for synchronization (e.g., a compare-and-swap to acquire a lock).
In general, the atomic execution of a read followed by a write to a key may fail if naively implemented with simple reads and writes. This is because a 
read followed by a write to a key is not guaranteed to be performed atomically, since another concurrent write to the same key with a smaller logical timestamp could be linearized between the read-write pair, hence violating the \CAP{RMW} semantics.

For this reason, an \CAP{RMW} update in Hermes is executed similarly to a write, 
 but it is conflicting.
Hermes may abort an \CAP{RMW} that is
concurrently executed with another \textit{update} operation (either a write or another \CAP{RMW}) to the same key. Hermes commits an \CAP{RMW} if and only if the \CAP{RMW} has the highest timestamp amongst any concurrent updates to that key. Moreover, it purposefully assigns higher timestamps to writes compared to their concurrent \CAP{RMW}s. As a result, any write racing with an \CAP{RMW} to a given key is guaranteed to have a higher timestamp, thus safely aborting the \CAP{RMW}. Meanwhile, if only \CAP{RMW} updates are racing, the \CAP{RMW} with the highest node \CAP{ID} will commit, and the rest will abort. 

More formally, Hermes always maintains safety and guarantees progress in the absence of faults
by ensuring two 
properties: (1) \textit{writes always commit}, and (2) \textit{at most one of the possible concurrent \CAP{RMW}s to a key commits}. 
To maintain these properties, the following protocol alterations are needed:

\begin{itemize}[leftmargin=*]
\item\textbf{\scalebox{0.90}{Metadata}}: To distinguish between \CAP{RMW} and write updates, an additional binary flag (\VAR{RMW\_flag}) is included in \CAP{INV} messages. The flag is also stored in the per-key metadata to accommodate \textit{update replays}. 
\item\textbf{C\textsubscript{\CAP{TS}}}: When a coordinator issues an update, the version of the logical timestamp is incremented 
by one if the update is an \CAP{RMW} and by two if it~is~a~write. 

\item\textbf{F\textsubscript{\CAP{RMW-ACK}}}: A follower \CAP{ACK}s an \CAP{INV} message for an 
\CAP{RMW} only if its timestamp is equal to or higher than the local one; otherwise, the follower responds with an \CAP{INV} based on its local state (i.e., the same message used for write replays).

\item\textbf{C\textsubscript{\CAP{RMW}-abort}}: In contrast to non-conflicting writes, an \CAP{RMW} with pending \CAP{ACK}s is aborted if its coordinator receives an \CAP{INV}
to the same key 
with a higher timestamp.


\item
\textbf{C\textsubscript{\CAP{RMW}-replay}}: After an \CAP{RM} reconfiguration, the coordinator resets any 
gathered \CAP{ACK}s of a pending \CAP{RMW} and replays the \CAP{RMW} to ensure that it is not conflicting.
\end{itemize}


\vspace{-15pt}
\tsubsection{Taming Asynchrony}
\vspace{-6pt}
\label{secH:async}
This thesis mainly considers a failure model with loosely synchronized clocks (\CAP{LSC}s), which is representative of intra-datacenter deployments~\cite{Shamis:2019, Li:2020}. 
Nevertheless, Hermes leverages \CAP{LSC}s only for \CAP{RM} lease management to 
\linebreak
ensure that a node with a lease is always part of the latest membership.
%
\linebreak
Updates in Hermes seamlessly work under asynchrony (i.e., without \CAP{LSC}s), since they commit only after all acknowledgments are gathered, which occurs only if the coordinator has the same membership as every other live follower.\footnote{Followers with a different membership value would have otherwise ignored the received \CAP{INV}s due to discrepancy in the message epoch\_ids (\cref{secH:reliable-protocols})}

Linearizable reads in Hermes can also be served under asynchrony. The basic idea is to use a committed update to {\em any} key after the arrival of a read request as a guarantee that the given node is still part of the replica group (thus validating the read). More specifically, observe that a node can establish that it is a member of the latest membership by successfully committing an update. Using this idea, a read at a given node can be speculatively executed but not immediately returned to the client. Once the node executes a subsequent update to any key and receives acknowledgments from a \textit{majority} of replicas, it can be sure that it was part of the latest membership when the read was executed. Once that is established, the read can be safely returned to the client. Note that a majority of acknowledgments suffices, as the membership 
is updated via a majority-based protocol and is maintained by the replicas rather than an external service.


If a subsequent update is not readily available (\eg due to low load), the coordinator replica of a read can send a \textit{membership-check} message to the followers containing only the membership epoch\_id. The followers acknowledge this message if they are in the same epoch. After a majority of acknowledgments have been collected, the coordinator returns the read. The membership-check is a small message and can be issued after a batch of read requests are speculatively executed by the coordinator.
Overall, although serving reads without \CAP{LSC}s increases the latency of reads until a majority of replicas respond, it 
incurs zero (if a subsequent update is timely) or minimal network cost to validate a batch of reads. 
We experimentally study the impact of this asynchronous variant in \cref{secH:async-eval}.

\tsubsection{Protocol Summary}
This section introduced Hermes, a reliable membership-based protocol that guarantees linearizability. Hermes' decentralized, broadcast-based design is engineered 
for high throughput and low latency. Leveraging invalidations and logical timestamps, Hermes enables 
efficient local reads and high-performance updates that are decentralized, fast, and inter-key concurrent. Writes (but not \CAP{RMW}s) in Hermes are also non-conflicting.
Hermes seamlessly recovers from a range of node and network faults thanks to
its update replays, enabled by early value propagation and logical timestamps.
Finally, Hermes can be deployed in a fully asynchronous setting without compromising safety via a simple read-validation scheme.

\tsection{Experimental Methodology}
\label{secH:methodology}

\tnsubsection{HermesKV}
\label{secH:system}

To evaluate the benefits and limitations of the Hermes protocol, we build HermesKV, an in-memory \CAP{RDMA}-based \CAP{KVS} with a single-object read/write \CAP{API}. 
HermesKV is replicated across all the machines comprising a deployment and relies on the Hermes protocol to ensure the consistency of the deployment.

\beginbsec{Overview and KVS} 
\label{secH:functional}
Each node in HermesKV is composed of a number of identical \emph{worker} threads. Each worker performs the following tasks: (1) decodes client requests; (2) accesses the local \CAP{KVS} replica; and (3) runs the Hermes protocol to complete requests. Client requests are distributed among the worker threads of the system. Requests can be either reads or writes. Worker threads communicate solely to coordinate writes (and write replays) as reads are completed locally. 

The implementation of HermesKV is based on cc\CAP{KVS}~\cite{A&V:2018}.
%
%
%
We extend cc\CAP{KVS} by replicating all objects for availability and
to accommodate the Hermes-specific protocol actions, state transitions, and request replies based on the replica state. 
To focus on the performance of Hermes, we strip the caching layer of cc\CAP{KVS}.
Note that the Hermes protocol is 
agnostic 
to the choice of a datastore and 
can be used with any datastore. We choose cc\CAP{KVS} because its minimalist design allows us to focus on the impact of the replication protocol itself, without regarding other irrelevant overheads of commercial-grade datastores. 

\beginbsec{Networking}
State-of-the-art \CAP{RDMA}-based \CAP{KVS} designs such as \CAP{HERD}~\cite{Kalia:2014} and cc\CAP{KVS}~\cite{A&V:2018} have shown remote procedure calls to be a highly effective design paradigm. Hence, we leverage two-sided \CAP{RDMA} primitives (over unreliable datagram sends) and all of the networking optimizations discussed in \cref{secC:communication}.
%
These include opportunistic coalescing of multiple messages into one network packet, application-level flow control, support for software broadcasts, and other low-level \CAP{RDMA} optimizations.
\vspace{-15pt}
\tsubsection{Evaluated Systems} 
\vspace{-5pt}
\label{secH:meth-protocols}
We evaluate Hermes by comparing its performance with majority-based and mem\-ber\-ship-based \CAP{RDMA}-enabled baseline protocols. To facilitate a fair 
\linebreak
protocol comparison, we study all protocols over a common multi-threaded \CAP{KVS} implementation based on HermesKV (as described in \cref{secH:system}). All protocols are implemented in C over the \CAP{RDMA} \emph{verbs} \CAP{API}~\cite{Barak:2015}.

\noindent The evaluated systems are as follows:
\squishlist
    \item \textbf{r\CAP{ZAB}}: In-house, multi-threaded, \CAP{RDMA}-enabled  
    \CAP{ZAB}~\cite{Reed:2008}.
    \item \textbf{r\CAP{CRAQ}}: In-house, multi-threaded, \CAP{RDMA}-based  \CAP{CRAQ}~\cite{Terrace:2009}.
    \item \textbf{HermesKV}: Implementation of Hermes as presented in \cref{secH:hermes}, without the latency optimization \textbf{[O\textsubscript{3}]} from \cref{secH:optimizations}.
\squishend

Our evaluation mainly focuses on comparing HermesKV to r\CAP{ZAB} and r\CAP{CRAQ}, since they share the same \CAP{KVS} and communication implementation, which allows us to isolate the effects of the protocol itself on performance. We also compare Hermes to \textbf{Derecho}~\cite{Jha:2019} (\cref{secH:eval-derecho}), the state-of-the-art \CAP{RDMA}-optimized open-source implementation of mem\-ber\-ship-based (i.e., virtually synchronous) Paxos.
\cref{tabH:eval} summarizes the read and write features of the evaluated systems.

\begin{table}[h!]
\centering

\begin{adjustbox}{max width=0.95\textwidth}

\begin{tikzpicture}

\node (table) [inner sep=-0pt] {
\begin{tabular}{l|c|c|c|c|c}
\rowcolor[HTML]{9B9B9B} 
\cellcolor[HTML]{9B9B9B} & \multicolumn{2}{c|}{\cellcolor[HTML]{9B9B9B}\textbf{Local reads}} & \multicolumn{3}{c}{\cellcolor[HTML]{9B9B9B}\textbf{Writes}} \\ \cline{2-6} 
\rowcolor[HTML]{9B9B9B} 
\multirow{-2}{*}{\cellcolor[HTML]{9B9B9B}\textbf{System}} & \multicolumn{1}{c|}{\cellcolor[HTML]{9B9B9B}Leases} & \multicolumn{1}{c|}{\cellcolor[HTML]{9B9B9B}Consistency} & \multicolumn{1}{c|}{\cellcolor[HTML]{9B9B9B}Concurrency} & Latency (RTT) & \multicolumn{1}{l}{\cellcolor[HTML]{9B9B9B}Decentralized} \\
\textbf{HermesKV} & one per RM & linearizable & inter-key & 1 & \cmark \\
\textbf{rCRAQ} & one per RM & linearizable & inter-key & O(\textit{n}) & \xmark \\
\textbf{rZAB} & none & SC & serializes all & 2 $\dagger$ & \xmark \\
\textbf{Derecho} & none & SC & serializes all & 1 $\ddagger$ & \cmark
\end{tabular}
};
\draw [rounded corners=.3em] (table.north west) rectangle (table.south east);

\end{tikzpicture}

\end{adjustbox}

\mcaption{Read and write features of the evaluated systems.}
{Comparison of read and write features for the evaluated systems. 
SC: sequentially consistent; \CAP{RM}: reliable membership; \textit{n}: number of replicas; $\dagger$1 \CAP{RTT} for master's writes; $\ddagger$lockstep commit.}
\label{tabH:eval}
\end{table}

\vspace{-20pt}
\subsection*{rZAB}
\vspace{-10pt}
In the \CAP{ZAB} protocol, one node is the leader and the rest are followers. A client can issue a write to any node, which in turn propagates the write to the leader. 
The leader receives writes from all nodes, serializes them, and proposes them by broadcasting atomically to all followers. The followers send \CAP{ACK}s back to the leader. Upon receiving a majority of \CAP{ACK}s for a given write, the leader commits the write locally and broadcasts commits to the followers.

A client's read can be served locally by any node without any communication as long as the last write of that client has been applied in that node. 
%
However, local reads in \CAP{ZAB} are sequentially consistent, which is weaker than linearizable. Problematically, the fact that \CAP{ZAB} is not linearizable leads to a performance issue on writes.
This is because, in contrast to the stricter linearizability, sequential consistency (\CAP{SC}) is not composable~\cite{Attiya:1994}. As a result, it is not possible to deploy independent (e.g., per-key) instances of SC protocols such as \CAP{ZAB} to increase the concurrency of writes because the composition of those instances would violate \CAP{SC}.
If a  \CAP{ZAB} client requires linearizable reads, it can issue a \emph{sync} command prior to the read. A sync completes similarly to a write, hence significantly increasing the read message cost and latency. 
In order to obtain the upper-bound performance of \CAP{ZAB},
we do not evaluate linearizable reads.

\beginbsec{rZAB optimizations}
We apply to r\CAP{ZAB} all HermesKV optimizations and use the \CAP{RDMA} multicast~\cite{Barak:2015} to tolerate \CAP{ZAB}'s asymmetric (i.e., leader-oriented) network traffic pattern.
Our highly optimized, \CAP{RDMA} implementation of \CAP{ZAB} outperforms the open-source implementation of Zook\-eep\-er (evaluated in ~\cite{Jin:2018}) by three orders of magnitude. Of course, Zookeeper is a production system that incorporates features beyond the \CAP{ZAB} protocol, such as client tracking and checkpointing to disk. By evaluating a lean and optimized version of \CAP{ZAB}, alone, we 
facilitate a fair protocol comparison.

\vspace{-15pt}
\subsection*{rCRAQ}
\vspace{-10pt}
\CAP{CRAQ} 
affords local reads and inter-key concurrent -- but not decentralized -- writes (\cref{secH:craq} details the \CAP{CRAQ} protocol). We identify two undesirable properties of \CAP{CRAQ}: (1) writes must traverse multiple hops before completing, adversely affecting the system's latency; and (2) the nodes of the chain are generally not well balanced 
in terms of the amount of work performed per packet, potentially affecting the system's throughput. 
To evaluate how these properties affect performance, we study our own \CAP{RDMA}-enabled version of \CAP{CRAQ} (r\CAP{CRAQ}), which takes advantage of all optimizations available in HermesKV.

\vspace{-10pt}
\tsubsection{Testbed}
\vspace{-5pt}
%
All of our experiments (except those described in \cref{secH:async-eval}) are conducted on a cluster of 7 servers interconnected via a 12-port Infiniband switch (Mellanox \CAP{MSX6012F}). Each machine runs Ubuntu 18.04 and is equipped with two 10-core \CAP{CPU}s (Intel Xeon E5-2630) with 64 \CAP{GB} of system memory and a single-port 56Gb \CAP{NIC} (Mellanox \CAP{CX4} \CAP{FDR} \CAP{IB} \CAP{PCI}e3 x16). Each \CAP{CPU} has 25 \CAP{MB} of L3 cache and two hardware threads per core. We disable turbo-boost, pin threads to cores, and use huge pages (2 \CAP{MB}) for the \CAP{KVS}. The \CAP{KVS} consists of one million key-value pairs, replicated in all nodes. Unless stated otherwise, we use keys and values of $8$ and $32$ bytes, respectively, which are accessed uniformly.

\tsection{Evaluation}
\label{secH:evaluation}

\tsubsection{Throughput on Uniform Traffic} 
\label{secH:throughput} 

\begin{figure}[t]
  \centering
  \includegraphics[width=0.75\textwidth]{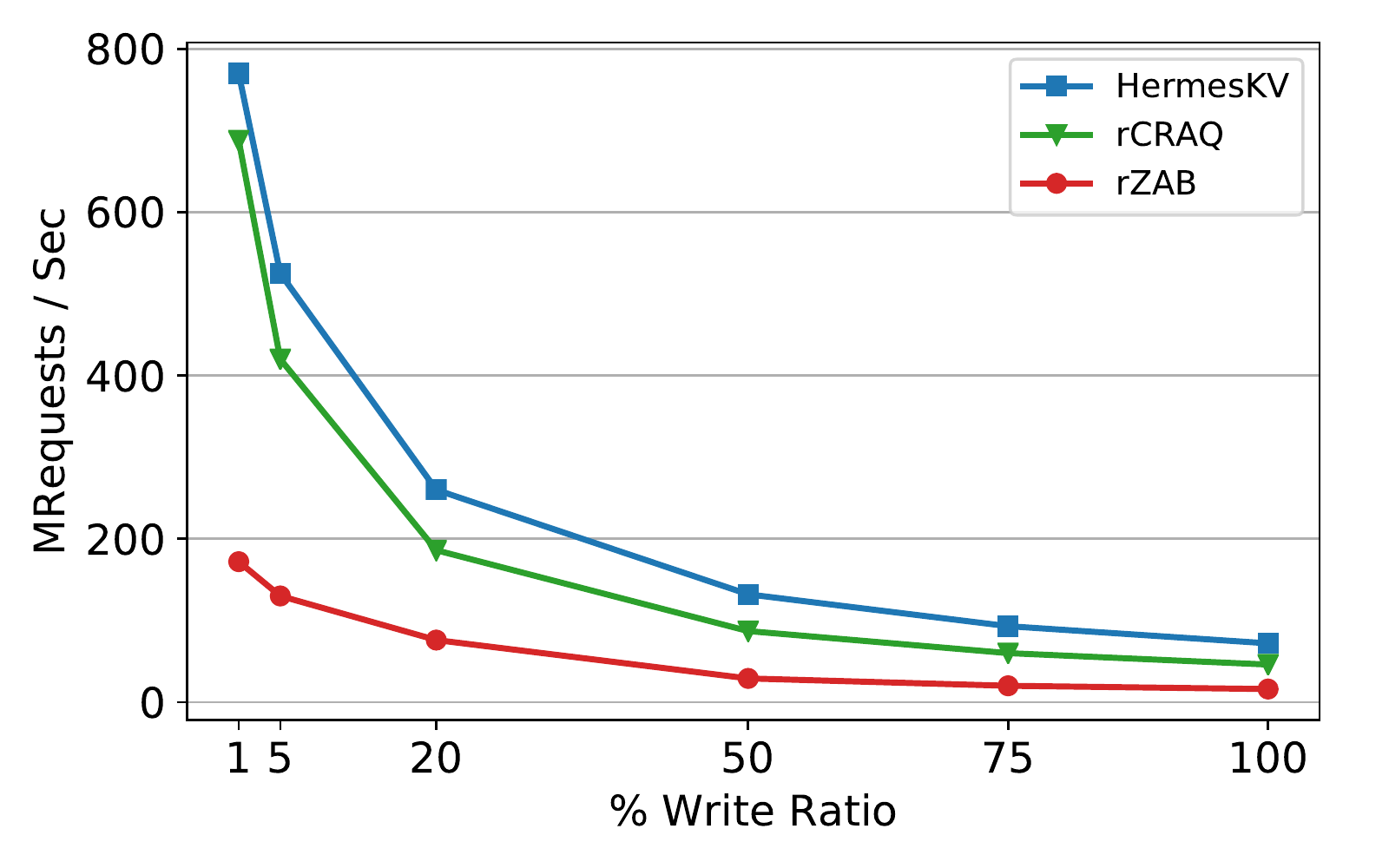}
  \mcaption{Throughput while varying write ratio with uniform accesses.}
  {Throughput for 1\% to 100\% writes with uniform accesses. [5 nodes]}
  \label{figH:write-rate}
\end{figure}

\cref{figH:write-rate} shows the performance of HermesKV, r\CAP{CRAQ}, and r\CAP{ZAB} while varying the write ratio under uniform traffic.\goodbreak\noindent 

\beginbseceval{Read-only} For read-only, all three systems exhibit identical behavior, achieving 985 million requests per second (MReqs/s), as all systems perform reads locally from all replicas. To reduce clutter, we omit the read-only from the figure.

\beginbseceval{HermesKV} 
At 1\% write ratio, HermesKV achieves 770 MReqs/s, outperforming both baselines (12\% better than r\CAP{CRAQ} and 4.5$\times$ better than r\CAP{ZAB}). 
As the write ratio increases, the throughput of HermesKV gradually drops, reaching 72 MReqs/s on a write-only workload. The throughput degradation at higher write ratios is expected, as writes require an exchange of messages over the network, which costs both \CAP{CPU} cycles and network bandwidth.

At 20\% write ratio, HermesKV significantly outperforms the baselines (40\% over r\CAP{CRAQ}, 3.4$\times$ over r\CAP{ZAB}). The reason for HermesKV's good performance compared to alternatives is that it combines local reads with high-performance writes. 

\beginbseceval{rCRAQ} 
The \CAP{CRAQ} protocol is well suited for high throughput, comprising both inter-key concurrent writes and local reads. Nevertheless, r\CAP{CRAQ} performs worse than HermesKV across all write ratios, with the gap widening as write ratios increase. That difference has its root in the design of \CAP{CRAQ}.

First, reads in \CAP{CRAQ} are not always local. If a non-tail node is attempting to serve a read for a key for which it has seen a write but not an \CAP{ACK}, then the tail must be queried to determine whether the write has been applied or not. Therefore, increasing the write ratio has an adverse effect on reads, as more reads need to be served remotely via the tail node.

This disadvantage hints at a more important design flaw: the \CAP{CRAQ} design is heterogeneous, mandating that nodes assume one of three different roles -- head, tail, or intermediate -- each of which has different responsibilities. As such, 
the load is not equally
balanced, so the system is always bottlenecked by the node with the heaviest responsibilities. For instance, at high write ratios, the tail node is heavily loaded, as it receives read queries from all nodes. Meanwhile, at low write ratios, the tail has fewer responsibilities than an intermediate node since it only propagates acknowledgments up the chain, while an intermediate must also propagate writes downstream. 

\beginbseceval{rZAB} 
As expected, \CAP{ZAB} fails to achieve high throughput at non-zero write ratios, as it imposes a strict ordering constraint on {\em all} writes at the leader. The strict ordering makes it difficult to extract concurrency, inevitably causing queuing of writes and delaying subsequent reads within each session. At 1\% write ratio, r\CAP{ZAB} achieves 172 MReqs/s, which drops to a mere 16 MReqs/s for a write-only workload. 

\begin{figure}[t]
  \centering
  
  \includegraphics[width=0.7\textwidth]{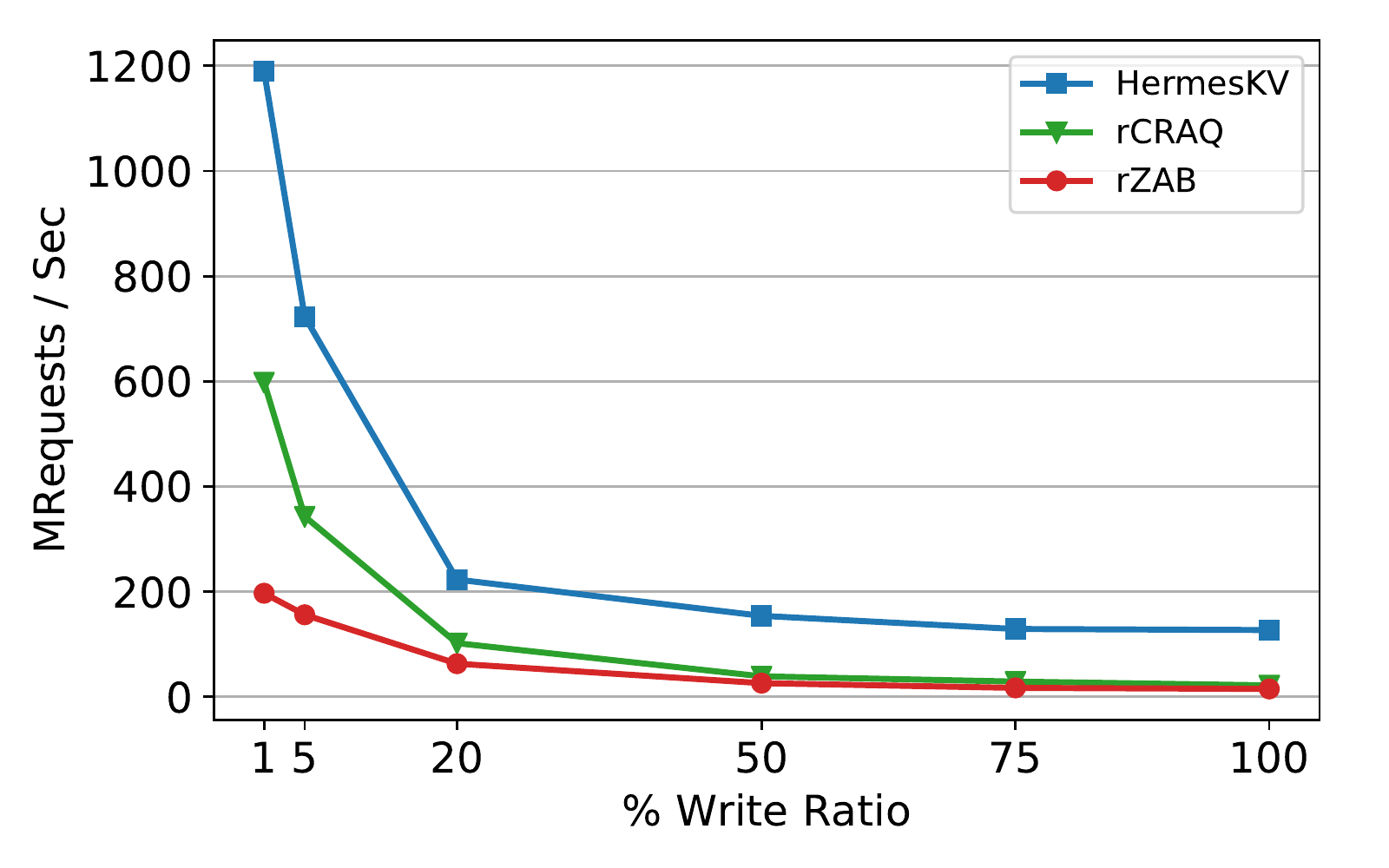}
  \mcaption{Throughput while varying write ratio under skew.}
  {Throughput for 1\% to 100\% writes under skew. [$\alpha = 0.99$, 5 nodes]}
  \label{figH:skew-line}
\end{figure}
\tsubsection{Throughput under Skew} 
\label{secH:skew}
We next explore how the evaluated protocols perform under access skew. We study an access pattern that follows a power-law distribution with a Zipfian exponent $a = 0.99$, as in \CAP{YCSB}~\cite{Cooper:2010} and recent studies~\cite{Dragojevic:2014, A&V:2018, RNovakovic:2016}. \cref{figH:skew-line} shows the performance of the three protocols when varying the write ratio from 1\% to 100\%. We discuss the read-only scenario separately.

\beginbseceval{Read-only} 
Similar to the uniform read-only setting, all three protocols achieve identical performance (4183 MReq/s) due to their all-local accesses. Unsurprisingly, the read-only performance under the skewed workload is higher than the uniform performance for all protocols. This is because, under a skewed 
workload, there is temporal locality among the popular objects, which is captured by the hardware caches.

\beginbseceval{HermesKV} 
HermesKV gracefully tolerates skewed access patterns, especially at low write ratios (achieving 1190 MReq/s at 1\% write ratio). Repeatedly 
accessing popular objects cannot adversely affect HermesKV's write throughput, as concurrent writes to popular objects can proceed without stalling (as explained in \cref{secH:conc-writes}). Meanwhile, read throughput thrives under a skewed workload because reads are always local in HermesKV, and as such can benefit from temporal locality.

\beginbseceval{rCRAQ} 
Similarly, r\CAP{CRAQ} benefits from temporal locality when accessing the local \CAP{KVS}, while write throughput is unaffected by the skew,
as multiple writes for the same key can concurrently flow through the chain. The problem, however, is that non-tail nodes cannot complete reads locally if they have seen a write for the same key but have not yet received an \CAP{ACK}. In such cases, the tail must be queried. Under skew,
such cases become frequent, with reads to popular objects often serviced by the tail rather than locally. Thus, at higher write ratios, the tail limits r\CAP{CRAQ}'s performance. 

\beginbseceval{rZAB} 
r\CAP{ZAB} is not affected by the conflicts created by the skewed access pattern, as it already serializes all writes irrespective of the object they write. In practice, r\CAP{ZAB} performs slightly better under skew, as hardware caches are more effective due to better temporal locality for popular objects.

\begin{figure*}[t]
  \centering
  \begin{subfigure}[b]{0.7\textwidth}
    \includegraphics[width=\textwidth]{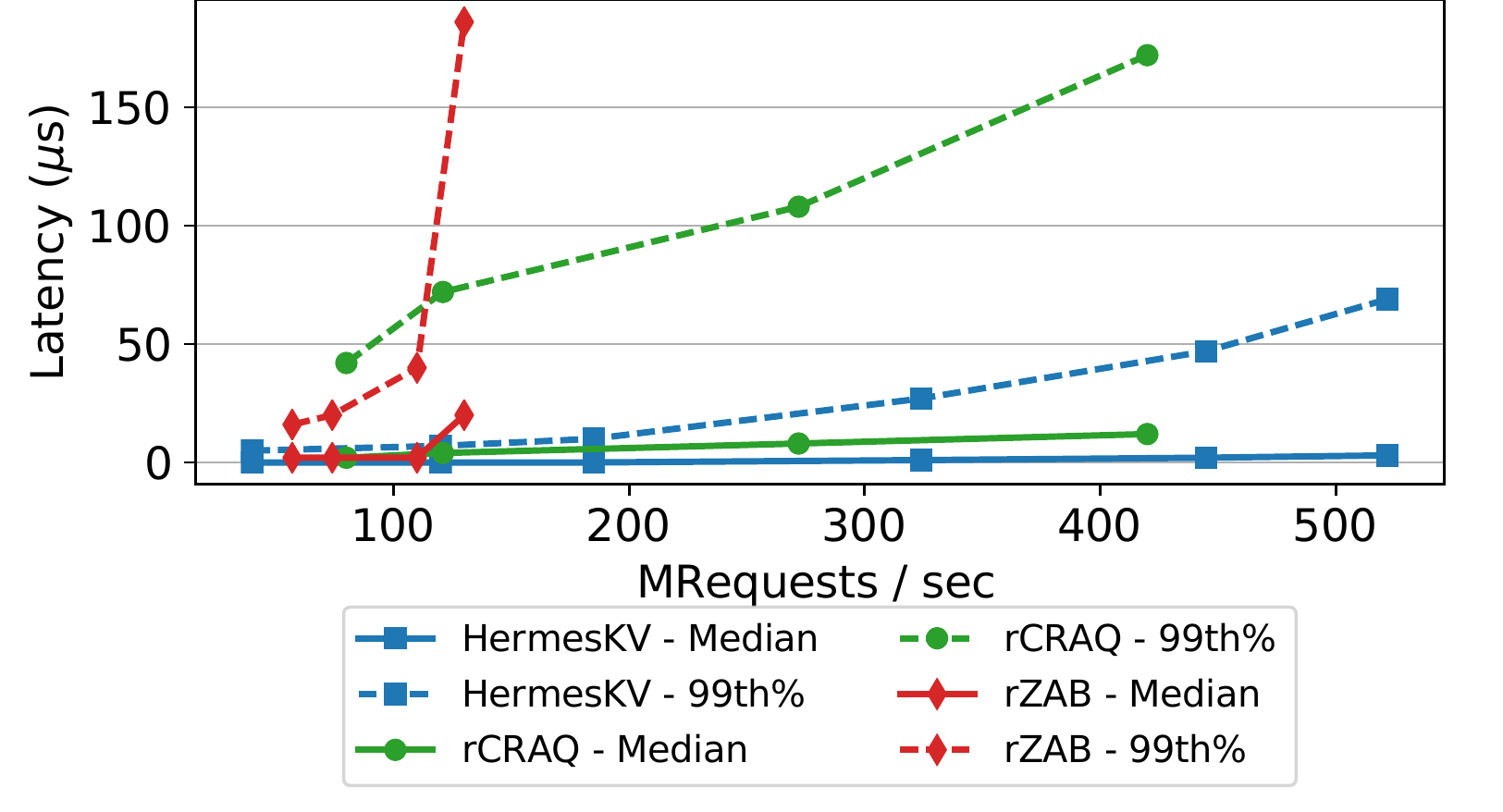}
    \caption{Latency vs. throughput. [uniform traffic, 5\% write ratio]}
    \label{figH:latency}
    \vspace{10pt}
  \end{subfigure}
  \begin{subfigure}[b]{0.7\textwidth}
    \includegraphics[width=\textwidth]{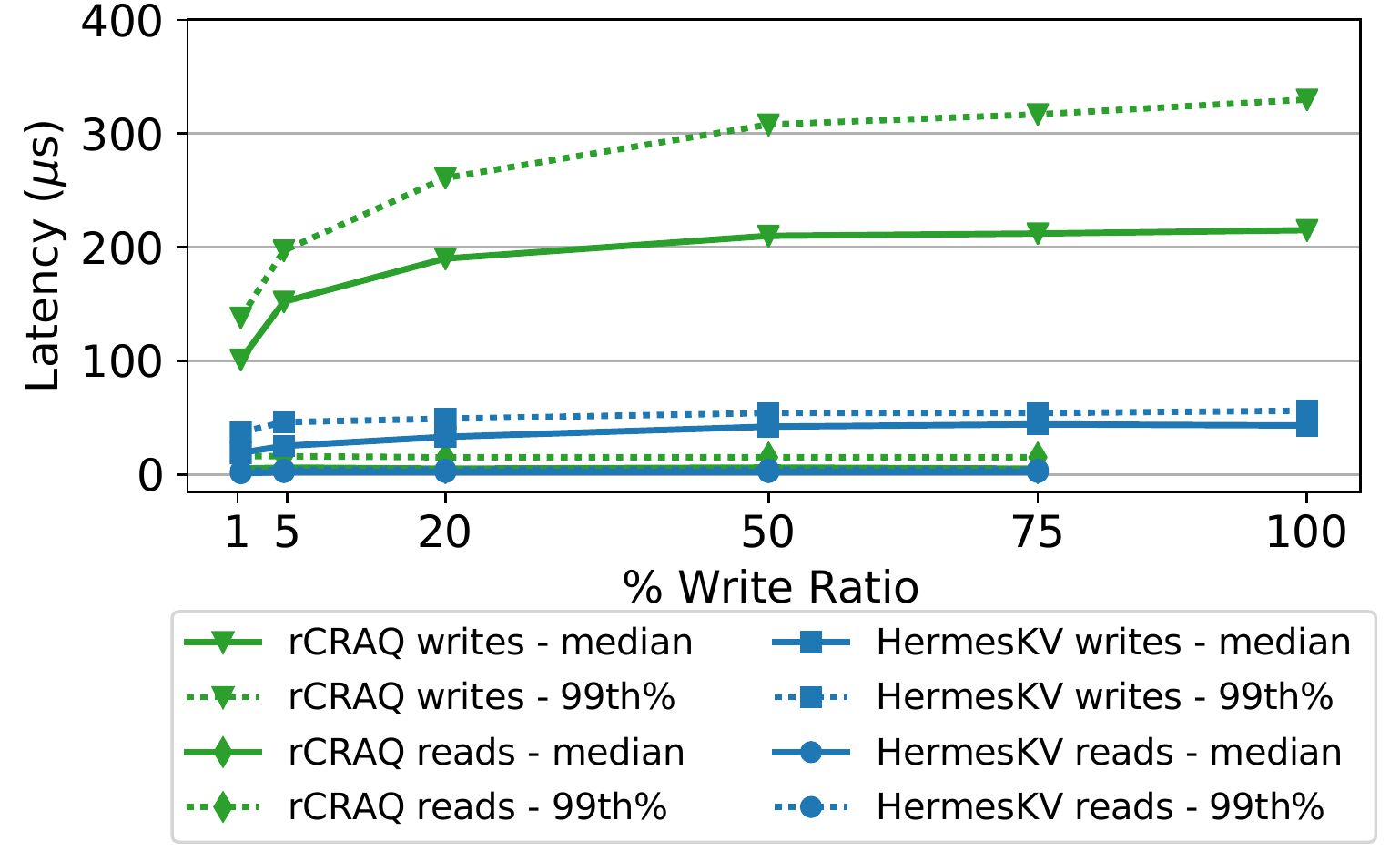}
    \caption{Median and 99th percentile. [uniform traffic]}
    \label{figH:latency-uni}
    \vspace{10pt}
  \end{subfigure}
  \begin{subfigure}[b]{0.7\textwidth}
    \includegraphics[width=\textwidth]{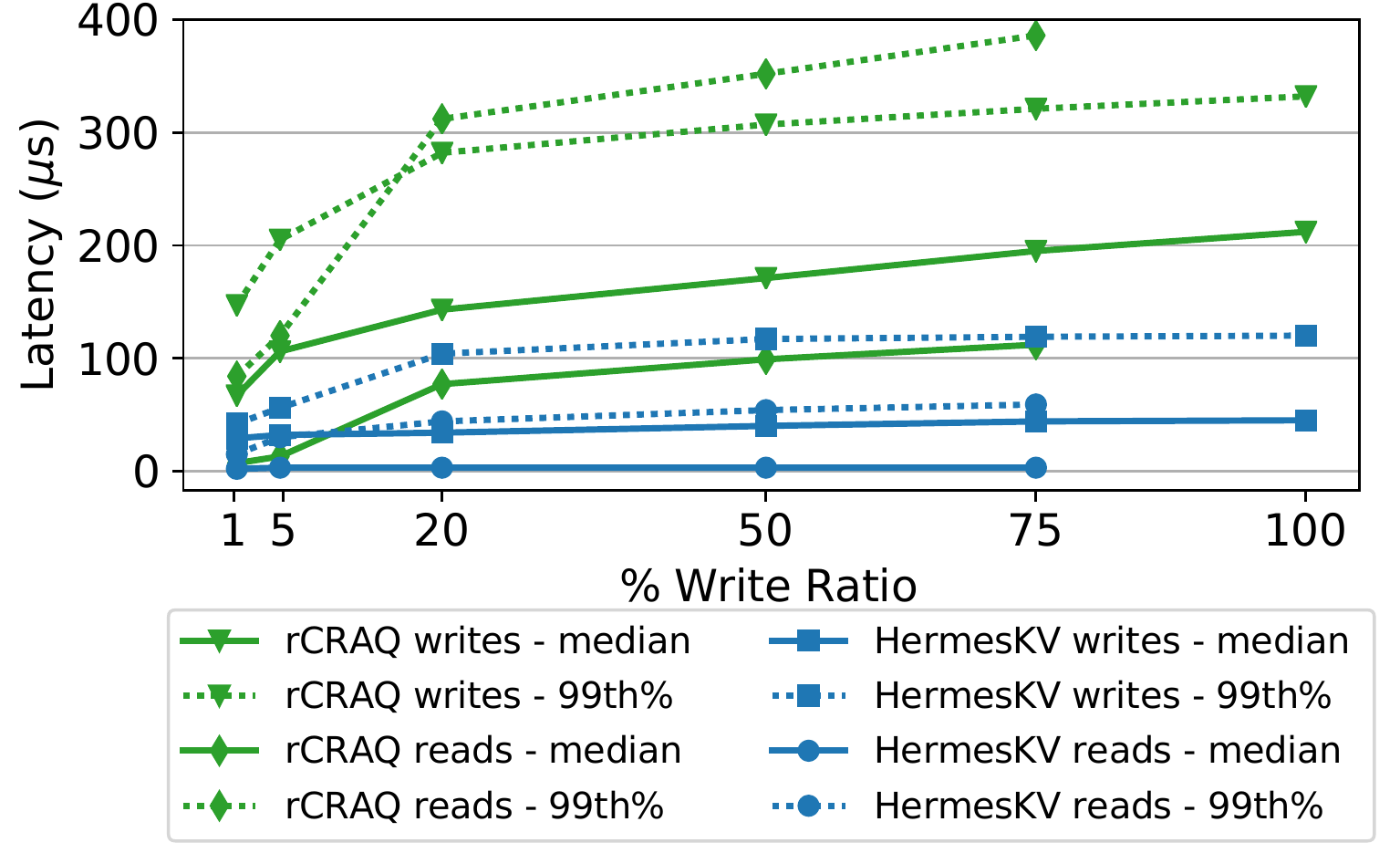}
    \caption{Median and 99th percentile. [$\alpha = 0.99$]}
    \label{figH:latency-skew}
  \end{subfigure}
  \mcaption{Latency analysis.}{Latency analysis. [5 nodes]}
\end{figure*}
\tsubsection{Latency Analysis}
\label{secH:latency}

\beginbsecBig{Latency vs. throughput}

\cref{figH:latency} illustrates the median (50th\%) and the tail (99th\%) latencies of the three protocols as a function of their throughput at 5\% write ratio. We measure the latency of each request from the beginning of its execution to its completion. 

All three systems execute reads locally, while writes incur protocol actions that include traversing the network. Therefore, at 5\% write ratio, we expect the median latency of all protocols to be close to the latency of a read and the tail latency to be close to that of a write. Consequently, the gap between the median and tail latency is to be expected for all systems and should not be interpreted as unpredictability.

\beginbseceval{HermesKV} 
The median latency of HermesKV is the latency of a read, and as expected, is consistently very low (on the order of 1$\mu s$) even at peak throughput because reads are local. Tail latency is determined by the writes. The tail latency increases with the load because writes traverse the network and thus can be subject to queuing delays as load increases. At peak throughput, the tail latency of HermesKV is 69$\mu s$.

\beginbseceval{rCRAQ} 
In r\CAP{CRAQ}, the median latency is the latency of a read, and as such, is typically on the order of a few microseconds. As expected, the tail latency, which corresponds to a write, is consistently high: at least 3.6$\times$ larger than HermesKV at the same throughput points (ranging from 42$\mu s$ at lowest load to 172$\mu s$ at peak load). The high write latency is directly attributed to the protocol design, as writes in r\CAP{CRAQ} need to traverse multiple network hops, incurring both inherent network latency and queuing delays in all the nodes.

\pagebreak 

\beginbseceval{rZAB} 
Like the other two protocols, r\CAP{ZAB} achieves a low median latency because of its local reads. However, even at 
moderate throughput,
its tail latency is much larger
(e.g., over 3.6$\times$ than that of Hermes at 75 MReq/s)
because of the high latency of the writes that must serialize on the leader. 

\vspace{10pt}
\beginbsecBig{Latency vs. write ratio} 

\cref{figH:latency-uni} and \cref{figH:latency-skew} depict the median and tail latencies of reads and writes separately, under both uniform and skewed workloads, when operating at the peak throughput of \CAP{CRAQ}, which corresponds roughly to 50--85\% of HermesKV's peak throughput. r\CAP{ZAB} cannot achieve sufficiently high throughput to be included in the figures. 

\beginbseceval{Uniform} 
HermesKV delivers very low, tightly distributed latencies across all write ratios, for both reads (2$\mu s$ -- 15$\mu s$) and writes (29$\mu s$ -- 42$\mu s$). As expected, r\CAP{CRAQ} exhibits a similar behavior for reads but not for writes. r\CAP{CRAQ} write latencies are at least 3.9$\times$ to 5.9$\times$ larger than the corresponding write latencies of HermesKV, with median latencies ranging from 101$\mu s$ to 215$\mu s$ while the tail latencies range from 138$\mu s$ to 330$\mu s$.

\beginbseceval{Skew} 
Under skew, the tail latencies of both reads and writes increase in HermesKV, as reads and writes are more likely to conflict on popular objects. The tail read latency is the latency of a read that stalls waiting for a write to return. Not surprisingly, that latency is roughly equal to the median latency of a write. 
Similarly,
the tail latency of a HermesKV write increases up to 120$\mu s$  because, in the worst case without failures, a write might need to wait for an already outstanding write (to the same key) issued from the same node.

In r\CAP{CRAQ}, the latencies of writes remain largely unaffected compared to the uniform workload. However, the behavior of reads changes radically because reads are far more likely to conflict with writes under skew; such reads are sent to the tail node. Consequently, the tail node becomes very loaded, which is reflected in both the median (up to 112$\mu s$) and tail (up to 386$\mu s$) read latencies. This is a very important result; while high write latencies are expected of r\CAP{CRAQ}, we show that read latencies can suffer as well, making \CAP{CRAQ} an undesirable protocol for systems that target low latency. 
\FloatBarrier

\begin{figure}[t]
  \centering
  \includegraphics[width=0.7\textwidth]{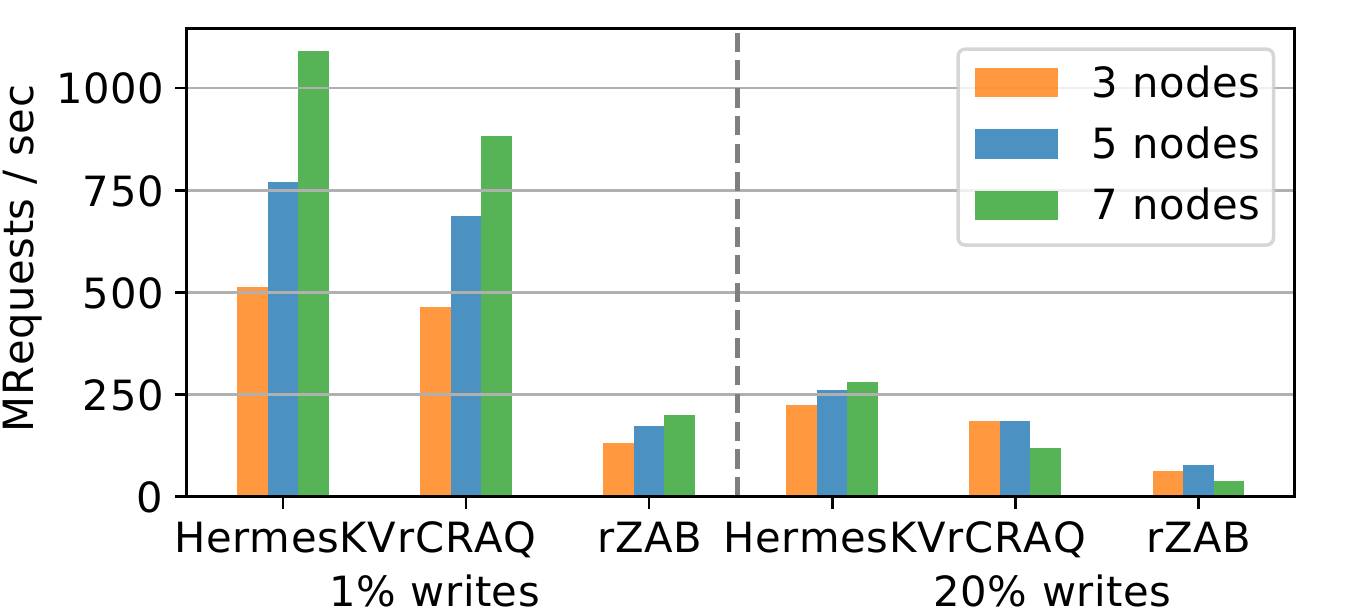}
  \mcaption{Scalability study.}
  {Scalability study. [uniform traffic] }
  \label{figH:scalability-line}
\end{figure}

\tsubsection{Scalability Study}
\label{secH:scalability}

To investigate the scalability of the evaluated protocols, we measure their performance by varying the replication degree. \cref{figH:scalability-line} depicts the throughput of the three protocols under 1\% and 20\% write ratios for 3, 5, and 7 machines.

\beginbseceval{HermesKV} 
Reads in HermesKV are always local and their overhead is thus independent of the number of replicas, allowing HermesKV to take advantage of the added replicas to increase its throughput. Therefore, HermesKV's scalability is dependent on the write ratio, achieving almost linear scalability with the number of replicas at 1\% writes while maintaining its performance advantage at 20\% write ratio.

\beginbseceval{rCRAQ} 
When scaling r\CAP{CRAQ}, the expectations are similar to HermesKV: reads are scalable, but writes are not. However, scaling the 
replicas in \CAP{CRAQ} implies extending the size of the chain. Consequently, more non-tail nodes redirect their reads to the tail node. Thus, the tail becomes loaded, degrading read throughput while also creating back pressure in the chain, which adversely affects write throughput. This phenomenon is apparent in \cref{figH:scalability-line}; at 20\% write ratio, r\CAP{CRAQ} throughput degrades when the chain is extended from 5 to 7 nodes. 

\beginbseceval{rZAB} 
r\CAP{ZAB} also performs reads locally and thus is expected to see a benefit from greater degrees of replication at low write ratios. However, write requests incur a large penalty in r\CAP{ZAB}, as the leader receives and serializes writes from all machines. When the leader cannot keep up with the write stream, the replicas inevitably fall behind, as the reads stall waiting for the writes to complete and the writes are queued on the leader. Indeed, in \cref{figH:scalability-line}, we observe that even though r\CAP{ZAB} scales well for a read-dominant workload, at a 20\% write ratio, increasing the replication degree from 5 to 7 cuts performance almost in half. Our results are in line with the original scalability analysis of Zookeeper~\cite{Hunt:2010}.

\vspace{-5pt}
\tnsubsection{Comparison to Derecho}
\vspace{-10pt}
\label{secH:eval-derecho}
In this section, we compare HermesKV's throughput
with the \CAP{RDMA}-optimized open-source Derecho~\cite{Jha:2019}, the state-of-the-art
mem\-ber\-ship-based 
variant of
Paxos. 
Derecho's codebase partitions work at each node across several threads (3--4) but does not support higher degrees of threading. 
To ensure the fairest possible comparison, we limit HermesKV to a single thread. 

\begin{figure}[t]
  \centering
  \includegraphics[width=0.7\textwidth]{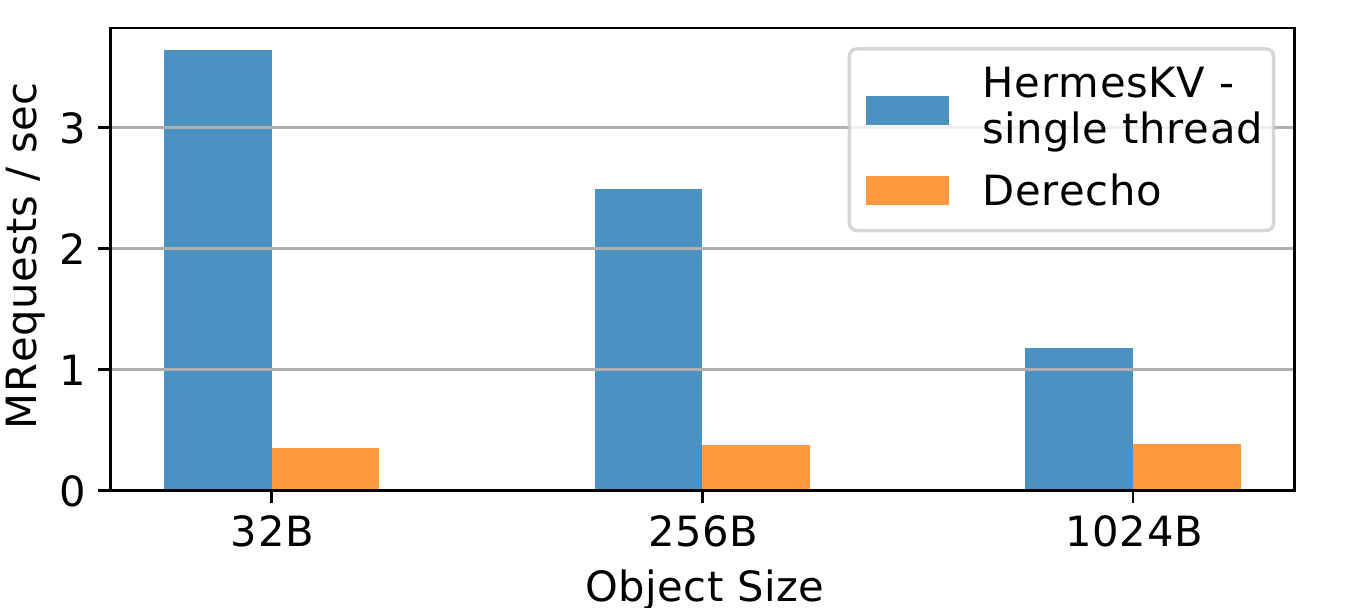}
  \mcaption{Comparison to Derecho.}
  {Comparison to Derecho. [uniform traffic, 5 nodes, write-only]}
  \label{figH:derecho}
\end{figure}

\cref{figH:derecho} shows the throughput of a write-only workload when the object size is varied from \CAP{32B} to \CAP{1KB}.
Such relatively small object sizes are typical for datastore workloads~\cite{Atikoglu:2012,Lim:2014}.
Despite being constrained to a single thread, HermesKV outperforms Derecho by an order of magnitude on small object sizes (\CAP{32B}) while maintaining its advantage even on larger objects (\CAP{3$\times$} at \CAP{1KB}). 
Derecho increases the performance of its totally ordered writes by exploiting monotonic predicates~\cite{Jha:2019}. Nevertheless, due to its lockstep delivery and its inability to offer inter-key concurrent writes, it fails to match the performance of Hermes.
We note that HermesKV's throughput naturally decreases as the object size increases and more bytes per request are transferred.

\begin{figure}[t]
  \centering
  \includegraphics[width=0.75\textwidth]{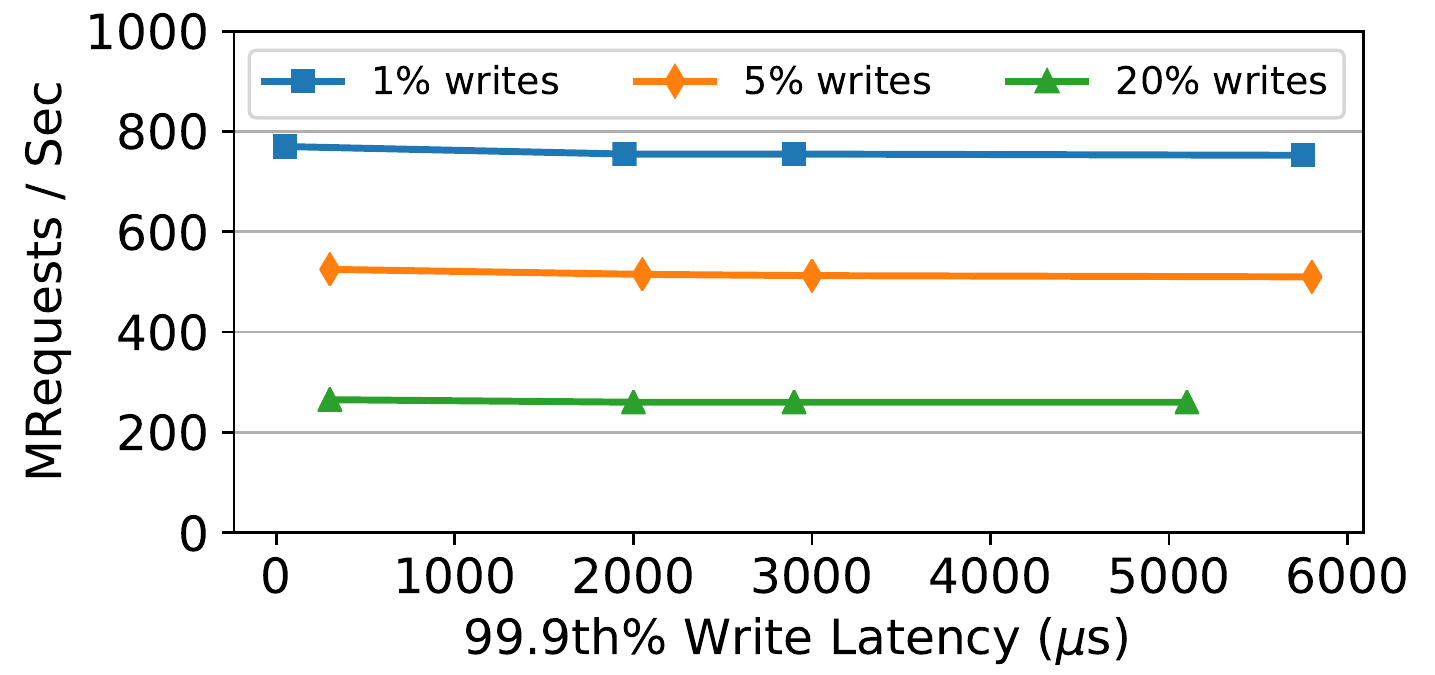} 
  \mcaption{Throughput while increasing the tail latency of writes.}
  {\newtext{Throughput of Hermes while artificially increasing write tail latency.
   [uniform traffic, 5 nodes, 5\% write ratio]}}
  \label{fig:tail}
\end{figure}

\tnsubsection{Tolerance to tail latency}
\label{sec:tail-tolerance}

\newtext{
Slowdowns (e.g., due to message losses) are less rare than node crashes and could affect a protocol's throughput.
In contrast to some majority-based replication protocols that may naturally tolerate such slowdowns, Hermes is a membership-based protocol and must gather acknowledgements from all the replicas when performing a write. Thus, raising the question of what impact this design decision has on its performance. More precisely, \textit{can an increase in tail latency adversely affect the throughput of Hermes?}
}

\newtext{
To answer this question, we artificially increase the tail latency by randomly choosing
writes and delay their commit. Figure~\ref{fig:tail} shows the throughput of Hermes when increasing the 99.9th\% latency of writes by an amount specified on the X axis. From the data, we conclude that Hermes is tail-tolerant because increasing the tail latency has a negligible effect on its throughput. The reason is that requests from different sessions and keys can be executed concurrently in Hermes and thus, if a write is stalled due to a slow tail effect, Hermes is able to process requests from other sessions to mask the impact of the slow request. The only time another session stalls on an outstanding request is if it performs a read to an in-progress write; however, as results show, this is rarely the case.}

\newtext{
Note that the above experiment focuses on inflated tail latencies (e.g., emulating messages losses). 
In contrast, if a replica \textit{always} responds slowly (e.g., due to an overloaded node), it would impact Hermes performance. Such a slow node should be removed from the membership to avoid reducing the throughput of Hermes.
However, notice that unlike centralized protocols, Hermes load-balanced design naturally minimizes such issues~\cite{Gavrielatos:21}.
}

\tsubsection{Throughput with Failures} 
\label{secH:failure-study}

Like \CAP{CRAQ} and other membership-based protocols, Hermes is designed to run in conjunction with a leased reliable membership (\CAP{RM}). 
In this section, we empirically identify the length of the lease timeout in order to minimize the likelihood of a false positive in failure detection. Given that timeout and industry-observed failure rates, we calculate the availability of Hermes. Finally, we evaluate the throughput of Hermes under failures.

\beginbsec{Timeout length}
To determine the required \CAP{RM} timeout in our setup, we 
conduct the following experiment. While Hermes is running at peak throughput, an additional process is spawned in each server. These processes exchange 40-byte messages over \CAP{RDMA}, emulating heartbeats through an inquire-response pattern, while logging the response latencies. 
After running the experiment for 10 hours, the highest observed latency was 6ms, likely arising from kernel (scheduler) interference.  

Ideally, the timeout would be sufficiently larger than the highest latency, minimizing 
the probability of a false positive. We quantify the meaning of "sufficiently larger" through a \emph{safety factor} $sf$, defining that the timeout must be $sf$ times larger than the highest observed latency. We set $sf$ equal to 25, which yields a timeout of 150ms.

\begin{figure}[t]
  \centering
  \includegraphics[width=0.75\textwidth]{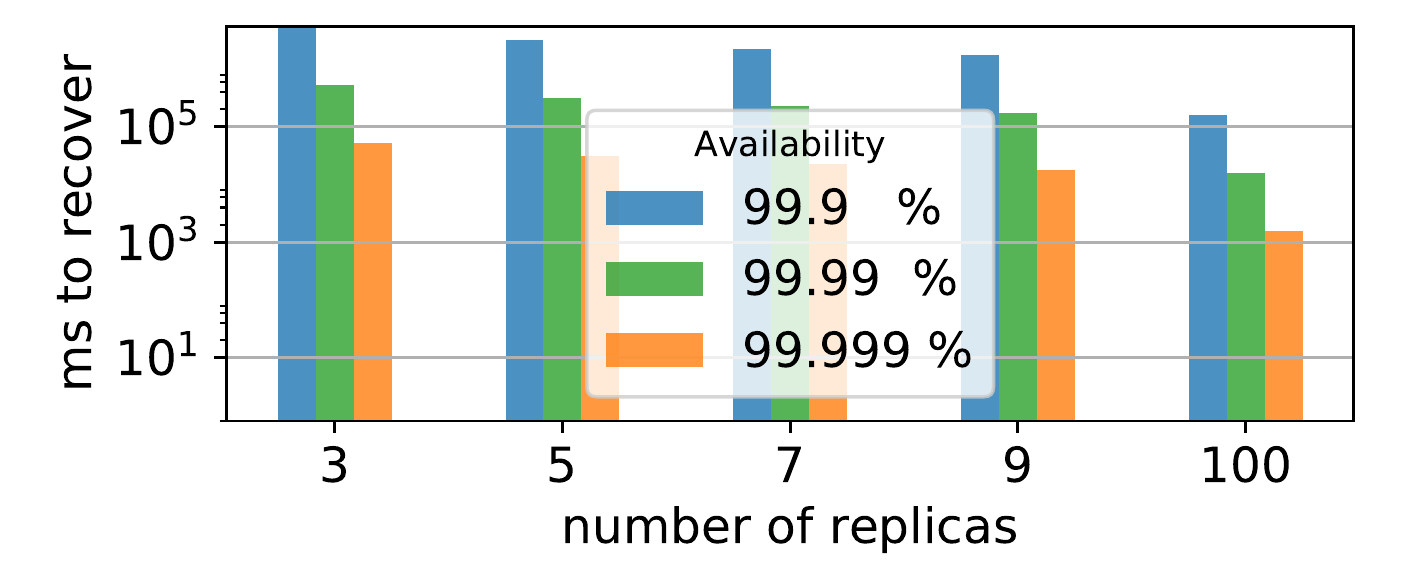} 
  \mcaption{Maximum recovery per crash based on availability targets.}
  {Maximum recovery time per failure while meeting various availability targets, based on a failure rate of two crashes per year per server.}
  \label{figH:latency-limit}
\end{figure}

\begin{figure}[t]
  \centering
  \includegraphics[width=0.7\textwidth]{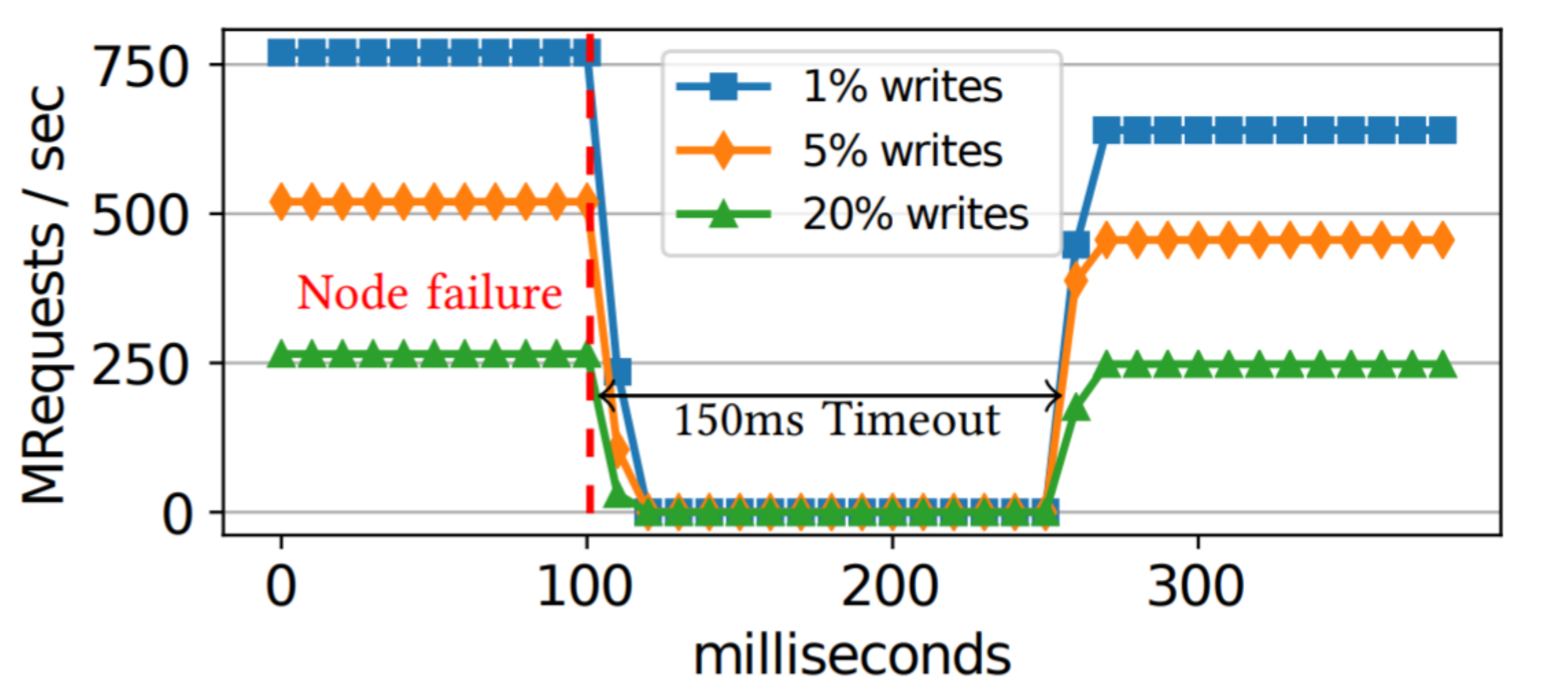} 
  \mcaption{Hermes throughput under a node crash.}
  {HermesKV under failure. [uniform traffic, 5 nodes, timeout=150ms]}
  \label{figH:failure-throughput}
\end{figure}

\beginbsec{Availability}
Using the observation from~\cite{Barroso:2018} that a server may crash up to two times per year, we plot in \cref{figH:latency-limit} the maximum recovery time per failure for various deployments with various availability targets. For example, a five-replica deployment achieves 99.999\% availability if it is able to recover from a server failure within 31 seconds, assuming each server fails twice per year.

Hermes is able to recover a few microseconds after the lease timeout (more details in the "Throughput under failure"), and thus the recovery time of Hermes is practically equal to its timeout. We note that the 150ms recovery time is two orders of magnitude smaller than the maximum allowed 31-second recovery time, enabling Hermes to comfortably achieve the common target of 99.999\% availability. Indeed with a recovery time of 150ms, a 5-replica deployment can achieve 99.999\% availability while tolerating 420 failures per server each year.

\beginbsec{Throughput under failure}
In order to study HermesKV's behavior when a failure occurs, we implement \CAP{RM}
similarly to~\cite{Kakivaya:2018} and integrate it with HermesKV. \cref{figH:failure-throughput} depicts the behavior of HermesKV when a failure is injected at 1\%, 5\%, and 20\% write ratios in a five-node deployment and a conservative timeout of 150ms. The throughput drops to zero almost immediately after the failure because all live nodes are blocked while waiting for acknowledgments from the failed node. After the timeout expires, the machines reach agreement
(via a majority-based protocol) to reliably
remove the failed node from the membership and subsequently continue operating with four nodes. The agreement part of the protocol entails exchanging a handful of small messages over an unloaded \CAP{RDMA} network, which takes only a few microseconds and is not noticeable in the figure. 
The recovered steady-state throughput is lower after the failure because one node is removed from the replica group.
\FloatBarrier

\begin{figure}[t]
  \centering
  \includegraphics[width=0.7\textwidth]{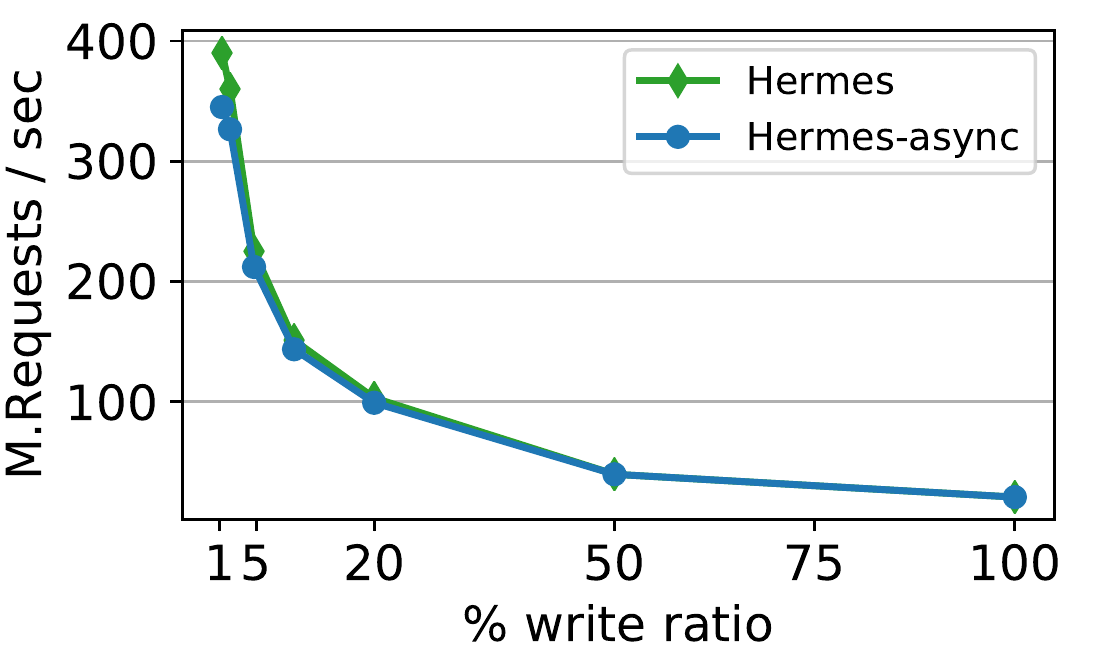}
  \mcaption{Hermes vs. Hermes-async throughput.}
  {Hermes vs. Hermes-async throughput. [5 nodes, varying writes, Cloudlab]}
  \label{figCA:hermes_xput}
\end{figure}

\tsubsection{Study under Asynchrony}
\label{secH:async-eval}
We conduct a brief study on Cloudlab~\cite{Duplyakin:19} to evaluate the performance of the asynchronous Hermes variant (Hermes-async), described in \cref{secH:async}. 
We use HermesKV and the same optimizations to compare the performance of Hermes-async with that of Hermes (i.e., the original variant with loosely synchronized clocks).
To that end, we perform two experiments on a cluster of 5 nodes equipped with a Xeon E5-2450 processor (8 cores, 2.1Ghz) and interconnected over 40Gb network cards (CX3 FDR IB). The \CAP{KVS} consists of one million key-value pairs, replicated in all five nodes. We use keys and values of $8$ and $32$ bytes, respectively, which are accessed uniformly.

\beginbsec{Throughput while varying the write ratio}
We first evaluate the throughput while varying the write ratio. As expected, \cref{figCA:hermes_xput} shows that Hermes outperforms Hermes-async for low write ratios (up to 5\%), as Hermes-async needs to send extra messages to validate reads. However, the difference is marginal because several reads can be verified by a constant lightweight message carrying just the epoch\_id. For higher write ratios, a write is almost always readily available and able to verify reads in Hermes-async, thus rendering the throughput difference negligible to non-existent.

\begin{figure}[t]
  \centering
  \includegraphics[width=0.7\textwidth]{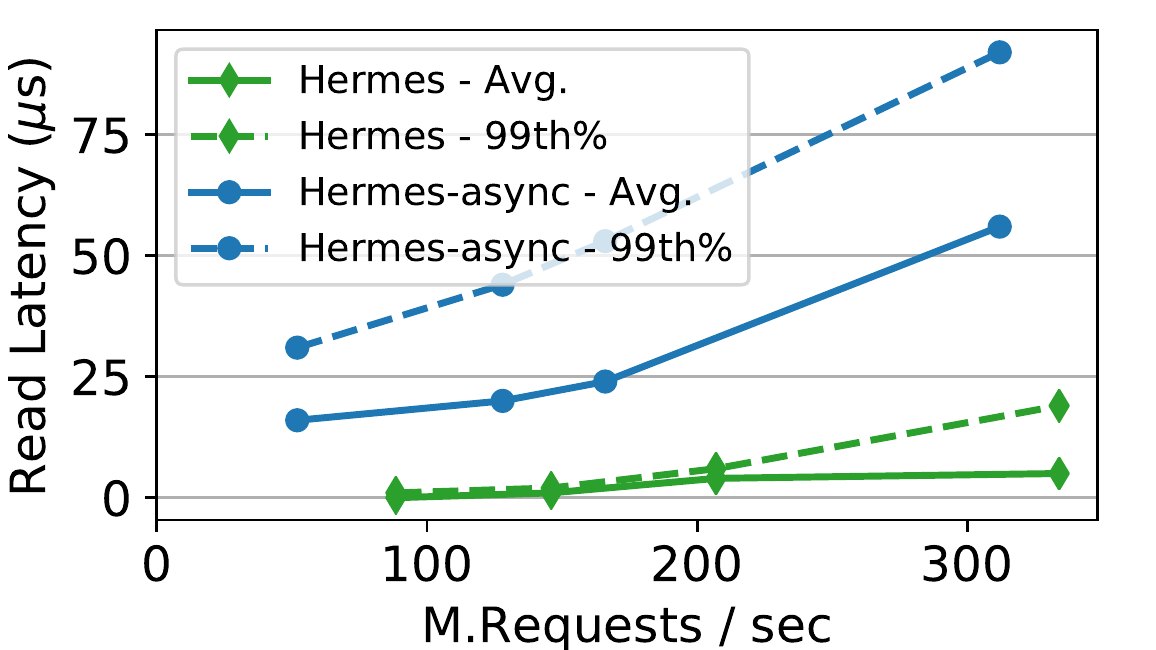} \mcaption{Hermes vs. Hermes-async latency.}{Hermes vs. Hermes-async latency. [5 nodes, 2\% writes, varying load, Cloudlab]}
  \label{figCA:hermes_latency}
\end{figure}
\beginbsec{Latency while varying the load}
We also study the latency of reads in Hermes and Hermes-async while varying the load.
We focus on a read-dominant workload with just 2\% write ratio, in which 
reads in Hermes-async will most likely need to send extra messages for validation.
The results are shown in \cref{figCA:hermes_latency}. As expected, both the average and the tail latency of Hermes-async are higher than that of Hermes, reaching about $50\mu s$ average latency with maximum load.
This difference is justified because -- unlike Hermes' reads, which immediately return the local value -- reads in Hermes-async must wait for a round-trip to a majority of replicas. 



\tsection{Related Work}
\label{secH:related-work}

\beginbsec{Consensus and atomic broadcast}
State machine replication (\CAP{SMR})~\cite{Schneider:1990}
provides linearizability by explicitly ordering all client requests (reads and writes) and requiring all replicas to execute requests in the determined order. \CAP{SMR} can be implemented using any fault-tolerant consensus or atomic broadcast algorithm to order the requests.  
Numerous such algorithms have been proposed~\cite{Chandra:1996, Birman:1987, Liskov:2012, Oki:1988}, the most popular of which are variants of Paxos~\cite{Lamport:1998}.
Recent works present optimized variants of these protocols that exploit commutative operations~\cite{Moraru:2013, Aguilera:2000, Lamport:2005} and rotating coordinators~\cite{Mao:2008}. 
Others leverage a ring-based topology~\cite{Marandi:2010, Amir:1995, Guerraoui:2007}, similar to \CAP{CRAQ}, to increase throughput but at the cost of latency.

Most of these
protocols are majority-based and sacrifice performance for a failure model without \CAP{RM} support.
Therefore, they typically enforce strong consistency at the cost of performance by sacrificing either local reads or concurrency.
An abundance of such protocols forfeits local reads~\cite{Lamport:1998, Moraru:2013, Attiya:1995, Lamport:2001, Lamport:2006, Lamport:2005, Mao:2008, Ekstrom:2016, Birman:1987, Bolosky:2011, Ongaro:2014, Li:2016-NoPaxos, Marandi:2011, Poke:2015}, thereby incurring a significant penalty on read-dominant datastore workloads. 

Meanwhile, protocols that allow local reads sacrifice performance on writes. A recent atomic broadcast protocol offers local reads but does so by relaxing consistency and applying writes in lockstep~\cite{Poke:2017}. Chandra \etal\cite{Chandra:2016} present a protocol with linearizable local reads 
through object leases that serializes writes on a leader.
\CAP{ZAB}~\cite{Reed:2008}, a characteristic example of such protocols, enables local reads and serializes writes on a leader, but without using object leases; thus increasing performance but at the cost of consistency. 
As shown in our evaluation, Hermes significantly outperforms \CAP{ZAB} with its decentralized and inter-key concurrent writes.

\beginbsec{Per-key leases} 
Linearizable protocols that use object leases for local reads, such as those in~\cite{Chandra:2016, Moraru:2014}, could be deployed on a per-key basis (\ie one protocol instance for each key) to match the inter-key concurrency -- but not latency -- of writes in Hermes. However, this mandates a lease for \textit{each} individual key, which is not scalable for realistic datastores with millions of keys. In this approach, for linearizable local reads, leases must be continuously renewed for each key, even in the absence of writes or reads. This renewal costs at least $\Theta(n)$ messages (where \textit{n} = number of replicas) per key and must occur 
before each lease expires,
causing significant network traffic. Moreover, the lease duration cannot be very long, since a long lease would translate into similarly long unavailability upon a fault. In contrast, Hermes, with its invalidating writes and only a single \CAP{RM} lease per replica, offers local reads while being fully inter-key concurrent at a message cost independent of the number of the keys stored in the datastore.  


\beginbsec{Hardware-assisted replication}
Some proposals leverage hardware support to reduce the latency of reliable replication, such as \CAP{FPGA} offloading~\cite{Istvan:2016} and programmable switches~\cite{Jin:2018, Dang:2015, Li:2016-NoPaxos, Kogias:20}. For instance, 
Zhu \etal\cite{zhu:2019} use program\-mable switches for in-network conflict detection to allow local reads from any replica. 
Other works tailor reliable protocols by exploiting \CAP{RDMA}~\cite{Poke:2015, Wang:2017, Behrens:2016}. 
Hermes offers local reads without hardware support. When evaluated over 
\CAP{RDMA}, Hermes significantly outperforms Derecho, which represents the state of the art of \CAP{RDMA}-based approaches.

\beginbsec{Optimized reliable replication}
A recent work~\cite{Park:2019} proposed a primary-backup optimization to reduce the exposed write latency for external clients. However, its correctness relies on
commutative operations.
Howard's optimization~\cite{Howard:2019} allows Paxos to commit after 1 \CAP{RTT} in conflict- and failure-free rounds, but reads are not local.
In contrast, Hermes is not limited to commutative operations and affords local reads. 

\beginbsec{Reliable transaction commit}
Hermes provides single-object linearizable reads, writes, and \CAP{RMWs}, but does not offer reliable multi-object transactions. 
The distributed transaction commit requires an agreement as to whether a transaction should atomically commit or abort. A transaction may only be committed if all parties agree on it. 
A popular protocol to achieve this is the two-phase commit (\CAP{2PC})~\cite{Gray:78}. However, the \CAP{2PC} is a blocking protocol and must be extended to three phases (\CAP{3PC}) to 
tolerate coordinator failures~\cite{Guerraoui:1995, Guerraoui:2002, Skeen:1981}. 
A more common way to achieve reliable transactions is layering a transactional protocol over a reliable replication protocol~\cite{Zhang:2013, Kraska:2013, Corbett:2013}. For instance, FaRM and Sinfonia use a primary-backup protocol~\cite{Dragojevic:2015, Aguilera:2007}.
In this type of setting, Hermes could be used as the underlying reliable replication protocol to increase locality and improve performance.

\tsection{Discussion}
\label{secH:discussion}
\beginbseq{Are local reads beneficial in a large-scale datastore} 
Throughout this chapter, we report the latency of operations with respect to a
node (replica) in a distributed replicated datastore.
In a large-scale datastore, clients might be external and not co-located with the replica they want to access. Although in this case reads in Hermes do not provide locality with respect to the client, they still ensure low latency and cost. This is because, 
in Hermes, a remote read from an external client would be solely served by one replica without 
additional message delays or coordination among replicas.

\beginbsec{Reducing write latency of external clients}For the protocols discussed in this work, if clients are external, an additional round-trip is required to reach and obtain a response from the replica ensemble. Thus, the common-case exposed latency for an external client when committing a write in Hermes is 2~\CAP{RTT}s. To reduce response time,
followers can send \CAP{ACK}s to both the coordinator of the write and the client.
This reduces the latency in completing linearizable writes from external clients to 1.5~\CAP{RTT}s.
The message cost of this optimization (approximately twice the number of \CAP{ACK}s 
of the baseline protocol) is linear with the replication degree.

\beginbsec{Reducing \CAP{RMW} conflicts with \textit{opportunistic leadership}}
\newtext{
Despite Hermes' decentralized design, concurrent writes from different coordinators need never abort, even if they target the same key, as they can be safely linearized directly at the end-points based on their logical timestamps. Thus, concurrent writes in Hermes are always non-conflicting. 
However, \CAP{RMW}s to the same key in Hermes are conflicting and may abort (as detailed in \cref{secH:hermes-RMWs}). 
}

\newtext{
Under high \CAP{RMW} conflicts to the same key a leader-based design could reduce conflicts and aborts. Such high conflicts would only arise under skewed distributions and for just a few hot keys which must also be dominantly updated via \CAP{RMW}s. For those keys\footnote{\newtext{Techniques to identify such popular keys are discussed in \cref{secC:cache}.}} one could apply what we call an \textit{opportunistic leader} optimization over Hermes, in which a hash function is used to steer update requests for such a hot key to the same Hermes replica. In short, such an opportunistic leader in Hermes reduces conflicts and enables batching of concurrent \CAP{RMW} requests to the same (hot) key, while being a best-effort optimization. 
In other words, unlike native leader-based protocols, Hermes would still safely handle any updates issued by a non-(opportunistic-)leader.
%
}

\tsection{Summary}
\label{secH:conclusion}

In this chapter, we introduced Hermes, 
a membership-based reliable replication protocol that offers both high throughput and low latency. Hermes utilizes 
invalidations and logical timestamps to achieve linearizability, with local reads and high-performance updates at all replicas. In the common case of no failures, Hermes broadcast-based writes are non-conflicting and 
always commit after a single round-trip.
Hermes also tolerates node and network failures through its safe write replays.
Our evaluation of Hermes against state-of-the-art protocols shows that it achieves superior throughput at all write ratios and considerably reduces tail latency.
Finally, we demonstrated that Hermes can be applied safely even under asynchrony, with a modest latency penalty on reads but without sacrificing its high throughput.

\markedchapterTOC
{Zeus}
{Zeus:\nextlinepdf Locality-Aware Replicated Transactions}
{Zeus: Locality-Aware Replicated Transactions}
\label{chap:Zeus}

\equote{-40}{10}{Think global, act local.}{Patrick Geddes}

The last two chapters focused on a single-object interface over a statically sharded datastore.
This chapter applies invalidation-based protocols in a datastore that dynamically shards data and offers a richer transactional interface with data availability and the performance benefits of locality awareness.


\tsection{Overview}
\label{secZ:introduction}

Cloud applications over commodity infrastructure are becoming increasingly popular. 
They require distributed, fast, and reliable datastores.
Recent 
\linebreak
in-memory datastores that operate within a datacenter and leverage replication for fault tolerance (FaRM \cite{Dragojevic:2014}, FaSST~\cite{F-Kalia:2016}, and DrTM~\cite{wei2015fast}) offer strongly consistent distributed transactions in the order of millions per second.
They do not make any assumptions about the workloads and rely on highly optimized remote access primitives (e.g., \CAP{RDMA}) to enable a variety of use cases.  

These datastores run \CAP{OLTP} workloads with transactions involving a small 
\linebreak 
number of objects. 
In addition, many applications have a high degree of 
\linebreak 
locality. 
For example, many transactions in a cellular control plane involve one user always accessing the same set of objects (e.g., the nearest base station or the same call forwarding number~\cite{TATP:2009}). 
Many Internet middleboxes mostly access the same state for all packets of a single flow (e.g., intrusion detection systems~\cite{Woo:18}).
Bank transactions often recur between the same parties~\cite{cahill2009serializable, Venmo17, Venmo20}. 
As Stonebraker \etal report~\cite{Harding:17}, a transactional concurrency control scheme can derive significant benefits from leveraging application-specific characteristics such as locality.

Existing works~\cite{F-Kalia:2016, Dragojevic:2014, wei2015fast} can exploit locality through \textit{static sharding} -- if and only if all objects involved in each transaction are stored on the same node.
Consequently, static sharding only helps if the optimal placement is known a priori and never changes. 
However, this is often not the case for two main reasons. 
First, the set of objects involved in a transaction may change over time. 
For instance, as a mobile phone user moves, her {\em cellular handover} transaction involves different base stations.
Second, the popularity of each object changes over time, be it a network service or a financial stock. 
If several popular objects are located on the same server, the server becomes a bottleneck, and the popular objects should be spread across servers.
In both cases, {\em the rate of changes in access locality is multiple orders of magnitude lower than the rate of processed transactions} (which is in the millions per second). 
We describe these cases in more detail in \cref{secZ:background}.

In contrast, \textit{dynamic sharding}, where objects are moved across nodes on demand, helps both
when the set of objects involved in a transaction changes or when object popularity shifts. 
In the first case, dynamic sharding ensures that all objects involved in a transaction are co-located, thereby reducing expensive remote accesses. 
In the second case, dynamic sharding allows the most popular objects to be quickly spread out, thus alleviating bottlenecks. 
However, state-of-the-art replicated datastores~\cite{F-Kalia:2016, Dragojevic:2014, wei2015fast} do not support dynamic object sharding.
Once the existing sharding is no longer optimal, they revert to remote transactions. Remote transactions are inherently slower because they impose the overhead of several round-trips, both to execute a transaction via remote accesses 
and to atomically commit it.
The overhead of the latter results from the complexity of distributed atomic commit for conflict resolution under the uncertainty of faults.

Several systems propose application-level load balancer designs that enable applications to make fine-grained decisions regarding which node each transaction should be routed to~\cite{Adya:2016, banerjee2015scaling, ahmad2020low, annamalai2018sharding}.
However, most of these systems rely on custom datastores that either do not provide strong consistency or are not as fast as the state-of-the-art datastores~\cite{F-Kalia:2016, Dragojevic:2014, wei2015fast}. 
As Adya \etal\cite{adya2019fast} argue, there is a need for a general distributed protocol that provides strongly consistent transactions and better exploits dynamic locality.

In this chapter, we address the problem of high-performance dynamic sharding for transactional workloads by presenting a novel distributed datastore called {\em Zeus}.
The key insight behind Zeus is that, for many workloads, the benefits of local execution outweigh the cost of (relatively infrequent) re-sharding. Zeus capitalizes on this insight through two novel 
reliable invalidation-based
protocols designed from the ground up to exploit locality in transactional workloads.
One protocol is responsible for reliable (atomic and fault-tolerant) object ownership migration, requiring at most 1.5 round-trips during common fault-free operation.
Using this protocol, when executing a transaction Zeus moves all objects to the server executing that transaction and ensures exclusive write access. 
Once this is done, and unless the access pattern changes, all subsequent transactions to the same set of objects are executed entirely locally, eschewing the need for costly distributed conflict resolution. 
The second protocol is a fast reliable commit protocol for the replication of localized transactions. 
By combining these two protocols, Zeus achieves the performance and simplicity of single-node transactions with the generality of distributed transactions.
To further exploit locality, Zeus reliable commit enables local yet consistent read-only transactions from all replicas.

Zeus' design provides an extra benefit in that it allows for easy portability of existing applications. 
Since most Zeus transactions are local, Zeus can pipeline executions without compromising correctness.
A subsequent transaction need not wait for the replication of the current one.
This is in contrast to the existing in-memory distributed transactional datastores~\cite{F-Kalia:2016, Dragojevic:2014, wei2015fast}, in which each transaction blocks until the replication is finished.
To mitigate the effects of blocking, these datastores use custom user-mode threading (e.g., co-routines) that requires substantial effort when porting existing applications. 
In contrast, Zeus' transaction pipelining enables easy porting of legacy applications onto it, making them distributed and reliable while reaping the performance benefits of locality with minimal developer effort.

We implement Zeus and evaluate it on several relevant benchmarks: Smallbank~\cite{cahill2009serializable}, Voter~\cite{oltp2013bench}, and \CAP{TATP}~\cite{TATP:2009}. 
We also introduce and implement a new benchmark that models handovers in a cellular network based on observed human mobility patterns. 
To demonstrate the ease of porting existing applications to Zeus, we port several networking applications that exhibit locality: a cellular packet gateway~\cite{OpenEPC}, an Nginx server~\cite{nginx_session_persisten}, and the \CAP{SCTP} transport protocol~\cite{usrsctp:2015}.

\noindent In brief, the main contributions of this chapter are as follows:

\squishlist
    \item \textbf{We introduce \textit{Zeus}, a reliable locality-aware transactional datastore} (\cref{secZ:design}) that replicates data in-memory to ensure availability.
    Unlike state-of-the-art strongly consistent transactional datastores, transactions in Zeus are fast by virtue of exploiting dynamic sharding and locality that exists in certain transactional workloads (as demonstrated in \cref{secZ:evaluation}).
    
    \item \textbf{We propose two invalidation-based reliable protocols} (\cref{secZ:sharding-protocol} and \cref{secZ:zeus-protocol}): an \textit{ownership protocol} for dynamic sharding that quickly alters object placement and access levels across replicas; and a transactional protocol for fast pipelined \textit{reliable commit} and local read-only transactions from all replicas. Both protocols, which ensure the strongest consistency under concurrency and faults, are verified in \CAP{TLA$^{+}$}. 
    
    \item \textbf{We implement and evaluate Zeus} (\cref{secZ:system} and \cref{secZ:evaluation}) 
    over \CAP{DPDK} on a six-node cluster, using three standard \CAP{OLTP} benchmarks and a new 
    \linebreak
    cellular handover benchmark. For workloads with high access locality, 
    Zeus achieves up to 2$\times$ the performance of 
    state-of-the-art \CAP{RDMA}-optimized
    \linebreak
    systems while using less network bandwidth and without relying on \CAP{RDMA}.
    On the handovers benchmark, Zeus' performance with dynamic sharding is just 4\% to 9\% from the ideal of all-local accesses.
    We also demonstrate the ease of portability by porting three legacy applications, showing scalability and reliability with little to no performance drop. 
\squishend
\tsection{Objectives and Motivation}
\label{secZ:background}

\vspace{-5pt}
We first describe high-level objectives that datacenter operators and application developers desire in a datastore. 
We next discuss the opportunities that arise with regard to local access patterns and analyze why they have not been fully explored before. 

\vspace{-10pt}
\tsubsection{Datastore Design Objectives}
\label{secZ:design_objectives}
\vspace{-5pt}

Our goal is to design an intra-datacenter shared-nothing transactional database for \CAP{OLTP} workloads that allows programmers to deploy their software on top of a distributed infrastructure without needing to re-architect the application. 
More specifically, we want to provide the following:

\vskip 1pt \noindent {\bf Performance and reliability. } 
Our target is to have a reliable datastore capable of processing millions of operations per second. 
Moreover, to remain available despite node failures,
each state update needs to be replicated across nodes.

\vskip 1pt \noindent {\bf Transactions. } 
A single operation may arbitrarily access or modify multiple objects. 
A notion of transaction guarantees that either all modifications are committed or none are. 
This is in contrast to many widely used in-memory key-value stores (e.g.,~\cite{redis:2020}), which essentially provide only single-object primitives and some generalizations as an afterthought.

\vskip 1pt \noindent {\bf Strong consistency. } 
We want to provide a simple programming model where a programmer has the intuitive notion of a single copy of state, despite the state being replicated for reliability.
This model requires strongly consistent distributed transactions guaranteeing strict serializability. Recall that, under strict serializability, all transactions appear as if they are atomically performed at a single point in real time to all replicas between their invocation and response.

\vskip 1pt \noindent {\bf Support for legacy applications. } 
The state-of-the-art in-mem\-ory datastores~\cite{F-Kalia:2016, Dragojevic:2014, wei2015fast} meet the above criteria. 
However, when executing remote transactions, they block the associated threads.  
To mask the performance cost of blocking, they rely on 
transaction multiplexing and 
user-mode threads~\cite{F-Kalia:2016}.
However, this makes porting existing applications on top of these frameworks difficult.
Our goal is to provide a datastore that allows legacy applications to run on top of it without mandating modifications to the existing architecture.

\tsubsection{A Case for Access Locality}

As noted in \cref{secZ:introduction}, many real-world applications exhibit transactional access patterns with a high degree of locality. 
In these cases, data are usually sharded for efficiency. 
However, the optimal sharding may change over time for two reasons:
changes in object popularity or changes in access locality. 
In this chapter, we use the term \textit{locality} 
to refer specifically to the temporal reuse of transactions between (spatially related) objects that reside on the same node.

Let us consider changes in locality via an example of call handovers in a cellular network. 
Every time a phone wakes up to process data traffic (a {\em service request}) or goes to sleep (a {\em release request}), the cellular control plane updates various objects related to the phone and to the base station to which the phone is attached. 
This is an example of data access locality, where each consecutive operation on the same phone accesses the same two objects (the phone and the base station contexts).

However, the access locality may slowly and gradually change over time due to mobility. 
Every time a cellular user moves from one base station to another, her phone performs a {\em handover} operation.
This is a transaction that involves three entities: the phone, the old base station that the user is leaving, and the new base station the user is connecting to. 
As the user travels (e.g., during a daily commute), her phone performs many such transactions, each involving one object that stays the same (the phone context) and two other objects that continuously change (contexts of the base stations along the trip). 
Once the user finishes her commute, the access locality resumes, and every subsequent {\em service request} and {\em release} for the user again involves a single base station (the one the user is currently attached to, which is different from the one she was attached to at the beginning of her commute).

This change is slow, as people are stationary most of the time. 
A study~\cite{bostom2013mobility} shows that, on average, a person makes five one-way trips per day with a total length of 100km for drivers and 20km for non-drivers. 
Consequently, handover requests are only between 2.5\% and 5\% of service and release requests~\cite{mohammadkhan2016considerations, cell_params}, while the vast majority of service and release requests repeatedly include the same base station. 
Another fact that further improves locality in this scenario is
that a base station will only take part in handovers with other base stations that are geographically close to it.

The optimal sharding should adapt to keep relevant objects together in the same node. 
In this example, it should strive to keep the contexts of a phone and the base station with which it is associated on the same node. 
However, based on the above observations regarding user mobility, 
re-sharding will occasionally need to happen, though only for a single-digit fraction of transactions.
We further discuss and evaluate this example in \cref{secZ:evaluation}.

Another example of access locality is peer-to-peer financial transactions. 
Several studies of the popular peer-to-peer mobile payment system Venmo~\cite{Venmo17, Venmo20} show that transactions mainly occur among groups of friends and that the transaction graph exhibits a greater local clustering than Facebook and Twitter graphs. 
Moreover, as noted by Unger~\etal\cite{Venmo20}, the network remains largely consistent across studies, indicating slow temporal changes in the interaction graph. 
We study this case using publicly available data from a recent Venmo study~\cite{venmo-dataset} and evaluate it on a popular financial transaction benchmark, Smallbank~\cite{cahill2009serializable} (discussed in \cref{secZ:evaluation}).

The optimal sharding may also change due to a shift in object popularity.
One example of this can be found in the Voter benchmark~\cite{oltp2013bench}, which we also evaluate in \cref{secZ:evaluation}. 
In a long-lasting online public contest (e.g., Eurovision), many users vote for a few contestants. 
The optimal sharding should spread the load evenly and would ideally put each of the most popular contestants on a separate server
while potentially grouping the least popular contestants together on a single server. 
However, the popularity of each contestant changes over time, and as she receives more or fewer votes, the optimal sharding changes, as well.
As in the previous example, each transaction involves only a few objects (a voter and a contestant), and the frequency of change in the optimal sharding is much lower than the frequency of the voting transactions. 

Another example is the stock exchange. 
Between 40\% and 60\% of the volume on the New York Stock Exchange occurs on just 40 out of 4000 stocks~\cite{taft2014store}.
Stock popularity changes at the granularity of hours or days, whereas daily trading volume is on the order of 5--10 billion shares~\cite{stocks}. Thus, while transaction volume is high, the change in popularity is slow. 
Similar to the case of handovers, re-sharding will need to happen but relatively infrequently. 

Existing works~\cite{taft2014store, serafini2016clay, elmore2015squall, curino2010schism, Rococo, Rocksteady} propose dynamic sharding to adapt to these kinds of changes. 
However, their datastore designs that support 
\linebreak
re-sharding and provide strong consistency operate at a sub-Mtps throughput. For instance, Squall~\cite{elmore2015squall} and Rococo~\cite{Rococo} report up to 100 Ktps per server and Rocksteady~\cite{Rocksteady} up to 700 Ktps per server. 

Meanwhile, the state-of-the-art reliable in-memory datastores (e.g., FaRM, FaSST) reach millions of tps per node but have limited support for changes in locality. 
For instance, FaRM only supports static location hints. 
If the access locality changes, both FaRM and FaSST must execute remote transactions. 
Some domain-specific datastores have been built that exploit locality, but they do not meet all design objectives. For example, S6~\cite{Woo:18} does not provide replication (a must for availability), while \CAP{FTMB}~\cite{Sherry:15} runs on only one node and replicates on disk. 
Overall, to the best of our knowledge, there is no in-memory datastore that meets all our design objectives and effectively exploits locality. 

\tsection{Zeus Design}
\label{secZ:design}

We start this section by outlining the Zeus datastore's system architecture. 
We then present a high-level overview of the core of Zeus: a pair of protocols that exploit locality for high-performance transaction processing with fault tolerance, strong consistency, and programmability.

\tsubsection{System Architecture}
Zeus exploits request locality and uses an application-level load balancer to enforce it. 
External requests issued to Zeus are routed through a load balancer. 
The load balancer can extract the application level information, locate relevant object keys, and always forwards requests with the same set of keys to the same server.
Application-level load balancers are not a new concept. 
Several previous systems have demonstrated such load balancers~\cite{Adya:2016, Nguyen:2018, ahmad2020low, annamalai2018sharding}. 
We implement a simple load balancer using a distributed, replicated key-value store based on Hermes. 
We extract a key from each request and look it up in the key-value store. 
If it is not found, we pick a destination Zeus node at random, store it in the load balancer's key-value store, and forward the request.
If the key is found, we forward the request to the corresponding destination. 

Zeus considers \newtext{a partially synchronous model with crash-stop node failures and network faults, including message losses (as described in \cref{secB:fault-model})}.
It implements a reliable messaging library with low-level retransmission to recover lost messages. 
Similar to Hermes, it uses a reliable membership with leases to deal with the uncertainty of detecting node failures. 
Each membership update is tagged with a monotonically increasing \epoch 
and is performed across the deployment only after all node leases have expired. 
For data reliability, Zeus maintains replicas of each object. 
The replication degree is configurable; however, the higher the degree of replication, the greater the \CAP{CPU} and network overhead, and the lower the throughput of transactions that modify the state.

\tsubsection{Overview of Protocols}
\label{secZ:protocol}

Zeus is efficient in executing distributed transactions by forcing them to become local. 
At the heart of Zeus are two separate, loosely-connected reliable protocols.
One is the {\em ownership protocol} responsible for the on-demand migration of the object data from one server to another and for changing the access rights (read or write) of servers storing the replica of an object.
The other one is the {\em reliable commit protocol} for committing the updates performed during a transaction to the replicas. 
As these two protocols are only loosely connected, they can be independently optimized, verified, and tested. 

Zeus, inspired by hardware transactional memory~\cite{htm}, executes and commits each transaction locally on a server designated to be the {\em coordinator} for that transaction. 
When executing a transaction, the coordinator must secure the appropriate ownership level for each object involved in the transaction.
This is the task of the ownership protocol.
Once the coordinator acquires the
required ownership levels and finishes execution, it commits the transaction locally.
Subsequently, it copies
the state of modified objects to backup servers, also called {\em followers}.
The latter is the task of the reliable commit protocol. Crucially, the ownership protocol is invoked only the first time a node accesses an object.
Subsequent transactions proceed without invoking it until another node takes over the ownership  (i.e., the locality changes).

\begin{figure}[t]
  \centering
  \includegraphics[width=0.8\columnwidth]{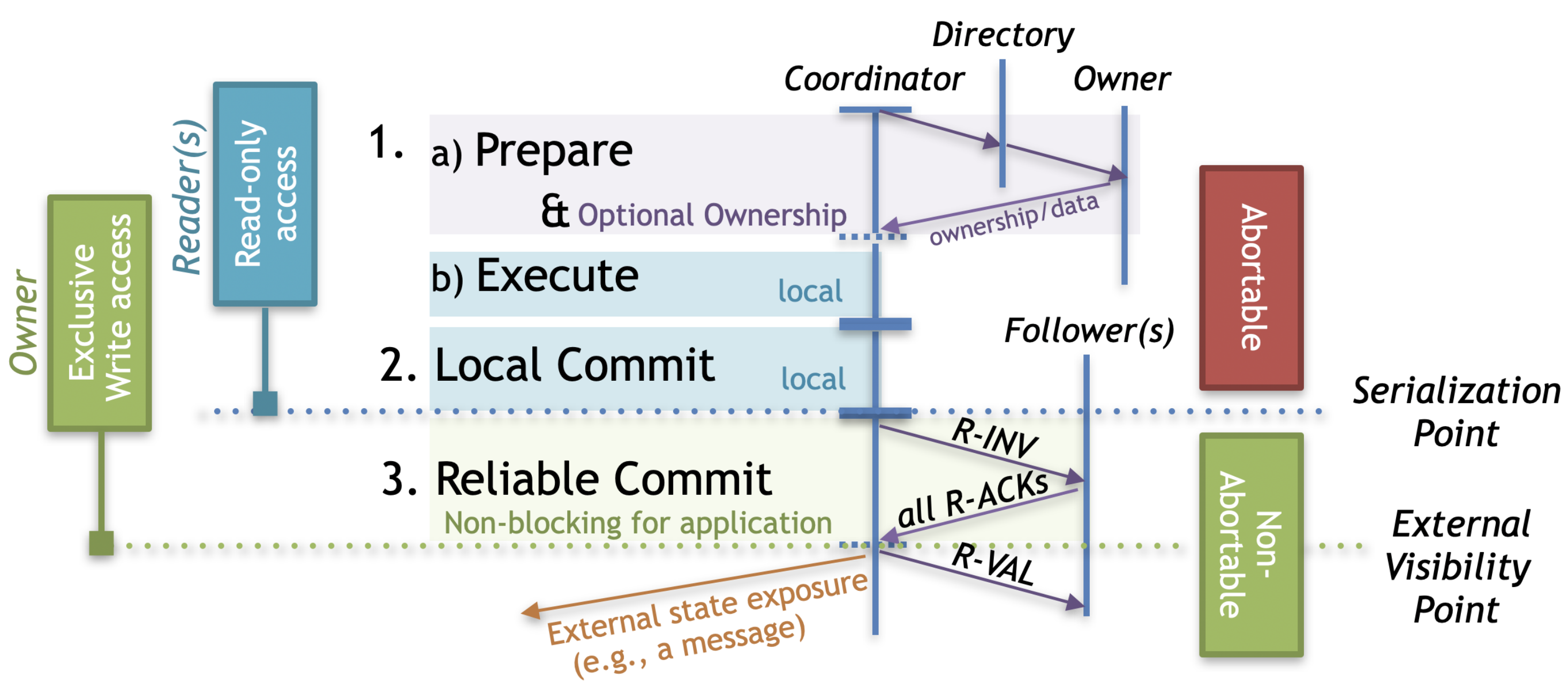}
  \mscaption{Locality-aware distributed transactions in Zeus.}
  \label{figZ:zeus}
\end{figure}

At a high level, a transaction in Zeus is carried out through the following three steps (also shown in \cref{figZ:zeus}):

\squishenum
\item \textbf{Prepare \& Execute}: 
When the coordinator executes a transaction, it verifies prior to accessing an object that it holds the appropriate ownership level (read or write) for that object.
If not, it acquires the appropriate ownership level via the {\em ownership protocol} (described in \cref{secZ:sharding-protocol}) and continues execution.  
Before performing its first update to an object,
the coordinator creates a local private (to the transaction) copy of the object. This private copy is then used for all accesses of the transaction to the object.

\item \textbf{Local Commit}: The coordinator attempts to serialize the transaction locally via a single-node commit.
This commit is local and unreliable but does not yet expose any updated values to other servers.
We implement a simple multi-threaded local commit that resolves contention across threads using a simplified, local version of the ownership protocol (detailed in \cref{secZ:system}). 

\item \textbf{Reliable Commit}: If the local commit is successful, the coordinator pushes all updates to the followers for data reliability.
In case the coordinator fails in the middle of this process,
the followers recover by safely replaying any pending reliable commit of the failed coordinator.
Both backup and recovery are performed by the {\em reliable commit protocol} (detailed in \cref{secZ:zeus-protocol}). 
\squishenumend

Zeus allows only a single server to modify an object at any given time. 
This server is called the {\em owner} and is the only node able to use the object to execute write transactions (i.e., transactions that modify at least one object).
Each object is replicated on one or more backup servers. 
These backups are active and are called the {\em readers} of the object;
they can perform read-only transactions but not write transactions using the object.\footnote{
  Note that a {\em reader} is per object, whereas a {\em follower} is per transaction (potentially spanning multiple objects).
} Only the owner and the readers store the content of the object. 
The owner (as a coordinator of write transactions) updates all readers during the reliable commit phase. 
A user can specify and dynamically change the number of readers (i.e., replicas) of each object, making a trade-off between reliability and replication overhead.

Zeus avoids the conventional distributed commit protocols~\cite{Mohan:86, Skeen:1981}, which are complex~\cite{Binning:16} because they need to deal with 
distributed conflict resolution and 
the uncertainty of commit or abort after faults. 
Zeus sidesteps these challenges through a simple invariant: an initiated reliable commit is idempotent and cannot be aborted by remote participants. 
This is accomplished via the exclusive write access of the coordinator and 
the use of {\em idempotent invalidations} (\cref{secZ:reliable_commit_protocol}), which are sent to all of the remote participants at the start of the reliable commit. 
In the case of a fault, any of the participants can replay the invalidation message, which contains enough data to complete the transaction. 

Zeus further introduces two key optimizations.
First, it supports efficient, strictly serializable read-only transactions.
Any node that is a \textit{reader} of all objects involved in a read-only transaction is able to execute that transaction without invoking the ownership protocol. 
A read-only transaction does not require a reliable commit phase; as such, it is lightweight and incurs no network 
\linebreak
traffic. 
The consistency of read-only transactions is enforced through 
\linebreak
invalidation messages, as a read-only transaction cannot execute on an object that is invalidated. 

\begin{figure}[t]
  \centering
  \includegraphics[width=0.8\columnwidth]{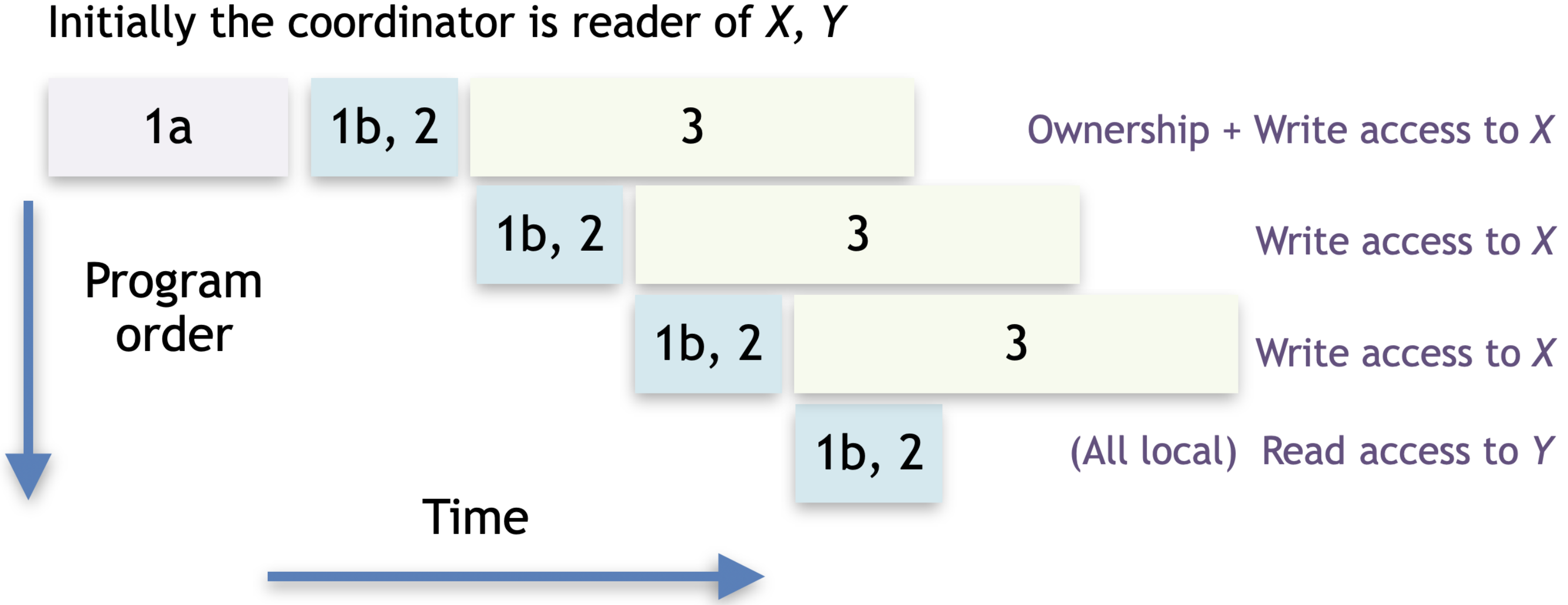}
  \mcaption{Pipelined execution of transactions in Zeus.}
  {Zeus’ pipelined execution of transactions for objects \emph{X} and \emph{Y} on the same coordinator (labels in boxes are the same as in \cref{figZ:zeus}).}
  \label{figZ:zeus_pipe}
\end{figure}

Second, a transaction coordinator in Zeus pipelines local execution and commit with the reliable commit, as shown in \cref{figZ:zeus_pipe}.
This is possible because no other server can update the objects at the same time. 
The latter is guaranteed by the ownership protocol, which ensures that only one node (the current owner) can modify an object.
It is thus safe for the coordinator to keep modifying the same object without waiting for the reliable commit to finish. 
As a consequence, any local transactions to objects for which permissions have already been acquired will not block the application execution. 

We also note that, in order to simplify application portability, we made a conscious design trade-off in making the ownership protocol blocking and Zeus transactions (the most frequent operations) non-blocking. 
In other words, the application thread stalls when executing an ownership request (phase 1(a) in \cref{figZ:zeus}). 
This design is justified because ownership requests are much less frequent than transactions, as discussed in \cref{secZ:background}. 
It would be straightforward to improve the performance of the ownership protocol, e.g., via a user-mode thread scheduling framework, as in~\cite{F-Kalia:2016}. However, doing so would increase the burden on the developer and likely require re-architecting the application, thus invalidating a key design requirement, as laid out in \cref{secZ:background}. 

Finally, we specified Zeus ownership and Zeus reliable commit in \CAP{TLA$^{+}$} and model-checked them. The details are provided in \cref{secZ:evaluation}.

\tsection{Zeus Ownership}
\label{secZ:sharding-protocol}

The reliable ownership 
atomically alters object access rights and transfers content between nodes. 
We start by introducing the main terminology used in the protocol. 
We then overview its operation without faults and contention, and follow by discussing these other cases. 

\vskip 6pt \noindent{\bf Access levels, directory, and metadata.}
A node can be the {\em owner}, a {\em reader}, or a {\em non-replica} of an object. 
Each object has at most one owner at any given time, which has exclusive write and (non-exclusive) read access to it. 
An object can also have several other readers with read access. 
Both the owner and the readers store a replica of the object. 
A non-replica node has neither the access rights nor the data for the object. 

Zeus maintains an {\em ownership directory} that stores ownership metadata about each object. 
This directory is replicated across three nodes for reliability (even if a Zeus deployment has more nodes). 
The nodes that store directory information are called the {\em directory} nodes. 

\noindent  The directory stores the following metadata for an object: 
\squishlist
\item $o\_state$: the ownership state of the object, which can be \textit{Valid}, \textit{Invalid}, \textit{Request}, or \textit{Drive}; 
\item $o\_ts =$ <$obj\_ver, node\_id$>: the ownership timestamp, which is a tuple of a mono\-ton\-ically-increasing number and a node \CAP{ID}; and
\item $o\_replicas$: a bit vector that denotes all nodes storing a replica of the object and their access rights (i.e., the owner and readers). 
\squishend
These ownership metadata are also stored by each object's owner node. 
A summary of the above is given in \cref{tabZ:shard_metadata}.

\begin{table}[t!]
\centering
\resizebox{0.8\columnwidth}{!}{

\begin{tikzpicture}

\node (table) [inner sep=-0pt] {
\begin{tabular}{l|cccc}
\rowcolor[HTML]{9B9B9B}

\multicolumn{1}{c|}{} & \multicolumn{1}{l}{\textbf{directory}} & \multicolumn{1}{l}{\textbf{owner}} & \multicolumn{1}{l}{\textbf{reader(s)}} & \multicolumn{1}{l}{\textbf{non-replica}} \\
\textbf{data}               &        & \cmark & \cmark &  \\
\textbf{ownership metadata}  & \cmark & \cmark &        &  \\
\textbf{ownership levels} & - & \textbf{w}/\textbf{r}    & \textbf{r}      & -                             
\end{tabular}

};
\draw [rounded corners=.3em] (table.north west) rectangle (table.south east);
\end{tikzpicture}

}
\mcaption{Nodes storing object (meta)data and their access permissions.}
{Object data and metadata stored by each node along with their read (r) and exclusive write (w) access permissions.}
\label{tabZ:shard_metadata}
\end{table}

\begin{figure}[t]
  \includegraphics[width=1\columnwidth]{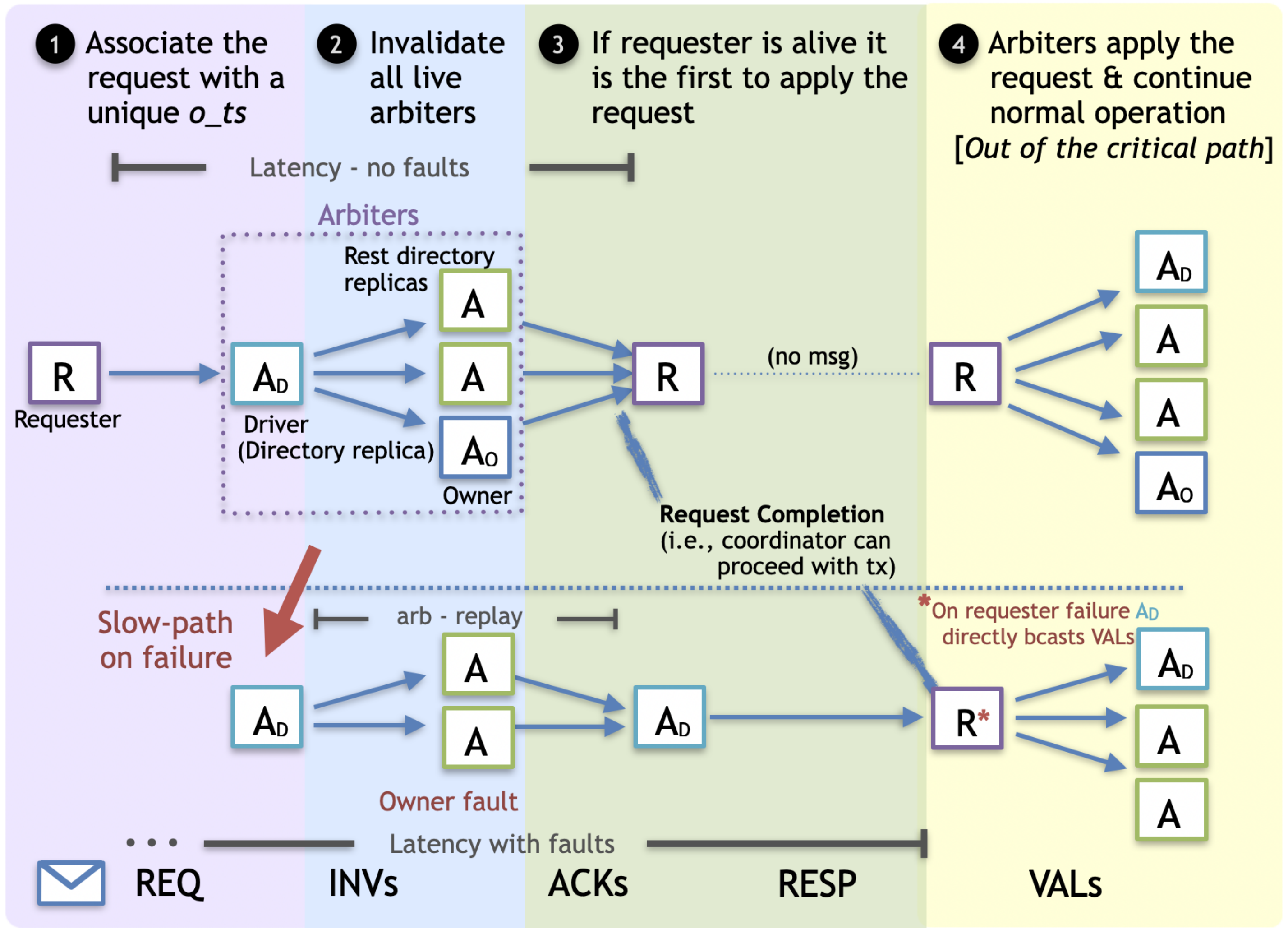}
  \mscaption{Zeus ownership protocol with and without faults.}
  \label{figZ:sharding-prot}
\end{figure}

\tsubsection{Reliable Ownership Protocol}
\vskip 0pt \noindent{\bf Failure- and contention-free operation.}
An ownership request is illustrated at the top of \cref{figZ:sharding-prot}.
The coordinator that starts a request is called a \textit{requester} node. 
The requester assigns a locally unique request \CAP{ID} to the request (to be able to match the response) and sets the object's local $o\_state = Request$.
It then sends a \textit{request} (\CAP{REQ}) message with the request \CAP{ID} to an arbitrarily chosen directory node, and this node becomes the {\em driver} of the request. 
The directory nodes and the object owner help arbitrate concurrent ownership requests to the same object and are called {\em arbiters}.

Upon receipt of a \CAP{REQ} message, the driver assigns an ownership timestamp $o\_ts$ to the object and sets its local state to $o\_state = Drive$ \scalebox{0.9}{\circledZ{1}}. 
It also 
sends an \textit{Invalidation} (\CAP{INV}) message containing both the request \CAP{ID} and ownership metadata 
to the remaining arbiters (including the current owner)~\scalebox{0.9}{\circledZ{2}}. 
Assuming no contention over the ownership of the object, each arbiter sets the object's local state to $o\_state = Invalid$, updates its local $o\_ts$ and $o\_replicas$, and responds directly to the requester with an \CAP{ACK} message. 
Note that we optimize the ownership latency by sending the responses directly to the requester rather than passing them through the driver. 
If the requester is a non-replica and does not have the data of the object, the current owner includes the data in her \CAP{ACK}.

When the requester receives all expected \CAP{ACK} messages, it \textit{applies} its request locally before responding to all arbiters with a \textit{Validation} (\CAP{VAL}) message~\scalebox{0.9}{\circledZ{3}}. To apply the request, it updates the $o\_replicas$ to specify itself as the new owner and sets its object's local $o\_state = Valid$.
Finally, upon reception of the \CAP{VAL} message, each arbiter applies the request in the same way, and the request is finished~\scalebox{0.9}{\circledZ{4}}.

Notice that, to keep $o\_replicas$ consistent with the replica placement and the access levels of the object, the requester must apply the request before any of the arbiters does. Moreover, once the requester receives all the \CAP{ACK} messages, it unblocks the application. The application thus resumes its transaction after 1.5 round-trips, as shown in the top part of \cref{figZ:sharding-prot}.

\vskip 2pt \noindent{\bf Contention resolution.}
Zeus ownership uses the $o\_ts$ timestamp to resolve contending requests. 
Multiple nodes may concurrently issue an ownership request for the same object through different drivers. 
Each driver creates a per-object unique timestamp for the request $o\_ts =$ <$obj\_ver + 1, node\_id$> using its previous local $obj\_ver$ and own $node\_id$\vspace{1pt}~\scalebox{0.9}{\circledZ{1}}. 
In case of contention, a driver of one of the contending requests will receive an \CAP{INV} message of another contending request (for the same object) \scalebox{0.9}{\circledZ{2}}. 
It will only process the \CAP{INV} message if the $o\_ts$ in the message is lexicographically larger than its own $o\_ts$ for the object.
This guarantees that there is one and only one winner of each contention. 
All the drivers whose requests fail send a \CAP{NACK} message to their requesters.
Similarly, the owner responds directly to the requester with a \CAP{NACK} message if the requested object is involved in a pending transaction (\cref{secZ:zeus-protocol}).
Upon receiving a \CAP{NACK}, the requester either aborts its ownership request or retries it at a later point. 

\vskip 2pt \noindent{\bf Failure recovery.}
The failure recovery procedure starts when the reliable membership is updated after fault detection and the expiration of leases. Each live directory node (and the live owners) update their $o\_replicas$ to remove any non-live nodes. The objects whose owners have died will be taken over by a new owner on the next write transaction. 
After the membership update, which increases the \epoch ($e\_id$), 
requests from previous epochs are ignored. 
This is achieved by including the $e\_id$ of the current epoch in the \CAP{INV} and \CAP{ACK} messages. The requester and arbiters ignore these types of messages when their $e\_id$s differ from their local ones.

A node fault followed by a membership update can leave the arbiters of a pending ownership request in an \textit{Invalid} $o\_state$.
Nevertheless, any arbiter has all the necessary information to replay the idempotent arbitration phase of the ownership request (termed \textit{arb-replay}) between the live arbiters and unblock. A blocked arbiter acts as the request driver and initiates an \textit{arb-replay} by constructing and transmitting the same \CAP{INV} message using its local state.
During \textit{arb-replays}, some arbiter may receive an \CAP{INV} message for a request it has already applied locally (with the same $o\_ts$). In this case, the arbiter simply responds with an \CAP{ACK}.
A basic recovery path from an owner failure is illustrated at the bottom of \cref{figZ:sharding-prot}.

Note that, in the recovery process, the arbitration phase of an ownership request is finalized with \CAP{ACK} messages sent from the arbiters to the driver instead of the requester, as shown in \cref{figZ:sharding-prot}. 
This is done in order to have a single recovery process that covers the failures of all nodes, including the requester. 
If the requester is not live, the driver directly sends \CAP{VAL} messages to unblock the other live arbiters. 
Otherwise, for safety reasons -- as in the failure-free case -- the requester must be the first to apply the request. To achieve this, we introduce a new \CAP{RESP} message which confirms the win of the arbitration to the requester, who can then apply the request prior to sending \CAP{VAL} messages to the live arbiters, as before.

\tsubsection{Fast Scalable Ownership}

The Zeus ownership protocol is \textit{scalable} since (1) it does not store directory metadata for each object at every transactional node
and (2) it does not broadcast to every transactional node to locate an object's owner.
The Zeus ownership protocol has a latency of at most 3 hops (without faults and contention) to reliably acquire the ownership, regardless of the node requesting the ownership.
We believe this to be the lowest possible latency for a scalable ownership protocol. 
The worst-case latency is incurred when an ownership request originates from a non-replica node where neither the owner nor the requester is co-located with the object's directory metadata.
To proceed in this case, the requester must receive the latest value of the object. In order to locate the object, the requester should first contact the directory. The directory forwards the request to the owner, which in turn sends the value to the requester, resulting in 3 hops. Note that if the requester is co-located with a directory replica, the first hop is eliminated, and the ownership is acquired after just one round-trip (2 hops) to the owner.
\tsection{Zeus Reliable Commit}
\label{secZ:zeus-protocol}

\begin{figure}[t]
    \centering
  \includegraphics[width=0.925\columnwidth]{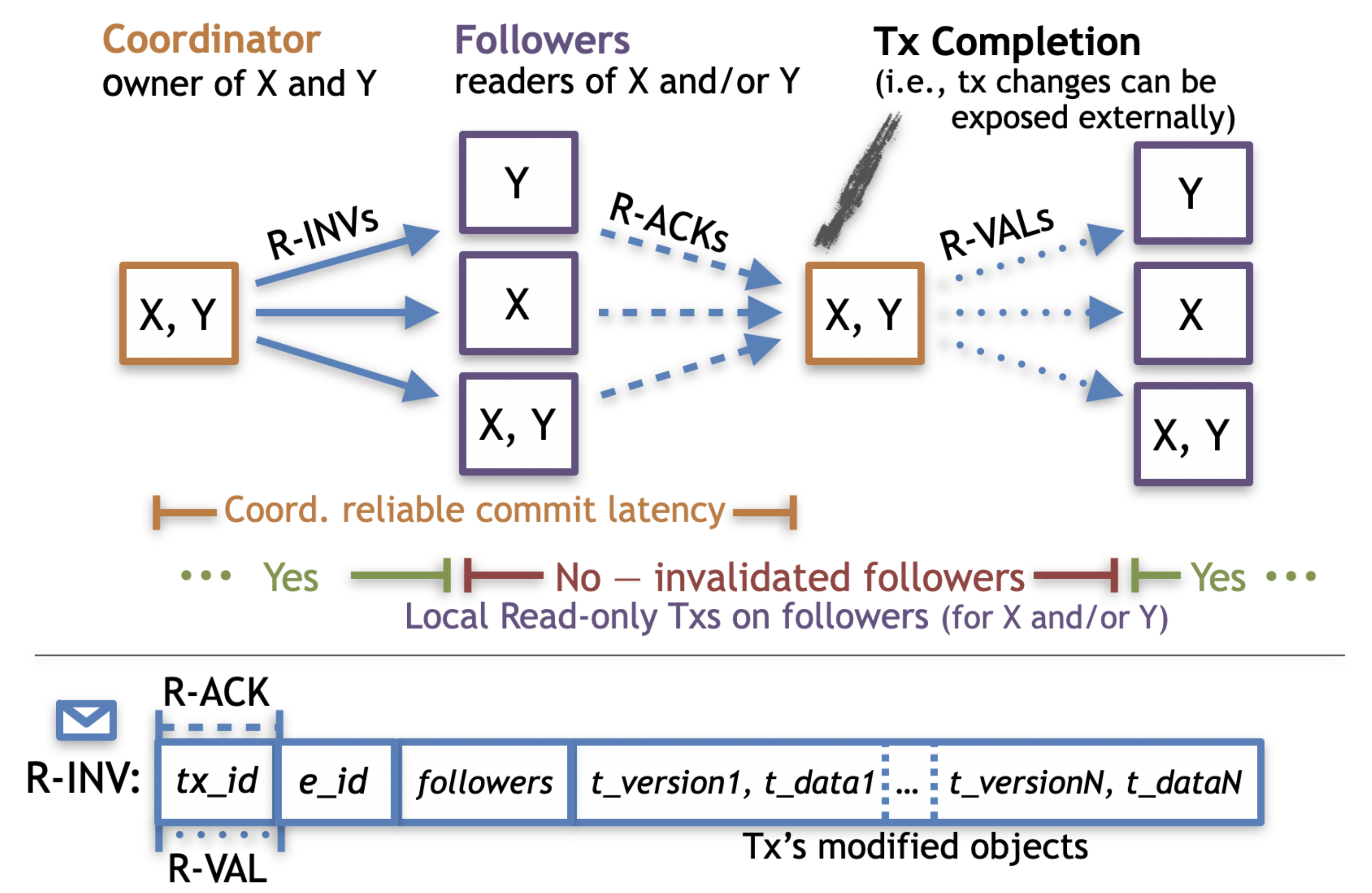}
  \mscaption{Zeus reliable commit protocol and its messages.}
  \label{figZ:reliable_commit}
\end{figure}

The Zeus reliable commit protocol is responsible for propagating the updates made by a local transaction to all followers (illustrated in \cref{figZ:reliable_commit}). 
For clarity, we begin by describing the information maintained by the protocol. 
We next overview the operation without faults and then discuss the case with failures. 
\linebreak
Finally, we present two optimizations: pipelining and local read-only transactions from all replicas.

\vskip 4pt \noindent{\bf(Meta)data.}
Each replica (i.e., the owner and readers) keeps the following information for an object:
\squishlist
\item $t\_state$: the state of the object, which can be \textit{Valid}, \textit{Invalid} or \textit{Write};
\item $t\_version$: the version of the object, which is incremented on every transaction that modifies the object; and
\item $t\_data$: the data of the object stored by the application.
\squishend

\noindent For every transaction, at the beginning of the reliable commit, the coordinator generates a unique $tx\_id =$ <$local\_tx\_id,$ $node\_id$>, where $node\_id$ is its own \CAP{ID} and $local\_tx\_id$ is a locally unique, monotonically-increasing transaction \CAP{ID}.

\tsubsection{Reliable Commit Protocol}
\label{secZ:reliable_commit_protocol}
\vskip 0pt \noindent{\bf Failure-free operation.} 
At the end of the Local Commit phase, the transaction coordinator updates the $t\_data$ of all modified objects with its private copies created during the Prepare \& Execute phase. 
It also increments their $t\_versions$ and sets $t\_state = Write$ --- for the pending reliable commit.

At the beginning of the Reliable Commit phase, the coordinator broadcasts an {\em Invalidation} (\CAP{R-INV}) message to all followers.
As shown at the bottom of \cref{figZ:reliable_commit},
this message contains the $tx\_id$, the current $e\_id$, and the $node\_id$s of all followers.
For each updated object, it also contains the new $t\_version$ and $t\_data$.
The coordinator temporarily stores the \ZINV message locally.

Upon receiving an \ZINV message, a follower checks whether the received and local $e\_id$ match. If not, the message is ignored.
If they match, the follower goes through each updated object and compares its local $t\_versions$ with that of the message. In the case that an object's local version is greater than or equal to the object version in the message, the follower skips the update of that object. Otherwise, it updates the local $t\_data$ (the actual content of the object) and $t\_version$ with their new counterparts from the message and sets its local $t\_state = Invalid$ --- denoting that the object has a pending reliable commit.
A follower then responds to the coordinator with an \ZACK message containing the same $tx\_id$ 
and temporarily stores the \INVNOSPACE.

Once the coordinator receives {\ACKNOSPACE}s from all the followers, 
it reliably commits the transaction locally by changing the $t\_state$ of each updated object to \textit{Valid}. 
Subsequently, the coordinator broadcasts a \textit{Validation} (\CAP{R-VAL}) message containing the $tx\_id$ to all followers and discards the previously-stored \ZINV message of the transaction.
When a follower receives an \ZVAL message for which it has already stored an \ZINV message (with the same $tx\_id$), it sets the $t\_state$ of all objects previously updated by the transaction to \textit{Valid} if and only if their $t\_version$ has not been increased. 
It then discards the stored \ZINV message.

\vskip 2pt \noindent{\bf Reliable replay under failures.}
A node failure triggers a membership reconfiguration wherein the \epoch ($e\_id$) is increased and the set of live nodes is updated. 
Subsequently, the ownership protocol temporarily stops accepting requests for objects whose owner node is not live in the current membership.

At this point, each locally stored \ZINV message on any live node represents a pending transaction in the Reliable Commit phase. 
A live node replays its own pending reliable commits as well as those from the failed nodes. 
This is accomplished by first updating the local pending \ZINV messages (issued or received) with the new $e\_id$ and removing all non-live nodes from followers.
The messages are then re-sent and handled as explained before. 
A follower who receives an \ZINV message with the latest $e\_id$ for a transaction that it has previously stored locally (i.e., with the same $tx\_id$) simply ignores its content and responds with an \CAP{R-ACK}.
Although multiple nodes may replay the reliable commit phase of the same transaction, all relevant \ZINV messages are idempotent, containing the same $tx\_id$ (and $t\_versions$), so the only one can apply updates. 

When a node has no more pending reliable commits (\ZINV messages) from nodes that are not live, it informs the ownership protocol that it has finished the recovery (\cref{secZ:sharding-protocol}). 
Once all live nodes finish the recovery, the ownership protocol again starts to accept all ownership requests as normal.

\begin{figure}[t]
  \centering
  \includegraphics[width=1\linewidth]{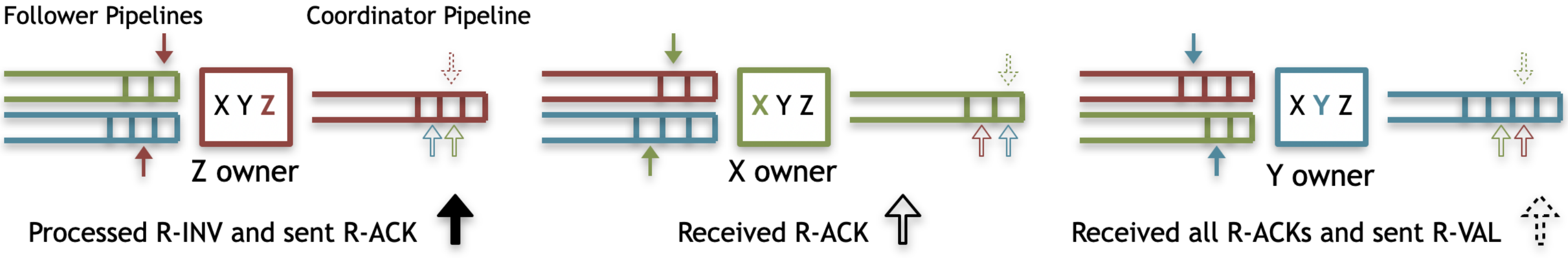}
  \mcaption{Per-node pipelines in Zeus.}{Zeus' per-node (in reality, per-thread; see \cref{secZ:system}) pipelines.}
  \label{figZ:pipe}
\end{figure}    
\tsubsection{Non-blocking Transaction Pipelining}
\label{secZ:tx-pipelining}
We further introduce transaction pipelining to avoid blocking the application at the coordinator during replication (illustrated in \cref{figZ:zeus_pipe}).
This is possible because a locally (unreliably) committed transaction at the coordinator cannot be aborted. 
Thus, the coordinator can proceed using its locally committed values with certainty. 

However, Zeus also needs to maintain strict serializability 
on each follower replica. 
Thus, followers must respect the pipeline order of the coordinators when applying updates.
For this, Zeus uses $tx\_id =$ <$local\_tx\_id,$ $node\_id$>, which is transmitted in every \ZINV message and contains both the local transaction order within the node $local\_tx\_id$ and the $node\_id$. 
As a result, although there could be several pending and causally related reliable commits, all will be applied in the correct order as specified by the $local\_tx\_id$.

Note that the ordering is enforced only within each different pipeline, as shown in \cref{figZ:pipe}.
This is because an object's owner change (i.e., when an 
\linebreak
object switches pipelines) is not approved until all pending reliable commits with that object have been completed (\cref{secZ:sharding-protocol}). 
Thus, an object cannot be involved in pending transactions from two different coordinator nodes, and the ordering across coordinators is irrelevant. 
We further optimize this by enabling per-thread (rather than per-node) pipelines via our choice of local commit, as explained in \cref{secZ:system}.
The pipelining optimization also reduces the 
number of \ZACK and \ZVAL messages, since sending a message with a 
\linebreak
$tx\_id$ implies the successful reception and processing of all previous messages in that pipeline.

A node may not be a follower of all \CAP{R-INV}s and thus may receive only a partial stream of a pipeline.
An extra condition is needed for when such followers can \textit{apply} an \CAP{R-INV}.
A follower applies an \CAP{R-INV} if for the previous $local\_tx\_id$ (slot) of the pipeline it has either applied an \CAP{R-INV} or has received an \CAP{R-VAL}. The latter occurs 
for a transaction follower \textit{F} who was not also a follower of the previous slot in the pipeline. 
To facilitate this, during the broadcast of an \CAP{R-INV}, the coordinator piggybacks a \textit{prev-\CAP{VAL}} bit if it has broadcasted \CAP{R-VAL}s for the previous slot.
Otherwise, it includes \textit{F} in the \CAP{R-VAL} broadcast of the previous slot.
Finally, after a coordinator's failure, an \CAP{R-INV} is considered a pending reliable commit and is replayed by a follower if and only if that follower has not only received but also applied the \CAP{R-INV} message. 

\begin{figure}[t]
  \centering
  \includegraphics[width=0.9\linewidth]{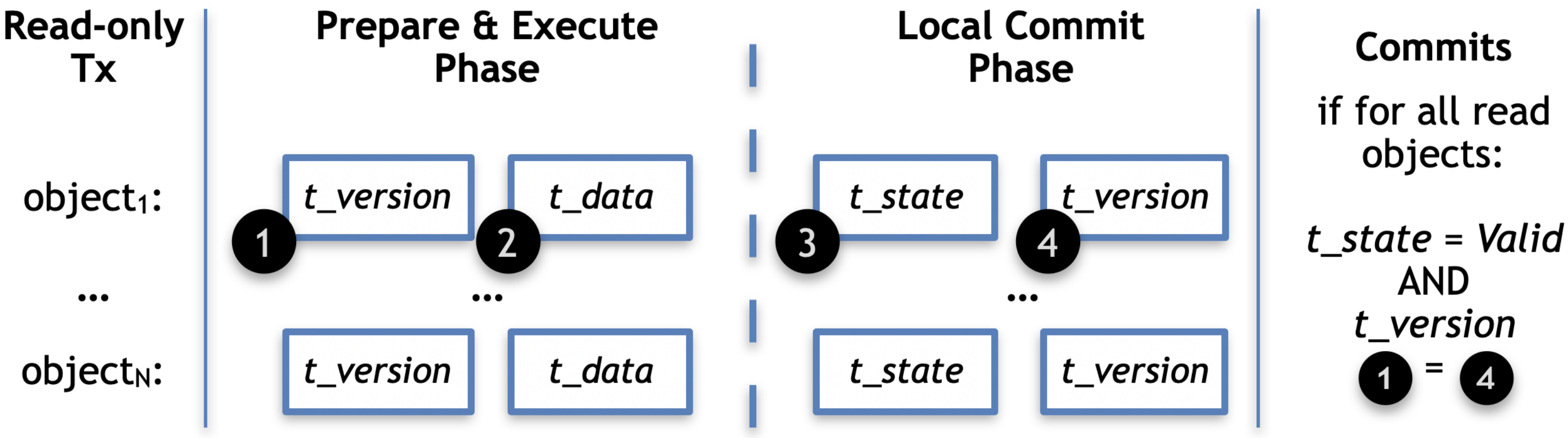}
  \mcaption
  {Consistent read-only transactions in Zeus.}
  {Zeus' consistent read-only transactions on readers.}
  \label{figZ:read-only}
\end{figure}
\tsubsection{Read-only Transactions}
Zeus optimizes read-only transactions by allowing them to be executed locally from any replica that stores all relevant objects, regardless of the ownership level (read or write) and without compromising strict serializability. 
This is enabled by three factors.
First, read-only transactions do not need to communicate any updates to other replicas. 
Second, a verification-based scheme can be applied to exploit the local object versioning
and ensure a consistent snapshot across all reads of a read-only transaction.
Third, the reliable commit guarantees that all replicas are invalidated before any updated state is exposed externally by the readers. 
We elaborate on the latter before discussing the read-only protocol.

\beginbsec{Invalidation-based reliable commit}
A locally committed write transaction does not reliably commit on the owner unless it has invalidated all its followers (i.e., the readers of modified objects).
As noted previously, a reader that applies an invalidation to its local object also updates its object's local value with the newly received value. Thus, it can return neither the old nor the new value, as the object has been invalidated.
The reader can return the new value only after it receives the \CAP{R-VAL} message and validates its local object.

Simply put, there is a transitioning period until a reader can safely return the new value.
That period ends once all readers of a modified object have stopped returning the old value and have received the new one.
If a reader was set to prematurely return the new value (i.e., prior to receiving the \CAP{R-VAL} message and before the end of that period), two problems could arise. First, another reader who has not yet invalidated the object could subsequently return the old value and compromise consistency. 
Second, if all nodes that had received the new (not yet reliably committed) value were to fail,\footnote{That is a smaller number of nodes than the replication degree.} then the prematurely returned value would be permanently lost. 

\beginbsec{Read-only protocol}
Consequently, in Zeus, a read-only transaction completes after only two phases, as shown in \cref{figZ:read-only} and described next.
In the Prepare \& Execute phase, the coordinator of a read-only transaction 
sequentially reads and buffers the $t\_version$ and the value ($t\_data$) of each local object as specified by the transaction.
In the Local Commit phase, the coordinator checks whether all accessed objects are in $t\_state = Valid$ before verifying that all $t\_versions$ have remained the same.
If yes, the transaction commits successfully. 
Otherwise, there is an ongoing conflicting (local or remote) reliable commit, and the read-only transaction is aborted or optionally retried.

\beginbsec{Use case}
Apart from the obvious performance benefits, one example where the read-only optimization is useful is control/data-plane applications, such as cellular network applications. There, write transactions are executed by a control-plane node (the Zeus owner) -- for instance, to configure routing -- while all data-plane nodes (i.e., Zeus readers) can perform consistent read-only transactions locally (e.g., for forwarding).

\tsection{Discussion}
\label{secZ:discussion}
\tnsubsection{Distributed commit vs. Zeus}

Traditional datastores statically shard objects and execute reliable transactions in a distributed manner across servers.
This poses two challenges. The first 
is accessing the objects. 
Static sharding schemes do not guarantee that all objects accessed by a 
transaction will reside on the same node. Frequently, one or more objects 
in a transaction are stored remotely. In such cases, the execution stalls until the objects are fetched, sometimes sequentially (e.g., for pointer chasing or control flow).

The second challenge is handling concurrent transactions on conflicting 
objects. If two nodes attempt to simultaneously commit transactions on conflicting
objects, one of them has to abort. Detecting and handling these conflicts under the uncertainty of faults requires extra signaling across nodes. Thus, transactional systems 
based on distributed commit 
necessitate numerous round-trips to commit each transaction (e.g., see FaSST). Moreover, a node cannot start the next transaction on the same set of objects until the commit is finished, as it cannot be sure that it will not need to abort. This introduces several round-trips of delay in the critical path of the commit and significantly reduces the transactional throughput.

Zeus replaces remote accesses and distributed commit with its (occasional) ownership, local accesses, and reliable commit to address the two main issues mentioned above and accelerate workloads with locality.
First, the ownership makes objects accessed by a transaction accessible locally most of the time, which avoids stalls during the execution. 
Second, only a single node (the owner) can execute a write transaction on an object at a given time. Therefore, a transaction cannot be aborted remotely, commits after a single round-trip, and is pipelined. 
Zeus reliable commit also affords local and consistent read-only transactions from all replicas.

Unlike distributed commit, the Zeus ownership protocol is specialized for single-object atomic operations (including migration).
Zeus resolves concurrent 
\linebreak
ownership requests in a decentralized way and applies an idempotent scheme to tolerate faults without extra overhead in the common failure-free case. This makes acquiring ownership reliable yet fast (1.5 round-trips) during fault-free operation.

\tsubsection{Further Details}
\beginbsec{Cost of ownership vs. remote access}
The object size influences the cost of acquiring ownership for it by a non-replica node similarly to a remote access, since in the fault-free case the value is included in a single ownership message, as in the response of a remote access.
A reader acquires the ownership without the value and thus is not influenced by its size.
The reliability of Zeus ownership comes with a higher message cost compared to a remote access. These are small constant messages whose cost is amortized over several local accesses in workloads with locality. 
Nevertheless, for workloads without sufficient locality, that cost renders Zeus less suitable than remote accesses and distributed commit.

\beginbsec{Deadlocks}
Zeus currently circumvents deadlocks via a simple backoff mechanism. 
For Zeus, such a situation may arise only early in a transaction (i.e., in the Prepare \& Execute phase)
-- when requesting ownership for an object. This manifests with repeated failed ownership requests, after which Zeus aborts and retries a transaction with an exponential backoff. In practice, deadlocks in Zeus are rare because transactions on the same object are mostly executed on the same server by virtue of load balancing.
For deployments where this is not the case, a more sophisticated scheme such as that proposed by Lin \etal\cite{lin2016towards} may be considered.

\beginbsec{Distributed directory}
For simplicity, Zeus 
uses a single directory 
for all objects in the deployment. 
The directory is replicated for fault tolerance, and the ownership protocol is lightweight and is designed to balance the load across all directory replicas.
However, a single replicated directory may become a scalability bottleneck at large deployment sizes or when locality is limited. In such cases, a distributed directory scheme (i.e., one using consistent hashing on an object to determine its directory nodes) should be used instead.

\beginbsec{Sharding request types}
Zeus exploits the ownership protocol for other types of sharding requests, such as reliably removing a reader. For example, when a non-replica acquires the ownership of an object, the total number of replicas increases. 
To maintain the initial replication degree and avoid increasing the cost of reliable commits,
we invoke the ownership protocol out of the critical path to discard a reader.

\beginbsec{Write transactions with opacity}
Apart from strict serializability, Zeus 
\linebreak
provides an additional guarantee that all write transactions will see a consistent snapshot of the database, even if they abort. This is also referred to as \textit{opacity}~\cite{Guerraoui:08}. Opacity further enhances Zeus' programmability; by preventing inconsistent accesses in write transactions, it relieves the programmer from the effort of handling those cases.
\tsection{System}
\label{secZ:system}

We built a custom in-memory datastore and implemented the Zeus protocols on top of it. 
In this section, we briefly discuss the details of the implementation.

An application communicates with the datastore through a transactional memory \CAP{API} (summarized in \cref{listing:rvnf_api}), which consists of a traditional key-value interface and primitives to create and manage memory objects of different sizes.
The latter includes implementations of \Zmylisting{malloc} to create an object, \Zmylisting{free} to destroy an object, and \Zmylisting{tr\_read} (\Zmylisting{tr\_write}) for marking an object as used in a transaction for reading (writing).
Each transaction starts with a create transaction call \Zmylisting{tr\_create}, followed by an arbitrary code that can invoke the above \CAP{API}s, and finishes with  \Zmylisting{tr\_commit} (or \Zmylisting{tr\_abort}), at which point the local commit starts (aborts). 
This is a low-level \CAP{API} very similar to the one used by FaRM, and it allows for great flexibility in building further abstractions on top of it.

The datastore is implemented in C over \CAP{DPDK} and consists of two parts. 
The first is the datastore module, which runs as a separate process implementing the main datastore functionality.
The other part is the Zeus library, which is linked to any application over shared memory without limiting its architecture (e.g., can be a separate process, a \CAP{VM}, or a container).

The datastore module implements the transactional memory \CAP{API} as well as the Zeus protocols.
Zeus nodes communicate with each other using a custom reliable messaging library we built on top of \CAP{DPDK}.
The datastore module also includes a customizable, application-aware load balancing functionality, as described in \cref{secZ:design}.

\lstset{language=C++,
                basicstyle=\ttfamily,
                keywordstyle=\color{blue}\ttfamily,
                stringstyle=\color{red}\ttfamily,
                commentstyle=\color{green}\ttfamily,
                morecomment=[l][\color{magenta}]{\#}
}

\begin{figure}[t]
    \begin{lstlisting}[basicstyle=\ttfamily\small]
trans*  tr_create(bool is_read_only);
void    tr_abort (trans* t);
void    tr_commit(trans* t);
tr_addr tr_malloc(trans* t, int size);
void    tr_free  (trans* t, tr_addr addr);
void*   tr_read  (trans* t, tr_addr addr, int size);
void*   tr_write (trans* t, tr_addr addr, int size);
int tr_del(trans* t, void* key, int len);
int tr_get(trans* t, void* key, int len, void* val);
int tr_set(trans* t, void* key, int len, void* val, int vlen);
    \end{lstlisting}
    \mscaption{Zeus transactional \CAP{API}.}
    \label{listing:rvnf_api} 
\end{figure}

Both the application and datastore modules can run in multiple threads. 
In the evaluation, we use up to 10 application and 10 datastore worker threads.
These threads are pinned to their own cores. We also use one core for \CAP{DPDK}.

We implement a simple multi-threaded Local Commit (\cref{secZ:design}) using the same intuition as for the overall Zeus. 
Each thread that executes a transaction needs to become the owner of each object. 
However, this ownership is local and is managed through standard locking. 
We leverage the aforementioned load balancer to enforce locality across the threads and increase concurrency.  
Apart from simplicity, this also enables transaction pipelining to be applied on a per-thread basis, which increases the overall concurrency of reliable commits.

Currently, porting an application to Zeus requires manual code modification on pointer accesses, similar to prior work (e.g., FaRM). 
However, this can be automatized at a compiler level, as performed by Sherry \etal\cite{Sherry:15}. 
\tsection{Evaluation} 
\label{secZ:evaluation}

\beginfsec{Formal verification}
We specified the ownership protocol and the reliable commit of Zeus in \CAP{TLA$^{+}$} and model-checked them in the presence of crash-stop failures, message reordering and duplication. 
We verified them against several key invariants, including the following:
\squishlist
\item Live nodes\footnote{By construction, non-live nodes cannot compromise safety, as $e\_ids$ prevent them from participating in either transaction or ownership requests.} in $t\_state$ = $Valid$ \newtext{store the latest reliably committed value}.
\item All live arbiters in $o\_state$ = $Valid$ agree and correctly reflect the owner and reader nodes of the object.
\item At any given time, there is at most one owner, who stores the most up-to-date value of the object.
\squishend 
The detailed protocol specifications and the complete list of model-checked invariants can be found online.\footnote{\href{https://zeus-protocol.com/}{https://zeus-protocol.com}}
\newtext{\cref{Apendix} dives deeper into these invariants and informally sketches why the protocol provides strict serializability.}

\vskip 2pt \beginbsec{Locality in workloads}
We begin by briefly analyzing the locality of access patterns in workloads. For this, 
we report the fraction of remote transactions of three workloads spanning the telecommunications, financial, and trade sectors. 

\squishlist
\item \textbf{Boston cellular handovers}:
As explained in \cref{secZ:background},
in a cellular 
\linebreak
workload, remote transactions are caused by remote handovers.
To evaluate the real-world frequency of remote handovers, we use the population and mobility model from the Boston metropolitan area~\cite{bostom2013mobility} with the reported 
average daily commute of 100km.
We assume that base stations are uniformly spread throughout the area at a distance of 1km, with a typical coverage of a macro cell~\cite{cellCoverage} and a common ratio of cells per population~\cite{mohammadkhan2016considerations}.
These are sharded across all nodes in a deployment. 
As the number of nodes increases, the percentage of remote handovers also increases, up to 6.2\% for six nodes.
In summary, for a setup in which 5\% of all transactions are handovers
and out of these 6.2\% of handovers are remote (in a six-node deployment), there are in total 0.31\% remote transactions. 

\item \textbf{Venmo transactions}:
We use the most recent public Venmo dataset~\cite{venmo-dataset}, which contains more than seven million financial transactions, to analyze the fraction of remote transactions. 
We partition the users to nodes but 
still observe that 0.7\% and 1.2\% of remote transactions are remote for 3 and 6 nodes, respectively.

\item \textbf{TPC-C}:
We mathematically analyze the number of remote transactions in the \CAP{TPC-C} benchmark, which is considered representative for industries that trade products.
In \CAP{TPC-C}, only a small fraction of new-order and payment transactions may result in remote accesses. We find that only 2.45\% of the transactions in the benchmark are remote.
\squishend
We empirically evaluate benchmarks related to cellular and financial transactions (i.e., Handovers and Smallbank). While promising in terms of locality, we leave the experimental evaluation of \CAP{TPC-C} for future work, as our current implementation of Zeus does not support range queries.

\begin{table}[t!]
\centering
\resizebox{0.9\columnwidth}{!}{

\begin{tikzpicture}

\node (table) [inner sep=-0pt] {
\begin{tabular}{l|lrrrr}
\rowcolor[HTML]{9B9B9B}

 
 &
  \textbf{characteristic} &
  \multicolumn{1}{l}{\textbf{tables}} &
  \multicolumn{1}{l}{\textbf{columns}} &
  \multicolumn{1}{l}{\textbf{txs}} &
  \multicolumn{1}{l}{\textbf{read txs}} \\ 
\textbf{Handovers} & large contexts  & 5 & 36 & 4 & 0\%  \\
\textbf{Smallbank} & write intensive & 3 & 6  & 6 & 15\% \\
\textbf{TATP}      & read intensive  & 4 & 51 & 7 & 80\% \\
\textbf{Voters}    & popularity skew & 3 & 9  & 1 & 0\% 
\end{tabular}

};
\draw [rounded corners=.3em] (table.north west) rectangle (table.south east);
\end{tikzpicture}

}

\mscaption{Summary of evaluated benchmarks.}
\label{tabZ:bench}
\end{table}

\vskip 2pt \beginbsec{Experimental testbed}
We run all of our experiments on a dedicated cluster with six servers.
Each server has a dual-socket Intel Xeon Skylake 8168 with 24 cores per socket, running at 2.7GHz, 192 GB of \CAP{DDR4} memory, and a Mellanox \CAP{CX3} network card. We use and pin all our threads into the first socket only, where the network card resides. All servers communicate through a Dell \CAP{S6100-ON} switch with 40 Gbps links.

We first evaluate Zeus on several benchmarks (summarized in \cref{tabZ:bench}), 
\linebreak
including the three benchmarks discussed in \cref{secZ:background}
and the \CAP{TATP} benchmark~\cite{TATP:2009}, to further study Zeus' limits in comparison with FaSST and FaRM.
For benchmarks, as in prior work~\cite{F-Kalia:2016}, we consider three-way replication and enough co-located clients to saturate each evaluated system. The initial sharding of all systems is the same. Unlike Zeus, baselines do not support dynamic sharding (i.e., ownership).
We were not able to run the baseline systems FaRM, FaSST and DrTM on our platform.
However, as the hardware used in their evaluations is similar, we report the numbers from their papers~\cite{F-Kalia:2016, Dragojevic:2014, wei2015fast}. 
We conclude by demonstrating the ease of porting legacy applications onto Zeus by porting and evaluating a cellular packet gateway, an Nginx server, and the \CAP{SCTP} protocol.

\vspace{-20pt}
\tsubsection{Handovers}
\vspace{-5pt}

We start our evaluation with a cellular handovers benchmark. 
We evaluate three operations described in \cref{secZ:background}: a handover (consists of two transactions, one at the start and one at the end), a service request and a release (each a single transaction). 
We implement them as defined in the \CAP{3GPP} specification 
on top of Zeus.
All transactions are write transactions. 
The typical cellular phone context for these operations is large and many parts of it are modified, so we need to commit about 400B of data per transaction. 

Recall that mobile users perform both handovers and all other requests, while stationary users only perform other requests (i.e., no handovers). 
In our evaluation, we vary the ratios of the total number of handovers versus the total number of requests (handovers, service requests, and releases), each modeling different mobility speeds in the network. 
A typical cellular network has 2.5\% handovers~\cite{mohammadkhan2016considerations}. We also evaluate the case of 5\% handovers, corresponding to doubling the mobility. 

We run a benchmark on a population of 2M users, of whom 400k are mobile. 
We use the typical cell network provisioning as reported in~\cite{mohammadkhan2016considerations, cell_params}, scaled to 2M users (requiring 1000 base stations). 
Not all handovers involve ownership transfers because some occur between objects of the same node.
For the ratio of remote handovers, we use the numbers we analyzed from the Boston metropolitan area.

\begin{figure}[t]
    \centering
    \includegraphics[width=0.75\textwidth]{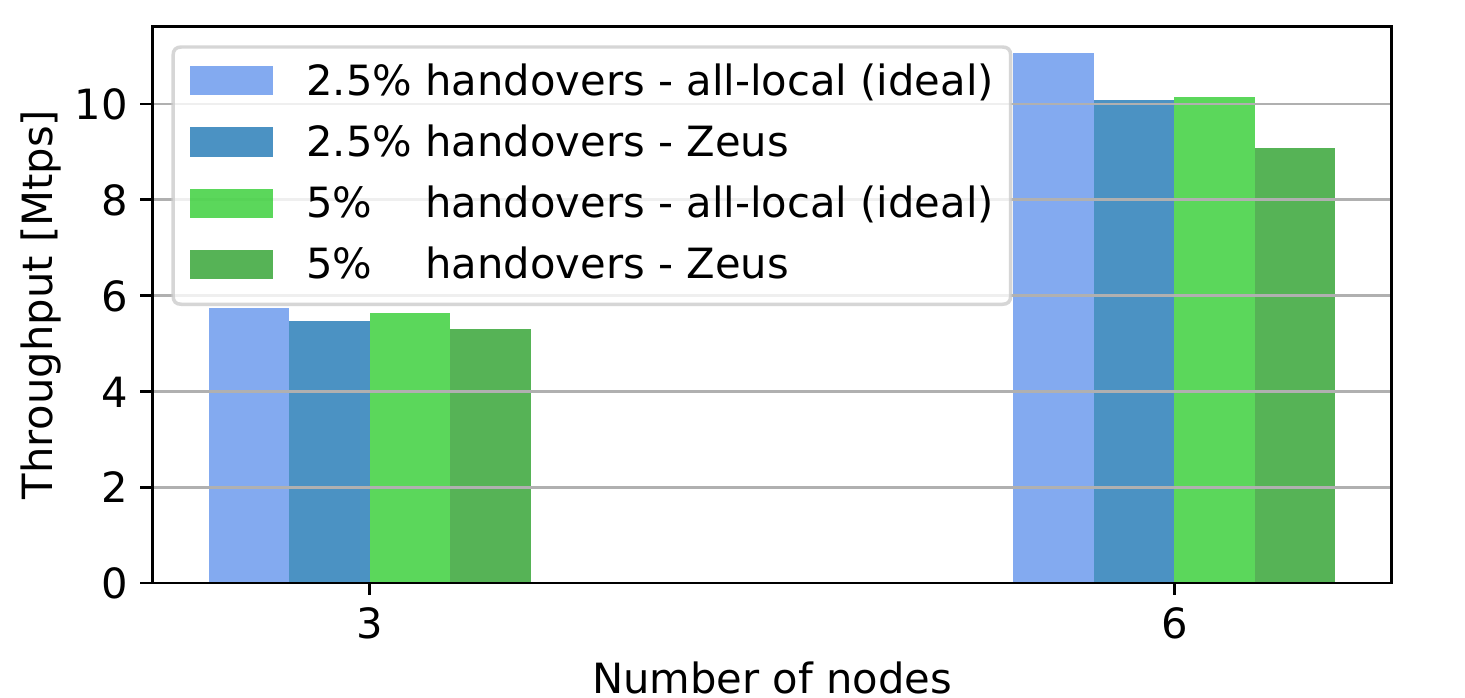}
    \mcaption
    {Handovers performance all-local (ideal) vs. Zeus.}
    {All-local vs. Zeus for 2.5\% and 5\% handovers on 3 and 6 nodes.}
    \label{figZ:ho_total}
\end{figure}

In our evaluation, we vary the number of nodes in the system and plot the total throughput for the two ratios as well as for all-local transactions (shown in \cref{figZ:ho_total}). 
The difference between Zeus and the perfect sharding is at most 9\%. 
This is because a large fraction of the transactions is local, and we have less than 0.5\% ownership requests.
We also see that the performance scales linearly with the number of nodes, even though there are more transactions with ownership transfers for a larger number of nodes. Lastly, we note that prior works have not studied the handover benchmark; as such, there are no published numbers for state-of-the-art systems to compare against.

\begin{figure}[t]
    \centering
    \includegraphics[width=0.75\textwidth]{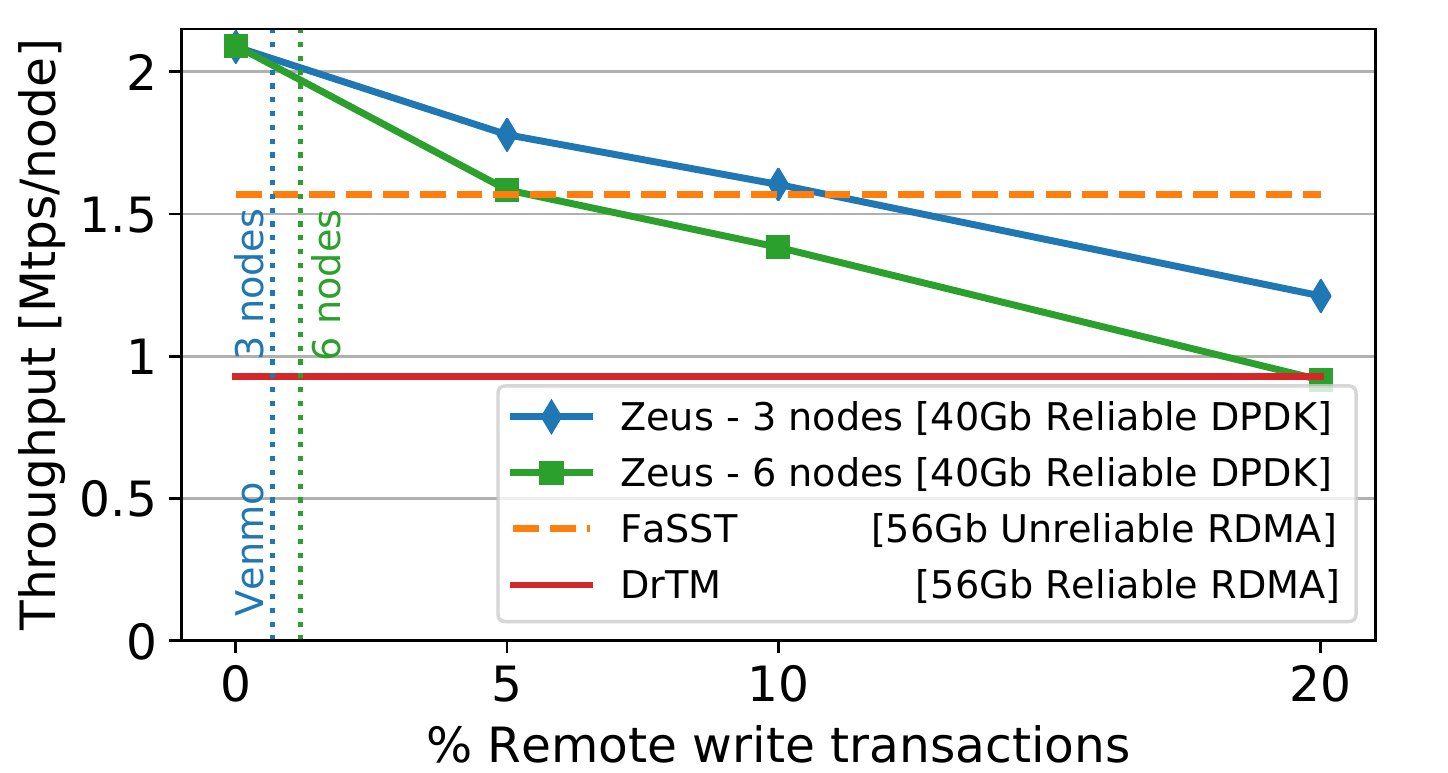}
    \mscaption{Smallbank performance while varying remote write transactions.}
    \label{figZ:eval_smallbank}
\end{figure}

\tsubsection{Smallbank}
Smallbank is a benchmark that simulates financial transactions~\cite{cahill2009serializable}.
It is write intensive, with 85\% write transactions. 
Of these, 30\% modify two objects, and the rest modify three or more objects per transaction.
All read transactions access three objects. 
We use the same access skew on objects as in FaSST.

Smallbank does not specify which pairs of users transact with each other and hence cannot be used to infer real-world transaction locality.
To understand how much the degree of locality affects Zeus, we increase the number of transactions that require an ownership change until Zeus breaks even with the baselines (shown in \cref{figZ:eval_smallbank}). 
We find that, when running Smallbank with the real-world remote transactions, as observed in Venmo, Zeus outperforms FaSST and DrTM by about 35\% and 100\%, respectively.
Recall that neither FaSST nor DrTM support dynamic sharding, so any gradual change in access pattern will eventually lead to an almost random placement and most requests being remote, which is what we show here.
As expected, Zeus' throughput drops as remote transactions increase and the trend remains the same between three and six nodes. As long as less than 5\% (20\%) of transactions require ownership change, Zeus provides an advantage over FaSST (DrTM).

\beginbsec{Reliable lower-end networking}
Note that, unlike FaSST, Zeus implements reliable messaging with its overheads. 
While this reduces Zeus' performance, it allows Zeus to gracefully tolerate message losses. In contrast, FaSST must kill and recover a node for each lost message.  Additionally, FaSST uses 56Gb \CAP{RDMA}. DrTM similarly leverages 56Gb \CAP{RDMA} and relies on hardware transactional primitives for performance. Zeus uses 40Gb non-\CAP{RDMA} networking and does not depend on hardware-assisted transactions for performance.

\begin{figure}[t]
    \centering
    \includegraphics[width=0.75\textwidth]{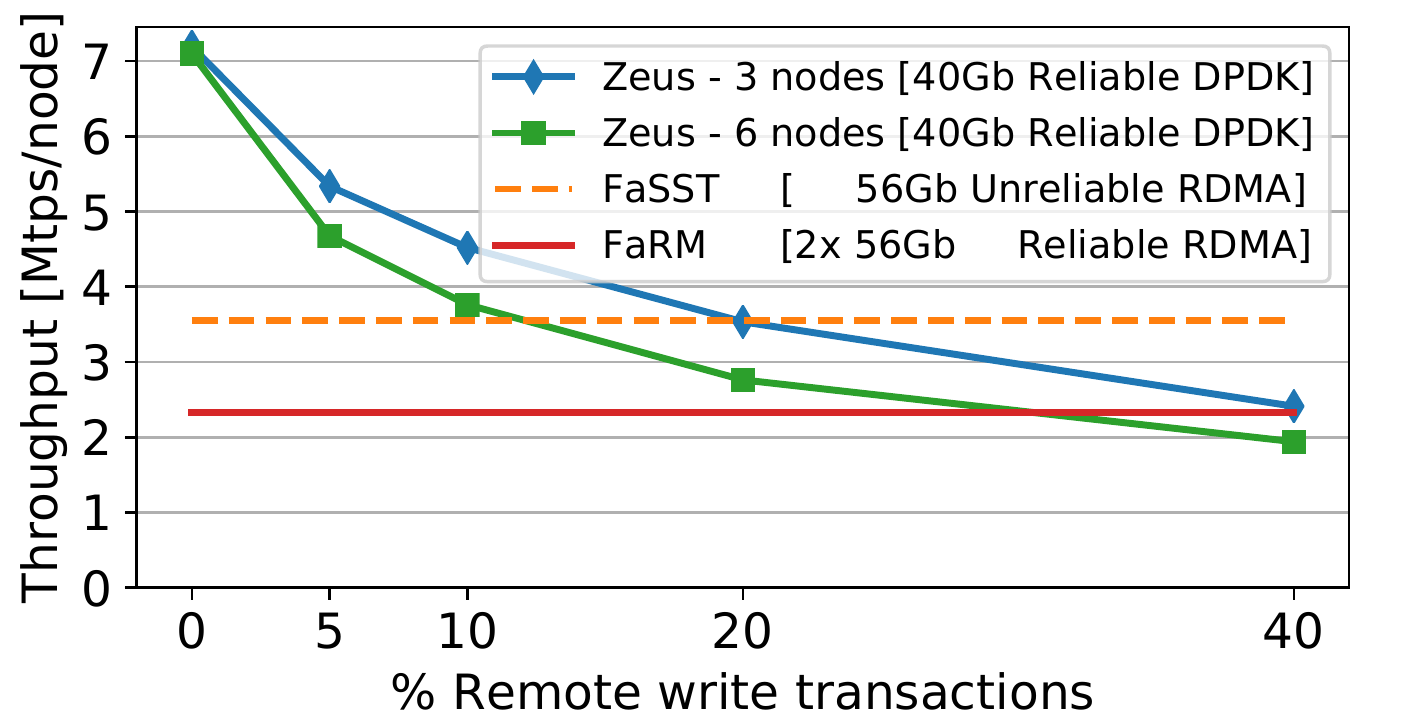}
    \mscaption{\CAP{TATP} performance while varying remote write transactions.}
    \label{figZ:tatp_local}
\end{figure}

\tnsubsection{TATP}

Next, we evaluate the \CAP{TATP} benchmark \cite{TATP:2009}, which provides a second point of comparison with other state-of-the-art systems~\cite{Dragojevic:2015, F-Kalia:2016}. 
It is read intensive, with 80\% read and 20\% write transactions. 
We use 1M subscribers per server, as in FaSST.
Similar to the Smallbank benchmark, we vary the fraction of transactions that require an ownership change. 
The total throughput is shown in \cref{figZ:tatp_local}. 
We see that when the fraction of remote requests is small, Zeus outperforms FaSST and FaRM by up to 2$\times$ and 3.5$\times$, respectively. 

As discussed in the Smallbank study, neither FaRM nor FaSST allow dynamic sharding, so they end up issuing remote requests whenever there is a changing access pattern.  
Zeus keeps the requests local by moving objects and is 
\linebreak
especially effective for a read-dominant benchmark like \CAP{TATP}, since there is little overhead on reads. 
In addition, as long as fewer than 20\% (40\%) of write transactions need ownership requests, Zeus outperforms FaSST (FaRM).
Again, these thresholds are higher than in the case of Smallbank due to the read-dominant workload.
The performance trend of Zeus for three and six nodes is the same as in Smallbank.

\begin{figure}[t]
    \centering
    \includegraphics[width=0.7\textwidth]{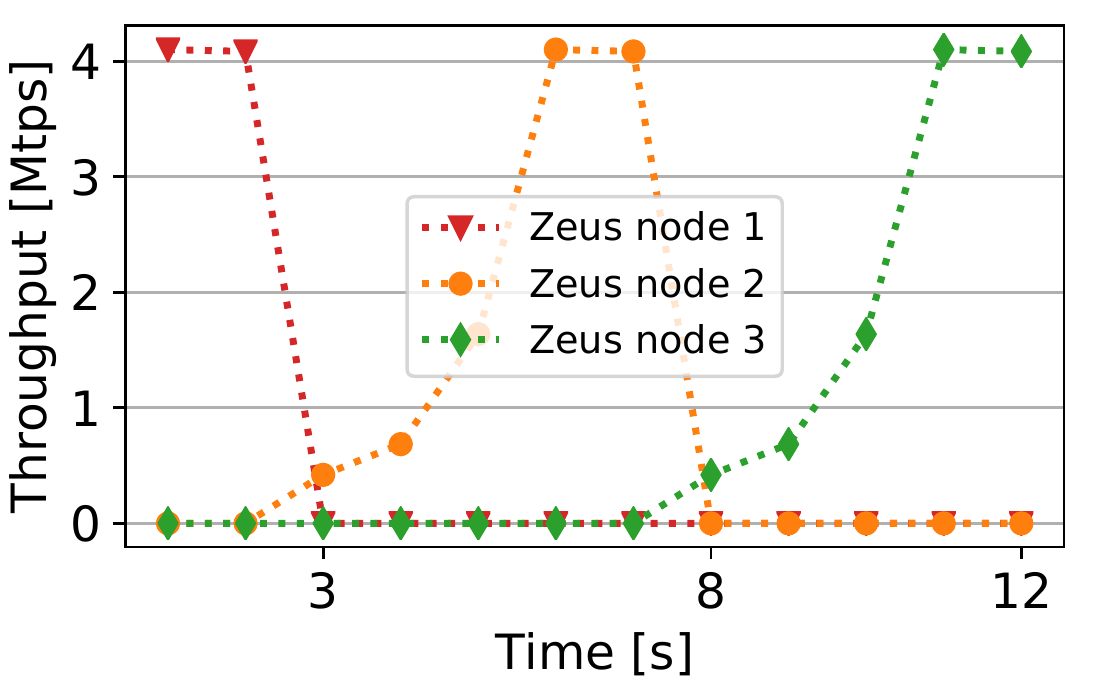}
    \mcaption{Voter performance when moving objects across nodes.}
    {Voter performance when moving 1M objects across nodes.}
    \label{figZ:voter_mov_1}
\end{figure}
\tsubsection{Voter}

Voter is a benchmark that represents a phone voting system~\cite{oltp2013bench}. 
Using three nodes, we simulate 20 contestants in a popularity show with 1M 
unique voters,
each identified by their phone number.
Each voter can vote for one contestant during one phone call, and there is a limit to how many times each voter may vote per unit of time. 
Therefore, each phone voting operation updates two objects: the total votes for a contestant and the voting history of the voter.

In this benchmark, we evaluate the ability of Zeus to move popular objects, as discussed in \cref{secZ:background}. 
In the first experiment, we evaluate the performance of the ownership transfer protocol in isolation. 
We have 1M voters that generate 4M transactions per second (in comparison, E-store~\cite{taft2014store} evaluates up to 200Ktps). 
At time 2s, we move all voter objects from node 1 to node 2, and at time 7s, we move them again to node 3. 
The results are shown in \cref{figZ:voter_mov_1}.
We see that the full move takes 4s, implying that a single worker thread (out of 10) can move 25k objects per second. 

\begin{figure}[t]
    \centering
    \includegraphics[width=0.7\textwidth]{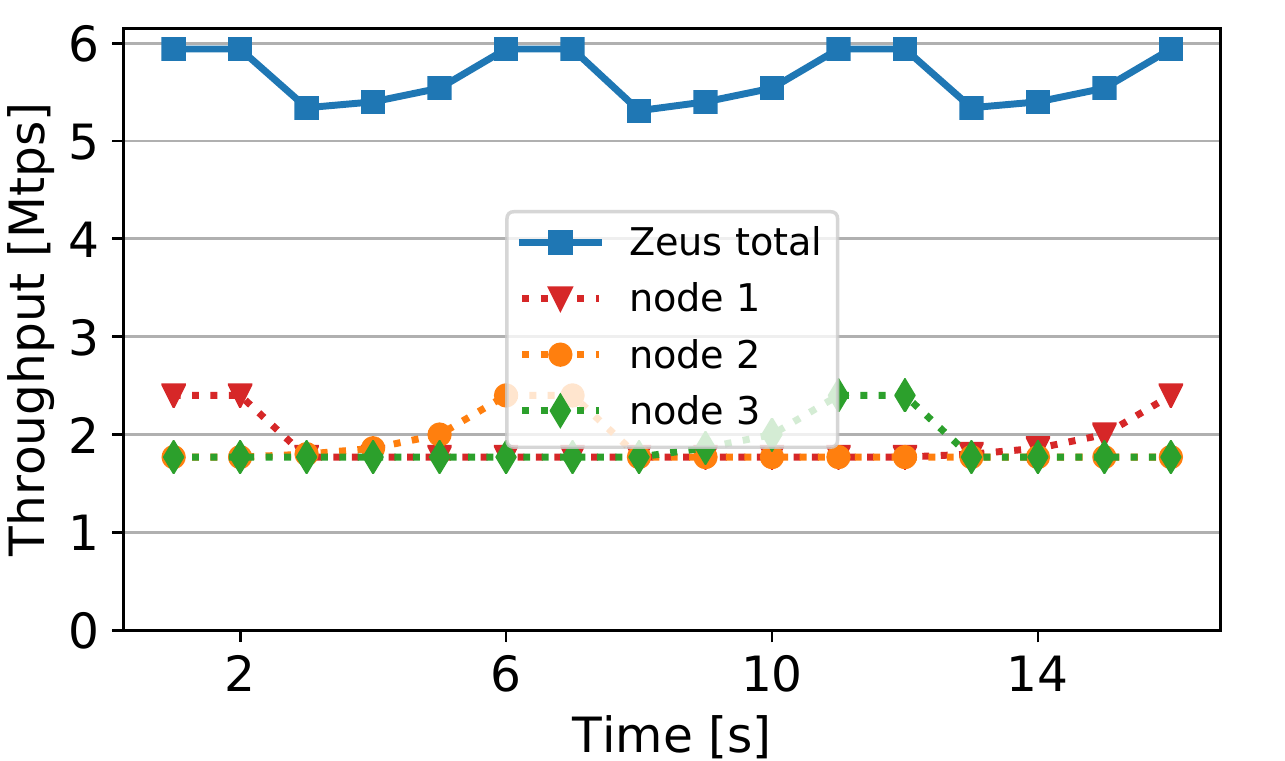}
    \mscaption{Voter performance when registering votes and moving objects.}
    \label{figZ:voter_mov_2}
\end{figure}
In the second experiment, we evaluate the performance of ownership transfers concurrently with transaction processing. 
We have one very popular contestant that has 100k voters voting for her, generating 700Ktps. 
All other voters vote for other contestants and generate about 5.3Mtps in aggregate. 
In this experiment, a single application and worker thread process the most popular contestant. 
As in the previous experiment, at times 2s, 6s, and 10s, we begin to move the object corresponding to the popular contestant to another node.
The results are shown in \cref{figZ:voter_mov_2}.
We see that the single worker thread still performs 25k ownership requests per second (moving 100k objects in 4s) while, at the same time, the rest of the system completes 5.3Mtps.
This shows that concurrent transactions do not impact the ownership performance. 

\begin{figure}[t]
    \centering
    \includegraphics[width=0.7\textwidth]{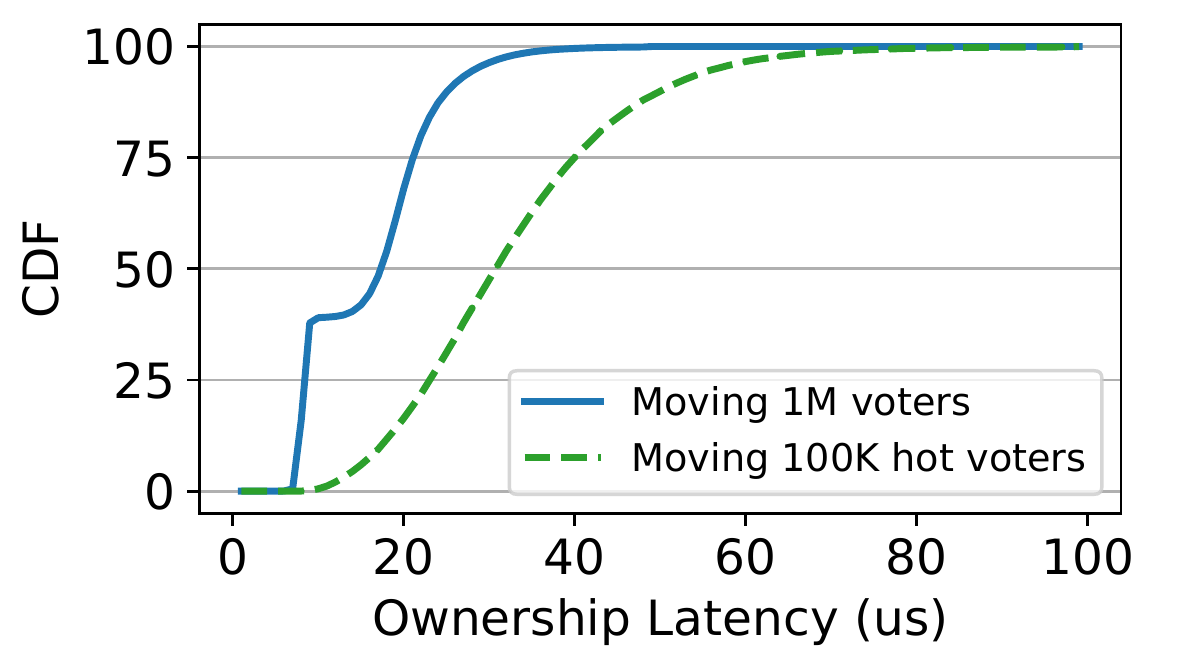}
    \mscaption{CDF of Zeus ownership request latency for Voter experiments.}
    \label{figZ:cdf}
\end{figure}
\cref{figZ:cdf} shows the latency distribution of ownership transfer. 
This metric is important because an application thread is stalled during an ownership transfer, which allows for easy porting of applications. 
The mean latency and the $99.9th$ percentile are close during the first voter experiment (17 and 36 $\mu$s, respectively). Under high load and while moving hot objects (during the second experiment), the mean latency is slightly higher at 29 $\mu$s, and the $99.9th$ 
percentile is 83 $\mu$s. This makes Zeus three times faster than Rocksteady\footnote{Evaluated in a similar setup with \CAP{DPDK} networking over 40Gb \CAP{CX3} \CAP{NIC}s.}~\cite{Rocksteady} in the $99.9th$ percentile, despite moving hot objects under load.

\tsubsection{Legacy Applications}
One of the advantages of Zeus is the ease of porting existing applications. 
Different applications assume different multi-threading or multi-process models, with a different role for each thread (process). 
They also often have dependencies on various external libraries and \CAP{OS} calls. 
FaRM, FaSST, and DrTM need to wait on each remote access.
To mitigate this latency, they
assume transaction multiplexing
via custom user-mode threading (e.g., co-routines or Boost user-threads in FaSST); however, this makes it difficult to integrate with many legacy applications.

As explained in \cref{secZ:design}, Zeus takes a different approach. 
Since most transactions are pipelined and do not block the application thread, there is no need to re-architect a legacy application. 
Zeus only blocks the application during ownership requests, which are infrequent. 

In order to verify our claim regarding portability, we port and evaluate three existing applications on top of Zeus: the control plane of a cellular packet gateway, the \CAP{SCTP} transport protocol, and an Nginx web server.

\begin{figure}[t]
    \centering
    \includegraphics[width=0.7\textwidth]{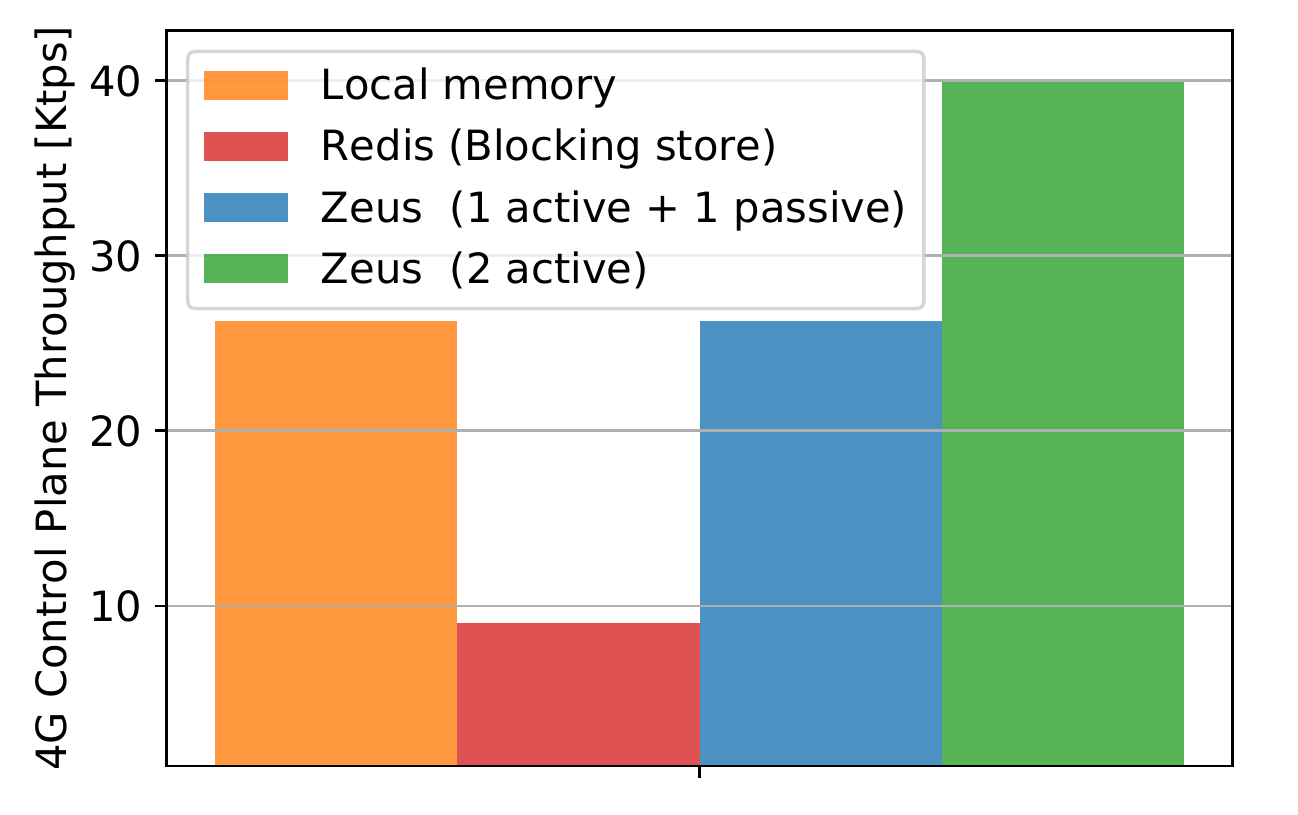}
    \mscaption{Cellular packet gateway control plane performance.} 
    \label{figZ:spgw_thr}
\end{figure}
\beginbsec{Cellular packet gateway}
A cellular packet gateway is a virtual network function in a cellular network that forwards all packets from mobile users. It has a control and data plane. 
The control plane performs service request and release operations, as described in the handover benchmark (but not the handovers themselves). 
Each of these operations is one transaction. 
We use the OpenEPCv8~\cite{OpenEPC} 4G implementation of the cellular core control plane. 
We remove the legacy datastore and instrument every access to use Zeus. 
We use a custom load generator to create test workloads with service and release requests. 
We test the gateway without any datastore (i.e., all data in local memory and no replication), using an off-the-shelf Redis datastore without replication, and Zeus. 

The results are shown in \cref{figZ:spgw_thr}. 
Requests to Redis are remote and, due to the OpenEPC design, the application thread blocks on every request. 
Redis' performance is thus lower than 10Ktps even without replication, which illustrates the challenges due to blocking when porting existing applications. 
With a single active node (and one passive replica), Zeus is as fast as the gateway with local accesses and no replication. 
This is because the bottleneck is in parsing and processing the signaling messages, not in the datastore access. 
When we treat both nodes as active (i.e., as each other's replicas), the throughput is 60\% higher. 
We are not able to scale beyond three nodes due to the limitations of our signal generator, which cannot saturate more than two Zeus nodes.

\begin{figure}[t]
    \centering
    \includegraphics[width=0.7\textwidth]{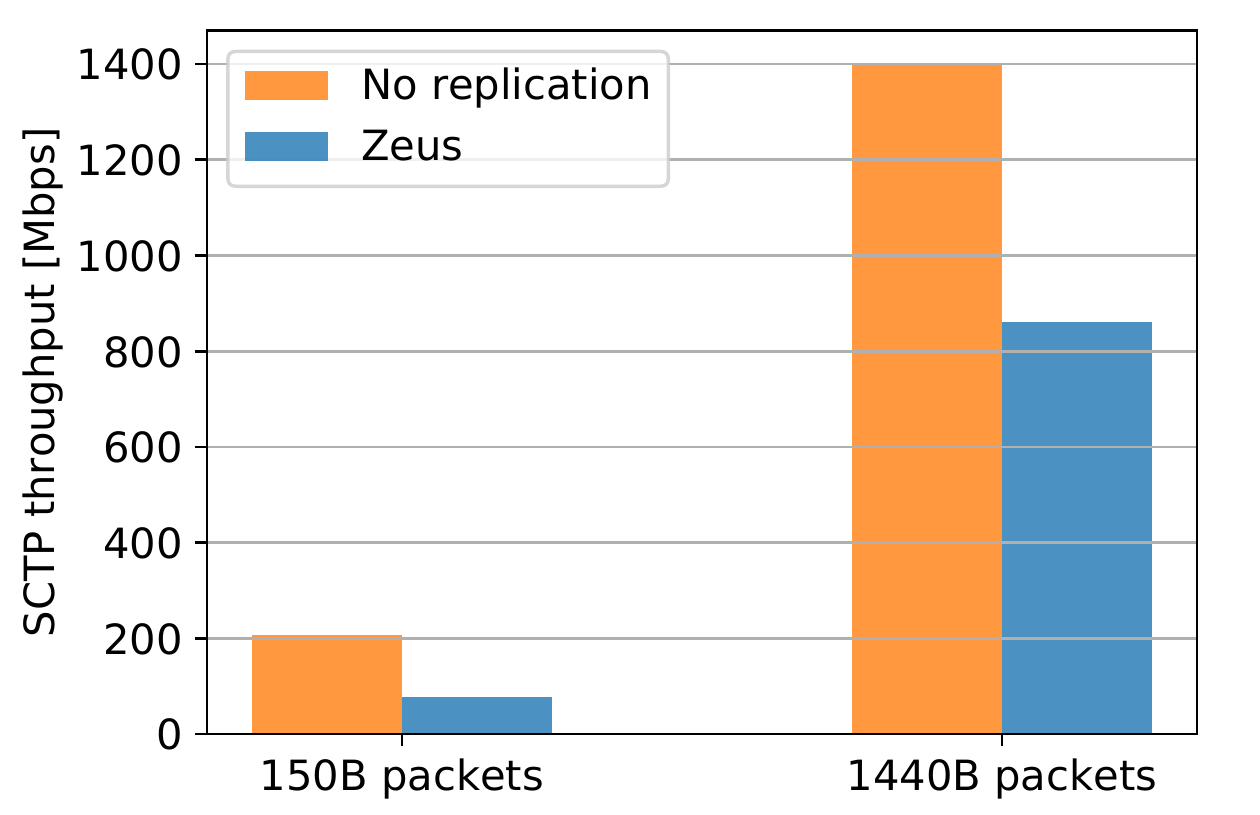}
    \mscaption{SCTP performance.}
    \label{figZ:sctp_thr}
\end{figure}
\beginbsec{SCTP transport protocol}
\CAP{SCTP} is commonly used in the cellular control plane to offer a degree of fault tolerance on network issues. 
For fault tolerance, \CAP{SCTP} natively supports multi-homing and is able to switch from one access network to another in case of a network failure without dropping a connection.\footnote{\newtext{A \textit{connection} is typically referred as an \textit{association} in \CAP{SCTP} parlance.}} 
However, current \CAP{SCTP} implementations cannot survive a node failure, as the connection state is not replicated.
If an \CAP{SCTP} connection fails, all active users drop calls.

To demonstrate Zeus' efficiency and ability to support legacy applications, we port an implementation of the \CAP{SCTP} protocol~\cite{usrsctp:2015} to Zeus. 
Each packet transmission, reception, and timer event is treated as a single transaction. 
Thus, any node failure is perceived by its peers as a network loss and dealt with by the protocol. 
\newtext{We replicate both the internal \CAP{SCTP} state and its buffer queues.}
\CAP{SCTP} uses standard \CAP{BSD} macros for basic data structures (\eg lists, hash tables) that are compatible with Zeus memory interfaces (described in \cref{secZ:system}).  
\newtext{Each pointer malloc, access, and free are converted to the Zeus equivalent.}
We are able to keep the original \CAP{SCTP} design (timer, \CAP{RX}, and \CAP{TX} threads) because we do not have to deal with thread blocking.
\newtext{In short, all operations are replicated, including, for example, a socket write that stores a packet to a shared queue. This approach does not necessarily provide optimal performance, but it requires minimum effort and is the least error-prone way to port \CAP{SCTP}.}

We use a standard iperf3 client to generate a single \CAP{SCTP} flow to a Zeus server running \CAP{SCTP}. 
All state is replicated on another Zeus server. 
\cref{figZ:sctp_thr} shows the throughput of the single flow for different packet sizes. 
For large packet sizes, Zeus is 40\% slower than vanilla \CAP{SCTP} with no modifications. 
This is because \CAP{SCTP} has a complex state that is modified for every packet and 6.8 KB of data must be replicated 
(note that we have not spent any time optimizing state access and providing read-only accesses). 
\newtext{The difference is greater for smaller packets because the replication rate is higher}. 
However, we argue that this is acceptable for the {\em control plane}, where reliability is more important than speed.
We also note that Zeus' pipelined transactions are important for the \CAP{SCTP} case with a few flows because many consecutive transactions access the same object and do not have to wait for the reliable commit of the previous transaction (\cref{secZ:tx-pipelining}).

\begin{figure}[t]
    \centering
    \includegraphics[width=0.7\textwidth]{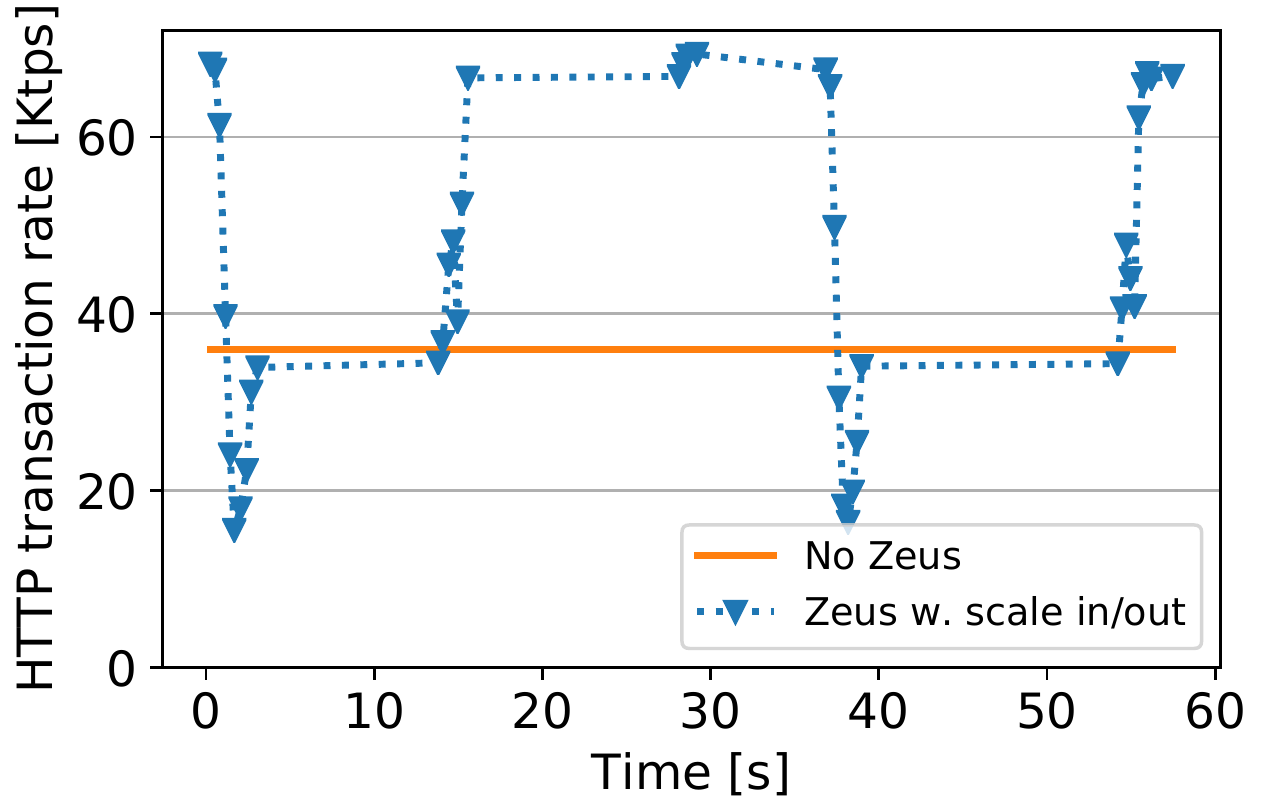}
    \mscaption{Nginx performance in a scale-in/scale-out scenario.}
    \label{figZ:nginx_time}
\end{figure}
\beginbsec{Nginx web server}
Finally, we evaluate the session persistence routing mode~\cite{nginx_session_persisten} of an Nginx web server on top of Zeus. 
%
In this mode, Nginx runs as an application-layer load balancer. 
It looks up a specific cookie in an \CAP{HTTP} request and chooses an end destination based on its value.
Session persistence is not available in the open-source version of Nginx, so we implement our own variant using the Zeus datastore. 
\newtext{
We store (and replicate) each cookie in a Zeus-supported key-value store that is accessed by all servers to determine subsequent request routing decisions.} If the requested cookie is found in the replicated datastore, we route the request to the destination stored in the entry. If not, we randomly select one of the two \CAP{HTTP} back-end servers and store it in the datastore (replicated over two nodes). 
\newtext{The rest of the state is not replicated. For instance, we do not replicate the \CAP{TCP} state as an \CAP{HTTP} connection will re-establish on a fault (i.e., a simpler failover mechanism than \CAP{SCTP}). 
}


A client creates a number of requests for a single small \CAP{HTTP} page \newtext{(the default page of Nginx -- 612B)}.
Initially, all packet requests are processed by the same Nginx server node using a \newtext{single software thread affinitized to a core}. 
We then emulate a scale-out and a scale-in by adding and removing another server node while spreading the load across all available nodes.
The number of forwarded \CAP{HTTP} requests processed by Nginx is shown in \cref{figZ:nginx_time}.
We see that the Nginx's performance with Zeus is the same as without Zeus, indicating that the bottleneck is in the application and not the datastore. 
We also see that it seamlessly scales in and out as the number of servers changes. Again, this illustrates the ease of porting an existing legacy application to Zeus. 

\tsection{Related Work}
\label{secZ:related}

\beginbsec{Modern transactional datastores}
Recent works on in-memory distributed transactions present distributed commit protocols that leverage modern 
\linebreak
hardware to achieve good performance with strong consistency but do not fully exploit locality~\cite{Dragojevic:2014,Dragojevic:2015,F-Kalia:2016,chen2016fast,wei2018deconstructing,sosp2015ramcloud}. Some systems expose object locality, which allows programmers to implement locality-aware optimizations~\cite{Dragojevic:2014, sosp2007sinfonia}, 
but, unlike Zeus, object relocation is costly and burdens the programmer.

\beginbsec{Cheaper distributed transactions}
There are also works that mitigate the cost of distributed transactions but impose other constraints.
For example, some mandate determinism~\cite{Lu:2020, Ren:2019, Le:2019, thomson2012calvin} and are limited to non-interactive transactions that require the read/write sets of all transactions to be known prior to execution~\cite{ren2014evaluation}. 
Others adopt epoch-based designs to amortize the cost of commit across several transactions but at the cost of increasing the transaction latency~\cite{lu:2021, lu2018star, Crooks:18}.
Unlike these, Zeus enhances programmability via non-deterministic transactions that need not wait until the end of epochs to commit.

\beginbsec{Object ownership}
Several works 
have used owner\-ship-related ideas, albeit in a single-node context~\cite{Marathe:08, Dice:06, Harris:14}. L-Store~\cite{lin2016towards} optimizes for locality using ownerships in a distributed local area setting but only supports durable transactions (i.e., without replicas and availability).  In contrast, Zeus enables strictly serializable transactions and ownerships over a replicated deployment that facilitates availability and local read-only transactions from any replica.

\beginbsec{Dynamic sharding}
Dynamic object sharding has been used to improve the performance of distributed transactions. Typically, objects are partitioned and migrated periodically to improve locality~\cite{curino2010schism, Psaroudakis:16, abebe2020dynamast, serafini2016clay, taft2014store, elmore2015squall, lee13asymmetric, Rocksteady}. 
In geo-distributed systems, object migration can significantly reduce \CAP{WAN} traffic~\cite{charapko2018adapting}. 
Facebook's Akkio~\cite{annamalai2018sharding} splits data in $\mu$-shards, which migrate across datacenters to leverage locality in workloads.
Similarly, \CAP{SLOG}~\cite{Ren:2019} deploys a periodic remastering scheme over a deterministic database to reduce across-datacenter round-trips but mandates coordination within a datacenter.
Other works also exploit locality to reduce across-datacenter round-trips~\cite{taft2020cockroachdb, fan2019ocean, zhang2018building}.
In contrast, Zeus infers locality and moves the object eagerly on the first access, supports non-deterministic transactions, and reduces coordination within the datacenter.

\beginbsec{Invalidating protocols}
Zeus protocols resemble cache coherence in 
\linebreak
multiprocessor systems. Cache coherence protocols move cache lines to the requesting node on access. Cache coherence protocols have been used to 
\linebreak
implement hardware transactions~\cite{htm}.
Zeus builds on the ideas in Hermes, which adapted concepts from cache coherence and applied them to enforce strong consistency for replicated in-memory datastores. Hermes allows for local reads and fast reliable updates to individual objects from all replicas; however, it does not support multi-object reliable transactions or reliable object ownership. 

\beginbsec{Distributed shared memory (DSM)}
A \CAP{DSM} provides the abstraction of a single shared memory space built on top of a collection of machines. Similar to Zeus, many \CAP{DSM}s use cache coherence protocols to move data to the accessing node, but, unlike Zeus, most focus on single-object consistency~\cite{munin,Stets:1997,Keleher:1994,Kermarrec:95}. A few support transactions (e.g., ~\cite{yu2018sundial, Cai:2018, Qing:21}) but relax consistency or forfeit data availability for performance. 



\tsection{Summary}
\label{secZ:conclusion}

Many real-world applications exhibit high access locality. 
Zeus leverages this to depart from conventional distributed transaction designs. 
Rather than executing a transaction across nodes, Zeus brings all objects to the same node and executes the transaction locally.
It does so via two new reliable invalidation-based protocols: one for fast localized transactions with replication and one for efficient object ownership.  
Another benefit of Zeus is the ease of porting existing applications on top of it without any re-architecting, as localized transactions can pipeline replication without blocking the application. 
Zeus is up to 2$\times$ faster than state-of-the-art systems on the \CAP{TATP} benchmark and up to 40\% faster on Smallbank while using lower-end networking. 
It can move up to 250K objects per server and process millions of transactions per second. 


\tchapter{Concluding Remarks}
\label{chap:conclusion}

\equote{-50}{25}{This is not the end.\\ It is not even the beginning of the end.\\ But it is, perhaps, the end of the beginning.}{Winston Churchill}

In this final chapter, we first summarize the main contributions of this thesis. We then discuss limitations and future work before concluding.
\tsection{Summary of Contributions}

\subsection*{Scale-out cc\uppercase{NUMA}: Replication for performance}
\vspace{-10pt}
In \cref{chap:cckvs}, we focused on skewed access patterns that are common in workloads of online services and drastically inhibit a datastore's performance.
State-of-the-art skew mitigation techniques resort to either (1) a front-end cache node to filter the skew or (2) exploit a \CAP{NUMA}-like shared memory abstraction that relies on remote access primitives to distribute the load across all servers.
The first approach is processing bound because a single cache node may not be able to keep pace with the load. Meanwhile, the second is network bound because the vast majority of requests require remote access.
%

In Scale-out ccNUMA, we addressed these shortcomings by combining the \CAP{NUMA} abstraction with caching and replication. Our symmetric caching strategy replicates a small cache that stores the hottest objects to all nodes. The request load is distributed among all nodes, allowing them to collectively serve the hottest objects through their replicated caches. Requests that miss in the symmetric cache are served via remote accesses, as in the \CAP{NUMA} abstraction. Thus, symmetric caching has two benefits. First, unlike a centralized front-end cache, the per-node cache scales its throughput with the size of the deployment. Second, by serving the hottest objects locally at any node, it significantly lowers the incidence of remote accesses compared with the pure \CAP{NUMA} abstraction.
A key challenge in symmetric caching is keeping the replicas of the hottest objects strongly consistent while avoiding a hotspot-prone write serialization.
To resolve this challenge, we proposed \textbf{Galene}, a novel invalidating protocol 
that uses logical timestamps for fully distributed write serialization.
Galene provides the strongest consistency while enabling any replica to drive a write to completion without imposing physical serialization points, hence eschewing hotspots and evenly spreading the cost of consistency actions across the deployment.
Under typical modest write ratios of skewed workloads, our 
evaluation shows that symmetric caching powered by Galene can more than double the performance of the state-of-the-art skew mitigation technique.

\subsection*{Hermes: Fast fault-tolerant replication}
\vspace{-10pt}
In \cref{chap:hermes}, we detailed the necessary protocol features for high-performance reads and writes and highlighted the performance limitations of existing replication protocols that guarantee strong consistency and fault tolerance. 
The state-of-the-art reliable protocol allows for efficient local reads from all replicas but still serializes writes on a dedicated node and requires numerous network hops to complete each write, thus harming the throughput and latency of writes.

%
To eliminate these performance limitations, we proposed \textbf{Hermes}, a strongly consistent protocol that extends Galene's combination of invalidations and logical timestamps to the fault-tolerant setting.
Hermes leverages logical timestamps for idempotence and propagates the value of an update early with the invalidation message. This simple strategy enables safe update replays that can tolerate node crashes and message failures.
Meanwhile, in the common fault-free operation, Hermes can achieve the holy grail of performance through its local reads and non-conflicting decentralized writes from all replicas, which complete quickly after a single round-trip to other replicas. 
We showed that five Hermes replicas can sustain hundreds of millions of reads and writes per second without delays, resulting in significantly better throughput and latency than the state-of-the-art protocols. Finally, we demonstrated that Hermes can be safely deployed under asynchrony with a negligible drop in throughput.

\subsection*{Zeus: Locality-aware replicated transactions}
\vspace{-10pt}
In \cref{chap:Zeus}, we pinpointed that the state-of-the-art datastores, which 
\linebreak
afford fault-tolerant transactions with strong consistency,
impose significant performance penalties to several workloads that exhibit locality in their access 
\linebreak
patterns.
The distributed commit protocols of these datastores rely on static sharding 
and cannot fully exploit locality in transactional accesses. 
Regardless of access pattern, each transaction in these datastores is likely to incur costly remote accesses during execution, while it also requires several communication rounds over the network for its distributed commit.
To address this issue, we proposed Zeus, an \CAP{HTM}-inspired transactional datastore that affords locality-aware reliable transactions. 
Zeus transforms expensive distributed transactions into efficient single-node transactions
using two novel reliable invalidating protocols.
First, the \textbf{Zeus ownership} protocol allows any node to 
quickly (at most three hops, in the typical case)
alter 
an object's 
access privileges and location 
without compromising on consistency or fault tolerance.
Using this protocol, 
Zeus moves all objects to the server executing a transaction and ensures exclusive write access (i.e., ownership).
Once this is done, and unless the access pattern changes, 
all subsequent transactions to the same set of objects are executed entirely locally, eschewing distributed conflict resolution.
The second protocol is \textbf{Zeus reliable commit}, 
which ensures data availability via a reliable replication of localized write transactions. Unlike a distributed commit, Zeus reliable commit is pipelined, completes quickly after only one round-trip to other replicas, and facilitates local read-only transactions from all replicas with strong consistency.
Our evaluation shows that, for workloads that exhibit locality in accesses, Zeus' locality-aware transactions can deliver up to twice the performance of state-of-the-art transactional datastores while using less network bandwidth.


\subsection*{Formal verification of protocols}
\vspace{-10pt}

To ensure that all four invalidating protocols we proposed in this thesis guarantee the strongest consistency under all circumstances, we formally specified them in \tla and verified their correctness.
We verified all invalidating protocols for safety 
under concurrency and conflicts.
For the three fault-tolerant protocols (i.e., Hermes, Zeus ownership, and Zeus reliable commit), we also verified their correctness in the presence of membership reconfigurations due to crash-stop failures as well as message reordering and duplicates. Recall that message losses are tolerated via retransmissions and are thus covered by duplicates.

\tsection{Limitations and Future Work}
In this section, we suggest possible research directions related to the ideas of this thesis which may inspire compelling future work. We first discuss 
\linebreak
improvements on the proposed protocols and schemes and then consider how our ideas might be expanded to different settings.





\tsubsection{Enhancing Proposed Protocols and Schemes}

\beginbsec{Scale-out ccNUMA} 
Although, in symmetric caching, requests for hot objects are served once they reach any node of the deployment, cold requests that miss in the cache must access the home node of the object. Consequently, most cold requests require an extra remote access. 
Even if this occurs over an uncongested network, as the symmetric caches filter the skew, it still results in substantial network traffic.
Future studies could consider a more involved request dispatching on the client side to facilitate a better path for cold requests.
For instance, if a client knows (or speculates) that its request is not for a hot object, it could send the request directly to the home node of the object, saving a hop and entirely bypassing the symmetric cache.

\pagebreak

\beginbsec{Hermes} 
A practical future work, perhaps as easy as a graduate exercise, might explore implementing Hermes as a fast path over existing state machine replication approaches (e.g., over the popular Raft~\cite{Ongaro:2014}). 
This has the potential to boost the performance of numerous datastores powered by these protocols.
A limitation of Hermes and all other membership-based protocols is that,
\linebreak
in favoring performance during fault-free operation, they temporarily disrupt operation when node faults occur.
To circumvent safety violations on false positives, when a node crashes, a membership reconfiguration takes place only after the membership leases expire. 
This suspends update operations on the affected shard, as they are waiting for a response from the failed node to complete.
In a modern datacenter with predictable low-latency communication, 
leases can be short-lived, and node faults within each shard are infrequent. 
However, even a small, rare disruption might be undesirable for certain 
applications.
The asynchronous variant of Hermes (Hermes-async) does not require membership leases and could reconfigure faster. 
Further exploring this direction, perhaps through a variant mixing Hermes with Hermes-async, could alleviate the performance drop when node failures occur.

\beginbsec{Zeus} 
%
A future study could focus more on the scalability of Zeus' locality-aware transactions. For instance, it might target to reduce metadata for dynamic sharding and accelerate the ownership directory (e.g., via partitioning) such that it can sustain a larger volume of requests, which would naturally arise in clusters with a large number of servers. Exploring prefetching of object ownership could also yield further performance gains. 

Another interesting research direction would be to compose locality-aware transactions and distributed commit in a hybrid transactional datastore. 
As expected and shown in our evaluation, traditional distributed commit can provide better performance when locality is low and vice versa. 
A hybrid transactional datastore could strive to deliver the best of both worlds. 
More precisely, it could perform transactions with locality through a Zeus-like locality-aware protocol and utilize a distributed commit for transactions without sufficient locality.


    %


\tsubsection{Expanding to a Different Context}
\label{secF:environment}


\beginbsec{Hardware offloading} 
%
Ultra-low latencies are becoming first-order concerns in datacenters, requirements partly driven by the emergence of machine actors such as online sensors and self-driving cars.
The need for low latency has ushered in an era of offloaded network stacks and high-bandwidth network gear. While our work takes advantage of these networking advances and offers dramatically lower latency than prior reliable replication, it is still subject to software overheads in the critical path of replication.
Given the emergence of programmable hardware in the datacenter, offloading the replication protocols of this thesis to hardware (e.g., to smart \CAP{NIC}s or programmable switches) is a promising way forward. 
%
%

Although works on reliable replication offloading already exist~\cite{Istvan:2016, Li:2016-NoPaxos, Jin:2018, Dang:2015, Kogias:20}, the protocols they offload forfeit linearizable local reads from all replicas and serialize updates on a single node (e.g., a switch).
In contrast, the multiprocessor-inspired common path of our work not only provides fast decentralized updates and local reads from all replicas, but we also believe that it offers a simpler scheme that is better suited to hardware implementations. Moreover, this line of work has not yet considered reliable dynamic ownership or distributed fault-tolerant transactions.

\beginbsec{Hybrid consistency}
The main purpose of this thesis was to maximize performance under strong consistency.
While strong consistency provides the most intuitive behavior and is inevitable for some use cases, it may impose unnecessary overheads in others. In the latter case, hybrid approaches that intuitively compose strong and weak consistency (e.g.,~\cite{Gavrielatos:20, Kraska:09, Li:12}) might be a better choice for developers willing to sacrifice the ease of programming with purely strong consistency for performance.
We believe that hybrid approaches can adopt our invalidating protocols and drastically boost their performance when delivering strong guarantees. 

\beginbsec{Byzantine failures}
Security is an ever-growing issue for cloud applications and datastores. 
The sharing and openness of these systems could be compromised by malicious participants. When considering failures in this thesis, we assumed a crash-stop fault model
wherein participants are
expected to be well-intentioned. It is worth researching whether techniques similar to those we have proposed in this thesis could provide performance benefits when applied to the \textit{Byzantine} fault model, where the system may experience faults due to participants' misbehavior. An interesting direction is to explore the interplay 
between the proposed invalidating protocols and trusted execution environments (e.g., Intel's \CAP{SGX}~\cite{Costan:16}), which are now prevalent in datacenter \CAP{CPU}s.

\beginbsec{Geo-replication}
In this dissertation, our primary objective was to improve data replication within a datacenter.
Invalidation-based protocols might also benefit geo-replication (i.e., replication across datacenters), where the latency of inter-replica communication is magnified. 
Unlike majority-based protocols, which are typically deployed in this setting, we expect that invalidating protocols would afford linearizable reads solely by accessing the closest (local) datacenter.
Recall that invalidating protocols can also provide decentralized updates that commit after only one round-trip.
However, these updates must contact all replicas before they complete, no matter how far they reside, rather than contacting only the closest majority, as in majority-based approaches~\cite{Moraru:2013, Enes:21}. 
Consequently, there is a trade-off to explore between (1) the benefits of strongly consistent local reads and (2) the latency of reaching all replicas once instead of reaching the closest majority of replicas potentially multiple times.
%



\tsection{Conclusion}
In this dissertation, we argued that multiprocessor-inspired invalidating 
\linebreak
protocols can advance data replication inside a datacenter by delivering strong consistency, data availability, and superior performance. 
To support our claim, we demonstrated significant performance gains in three typical use cases of data replication within replicated datastores. First, we examined replication to 
increase performance under skewed data accesses. We showed that symmetric caching powered by a fully distributed invalidating protocol delivers high performance and the strongest consistency despite aggressive replication. Second, we demonstrated that a strongly consistent invalidating protocol with logical timestamps can maximize the performance of reads and writes in the common fault-free case while also maintaining data availability when node crashes or network failures occur. Finally, we introduced invalidating protocols that enable fast dynamic sharding and distributed transactions with replication to ensure availability and high performance through locality awareness.
We expect that the performance demands will continue to grow, and we hope that this thesis will motivate the uptake of invalidating protocols in the next generation of replicated datastores.

\setlength{\parskip}{0em} 
\appendix
\tchapter{Invariants and Strong Consistency}
\label{Apendix}

\section*{Informal sketch of protocol correctness}

This appendix informally sketches the correctness of the single- and multi-object protocols proposed in the main body of this thesis by diving deeper into the model-checked protocol invariants and their linearization/serialization points.

\begin{figure}[t]
  \centering
  \includegraphics[width=0.7\textwidth]{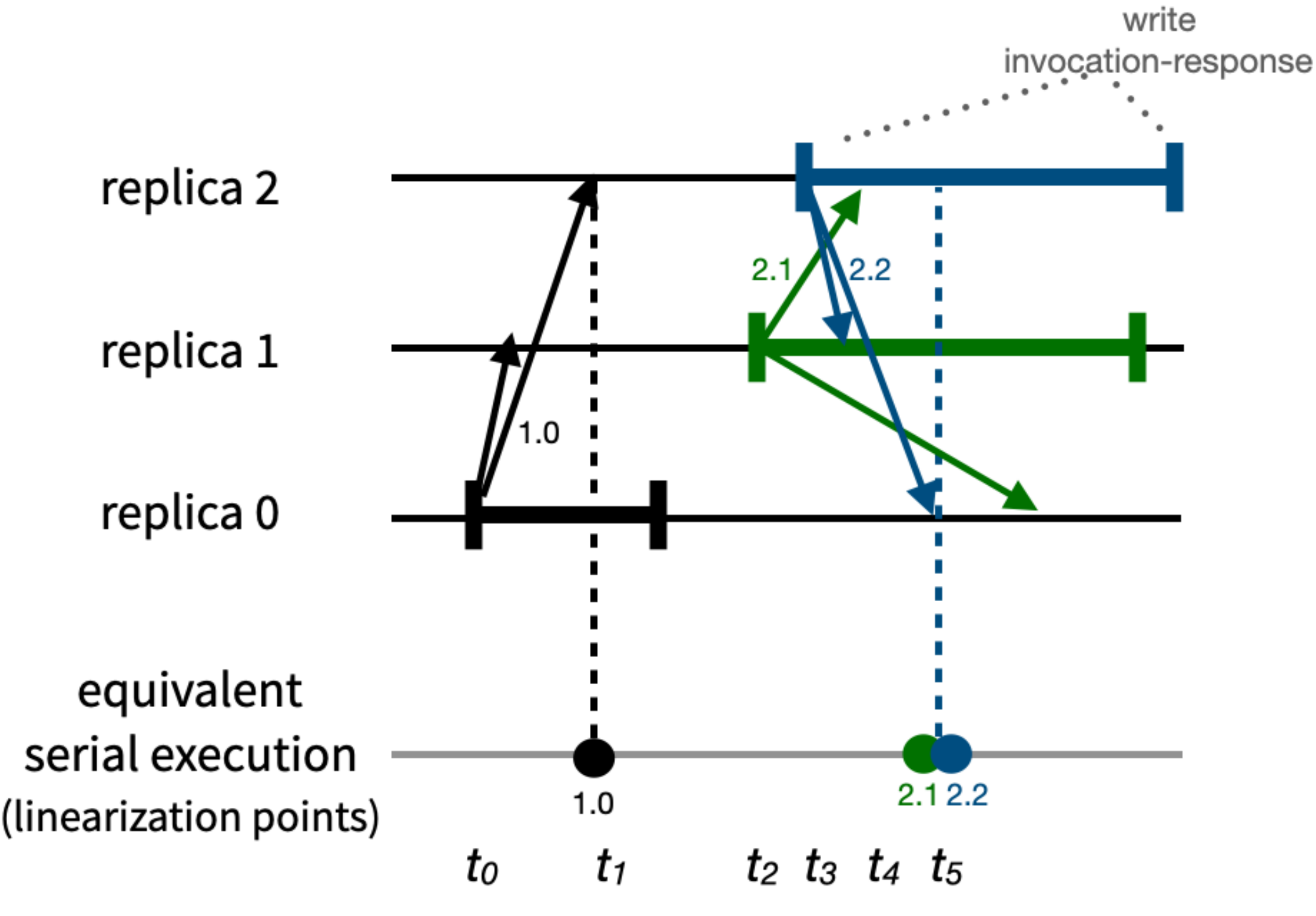} \mcaption{Example of linearization points in writes.}
  {Linearization points for a write executing solo followed by two concurrent writes (colored green and blue) to the same object. Arrows represent invalidation messages; timestamps are shown in the form of \textit{version.\CAP{node\_ID}}.}
  \label{figA:write_lin}
\end{figure}

\vspace{15pt}
\beginbsec{Single-object protocols}
As discussed in \cref{secB:consistency} we target the strongest consistency, which for single-object operations is captured by linearizability. Recall that under linearizability it appears \textit{as if} each (non-aborted) operation is executed without contention and instantaneously at a single point (i.e., the \textit{linearization point}) to all replicas between its invocation and response.

\vspace{10pt}
\noindent $\succ$ \underline{{\em Updates}}:
The protocol specifications of Galene and Hermes ensure that for all writes, there is an equivalent unreplicated serial execution in accordance with real time, which is established by linearization points.
Each write has a linearization point within its invocation and response and is associated with a unique per-object, monotonically-increasing, logical timestamp. The write is \textit{linearized} (i.e., its linearization point is established) when \textit{all} live replicas of the targeted object apply an invalidation with a timestamp greater or equal than the write's timestamp for the first time. More precisely, the linearization point of a write to an object \textit{o} with timestamp \textit{t} occurs once the following three conditions are satisfied:
\squishlist
    \item[\hspace{15pt} \textbf{1.}] 
    there is only one live replica (\textit{R}) storing \textit{o} with timestamp lower than \textit{t};
    \item[\hspace{15pt} \textbf{2.}]
    \textit{R} applies an invalidation to the object with a timestamp \textit{t'} $\geq$ \textit{t}; and
    \item[\hspace{15pt} \textbf{3.}]
    concurrent\footnote{Two operations are concurrent if their invocation-response periods overlap in time.} writes to \textit{o} with smaller timestamp than \textit{t} have been linearized
\squishend

\vspace{10pt}
Note that the above conditions are just a systematic way to produce an equivalent serial execution and not actual steps performed by the protocols.
Moreover, when considered in isolation, the first two conditions do not suffice to create an equivalent serial execution because they may result in overlapping linearization points for concurrent writes to the same key. 
Condition \textbf{3.} resolves these cases by enforcing the linearization points of such concurrent writes to be \textit{consecutive} (i.e., no other operations can be linearized in between) and ordered based on their timestamps.
For instance, the example execution illustrated by \cref{figA:write_lin} shows two concurrent writes, which are linearized at time $t_{5}$ with consecutive linearization points that respect their timestamp order (i.e., 2.1 is linearized before 2.2).

\vspace{10pt}
The above rationale can also be utilized to linearize more complex single-object updates such as Hermes RMWs and Zeus ownership requests. However, because when such updates commit (i.e., are not aborted), it is guaranteed that no other concurrent updates will also commit (we have model checked such an invariant as well), there is no need for the condition \textbf{3.} to establish their linearization points.

\vspace{10pt}
Besides ordering concurrent updates based on their timestamps, 
the linearization points of non-concurrent updates also follow the timestamp order.
Each update is tagged with a lexicographically higher timestamp than any preceding update.\footnote{An operation \textit{o$_{1}$} precedes an operation \textit{o$_{2}$} if \textit{o$_{1}$}'s response occurs before \textit{o$_{2}$}'s invocation.}  
This is because updates are only issued in the Valid state during which all preceding updates have propagated their version of their timestamp,\footnote{Recall that updates cannot return before sending a timestamped invalidation to all replicas.} which is subsequently incremented to create the timestamp of the current update.

\vspace{10pt}
\noindent $\succ$ \underline{{\em Reads}}:
In Galene, Hermes, and Zeus ownership, we have model-checked the data value invariant (among others). In short, this invariant ensures that at any given time, all replicas of an object in a Valid state reflect the most recent update (i.e., the timestamp and value of either a write, an \CAP{RMW}, or an ownership request).
The most recent is established based on the ordering formed by the linearization points of updates as described above.
Therefore, a read that encounters its requested object in a Valid state can be linearized (i.e., establish its own linearization point) before returning the local value and would never violate linearizability.
This is because each read is guaranteed to return the value associated with an update that is at least as recent as the value: 
\squishlist
    \item[\hspace{15pt} \textbf{1.}] 
    written by any preceding update; and
    \item[\hspace{15pt} \textbf{2.}]
    returned by any preceding read
\squishend

\vspace{10pt}
In a nutshell, for single-object reads and updates, we can establish linearization points within each operation's invocation and response boundaries to construct an equivalent serial execution and thus guarantee linearizability (i.e., by ensuring all of the real time orderings~\cite{Gavrielatos:21Papoc}).

\vspace{20pt}
\beginbsec{Multi-object protocols}
We also target the strongest consistency when it comes to multi-object transactions.
This means that transactions in Zeus must guarantee strict serializability.
Similarly to linearizability, in strict serializability, it appears \textit{as if} each (non-aborted) transaction is executed without contention and instantaneously at a single point (i.e., the \textit{serialization point}) to all replicas and relevant shards within its invocation and response.

\vspace{10pt}
\noindent $\succ$ \underline{{\em Write transactions}}:
As detailed in \cref{chap:Zeus}, write transactions in Zeus can be executed and committed only by the node that is the owner of all the objects involved in the transaction.
If the coordinator of a write transaction accesses an object for which it is not the owner, it leverages the Zeus ownership protocol to acquire the ownership.
For the Zeus ownership protocol, we have verified several invariants, including the following:
\squishlist
    \item[\hspace{15pt} \textbf{1.}] 
    At any time, there is at most one owner of an object who has exclusive write access to it.
    \item[\hspace{15pt} \textbf{2.}]
    The object owner holds the most up-to-date data and version of the object (i.e., the object has no pending committed updates from remote nodes).
\squishend

\vspace{10pt}
During the execution of a write transaction, the coordinator has ensured that it holds the ownership, which implies two things. First, the transaction accessed or modified the most \textit{recent} values all of the objects involved in the transaction. Second, the coordinator holds exclusive write access to all these objects.
Succinctly put, there can be no other concurrent write transaction on these objects from other nodes (or local threads in our case -- \cref{secZ:system}). Therefore, the transaction can always be committed locally without requiring further steps for distributed conflict resolution.
However, to guarantee that the values modified by the transaction remain accessible despite faults, Zeus has to replicate the modified state of the transaction to all the relevant replicas (i.e., the followers).
Similar to the single-object protocols, Zeus first invalidates the live replicas of all the objects involved in the transaction.
This invalidation message also holds the new values, which the replicas cannot immediately serve, and it is tagged with a unique transaction \CAP{ID}. As such, the serialization point of a non-aborted write transaction with transaction \CAP{ID} \textit{tx\_id} is naturally established at the point when both of the following conditions hold:
\squishlist
    \item[\hspace{15pt} \textbf{1.}] 
    there is only one live follower (\textit{F}) who has not yet applied the invalidation of the transaction with \textit{tx\_id}; and
    \item[\hspace{15pt} \textbf{2.}]
    \textit{F} applies an invalidation with a \textit{tx\_id}
\squishend

\vspace{10pt}
Because the serialization points take place within the invocation-response boundaries of each transaction, they form an equivalent serial execution of write transactions that respects real time. Note that the most recent values above are again established based on these serialization points.

\vspace{10pt}
\noindent $\succ$ \underline{{\em Read-only transactions}}:
In Zeus, read-only transactions can execute locally from any node replica that stores the relevant data (e.g., a reader).
Once again, we have verified the data value invariant. Thus, an object found in the Valid state on a Zeus reader always holds the most recent data based on the serialization points established by write transactions. This means that similarly to the single-object protocols, each individual access to an object in the Valid state is guaranteed to hold the latest value and be linearized, as explained in the section above. However, we need to ensure that multi-object read-only transactions are executed atomically in their entirety to establish their serialization points. For this reason, we perform a validation phase before committing a read-only transaction. 
The validation phase verifies that no conflicting write transaction got a serialization point in between the series of reads performed by the read-only transaction.
Recall that a read-only transaction proceeds as follows:
\squishlist
    \item[\hspace{15pt} \textbf{1.}] 
    The transaction is executed by recording the values and versions of all the requested objects.
    \item[\hspace{15pt} \textbf{2.}]
    The transaction commits and returns these values if it accesses all the objects again and finds them in the Valid state with their version unchanged.
\squishend

\vspace{10pt}
The serialization point of a successfully committed read-only transaction is established during the execution phase of the transaction, right after accessing the version of the last object to be read. 
In summary, Zeus guarantees strict serializability, as we can establish serialization points for both its write and read-only transactions within their invocation and response such that each operates on the most recent value according to serial execution provided by those serialization points.


\setcitestyle{numbers}
\bibliographystyle{plain}

\singlespace

\bibliography{main}

\end{document}